%% file: draft_27may.tex
\documentclass[a4paper,11pt]{article}
\pdfoutput=1 

\usepackage{jheppub} 
\usepackage[T1]{fontenc} 
\usepackage{subcaption}
\def\sgn{\text{sgn}}

\newcommand{\tl}[1]{\tilde{#1}}

\newcommand{\mc}[1]{\mathcal{#1}}
\newcommand{\CF}{{\rm CF}}
\newcommand{\RB}{{\rm RB}}
\newcommand{\CB}{{\rm CB}}
\newcommand{\RF}{{\rm RF}}
\newcommand{\Op}{{\rm O}}

\def\[{\left[}
\def\]{\right]}
\def\({\left(}
\def\){\right)}
\def\cC{{\cal C}}

\def\cS{{\cal S}}

\def\cV{{\cal V}}

\newcommand{\be}{\beta}

\def\sgn {\text{sgn}}

\def\tr{\mathrm{tr}}
\def\Tr{\mathrm{Tr}}

\def\nn{\nonumber\\}

\def\sgn{\text{sgn}}

\def\tr{\mathrm{tr}}
\def\Tr{\mathrm{Tr}}

\def\nn{\nonumber\\}

\def \be {\begin{equation}}
\def \ee {\end{equation}}
\def \bea {\begin{eqnarray}}
\def \eea {\end{eqnarray}}
\def \beal#1 {\begin{align}#1\end{align}}

\newcommand{\td}{\mathcal{D}}
\newcommand{\mbb}[1]{\mathbb{#1}}

\preprint{TIFR/TH/19-10} \title{\boldmath The large $N$ phase diagram
  of ${\cal N}=2$ $SU(N)$ Chern-Simons theory with one fundamental
  chiral multiplet}

 \author[a,1]{Anshuman
  Dey,\note{anshuman@theory.tifr.res.in}} \author[a,2]{Indranil
  Halder,\note{indranil.halder@tifr.res.in} } \author[b,3]{Sachin
  Jain,\note{sachin.jain@iiserpune.ac.in}} \author[a,4]{Shiraz
  Minwalla,\note{minwalla@theory.tifr.res.in}}\author[a,5]{Naveen
  Prabhakar \note{naveensp@theory.tifr.res.in}}
\affiliation[a]{Department of Theoretical Physics, Tata Institute of Fundamental Research, Homi Bhabha Rd, Mumbai 400005, India}
\affiliation[b]{Indian Institute of Science Education and Research, Homi Bhabha Rd, Pashan, Pune 411 008, India}

\abstract{We study the theory of a single fundamental fermion and
  boson coupled to Chern-Simons theory at leading order in the large
  $N$ limit. Utilizing recent progress in understanding the Higgsed
  phase in Chern-Simons-Matter theories, we compute the quantum
  effective potential that is exact to all orders in the 't Hooft
  coupling for the lightest scalar operator of this theory at finite
  temperature. Specializing to the zero temperature limit we use this
  potential to determine the phase diagram of the large $N$
  ${\cal N}=2$ supersymmetric theory with this field content. This
  intricate two dimensional phase diagram has four topological phases
  that are separated by lines of first and second order phase
  transitions and includes special conformal points at which the
  infrared dynamics is governed by Chern-Simons theory coupled
  respectively to free bosons, Gross-Neveu fermions, and to a theory
  of Wilson-Fisher bosons plus free fermions. We also describe the
  vacuum structure of the most general $\mc{N} = 1$ supersymmetric
  theory with one fundamental boson and one fundamental fermion
  coupled to an $SU(N)$ Chern-Simons gauge field, at arbitrary values
  of the 't Hooft coupling. }

\begin{document} 
\maketitle

\section{Introduction}

In this paper we continue the study, initiated in
\cite{Jain:2013gza,Gur-Ari:2015pca,Aharony:2018pjn}, of $U(N)$ (or
$SU(N)$) Chern-Simons theories coupled to a single fundamental boson
$\phi$ and a single fundamental fermion $\psi$ in the large $N$
limit. The theory we study is governed by the most general `power
counting renormalizable' Lagrangian\footnote{i.e. the most general
  Lagrangian built out of operators of free scaling dimension
  $\leq 3$. We use the following notation for Chern-Simons levels. $k$
  is the (integer valued) level of the pure, topological Chern-Simons
  theory obtained when the fermion in \eqref{generalaction} is given a
  mass with the same sign as the level (and the scalar given any mass)
  and both fields are integrated out. (Here we use standard
  terminology; the level of a pure Chern-Simons theory always refers
  to the level of the dual WZW theory). $\kappa$ is the `renormalized'
  level of this Chern-Simons theory
$$\kappa={\rm sgn}(k) \left( |k|+ N \right)$$
We use the dimensional regularisation scheme in this paper; this
entails that the coupling $\kappa$ satisfies $|\kappa| \geq N$. The
field theories we study in this paper are defined by the Lagrangian
\eqref{generalaction} and the dimensional reduction regularisation
scheme. As we are interested in the large $N$ limit in this paper, we
ignore potential order one corrections to the coefficient of the
Chern-Simons term in \eqref{generalaction}.}
\begin{align}
S  &= \int d^3 x  \biggl[ \frac{i\kappa}{4 \pi}\epsilon^{\mu\nu\rho}
\Tr( X_\mu\partial_\nu X_\rho -\tfrac{2 i}{3}  X_\mu X_\nu X_\rho) \nonumber\\
&+  \overline{D_\mu  \phi}  D^\mu\phi + \bar\psi \gamma^\mu D_\mu \psi 
+m_B^2 \bar\phi \phi + m_F \bar\psi \psi + \frac{4\pi b_4}{ \kappa} (\bar\phi \phi)^2 
+ \frac{4 \pi^2 (x_6+1)}{\kappa^2} (\bar\phi\phi)^3 \nonumber\\
&
+ \frac{4 \pi x_4}{\kappa} (\bar\psi \psi) (\bar\phi\phi)
+ \frac{2 \pi (y_4'-3)}{\kappa} (\bar\psi\phi)( \bar\phi \psi)
+ \frac{2 \pi y_4''}{\kappa} \left((\bar\psi \phi)( \bar \psi \phi ) 
+(\bar \phi \psi)( \bar \phi \psi )\right) \biggl]\ ,
\label{generalaction}
\end{align}
\footnote{The order one shifts $+1$ and $-3$ of the $(\bar\phi\phi)^3$
  and $(\bar\psi\phi)(\bar\phi\psi)$ couplings $x_6$ and $y_4'$ are
  present to make sure that these couplings transform without
  additional order one constant shifts under duality. Usually, under
  duality, additional Chern-Simons contact terms are generated for the
  background gauge fields that couple to the global symmetries of
  \eqref{generalaction}. These lead to order one shifts in some of the
  couplings in \eqref{generalaction} under duality. We put in
  appropriate order one constants in the action \eqref{generalaction}
  to remove the shifts that appear under duality.}. The theories
\eqref{generalaction} are of interest partly because they have been
conjectured \cite{Jain:2013gza, Gur-Ari:2015pca} to enjoy invariance
under a level-rank type strong-weak coupling duality under which
fermions are interchanged with bosons\footnote{Recently, a
  generalization of this duality to Chern-Simons theories coupled to
  arbitrary numbers of fundamental bosons and fermions has been
  proposed and analysed in \cite{Jensen:2017bjo, Benini:2017aed}.}. In
the large $N$ 't Hooft limit
\begin{equation}\label{tHooftlt}
\kappa \to \infty\ ,\quad N \to \infty\ ,\quad \lambda = \frac{N}{\kappa}\quad\text{held fixed}\ , 
\end{equation}
within which we work in this paper, the parameters of the theory
transform under the duality according to
\begin{align}\label{dualitymap}
  & N'=|\kappa|-N\ , \quad \kappa' =-\kappa\ ,\quad \lambda' = \lambda - \sgn(\lambda)\ ,\nonumber\\
  & x_4' = \frac{1}{x_4}\ ,\quad x_6'=-\frac{x_6}{x_4^3}\ , \quad (y_4')' = \frac{16 y_4'}{y_4'^2 - 4 y_4''^2}\ ,\quad (y_4'')' = -\frac{16 y_4''}{y_4'^2 - 4 y_4''^2}\ ,\nonumber\\
  &b_4'=-\frac{b_4}{ x_4^2} + \frac{3}{ 4} \frac{x_6}{ x_4^3}  m_F\ ,\quad m_B'^2=- \frac{1}{ x_4}  m_B^2 - \frac{3}{ 4} \frac{x_6}{ x_4^3}  m_F^2 + \frac{2}{ x_4^2}  b_4  m_F\ ,\quad  m_F'= -\frac{ m_F}{ x_4}\ .
\end{align}
as derived in \cite{Jain:2013gza,Gur-Ari:2015pca}. This duality is a
generalisation of the recently much-studied dualities between purely
fermionic and purely bosonic Chern-Simons matter theories
\cite{Jain:2013gza}-\cite{Cordova:2017vab} , and turn out (see below,
generalising \cite{Jain:2013gza, Aharony:2018pjn}) to imply these
earlier dualities in special scaling limits.

There is another angle from which one may view the dualities
\eqref{dualitymap}. Note that the theory \eqref{generalaction} is
${\cal N}=2$ superconformal when
\begin{equation}\label{parameters}
m_F = m_B^2 = b_4 = 0\ ,\quad x_4 = 1\ ,\quad x_6 = 0\ ,\quad y_4' = 4\ ,\quad y_4'' = 0\ .
\end{equation}
(the transformations in \eqref{dualitymap} leaves \eqref{parameters}
invariant).  It follows that the dualities of \eqref{generalaction}
may also be viewed as a generalisation of the Giveon-Kutasov type
supersymmetric duality of the $\mc{N} = 2$ theory\footnote{One reason
  this is of interest is the following. At the special point
  \eqref{parameters}, there is good evidence (from the computations of
  the $S^3$ partition function and the superconformal index using the
  technique of supersymmetric localisation) that the level-rank type
  duality of this theory holds true even at finite values of $N$
  \cite{Benini:2011mf}. The generalised duality \eqref{dualitymap}
  then implies that the same is true of the dualities between at least
  the subset of theories defined by \eqref{generalaction} that can be
  obtained from RG flows starting from this ${\cal N}=2$ fixed
  point.}.  The study of the theories \eqref{generalaction} thus seems
particularly interesting, as it holds the possibility of tying
together two rich but largely independent streams of work (so far),
namely the large $N$ studies of (non-supersymmetric in general)
Chern-Simons-Matter theories and the exact finite $N$ studies of
supersymmetric Chern-Simons-Matter theories.

In this paper we calculate an exact (i.e. all orders in the 't Hooft
coupling $\lambda$) analytic expression for the finite temperature
quantum effective potential for the lightest gauge invariant scalar -
${\bar \phi \phi }$ - of the theory \eqref{generalaction}. In
fact, we give a five-variable off-shell free energy for the theory
\eqref{generalaction} at finite temperature in Section \ref{off-shell},
which, upon integrating out four of the five variables, results in the
aforementioned quantum effective potential. The precise relation
between our off-shell free energy and the quantum effective potential
is described in Section \ref{qea}.

The computations presented in this paper can be motivated by the
following observation. Setting all kinetic and fermion terms to zero,
the action \eqref{generalaction} reduces to a cubic potential in the
variable $\xi =\frac{2 \pi {\bar \phi} \phi}{\kappa}$:
\begin{equation}
  U_{\rm cl}(\xi) = 2 \pi \kappa \left( m_B^2 \xi + 2 b_4 \xi^2 + (x_6+1) \xi^3 \right)\ \ .
\end{equation}
Note that $\kappa\, \xi > 0$ classically. Clearly this classical
potential is then bounded from below if and only if
\begin{equation}\label{classcondstab}
  x_6 > -1\ .
\end{equation}
In analogy with its classical counterpart, the quantum theory will
also be unstable to decay to $\kappa\, \xi \to \infty$ - and so will be
ill-defined - when $x_6$ is sufficiently negative. One of the goals of
this paper is to determine the (all orders in $\lambda$) quantum
version of the stability condition \eqref{classcondstab}. In order to
accomplish this we evaluate the exact quantum effective potential for
the variable
\begin{equation}\label{sigdef}
  \sigma  = \frac{\xi}{\lambda} = \frac{2 \pi \bar \phi \phi}{N}\ ,
\end{equation}
(i.e.~a quantum effective potential for the lightest gauge-invariant
scalar operator $\bar\phi \phi$) and work out the condition that
ensures that this effective potential is bounded from below at large
$\sigma$.

Our result for the exact quantum effective potential for the variable
$\sigma$ defined in \eqref{sigdef} has a surprise. Quantum
mechanically $\sigma$ is not necessarily positive definite (the
subtraction needed to define the composite operator ${\bar \phi} \phi$
could be negative). As a consequence, we shall see that the quantum
effective potential for $\sigma$ is well-defined also for negative
$\sigma$, apart from being well-defined for positive $\sigma$. The
detailed form of the quantum effective potential is listed in
\eqref{Feffofsh} at finite temperature and in \eqref{Ustab} at zero
temperature.

Thus, there will be two conditions for our theory to be stable:
firstly, a quantum-corrected version of \eqref{classcondstab} which
arises from requiring the quantum effective potential to be bounded
from below for large and positive $\sigma$; a second condition from
the requirement that the quantum effective potential must be bounded
from below at large negative $\sigma$ as well. As in the recent paper
\cite{Dey:2018ykx}, this second condition - which has no classical
counterpart - results in a second inequality for the variable
$x_6$. This inequality defines an upper bound for $x_6$ and hence it
is \emph{necessary} for the stability of the theory that $x_6$ is
smaller than a minimum value. It follows that the theory
\eqref{generalaction} is well-defined if and only if $x_6$ lies within
an interval of values. The lower and upper limits of this interval
turn out to depend on $x_4$ as well as the 't Hooft coupling $\lambda$
and are listed in detail in equation \eqref{stabcond} of Section
\ref{stab} below.

Of course the large $N$ exact quantum effective potential computed in
this paper has many applications beyond the analysis of vacuum
stability.  For instance, we demonstrate in Section \ref{dualfree}
that the quantum effective potential presented in this paper enjoys
invariance under the conjectured strong-weak coupling duality
\eqref{dualitymap}, yielding nontrivial new evidence for this duality
(generalising earlier results of \cite{Jain:2013gza,
  Aharony:2018pjn}).

However, the principal results of this paper concern the use of the
quantum effective potential to quantitatively (and exactly at large
$N$) determine the zero-temperature phase diagram of the theories
\eqref{generalaction}. We obtain this phase diagram by minimising the
quantum effective potential - which turns out to be piecewise cubic at
zero temperature - as a function of the parameters of
\eqref{generalaction} - and thereby determining the dominant phase of
our theory at zero temperature.

The theory \eqref{generalaction} has four dimensionless parameters
($x_6$, $x_4$, $y_4'$, $y_4''$) and three dimensionful parameters
$m_B^2, m_F$ and $b_4$ (and so two additional dimensionless ratios).
By varying these six parameters we could, in principle, obtain a six
dimensional phase diagram. In this paper we do not explore the full
six dimensional phase diagram but study only two relatively simple
slices of it. The first of these is the `phase diagram of the large
$N$ ${\cal N}=2$ theory' (see below for an explanation of these
words), i.e.~the phase diagram obtained by setting the four
dimensionless parameters $x_4, x_6, y_4', y_4''$ to the values
\eqref{parameters} but allowing the dimensionful variables
$m_B^2, b_4, m_F$ in \eqref{generalaction} to be arbitrary. Trading
one of the dimensionful parameters for a mass scale, the phase diagram
thus obtained is two dimensional. The second slice we study is
obtained by restricting our attention to the class of theories in
\eqref{generalaction} that preserve at least ${\cal N}=1$
supersymmetry:
\begin{equation}\label{susyN1intro}
  m_F=\mu\ ,\ \ m_B^2=\mu^2\ ,\ \ b_4= \mu w\ ,\quad x_4=\frac{1+w}{2}\ ,\quad x_6=w^2-1\ ,\quad y_4' = 3 + w\ ,\quad y_4'' = w-1\ .
\end{equation}
There is one dimensionful parameter $\mu$ on this slice which can be
traded for a mass scale and the remaining one dimensionless parameter
$w$ describes the one dimensional phase diagram.

Our motivation for studying special slices of \eqref{generalaction} -
rather than the whole shebang at once - are both practical as well as
conceptual in nature. At the practical level, a two (or one)
dimensional phase space is much easier to visualise than a six
dimensional phase space.  The conceptual reason is more important, and
we pause, over the next three paragraphs, to give provide a detailed
explanation.
 
Recall that a quantum field theory is defined in the UV as a fixed
point of the renormalization group. The phase diagram of a given
quantum field theory is defined as the set of phases obtained by
deforming the particular fixed point of interest with all possible
relevant deformations. In order to understand the phase diagram of
particular theories of the form \eqref{generalaction} we need to first
identify the set of fixed points (in the space of RG flows of the four
dimensionless couplings in \eqref{generalaction}).  With this
understanding in hand we can then study the phase diagram of any given
fixed point.

The study of fixed points of the Lagrangian \eqref{generalaction} is
complicated by the following fact; the beta function for all
dimensionless parameters in \eqref{generalaction} vanishes in the
strict large $N$ limit.  At leading order in large $N$ it follows that
the Lagrangian \eqref{generalaction} describes a four dimensional
hyperplane of conformal field theories\footnote{The hyperplane is
  obtained by setting the three dimensionful variables $m_B^2$, $b_4$
  and $m_F$ in \eqref{generalaction} to zero and is parametrized by
  the four dimensionless variables $x_4$, $x_6$, $y_4'$ and $y_4''$.}
rather than the more usual situation of a collection of isolated fixed
points. This picture is, of course, an artefact of the large $N$
limit. At any finite $N$, no matter how large, this fixed hyperplane
presumably breaks up into a set of isolated fixed points connected by
a presumably intricate pattern of RG flows. Each of these fixed points
defines a new conformal field theory; the phase diagram of this theory
is obtained by studying its relevant deformations.

It follows that there are as many physically interesting phase diagram
questions in \eqref{generalaction} as there are fixed points under the
RG flow (of the dimensionless variables) of
\eqref{generalaction}. Unfortunately the aforementioned RG flows have
not yet been studied, and their fixed points have not yet been
classified.  Despite this general state of ignorance, we do know of
one fixed point for the class of theories \eqref{generalaction}. This
is the ${\cal N}=2$ supersymmetric point defined by \eqref{parameters}
which lies on the fixed hyperplane of the RG flow at leading order in
the large $N$ limit. It also seems likely that deformations about this
point that are tangent to the four dimensional fixed hyperplane are
actually irrelevant once subleading corrections in $1/N$ are taken
into account\footnote{See Section 5.2 of \cite{Aharony:2018pjn} for a
  discussion for small values of the 't Hooft coupling $\lambda$.}.
Assuming this to be the case, it follows that the phase diagram
associated with this special fixed point - i.e.~the phase diagram of
the ${\cal N}=2$ theory - is obtained at large $N$ by restricting
attention to the special point \eqref{parameters} in the manifold of
parameters.

\begin{figure}
	\centering
	\scalebox{0.43}{\includegraphics{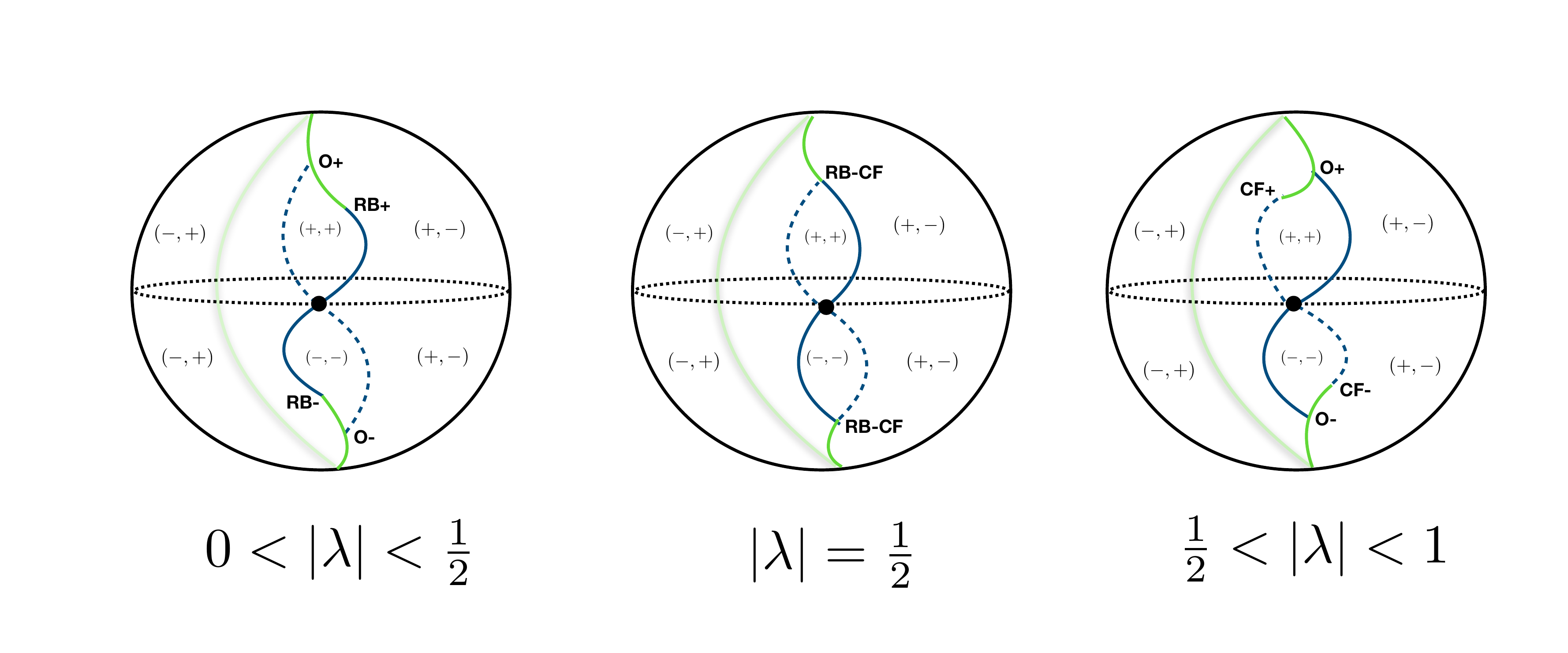}}
	\caption{A schematic phase diagram for the $\mc{N} = 2$ theory
          for three representative values of $|\lambda|$. The phase
          diagram is an ellipsoid (displayed here as a sphere for
          simplicity) in the three dimensional space spanned by the
          relevant parameters $m_B^2$, $\lambda b_4$ and
          $m_F \sgn(\lambda)$. The northern (southern) hemisphere
          corresponds to $m_F\sgn(\lambda) > 0$
          ($m_F\sgn(\lambda) < 0$) while the left (right) hemisphere
          corresponds to $m_B^2 > 0$ ($m_B^2 < 0$). The notation
          $(\pm,\pm)$ corresponds to the four low energy topological
          phases described in Table \ref{tabo}. The blue lines are
          second order transition lines and are governed by either CS
          gauged Wilson-Fisher bosons (solid lines) or CS gauged free
          fermions (dashed lines). The green line in each of the above
          figures is a first order phase transition line -- part of it
          lies on the far side of the ellipsoid in the viewpoint
          depicted above. For $|\lambda| < \tfrac{1}{2}$ the first
          order line meets the solid blue lines at one point each in
          the two hemispheres of the ellipsoid governed by CS gauged
          free boson CFTs ($\RB\pm$). For $|\lambda| > \tfrac{1}{2}$
          these points are governed by CS gauged Gross-Neveu fermions
          ($\CF\pm$). For $|\lambda| = \tfrac{1}{2}$, the first order
          line meets the solid and dashed lines at one point each in
          the two hemispheres which are governed by a theory of CS
          gauged free bosons plus Gross-Neveu fermions (RB-CF). The
          blue lines all intersect at the black dot on the equator and
          is governed by a theory of CS gauged Wilson-Fisher bosons
          plus free fermions (CB-RF). The phase diagrams for
          $|\lambda| < \tfrac{1}{2}$ and $|\lambda| > \tfrac{1}{2}$
          are related to each other by the duality map
          \eqref{dualitymap}. The diagram for
          $|\lambda| = \tfrac{1}{2}$ is self-dual.}
	\label{3dplot}
\end{figure}

Returning to the main flow of this introduction, the phase diagram of
the ${\cal N}=2$ supersymmetric theory is parametrized by two
dimensionless numbers that live on a space with the topology of a
sphere, and turns out to be rather intricate. A major part of this
phase diagram consists of regions of four distinct massive (more
precisely, topological) phases. The long distance dynamics of these
phases is governed by pure Chern-Simons theories with gauge group
either $SU(N)$ or $SU(N-1)$; the rank is $N$ or $N-1$ depending on
whether the massive bosons are unHiggsed or Higgsed. We denote these
as the $+$ and $-$ phases respectively of the boson. The level of the
low energy topological Chern-Simons theory is either $k$ or
$k-\sgn(k)$ depending on whether the dynamical massive fermions that
we integrate out to obtain the topological theory have the same sign
or opposite sign w.r.t.~$k$. These we term as the $+$ and $-$ phases
of the fermion. The four massive phases are then described by one sign
for the phase of the boson and one sign for the phase of the
fermion. In the rest of this paper we use the notation explained in
Table \ref{tabo} for these four massive phases of our theory.
\begin{table}[!htbp]
	\begin{center}
		\begin{tabular}{c|c|c|c}
			\hline
			Phase & Fermion & Boson & Low-energy TQFT \\ \hline
			$(+,+)$ & $\sgn({\tl m}_F) = \sgn(k)$ & unHiggsed & $SU(N)_k$ \\ \hline
			$(-,+)$ & $\sgn({\tl m}_F) = -\sgn(k)$  & unHiggsed & $SU(N)_{k-\sgn(k)}$ \\ \hline
			$(+,-)$ & $\sgn({\tl m}_F) = \sgn(k)$ & Higgsed & $SU(N-1)_k$ \\ \hline
			$(-,-)$ & $\sgn({\tl m}_F) = -\sgn(k)$ & Higgsed & $SU(N-1)_{k-\sgn(k)}$ \\ \hline
		\end{tabular}
	\end{center}
	\caption{The four massive phases of the one boson-one fermion
          theory. The notation is $(F,B)$ with $F$ and $B$ being the
          fermionic and bosonic phases respectively. The effective
          fermion mass is denoted by ${\tl m}_F$ and $\sgn({\tl m}_F)$
          and $\sgn(k)$ are the signs of ${\tl m}_F$ and $k$
          respectively. We list the Chern-Simons TQFTs as appropriate
          for an $SU(N)$ gauge group for the microscopic
          theory.} \label{tabo}
\end{table}
The phase diagram of the ${\cal N}=2$ theory turns out to be
qualitatively different depending on whether the absolute value
$|\lambda|$ of 't Hooft coupling is less than or greater than
half. The schematic phase diagram in each case is sketched in Figure
\ref{3dplot} (see Figures \ref{3dplot14},\ref{3dplot34},\ref{3dplot12}
for the actual accurate phase diagram at three sample values of
$\lambda$). Notice that the regions of the four topological phases
above are separated by blue or green lines. These are lines along
which the theory undergoes a second order or a first order transition
respectively. Along the second order phase transition lines the
dynamics is conformal and is generically governed by either the
critical boson (CB) or the regular fermion (RF) theories. At one point
on one of these phase transition lines, the order of the phase
transition jumps from second to first order. At this transition point
the theory reduces to the conformal Regular Boson (RB) theory (when
$|\lambda| <\frac{1}{2}$) or the conformal Critical Fermion (CF)
theory (when $|\lambda| > \frac{1}{2}$). Note also that in both cases
there is a point on the phase diagram at which the four second order
phase transition lines meet. At this point the dynamics is governed by
Chern-Simons gauged Wilson-Fisher bosons and regular fermions (the
CB-RF theory). This theory was first encountered 
(in the same context) in \cite{Jain:2013gza}, and 
has recently been intensively studied in their own
right at finite values of $N$ in \cite{Jensen:2017bjo, Benini:2017aed}.
Note also that the phase diagram at $|\lambda|= \frac{1}{2}$ has a
qualitatively new feature; in each of the northern and the southern
hemispheres this phase diagram has a special critical point that marks
the simultaneous end point of both the second order CS-gauged
Wilson-Fisher and CS-gauged free fermion phase transition lines. We
present a brief qualitative discussion about this interesting sounding
critical theory (and a related theory obtained by orbifolding this
theory by its duality symmetry) at the end of Section \ref{phasedN2}.

\begin{figure}
	\centering
	\scalebox{0.84}{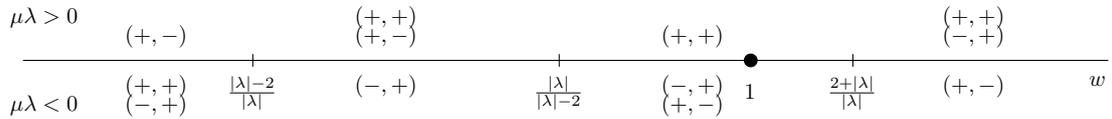}
	\caption{The vacuum structure of the $\mc{N} = 1$ theory
          \eqref{susyN1intro} as a function of the parameter $w$. The
          two cases $\sgn(\mu) = \pm \sgn(\lambda)$ are given above
          and below the $w$ line respectively. The quantum effective
          potential either has one or two $\mc{N} = 1$ supersymmetric
          vacua for a given range of $w$ and sign of $\mu\lambda$. In
          the cases where there are two such vacua, we have indicated
          both in the above phase diagram. The $w=1$ point corresponds
          to the $\mc{N} = 2$ theory.}
              \label{N1susyp}
\end{figure}

In addition to our analysis of the phase structure of the ${\cal N}=2$
theory, in Section \ref{N1phase} we have also presented a separate
analysis of the phase structure of the ${\cal N}=1$ subset of theories
\eqref{susyN1intro}. Our final results are summarised in Figure
\ref{N1susyp}.
Recall that the distinct ${\cal N}=1$ theories \eqref{susyN1intro} are
labelled by a single dimensionless number $w$ and a dimensionful scale
$\mu$. The phase diagram of our system depends on $w$ and
\footnote{Recall that $\mu$ and $\lambda$ each individually flip sign
  under a parity transformation; however the sign of $\mu\lambda$ is
  left invariant under this transformation and so is
  physical.}${\rm sgn}(\mu \lambda)$.  In Figure \ref{N1susyp} we
present the phase of our system at all allowed values of $w$ and for
both possible signs of $\mu \lambda$.

The results of Figure \ref{N1susyp} overlap with those of Figure
\ref{3dplot} in the special case $w=1$. When $\mu \lambda>0$, the
${\cal N}=1$ theory lies at a particular point in the $(+,+)$ region
in the northern hemisphere of either of the diagrams in Figure
\ref{3dplot}. On the other hand when $\mu \lambda<0$ the ${\cal N}=1$
theory lies at a particular point on the southern hemisphere of either
of the diagrams in Figure \ref{3dplot}. Remarkably enough, the point
in question turns out to lie exactly on the first order phase
transition line between the $(+,-)$ and $(-,+)$ phases in the southern
hemisphere. This explains why we have two possible phases - namely
$(+,-)$ and $(-,+)$ - for the ${\cal N}=1$ theory at $w=1$.

We emphasize that, at the conceptual level, the results of Figure
\ref{N1susyp} are only physical at those values of $w$ at which the
beta function has fixed points. While we know this is the case at
$w=1$ (i.e.~the ${\cal N}=2$ fixed point) we do not currently know
other values of $w$ where such fixed points occur. Precisely this question is the subject of investigation of the soon to appear paper \cite{ofer}.

Before ending this introduction we should emphasize that the finite
temperature free energy of the same theory \eqref{generalaction} was
already computed in \cite{Jain:2013gza} for the special case of the
$(+,+)$ and $(-,+)$ phases. About half of the phase diagram presented
in Figure \ref{3dplot} (i.e. the parts of the phase diagram covered by
the $(+,+)$ and $(-,+)$ phases) could already have been constructed
using the results of \cite{Jain:2013gza}.  The analysis of the phase
diagram of the theory presented in this paper has two advantages over
the approach of \cite{Jain:2013gza}: one methodological and the second
of principle. At the methodological level, the use of the simple
quantum effective potential, as opposed to simply the value of the
free energy at extrema (which is the only information we have access
to using the methods of \cite{Jain:2013gza}), makes the analysis of
the phase diagram more intuitive and much easier. The more important
in-principle advantage of the approach of this paper is that the
quantum effective potential allows us to access the $(+,-)$ and
$(-,-)$ phases (the analysis of \cite{Jain:2013gza} was blind to these
phases) and thus allows us to compute the complete phase diagram
schematically depicted in Figure \ref{3dplot}, a task that could not
have been accomplished using only the analysis of \cite{Jain:2013gza}.

\section{An off-shell free energy}\label{off-shell}

As we have explained in the introduction, one of the principal
technical results of this paper is an explicit formula for the
off-shell finite temperature free energy, analytic in all its
variables, for the class of field theories \eqref{generalaction}.

The off shell free energy presented in this paper is a close analogue
of the `three-variable off-shell free energy' for the regular boson
theory presented in equation 4.2 of \cite{Dey:2018ykx}. As in
\cite{Dey:2018ykx}, we obtain the thermodynamic finite temperature
free energy of our system by extremizing the off-shell free energy
w.r.t.~its variables and choosing the dominant extremum\footnote{The
  extremization over holonomies at finite temperature can be quite
  complicated, as it has to be done accounting for measure effects
  (see e.g. section 2.2. of \cite{Dey:2018ykx} for a brief discussion
  and \cite{Jain:2013py} for many more details). In this paper we will
  consider explicit results for the free energy only at zero
  temperature where the holonomy variables drop out thus considerably
  simplifying the analysis.}. The derivation of the off-shell free
energy for the theory \eqref{generalaction} is a straightforward
combination of the results and methods described in detail in
\cite{Jain:2013gza, Choudhury:2018iwf, Dey:2018ykx}. We relegate the
derivation of this off-shell free energy to Appendix \ref{federapp}.
In this section we simply present our final answer.

The finite temperature off-shell free energy of the theory
\eqref{generalaction} is given by
\begin{align}\label{Feffofsh}
  &F[c_B, c_F, {\tl \cS}, {\tl \cC}, \sigma]\nonumber\\
  &= \frac{N}{6\pi}\bigg[-3 {\hat c}_B^2 {\hat \sigma} + \lambda^2 {\hat \sigma}^3 +  3\left( {\hat m}_B^2 {\hat \sigma}  + 2\lambda {\hat b}_4 {\hat \sigma}^2 +  (x_6 + 1)  \lambda^2 {\hat \sigma}^3\right) \nonumber\\
  &\qquad\quad - 4\lambda^2 ({\tl \cS} + {\hat \sigma})^3 + 6|\lambda|{\hat c}_B ({\tl \cS} + {\hat \sigma})^2\nonumber\\
  &\qquad\quad -{\hat c}_B^3 + 3 \int_{-\pi}^{\pi}d\alpha\, \rho(\alpha)\,\int_{{\hat c}_B}^{\infty} dy\,y\,\left( \log\left(1-e^{-y-i\alpha}\right) + \log\left(1-e^{-y+i\alpha}\right)\right)  \nonumber\\
  &\qquad\quad -8 \lambda^2 {\tl \cC}^3 - 6\lambda {\tl \cC}^2 {\hat {\tl m}}_F - 3{\tl \cC} \left({\hat c}_F^2 - ({\hat {\tl m}}_F + 2 \lambda {\tl \cC})^2\right)\nonumber\\
  & \qquad\quad + {\hat c}_F^3  - 3 \int_{-\pi}^{\pi}d\alpha\, \rho(\alpha)\,\int_{{\hat c}_F}^{\infty} dy\,y\,\left( \log\left(1+e^{-y-i\alpha}\right) + \log\left(1+e^{-y+i\alpha}\right)\right)\nonumber\\
  &\qquad\qquad\qquad\qquad\qquad\qquad\qquad\quad -\frac{1}{2\lambda}\frac{3 {\hat m}_F (4 {\hat b}_4 {\hat m}_F x_4 - 4{\hat m}_B^2 x_4^2 - {\hat m}_F^2 x_6)}{8x_4^3}\Bigg]\ ,   
\end{align}
where
\begin{equation}\label{tlmfdef}
  {\tl m}_F = m_F + 2 x_4 \lambda \sigma\ ,
\end{equation}
and $\rho(\alpha)$ is the eigenvalue distribution function defined
e.g. in equation (1.7) of \cite{Choudhury:2018iwf}. The hats on the
variables in \eqref{Feffofsh} indicate that they have been scaled with
appropriate powers of the temperature $T$ to make them
dimensionless\footnote{The free energy in \eqref{Feffofsh} is
  dimensionless as well; the partition function of the theory is given
  by $\mc{Z} = e^{-\mc{V}_2 T^2 F}$ where $\mc{V}_2$ is the volume of
  two-dimensional space transverse to the thermal circle and $F$ is
  the value of the free energy at any of its saddle points.}. Note
that the large $N$ free energy does not depend on the dimensionless
parameters $y_4'$ and $y_4''$ in \eqref{generalaction}.

As we have mentioned on several occasions, \eqref{Feffofsh} is a
function of the five dynamical variables $c_B$, $c_F$, ${\tilde \cS}$,
${\tilde \cC}$ and $\sigma$ (in addition to the holonomy distribution
$\rho(\alpha)$) and needs to be extremized w.r.t.~these
variables. After extremization (i.e.~around any saddle point) these
variables have the following physical interpretation:
\begin{enumerate}
\item $c_{B}$ and $c_F$ are the thermal masses of the bosonic and
  fermionic excitations respectively and are positive by definition.

\item ${\tilde \cC}$ and ${\tilde \cS}$ are related to appropriate
  moments $\cC(c_F)$ and $\cS(c_B)$ of the holonomy distribution
  $\rho(\alpha)$:
  \begin{align}\label{moments}
    \cC(c_F) &= \frac{1}{2} \int_{-\pi}^\pi d\alpha\,\rho(\alpha)   \( \log(2 \cosh (\tfrac{\hat{c}_F +i\alpha}{2}))+ \log(2 \cosh (\tfrac{\hat{c}_F - i\alpha}{2})) \)\ ,\nonumber\\
    \cS(c_B) &= \frac{1}{2} \int_{-\pi}^\pi d\alpha\,\rho(\alpha)   \( \log(2 \sinh (\tfrac{\hat{c}_B + i\alpha}{2}))+ \log(2 \sinh (\tfrac{\hat{c}_B - i\alpha}{2})) \)\ .
  \end{align}

\item The variable $\sigma$ is related to the expectation value of
  ${\bar \phi } \phi$ via
\begin{equation}\label{sigd} 
\sigma= \frac{2 \pi}{N} \langle {\bar \phi} \phi \rangle\ .
\end{equation}
See Section 5 of \cite{Dey:2018ykx} for more details of this
interpretation in the closely related context of the regular boson
theory.
\end{enumerate}

The last term of \eqref{Feffofsh} is independent of all field
variables; this term shifts the finite temperature free energy of our
theory by a term proportional to $\frac{1}{T}$ and can be absorbed
into a cosmological constant (equivalently vacuum energy) of the
theory. We have added this `counterterm' to \eqref{Feffofsh} by hand;
this particular choice has the virtue that it renders the full
off-shell free energy invariant under the duality \eqref{dualitymap}
(rather than duality invariant upto a shift of the cosmological
constant counterterm). As we will show below the counterterm in fact
vanishes when evaluated on supersymmetric vacua (and so agrees with
the convention that assigns zero energy to susy vacua).

\subsection{Duality invariance of the off-shell free
  energy}\label{dualfree}

In this brief subsection we discuss the invariance of the off-shell
free energy of the theory governed by the action \eqref{generalaction}
under the duality transformations \eqref{dualitymap}. It is
straightforward to check that the off-shell free energy
\eqref{Feffofsh} is invariant under the duality transformations
\eqref{dualitymap}:
\begin{align}\label{dualitymaps2}
  & N'=|\kappa|-N\ , \quad \kappa' =-\kappa\ ,\quad \lambda' = \lambda - \sgn(\lambda)\ ,\nonumber\\
  & x_4' = \frac{1}{x_4}\ ,\quad x_6'=-\frac{x_6}{x_4^3}\ , \quad (y_4')' = \frac{16 y_4'}{y_4'^2 - 4 y_4''^2}\ ,\quad (y_4'')' = -\frac{16 y_4''}{y_4'^2 - 4 y_4''^2}\ ,\nonumber\\
  &b_4'=-\frac{b_4}{ x_4^2} + \frac{3}{ 4} \frac{x_6}{ x_4^3}  m_F\ ,\quad m_B'^2=- \frac{1}{ x_4}  m_B^2 - \frac{3}{ 4} \frac{x_6}{ x_4^3}  m_F^2 + \frac{2}{ x_4^2}  b_4  m_F\ ,\quad  m_F'= -\frac{ m_F}{ x_4}\ ,
\end{align}
provided the variables $c_B$, $c_F$, ${\tl \cS}$,
${\tl \cC}$ and $\sigma$ in \eqref{Feffofsh} are redefined according
to
\begin{align}\label{fieldredef}
  &c_B' = c_F\ ,\quad c_F' = c_B\ ,\quad \lambda' {\tilde \cS}' = \frac{1}{2}\left(- \sgn(\lambda) \hat{c}_F + 2\lambda {\tilde \cC}\right)\ , \nonumber\\
  & \lambda'{\tilde \cC}' = \frac{1}{2}\left( -\sgn(\lambda) \hat{c}_B + 2 \lambda \tilde\cS\right)\ ,\quad \lambda' \sigma' = \frac{1}{2}(m_F + 2\lambda x_4 \sigma)\ ,
\end{align}
in addition to the finite temperature holonomy distribution
$\rho(\alpha)$ transforming as
\begin{equation}\label{holoredef}
  |\lambda'|\rho'(\alpha) = \frac{1}{2\pi} - |\lambda|\rho(\pi - \alpha)\ .
\end{equation}
The field redefinition rules for the off-shell fields $\tilde\cC$ and
$\tilde\cS$ presented in the second line of \eqref{fieldredef} are
inspired by - and reduce on-shell to - the transformations rules of
the eigenvalue moments $\cS(c_B)$ and $\cC(c_F)$ \eqref{moments}
computed using \eqref{holoredef}\footnote{See around eq 4.9 of
  \cite{Dey:2018ykx} for a similar discussion of the duality
  invariance of the off-shell thermal free energy of the regular
  boson/critical fermion theories.}.

\subsection{Extremization of the off-shell free energy}

Extremizing \eqref{Feffofsh} w.r.t.~$\tilde{\cS}$, $c_B$,
$\tilde{\cC}$, $c_F$ and $\hat{\sigma}$ respectively yields the
following equations:
\begin{align} \label{offgap}
  & ({\tilde\cS}+{\hat \sigma}) (-{\hat c}_B + |\lambda| ({\tilde\cS}+{\hat \sigma}))=0\ ,\quad {\hat c}_B (\cS(c_B)+{\hat \sigma})-|\lambda| ({\tilde\cS}+{\hat \sigma})^2=0\ ,\nonumber\\
  & \hat{c}_F^2=(\hat{\tl m}_F + 2\lambda \tilde{\cC})^2\ ,\quad  \tilde{\cC}=\cC(c_F) \ ,\nonumber\\
  &{\hat c}_B^2 - {\hat m}_B^2-4 {\hat c}_B |\lambda|
    ({\tilde\cS}+{\hat \sigma}) + 4\lambda^2 {\tilde\cS}^2- 4 \lambda
    {\hat b}_4 {\hat \sigma}+ 8 \lambda^2 {\tilde\cS} {\hat \sigma} -
    3 \lambda^2 x_6 {\hat \sigma}^2 -4x_4 \lambda \tilde{\cC}(\hat{\tl
    m}_F +\lambda\tilde{\cC})=0\ ,
\end{align}
where $\cS(c_B)$ and $\cC(c_F)$ are defined in \eqref{moments}. Note
that the equations can be written a bit more symmetrically between the
bosonic and fermionic variables by making use of the equations in the
first line of \eqref{offgap} to solve for ${\tilde \cS}$. We find the
solution
\begin{equation}
  {\tilde \cS} = \cS(c_B)\ .
\end{equation}
Then, we have the following set of equations that follow from the
extremization of \eqref{Feffofsh}:
\begin{align} \label{offgapn}
  & ({\tilde\cS}+{\hat \sigma}) (-{\hat c}_B + |\lambda| ({\tilde\cS}+{\hat \sigma}))=0\ ,\quad {\tilde \cS} = \cS(c_B)\ ,\nonumber\\
  & \hat{c}_F^2=(\hat{\tl m}_F + 2\lambda \tilde{\cC})^2\ ,\quad  \tilde{\cC}=\cC(c_F) \ ,\nonumber\\
  &{\hat c}_B^2 - {\hat m}_B^2-4 {\hat c}_B |\lambda|
    ({\tilde\cS}+{\hat \sigma}) + 4\lambda^2 {\tilde\cS}^2- 4 \lambda
    {\hat b}_4 {\hat \sigma}+ 8 \lambda^2 {\tilde\cS} {\hat \sigma} -
    3 \lambda^2 x_6 {\hat \sigma}^2 -4x_4 \lambda \tilde{\cC}(\hat{\tl
    m}_F +\lambda\tilde{\cC})=0\ .
\end{align}
The equations in the first line in \eqref{offgapn} determine the
bosonic thermal mass ${\hat c}_B$ in terms of the
thus-far-undetermined variable ${\hat \sigma}$. There are two
solutions for ${\hat c_B}(\sigma)$ corresponding to unHiggsed and the
Higgsed phases\footnote{See Section \ref{offsh5} in Appendix
  \ref{federapp} for details of the gap equations in the different
  phases.}
\begin{align}\label{bosgapeq}
  \text{unHiggsed}(+):&\quad {\cS}(c_B) + {\hat \sigma} = 0\ ,\nonumber\\
  \text{Higgsed}(-):&\quad { \cS}(c_B) + {\hat \sigma}  = \frac{{\hat c}_B}{|\lambda|}\ .
\end{align}
The equations in the second line of \eqref{offgapn} determine the
modulus of the fermionic thermal mass $\hat{c}_F$ in terms of
${\hat \sigma}$ as
\begin{equation}\label{fergapeq}
  \hat{c}_F = \sgn(X_F) X_F\ ,
\end{equation}
where
\begin{equation}
X_F = {\hat { m}}_F + 2\lambda x_4 {\hat \sigma} + 2\lambda \cC(c_F)\ .
\end{equation}
Note that $\hat{c}_F=|X_F|$. $X_F$ is physically interpreted as the
true fermionic thermal mass including its sign. The fermionic phase is
decided by the sign $\varepsilon = \sgn(X_F) \sgn(\lambda)$ with
$\varepsilon = \pm$ sign corresponding to the $\pm$ phase of the
fermion.

Finally, plugging the equations \eqref{bosgapeq} and \eqref{fergapeq}
into the equation in the last line of \eqref{offgapn} gives an
implicit equation for ${\hat \sigma}$.

\subsubsection{The \texorpdfstring{$(\pm, +)$}{(pm,+)} phases}\label{unhiggapeq}
In the unHiggsed phase of the boson and either phase of the fermion,
${\hat \sigma}$ is simply given by $-\cS(c_B)$. Hence, the equations in
\eqref{offgapn} finally give equations for $c_B$, $c_F$ and $\sigma$
in the unHiggsed phase of the boson:
\begin{align}\label{unhigboseq}
  & {\hat c}_F = |{\hat m}_F - 2\lambda x_4 \cS(c_B) + 2\lambda \cC(c_F)|\ ,\quad {\hat \sigma} = -\cS(c_B)\ ,\nonumber\\
&{\hat c}_B^2 - 4\lambda^2 {\cS}^2(c_B)  - {\hat m}_B^2 + 4 \lambda
  {\hat b}_4 \cS(c_B) - 3x_6 \lambda^2
  \cS^2(c_B)\nonumber \\ &\hspace{150pt} - 4x_4 \lambda \cC(c_F)({\hat m}_F - 2\lambda x_4 \cS(c_B)  +\lambda {\cC}(c_F))=0\ .
\end{align}
Once these (effectively two) equations are solved for the two 
variables $c_B$ and $c_F$, the sign $\varepsilon$ of
$\lambda X_F = \lambda {\hat m}_F - 2\lambda^2 x_4 \cS(c_B) + 2\lambda
\cC(c_F)$ decides the phase of the fermion.

\subsubsection{The \texorpdfstring{$(\pm, -)$}{(pm,-)} phases}\label{higgapeq}
In the Higgsed phase of the boson and either phase of the fermion the
gap equations take the form
\begin{align}\label{higboseq}
& {\hat c}_F = |{\hat m}_F - 2\lambda x_4 \cS(c_B) + 2\lambda \cC(c_F)|\ ,\quad {\hat c}_B = |\lambda|(\cS(c_B) + {\hat \sigma})\ ,\nonumber\\
&  -3\lambda^2 (\cS(c_B) + {\hat \sigma})^2 + 4\lambda^2 \cS^2(c_B) - {\hat m}_B^2 - 4 \lambda {\hat b}_4 {\hat \sigma} - 3x_6 \lambda^2
{\hat \sigma}^2 \nonumber \\ &\hspace{150pt} - 4x_4 \lambda \cC(c_F)({\hat m}_F + 2\lambda x_4 {\hat \sigma}  +\lambda {\cC}(c_F))=0\ .
\end{align}
In this case we will find it convenient to choose ${\hat \sigma}$ as
our basic dynamical variable, i.e. to use the equations on the first
two lines of \eqref{higboseq} to solve for $c_F$ and $c_B$ in terms of
${\hat \sigma}$ and to then use the last equation to determine
${\hat \sigma}$. Again, once these equations are solved, the sign of
$\varepsilon$ of
$\lambda X_F = \lambda {\hat m}_F - 2\lambda^2 x_4 \cS(c_B) + 2\lambda
\cC(c_F)$ decides the phase of the fermion.
  
\subsection{Expressions for the free energy in different phases}

The expression \eqref{Feffofsh} is an elegant object in that it is a
single expression, analytic in all its variables, that simultaneously
captures the free energy of all distinct phases of the theory
\eqref{generalaction}. As we have seen above \eqref{generalaction} is
very useful for establishing formal properties like the invariance of
the free energy under duality. In order to find explicit expressions
for the free energy in each of the distinct `phases' of the theory
(and especially to make contact with the earlier results of
\cite{Jain:2013gza} valid for the $(\pm,+)$ phases) it is useful to
`simplify' the expression \eqref{Feffofsh} by eliminating some of its
variables using the equations \eqref{offgapn}.

The procedure is as follows. First we choose one of the two solutions
of the first equation in the first line of \eqref{offgap}:
\begin{align}\label{bosgapeqn}
  \text{unHiggsed}(+):&\quad {\tl \cS} + {\hat \sigma} = 0\ ,\nonumber\\
  \text{Higgsed}(-):&\quad {\tl \cS} + {\hat \sigma}  = \frac{{\hat c}_B}{|\lambda|}\ .
\end{align}
For the unHiggsed case, we substitute $\hat \sigma$ in terms of
$\tl \cS$ in the off-shell free energy in \eqref{Feffofsh} while for
the Higgsed case, we substitute $\tl \cS$ in terms of $\hat \sigma$
and $\hat c_B$. This gives two different expressions for the free
energy in the unHiggsed / Higgsed phases of the boson. Secondly, we
solve for ${\tl \cC}$ in terms of ${ c}_F$ and $\sigma$ from the first
equation in the second line of \eqref{offgapn}:
\begin{equation}
  {\tl \cC} = \sgn(X_F) {\hat c_F} - ({\hat m}_F + 2\lambda x_4 {\hat \sigma}) = \varepsilon\, \sgn(\lambda_F) {\hat c_F} - ({\hat m}_F + 2\lambda x_4 {\hat \sigma})\ ,
\end{equation}
where $\varepsilon = \sgn(X_F) \sgn(\lambda_F)$ decides the phase of
the fermion (cf.~Sections \ref{unhiggapeq} and \ref{higgapeq}).

Implementing this procedure we find the following expressions for the
free energies in the $(\varepsilon, \pm)$ phases i.e.~ in the
unHiggsed / Higgsed phase of the boson and either phase
($\varepsilon$) of the fermion:

\begin{align}\label{feunhigmain}
  &F^{(\varepsilon, +)}[c_B, c_F, \tl\cS]\nonumber\\
  &= \frac{N}{6\pi}\bigg[3{\hat c}_B^2 {\tl \cS}  - \lambda^2 {\tl \cS}^3 +  3\left( -{\hat m}_B^2 {\tl \cS}  + 2\lambda {\hat b}_4 {\tl \cS}^2 - (x_6 + 1)  \lambda^2 {\tl \cS}^3\right) \nonumber\\ 
                       &\qquad\quad -{\hat c}_B^3 + 3 \int_{-\pi}^{\pi}d\alpha\, \rho(\alpha)\,\int_{{\hat c}_B}^{\infty} dy\,y\,\left( \log\left(1-e^{-y-i\alpha}\right) + \log\left(1-e^{-y+i\alpha}\right)\right)  \nonumber\\
                       &\qquad\quad - \frac{\varepsilon}{|\lambda|} {\hat c}_F^3 + \frac{3}{2\lambda}  {\hat c}_F^2 ({\hat m}_F - 2\lambda x_4 {\tl \cS}) -\frac{({\hat m}_F - 2\lambda x_4 {\tl \cS})^3}{2\lambda} \nonumber\\
                       & \qquad\quad + {\hat c}_F^3  - 3 \int_{-\pi}^{\pi}d\alpha\, \rho(\alpha)\,\int_{{\hat c}_F}^{\infty} dy\,y\,\left( \log\left(1+e^{-y-i\alpha}\right) + \log\left(1+e^{-y+i\alpha}\right)\right)\Bigg]\ .   
\end{align}
\begin{align}\label{fehigmain}
  &F^{(\varepsilon, -)}[c_B, c_F, \sigma]\nonumber\\
  &= \frac{N}{6\pi}\bigg[\frac{2}{|\lambda|} {\hat c}_B^3 - 3 {\hat c}_B^2 {\hat \sigma} + \lambda^2 {\hat \sigma}^3 +  3\left( {\hat m}_B^2 {\hat \sigma}  + 2\lambda {\hat b}_4 {\hat \sigma}^2 +  (x_6 + 1)  \lambda^2 {\hat \sigma}^3\right) \nonumber\\ 
                       &\qquad\quad -{\hat c}_B^3 + 3 \int_{-\pi}^{\pi}d\alpha\, \rho(\alpha)\,\int_{{\hat c}_B}^{\infty} dy\,y\,\left( \log\left(1-e^{-y-i\alpha}\right) + \log\left(1-e^{-y+i\alpha}\right)\right)  \nonumber\\
                       &\qquad\quad - \frac{\varepsilon}{|\lambda|} {\hat c}_F^3 + \frac{3}{2\lambda}  {\hat c}_F^2 ({\hat m}_F + 2\lambda x_4 {\hat \sigma}) -\frac{({\hat m}_F + 2\lambda x_4 {\hat \sigma})^3}{2\lambda} \nonumber\\
                       & \qquad\quad + {\hat c}_F^3  - 3 \int_{-\pi}^{\pi}d\alpha\, \rho(\alpha)\,\int_{{\hat c}_F}^{\infty} dy\,y\,\left( \log\left(1+e^{-y-i\alpha}\right) + \log\left(1+e^{-y+i\alpha}\right)\right)\Bigg]\ .   
\end{align}
Note that the above two expressions are functions of three variables
($c_B$, $c_F$ and $\tl \cS$ in the unHiggsed case and $c_B$, $c_F$ and
$\sigma$ in the Higgsed case). The corresponding thermodynamic free
energies are obtained by extremizing the above three-variable
off-shell free energies and evaluating these free energies at the
respective extrema. These three-variable off-shell free energies
exactly match the free energies computed in the individual phases as
is summarised in \eqref{feunhig} / \eqref{fehig} in Appendix
\ref{federapp}.  Moreover the expression for $F^{(\epsilon, +)}$ given
in \eqref{feunhigmain} also agrees with the off-shell free energy
reported in \cite{Jain:2013gza}.

Since the off-shell free energy \eqref{Feffofsh} and the saddle point
equations \eqref{offgapn} are invariant under the duality map
\eqref{dualitymap}, the four expressions above correctly transform
into each other under duality.

\subsection{The quantum effective potential for \texorpdfstring{${\bar \phi}
  \phi$}{phibar-phi}} \label{qea}

In this subsection we will explain the relationship between the five
variable off-shell free energy \eqref{Feffofsh} and the quantum
effective potential for the field ${\bar \phi} \phi$. The discussion
of this subsection closely parallels that around equations 4.6 and 4.7
of \cite{Dey:2018ykx}.

In order to compute the quantum effective potential for the field
${\bar \phi}\phi$, we are instructed first to add the terms
$$\int d^3 x\, J \left( {\bar \phi} \phi - ({\bar \phi} \phi )_{\rm cl}
\right) $$ to the classical action \eqref{generalaction}, then perform
the path integral over $\phi$. The result of this path integral takes
the form
\begin{equation}\label{effact}
  \exp\bigg(-S_{\rm eff}[J\, , ({\bar \phi} \phi)_{\rm cl}]\bigg)
\end{equation} 
We are then instructed to extremize
$S_{\rm eff}[J,({\bar \phi} \phi)_{\rm cl}]$ over the field $J$. The
result of this extremization, $\Gamma[({\bar \phi} \phi)_{\rm cl}]$,
is the quantum effective action of our theory as a function of the
field $({\bar \phi} \phi)_{\rm cl}$.

Conceptually, the procedure described above works for any value of the
`classical' field $({\bar \phi} \phi)_{\rm cl}(x)$, and could, in
principle, be implemented to determine the full quantum effective
action for this operator.  In this paper, however, we specialise to
the case in which $({\bar \phi} \phi)_{\rm cl}$ is constant.  In other
words we focus attention on only the exact quantum effective
\emph{potential} for the field $(\bar\phi \phi)_{\rm cl}$
\footnote{Later in this paper we will make the very plausible
  assumption that the actual global minimum of the full quantum
  effective action $\Gamma[(\bar\phi \phi)_{\rm cl}]$ is indeed a
  constant field configuration and so the extremization of the quantum
  effective potential w.r.t.~the number $({\bar \phi} \phi)_{\rm cl}$
  correctly reproduces the phase diagram of our theory.}.

The first step in the programme outlined above is now easily
accomplished. The action \eqref{generalaction} already has a term
$m_B^2 {\bar \phi} \phi$. Consequently the result of the path integral
with the additional term
$J \left( {\bar \phi} \phi - ({\bar \phi} \phi)_{\rm cl} \right)$
added to the action is simply given by adding
$- \beta^3 J ({\bar \phi} \phi )_{\rm cl}$ to the extremized version
of the off-shell free energy \eqref{Feffofsh} along with replacement
$m_B^2 \rightarrow m_B^2 + J$. Now $m_B^2$ appears in \eqref{Feffofsh}
only in the term
$$ \frac{N}{2 \pi}{m_B^2 \sigma} . $$
It then follows that
\begin{equation}\label{effactimp} 
S_{\rm eff}[J,({\bar \phi} \phi)_{\rm cl}]
= \mc{V}_2 T^2 F + \mc{V}_2 \beta J \left( \frac{N}{2\pi}{ \sigma} - ({\bar \phi} \phi)_{\rm cl} \right)\ ,
\end{equation} 
where $\mc{V}_2$ is the volume of two dimensional space and $F$ is the
(dimensionless) five variable off-shell free energy \eqref{Feffofsh}
extremized over its dynamical variables.

In order to obtain the quantum effective potential, the expression in
\eqref{effactimp} must now be further extremized over $J$ as well as
holonomies and the five dynamical variables of \eqref{Feffofsh}. It is
convenient to perform the extremization over $J$ first; this yields \
\begin{equation}\label{sigph}
\sigma = \frac{ 2 \pi}{ N} ({\bar \phi} \phi)_{\rm cl} \ .
\end{equation}
Extremization over $\sigma$ fixes the value of $J$, but as the value
of the action is independent of $J$ (using \eqref{sigph}), this
extremization is unimportant and can be ignored. The extremization
over the other four dynamical variables $c_B$, $c_F$ $\tilde\cC$ and
$\tilde\cS$ still needs to be performed; the result of this
extremization is the quantum effective potential as a function of
$({\bar \phi} \phi)_{\rm cl}$.

The final prescription for computing the quantum effective action for
$ ({\bar \phi} \phi)_{\rm cl}$ from the off-shell free energy is
extremely simple; all we have to is to start with the expression
\eqref{Feffofsh}, extremize it w.r.t. the four variables $c_B$, $c_F$
$\tilde\cC$ and $\tilde\cS$. The resultant expression is a function of
$\sigma$. The substitution \eqref{sigph} inserted into this final
expression yields the required quantum effective potential. Since the
quantum effective potential is a function of some effective order
parameters $c_B$, $c_F$, ${\tl \cS}$, ${\tl \cC}$ and
$(\bar\phi \phi)_{\rm cl}$, it is apt to call it a
\emph{Landau-Ginzburg} effective potential. We compute this
Landau-Ginzburg effective potential in the next subsection in the zero
temperature limit.

\subsection{Explicit Landau-Ginzburg effective potential 
at zero temperature}

In this subsection we will explicitly implement the 
procedure described in Section \ref{qea} to 
find an explicit expression for the quantum effective 
potential of our theory in the zero temperature limit. 

In order to accomplish this we need to eliminate the variables
$\tilde\cS$, $\tilde\cC$, $c_B$ and $c_F$ using their equations of
motion. The first two variables listed above are particularly easy to
eliminate. Recall that the equation of motion for these two variables
can be cast in the form $\tilde\cS = \cS(c_B)$ and
$\tilde\cC = \cC(c_F)$. In the zero temperature limit, however, the
quantities $\cS(c_B)$ and $\cC(c_F)$ simply become
\begin{equation}
  \cS(c_B) = \frac{c_B}{2}\ ,\quad \cC(c_F) = \frac{c_F}{2}\ .
\end{equation}
It follows that $\tilde\cS$ and $\tilde\cC$ can be eliminated by
making the replacements
$$ \tilde \cS \rightarrow \frac{c_B}{2}\ ,\quad \tilde \cC \rightarrow \frac{c_F}{2}$$
in \eqref{generalaction} yielding the zero temperature free energy density
\begin{align}\label{FeffofshT0}
F & =  \frac{N}{6\pi} \Bigg[\frac{1}{2}(|\lambda| - 2)(1 - |\lambda|) {c}_B^3-3 { c}_B^2(1 - |\lambda|)^2  { \sigma} + 6{c}_B |\lambda|(1-|\lambda|) { \sigma}^2  \nonumber\\
&\qquad\qquad  + 3  \left({ m}_B^2 { \sigma} + 2\lambda   { b}_4 { \sigma}^2 +  x_6 \lambda^2 { \sigma}^3\right)-\frac{1}{2}(1 - \lambda^2){ c}_F^3   + \frac{3}{2}\lambda{c}_F^2{\tl m}_F + \frac{3}{2}c_F {\tl m}_F^2 + c_0\Bigg]\ ,
\end{align}
where ${\tl m}_F = m_F + 2\lambda x_4 \sigma$ and
\begin{equation}\label{const}
  c_0 = -\frac{1}{ 2\lambda}\frac{ m_F \left(4b_4 m_F x_4-4m_B^2 x_4^2-m_F^2 x_6\right)}{8x_4^3}\ .
\end{equation}
The above free energy density is analytic in the three variables
$c_B$, $c_F$ and $\sigma$. It remains to eliminate $c_B$ and $c_F$ and
thus finally obtain an expression for the exact effective potential as
a function of $(\bar\phi \phi)_{\rm cl}$ (making use of the relation
\eqref{sigph} between $\sigma$ and $(\bar\phi \phi)_{\rm cl}$ as
explained earlier in Section \ref{qea}).

\begin{figure}
	\centering
	\scalebox{0.7}{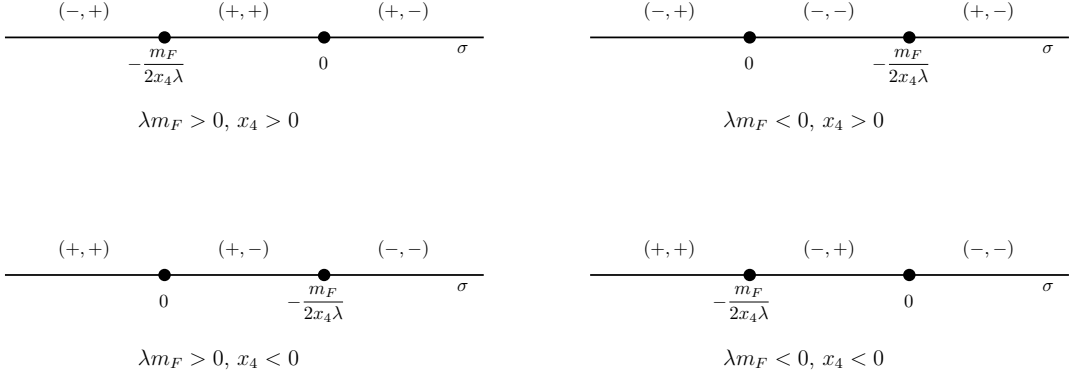}
	\caption{Regions of validity for different branches of the potential
		for various choices of $\sgn(m_F \lambda)$ and
		$\sgn(x_4)$.}
	\label{phaseseq}
\end{figure}

Extremizing \eqref{FeffofshT0} with respect to $c_B$ and $c_F$ we find
the gap equations
\begin{equation} 
  -(2 - |\lambda|)c_B^2 - 4 (1 - |\lambda|) \sigma  c_B + 4  |\lambda|\sigma^2 = 0\ ,\quad  c_F^2  = (\lambda  c_F +  {\tl m}_F)^2\ .
\end{equation}
The equation for $c_B$ is quadratic and has two roots given by
\begin{equation} \label{solcb}
  c_B = -2\sigma\ ,\quad\text{or}\quad c_B = \frac{2 |\lambda| \sigma}{2 - |\lambda|}\ ,
\end{equation}
corresponding to the unHiggsed and the Higgsed phases
respectively. Since $c_B$ is a positive quantity by definition, the
range of validity of $\sigma$ for the above roots are $\sigma < 0$ for
the unHiggsed phase and $\sigma > 0$ for the Higgsed
phase\footnote{Recall that $c_B$, by definition, is a positive
  quantity and is and is generically of order unity in the `classical'
  limit (in which we take $\lambda \to 0$ holding all other parameters
  in the action \eqref{generalaction} fixed, see Eq 2.10 of
  \cite{Dey:2018ykx} ). It follows first that $\sigma$ and so
  ${\bar \phi} \phi$ are negative on the first root in
  \eqref{solcb}. Note also that, in the small $\lambda$ limit,
  $\sigma$ is of order unity and so the square of `classical' field
  (see 2.8 of \cite{Dey:2018ykx}) is of order $\lambda$. It follows
  that this root is a quantum `blow up' of the classical vacuum
  $\phi=0$ and so represents the unHiggsed branch. The fact that
  $\bar \phi \phi$ on this root is quantum rather than classical
  allows its value to be negative (recall that products of local
  quantum fields are well-defined only after a subtraction). On the
  other hand on the second root of \eqref{solcb} $\sigma$ is positive
  and of order $\frac{1}{\lambda}$. It follows that the square of the
  classical field (again see Eq 2.8 of \cite{Dey:2018ykx}) is of order
  unity on this root. Consequently this root corresponds to expanding
  the scalar field around a nonzero classical value of $\phi$ and so
  lies on the Higgsed branch.}:
\begin{align} \label{bosecond}
&\text{Bosonic}\ +:\quad   \sigma<0 ,\nonumber\\
&\text{Bosonic}\ -:\quad \sigma >0 .
\end{align}
The $c_F$ equation can be rewritten as follows, keeping in mind that
$c_F$ is a positive quantity:
\begin{equation}
  c_F = \sgn({\tl X}_F) {\tl X}_F\ ,\quad\text{with}\quad {\tl X}_F = \lambda c_F + {\tl m}_F\ .
\end{equation}
It is easy to see that $\sgn({\tl X}_F) = \sgn({\tl m}_F)$. Thus, we have
\begin{equation}
  c_F = \sgn({\tl m}_F) (\lambda c_F + {\tl m}_F) \ .
\end{equation}
We now define the sign $\varepsilon = \sgn(\lambda) \sgn({\tl m}_F)$
such that $\varepsilon = \pm 1$ corresponds to the `$+$' or `$-$'
phases of the fermion. Then, the fermionic gap $c_F$ is given by
\begin{equation}
  c_F = \frac{|{\tl m_F}|}{1 - \varepsilon|\lambda|}\ .
\end{equation}
Recall that ${\tl m}_F = m_F + 2 x_4 \lambda \sigma$. The condition
$\varepsilon = \pm 1$ implies that the range of $\sigma$ for which the
$\pm$ phase of the fermion may occur is given by
\begin{align} \label{fermcond}
  &\text{Fermionic}\ +:\quad  \sgn(x_4) \sigma > -\frac{m_F}{2|x_4|\lambda}\ ,\nonumber\\
  &\text{Fermionic}\ -:\quad  \sgn(x_4) \sigma < -\frac{m_F}{2|x_4|\lambda}\ .
\end{align}
The conditions above unambiguously (and uniquely) determine the
bosonic and fermionic phase that any given value of $\sigma$ lies
in. When $x_4$ is positive, large negative values of $\sigma$ lie in
the $(-,+)$ phase, while large positive values of $\sigma$ lie in the
$(+,-)$ phase. As long as $\frac{m_F}{|x_4| \lambda}$ is nonzero, we
also have an intermediate range of $\sigma$ that lies either in the
$(-,-)$ or the $(+,+)$ phase depending on the sign of $m_F \lambda$
(see Figure \ref{phaseseq} for details). In a similar manner, when
$x_4$ is negative, large negative values of $\sigma$ lie in the
$(+,+)$ phase, while large positive values of $\sigma$ lie in the
$(-,-)$ phase. As long as $\frac{m_F}{|x_4| \lambda}$ is nonzero, we
also have an intermediate range of $\sigma$ that lies either in the
$(+,-)$ or the $(-,+)$ phase depending on the sign of $m_F \lambda$
(again see Figure \ref{phaseseq} for details).

The gap equations in the four phases along with their ranges of validity are as follows:
\begin{alignat}{3} \label{cbcf}
(+,+)\text{ Phase }:&\quad { c}_F =\frac{| {\tl m}_F|}{1-|\lambda|}\ ,\quad  &&{ c}_B =-2{ \sigma}\ ,\quad && \sgn(x_4)\sigma > -\frac{m_F}{2|x_4|\lambda}\ ,\quad \sigma < 0\ , \nonumber\\
  (-,+)\text{ Phase }:&\quad { c}_F =\frac{|{\tl m}_F|}{1+|\lambda|}\ ,\quad  &&{ c}_B =-2{ \sigma}\ , \quad && \sgn(x_4)\sigma < -\frac{m_F}{2|x_4|\lambda}\ ,\quad \sigma < 0\ ,\nonumber\\
  (+,-)\text{ Phase }:&\quad { c}_F = \frac{|{\tl m}_F|}{1-|\lambda|}\ ,\quad  &&{ c}_B =\frac{2|\lambda|\sigma}{2-|\lambda|} \ ,\quad && \sgn(x_4)\sigma > -\frac{m_F}{2|x_4|\lambda}\ ,\quad \sigma > 0\ , \nonumber\\
(-,-)\text{ Phase }:&\quad { c}_F =\frac{|{\tl m}_F|}{1+|\lambda|}\ ,\quad  &&{ c}_B =\frac{2|\lambda|\sigma}{ 2-|\lambda|}\ ,\quad && \sgn(x_4)\sigma < -\frac{m_F}{2|x_4|\lambda}\ ,\quad \sigma > 0\ .
\end{alignat}
As we have emphasised above (see Figure \ref{phaseseq}) at any
particular values of $x_4$ and $\lambda m_F$, the quantum effective
potential $U(\sigma)$ never accesses more than three (and generically,
when $m_F / x_4 \lambda$ is non-zero, exactly three) of these phases.

Plugging \eqref{cbcf} into \eqref{FeffofshT0}, we obtain the following
explicit expressions for the quantum effective potential valid in each
of the four possible phases:
\begin{align}\label{LGpot}
  U^{(\varepsilon,\pm)}(\sigma) &= \frac{N}{2\pi} \bigg[\psi_\varepsilon\frac{{\tl m}_F^3}{8\lambda} + m_B^2 \sigma + 2\lambda b_4 \sigma^2 + (x_6 - \phi_\pm)\lambda^2\sigma^3 + c_0\bigg]\ ,
\end{align}
with $\varepsilon = \pm$ denoting the fermionic phase and the
explicit $\pm$ denoting the bosonic phase.
The quantities $\psi_\pm$ and $\phi_\pm$\footnote{The functions( of
  $\lambda$) $\phi_+$ and $\phi_-$ are same as the functions $\phi_2$
  and $\phi_1$ that were encountered in the study of the regular boson
  theory. The functions $\psi_\pm$ map to $\phi_\pm$ under the duality
  map \eqref{dualitymap}.} are defined as
\begin{align} 
  &\psi_- = \frac{4}{3}\left(\frac{1}{(1+|\lambda|)^2} - 1\right)\ ,\quad \psi_+ = \frac{4}{3} \left(\frac{1}{(1 - |\lambda|)^2} - 1\right)\ ,\label{psipm}\\
  &\phi_- = \frac{4}{3}\left(\frac{1}{(2-|\lambda|)^2} - 1\right)\ ,\quad \phi_+ = \frac{4}{3} \left(\frac{1}{\lambda^2} - 1\right)\ .\label{phipm}
\end{align}
Plugging in ${\tl m}_F = m_F + 2x_4 \lambda \sigma$ in \eqref{LGpot}
we obtain the more explicit expression for the potential
\begin{multline}\label{Ustab}
U^{(\varepsilon,\pm)}(\sigma) = \frac{N}{2\pi}\Big[ \left(x_6 - \phi_\pm + x_4^3\psi_\varepsilon\right)  \lambda^2 { \sigma}^3  +  \left( { b}_4 + \tfrac{3}{4}  x_4^2\psi_\varepsilon m_F \right) 2 \lambda  { \sigma}^2\\
+ \left({ m}_B^2 + \tfrac{3}{4} x_4 \psi_\varepsilon m_F^2\right){ \sigma} + \psi_\varepsilon \frac{m_F^3}{8\lambda}+ c_0\Big]\ .
\end{multline}

Using the fact that $|\lambda|<1$ it is immediately obvious from
\eqref{phipm} that $\phi_+>\phi_-$ and from \eqref{psipm} that
$\psi_+>\psi_-$. The relative orderings between the $\psi$'s and the
$\phi$'s depends on whether $|\lambda|$ is less than or greater than
half. Explicitly, we have
\begin{align}\label{phipsiord}
 |\lambda| \geq \frac{1}{2}:&\quad \psi_+ \geq \phi_+ \geq 0 \geq \phi_- \geq \psi_-\ ,\nonumber\\
  |\lambda| < \frac{1}{2}:&\quad \phi_+ > \psi_+ \geq 0 \geq \psi_- > \phi_-\ .
\end{align}

Note that the quantum effective potential is a cubic function of
$\sigma$ in every phase. At any given value of microscopic parameters,
the full graph of the quantum effective potential is given by patching
together the expressions \eqref{LGpot} in the various regions depicted
in Figure \ref{phaseseq}. Recall from \eqref{cbcf} that $c_F$, and
hence ${\tilde m}_F$, vanishes at the value of $\sigma$ at which we
transit between fermionic phases, and that $c_B$, and hence $\sigma$,
vanishes (i.e.~$\sigma =0$) when we transit between bosonic phases. It
follows immediately from the explicit expressions \eqref{LGpot} that
the full potential $U(\sigma)$ is continuous across the `transition'
values of $\sigma$ depicted in Figure \ref{phaseseq}, even though it
is non-analytic at those points.

\section{Stability of the theory at zero temperature}\label{stab}

As we have explained above, when $x_4$ is positive, large values of
$\sigma$ lie in the $(+,-)$ phase and large negative values of
$\sigma$ lie in the $(-,+)$ phase. It follows that the effective
potential is bounded from below (and so the theory has a stable
vacuum) if and only if the coefficient of the cubic term in $\sigma$
is positive in the $(+, -)$ phase and negative in the $(-,+)$
phase. Similarly when $x_4<0$ the effective potential is bounded from
below if and only if the coefficient of $\sigma^3$ is positive in the
$(-,-)$ phase, and negative in the $(+,+)$ phase.  Using the explicit
formulae \eqref{Ustab}, it follows immediately that our theories have
a stable potential if and only if
\begin{align}\label{stabcond}
&  x_6 - \phi_+ + x_4^3 \psi_- < 0\quad{\rm and}\quad x_6 - \phi_- + x_4^3 \psi_+ > 0\ ,\quad{\rm when}\quad x_4>0\ ,\nonumber \\
&x_6 - \phi_+ + x_4^3 \psi_+ < 0\quad{\rm and}\quad x_6 - \phi_- + x_4^3 \psi_- > 0\ ,\quad{\rm when}\quad x_4<0\ .
\end{align}
\begin{figure}
	\centering
	\includegraphics[width=4in,height=3in]{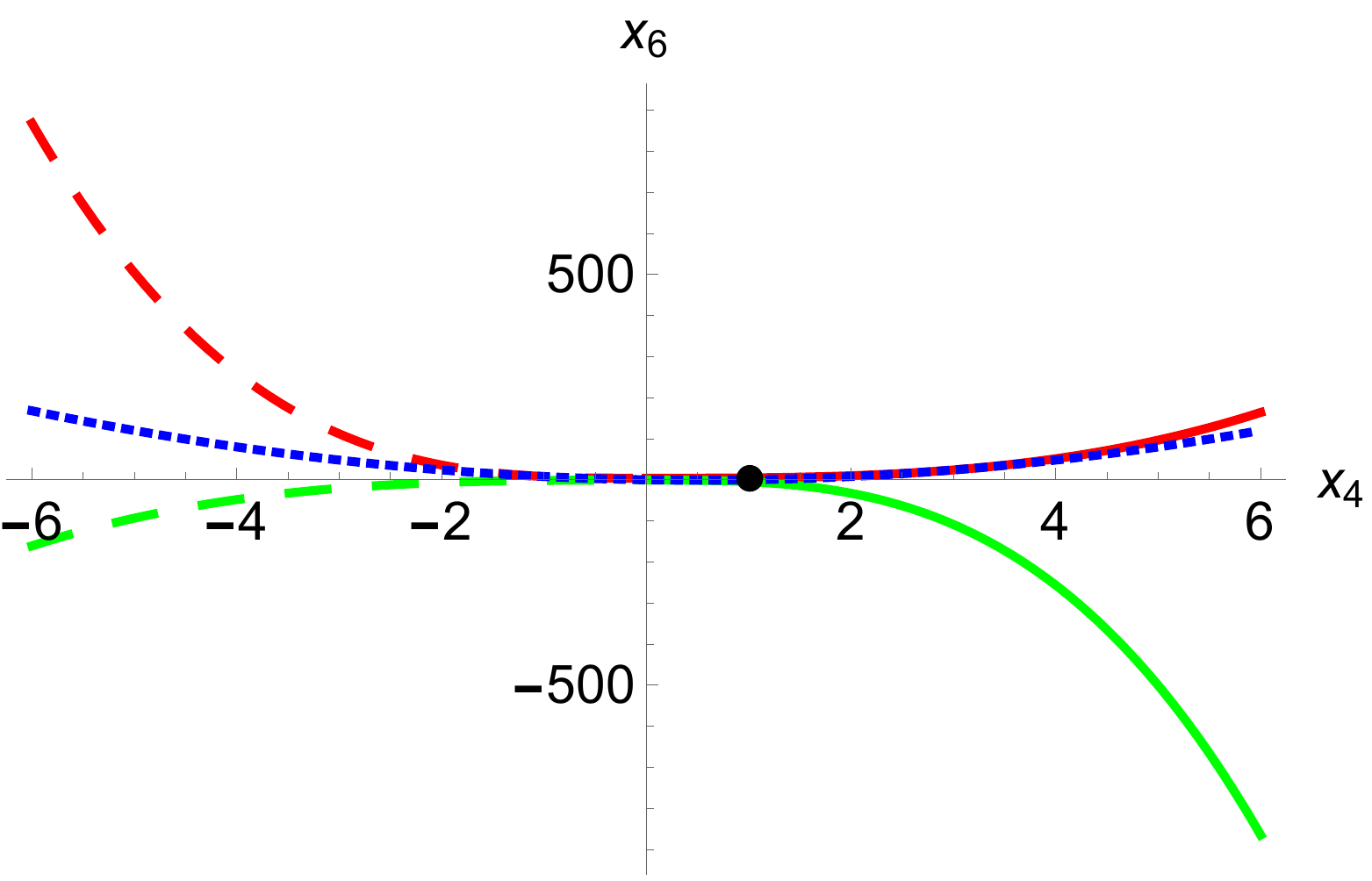}
	\caption{Region of stability in the $x_4-x_6$ plane at
          $|\lambda|=\frac{1}{ 2}$. The solid red curve represents
          $ x_6 - \phi_+ + x_4^3 \psi_- = 0$ while the solid green one
          corresponds to $ x_6 - \phi_- + x_4^3 \psi_+ =0$. The dashed
          red curve represents the line
          $x_6 - \phi_+ + x_4^3 \psi_+ < 0$ while the green dashed one
          denotes $x_6 - \phi_- + x_4^3 \psi_- = 0$. The region
          between the solid red and solid green curves, and the region
          between the dashed red and dashed green curves define the
          region of stability. The blue line is the set of
          ${\cal N}=1$ theories while the black dot is the
          ${\cal N}=2$ theory. Note that all susy theories lie within
          the stability regime.}
	\label{Stabilityfig}
\end{figure}
In Figure \ref{Stabilityfig} we have shown the lines (of
\eqref{stabcond}) constraining the stability of the potential for a
representative value of $|\lambda|=\frac{1}{2}$. With $x_4>0$, the
solid red curve corresponds to $ x_6 - \phi_+ + x_4^3 \psi_- = 0$,
while the solid green one corresponds to
$ x_6 - \phi_- + x_4^3 \psi_+ =0$. On the other hand, with $x_4<0$,
the dashed red curve denotes $x_6 - \phi_+ + x_4^3 \psi_+ < 0$ while
the green dashed one denotes $x_6 - \phi_- + x_4^3 \psi_- = 0$. The
conditions \eqref{stabcond} imply that the potential is stable only if
the $x_4-x_6$ parameters lie between the solid red, solid green,
dashed red and dashed green curves in Figure \ref{Stabilityfig}. Since
the quantum effective potential at leading order in $1/N$ does not
depend on the parameters $y_4'$ and $y_4''$ in \eqref{generalaction},
our analysis does not say anything about regions of stability for
those parameters.

\subsection{Stability of supersymmetric theories}
In this paper we are particularly interested in the $\mc{N} = 2$
superconformal theory and deformations by its classically relevant
parameters $m_B^2$, $b_4$ and $m_F$. This corresponds to setting the
marginal parameters $x_4$ and $x_6$ to
\begin{equation}
  x_4 = 1\ ,\quad x_6 = 0\ .
\end{equation}
In Figure \ref{Stabilityfig}, the black dot at $x_4=1, x_6=0$ denotes
the $\mathcal{N}=2$ SUSY point and it exists inside the region of
stability. 

We perform the stability analysis for general $|\lambda|$ now. The
following inequalities must be satisfied for the $\mc{N} = 2$ theory
to be stable:
\begin{equation}\label{N2stab}
  \psi_- < \phi_+\ ,\qquad \psi_+ > \phi_-\ .
\end{equation}
The inequalities in \eqref{N2stab} are clearly satisfied for all
values of $\lambda$ (see \eqref{phipsiord}) and so the ${\cal N}=2$
theory is stable.

In a similar manner, the $\mc{N} = 1$ theory corresponds to setting
\begin{equation}
  x_4 = \frac{1+w}{2}\ ,\quad x_6 = w^2-1\ .
\end{equation}
For these values of $x_4$ and $x_6$, it is not hard to verify that the
stability conditions follow from \eqref{phipsiord}. For
$|\lambda| = \tfrac{1}{2}$, the blue dotted line in Figure
\ref{Stabilityfig} is the $\mc{N} = 1$ locus. As expected, it lies
within the region of stability.

Thus, we see that the supersymmetric subset of theories in
\eqref{generalaction} are stable, as is expected to follow from
general supersymmetry arguments.

\section{Phase diagram of the \texorpdfstring{$\mc{N} = 2$}{N=2} theory}\label{N2phase}

As we have explained in the introduction, the $\mc{N} = 2$ theory is
defined in the UV by the choice of dimensionless parameters $x_4=1$,
$x_6=0$, $y_4'=4$ and $y_4''=0$ in the action \eqref{generalaction}.
These choices define a supersymmetric fixed point of the
renormalization group.  This fixed point admits three relevant
deformations parametrized by the massive parameters $b_4$, $m_B^2$
and $m_F$. The IR behaviour of the theory is a function of the two
dimensionless ratios of these parameters. It follows that the
${\cal N}=2$ theory has a two dimensional phase diagram. In this
section we will quantitatively work out this phase diagram in the
large $N$ limit. Our main tool in this section is the
Landau-Ginzburg effective potential is given by \eqref{LGpot}:
\begin{multline}\label{LGpotrel}
U^{(\varepsilon,\pm)}(\sigma) = \frac{N}{2\pi}\Big[ \left(\psi_\varepsilon - \phi_\pm\right)  \lambda^2 { \sigma}^3  +  \left( { b}_4 + \tfrac{3}{4}\psi_\varepsilon m_F \right) 2 \lambda  { \sigma}^2\\
+ \left({ m}_B^2 + \tfrac{3}{4}\psi_\varepsilon m_F^2\right){ \sigma} + \psi_\varepsilon \frac{m_F^3}{8\lambda}  -\frac{ m_F \left(b_4 m_F - m_B^2\right)}{4\lambda}\Big]\ .
\end{multline}
The parameter $\varepsilon = \sgn(m_F + 2\lambda\sigma)\sgn(\lambda)$
measures the sign of the effective fermionic mass expanded around a
vacuum with a particular value of $\sigma$; the fact that our
effective potential changes discontinuously as this sign flips is a
reflection of the fact that `phase' as a function of $\sigma$
undergoes a continuous second order `phase transition' \footnote{We
  have put quotes around the `phase' because all our statements here
  are true about any extremum of the effective potential
  \eqref{LGpot}, whether or not this extremum is global minimum of the
  potential and hence a true phase of the theory.} as
$m_F + 2\lambda \sigma$ goes through zero (the level of the low energy
Chern-Simons theory in the massive phase is an order parameter for
this phase transition). In a similar manner, the sign in $\phi_\pm$ is
either $+$ or $-$ depending on whether $\sigma$ is negative or
positive; so, as $\sigma$ changes sign, the coefficient of the
$\sigma^3$ term in \eqref{LGpotrel} changes. This non-analyticity
reflects the fact that our theory undergoes a second order Higgsing
`phase transition' as $\sigma$ goes from negative to positive values
(the rank of the low energy Chern-Simons theory in the massive phase
is an order parameter for this phase transition).
 
The 
quantities $\phi_\pm$ and $\psi_\pm$ are defined in \eqref{phipm} and
\eqref{psipm} respectively and are reproduced below:
\begin{alignat}{2}
  &\phi_+ = \frac{4}{3}\left(\frac{1}{\lambda^2} - 1\right)\ ,\quad&&\phi_- = \frac{4}{3}\left(\frac{1}{(2 - |\lambda|)^2} - 1\right)\ ,\nonumber\\
&\psi_+ = \frac{4}{3}\left(\frac{1}{(1 - |\lambda|)^2} - 1\right)\ ,\quad &&\psi_- = \frac{4}{3}\left(\frac{1}{(1 + |\lambda|)^2} - 1\right)\ ,
\end{alignat}
along with their ordering:
\begin{align}\label{phipsiordn}
1 \geq  |\lambda| \geq \frac{1}{2}:&\quad \psi_+ \geq \phi_+ \geq 0 \geq \phi_- \geq \psi_-\ ,\nonumber\\
0 \leq  |\lambda| < \frac{1}{2}:&\quad \phi_+ > \psi_+ \geq 0 \geq \psi_- > \phi_-\ .
\end{align}
Note that this ordering of $\phi_\pm$ and $\psi_\pm$ ensures that the
coefficient of $\sigma^3$ in \eqref{LGpotrel} is manifestly
positive/negative at asymptotically large positive/negative values of
$\sigma$; This ensures that the potential \eqref{LGpotrel} is bounded
from below for all values of $|\lambda|$.

From the expression for the potential above in \eqref{LGpotrel} it is
apparent that an odd power of $m_F$ always appears with a $\lambda$
and similarly, every occurrence of an odd power of $b_4$ is accompanied
by a $\lambda$. Thus, our theory depends only on the combinations
$m_F \sgn(\lambda)$ and $\lambda b_4$.

Our Landau-Ginzburg potential depends on three dimensionful parameters
in addition to the dimensionless coupling $\lambda$.  Of course, all
information about the phase diagram is invariant under changes of
units under which dimensionful parameters transform as
\begin{equation}\label{untr}
 m _F \rightarrow \alpha m_F\ ,\quad b_4 \rightarrow \alpha b_4\ ,\quad
m_B^2 \rightarrow \alpha^2 m_B^2\ ,
\end{equation}
where $\alpha$ is any positive number. In other words the three
dimensional parameter space parametrized by $m_F$, $b_4$ and $m_B^2$
can be foliated by the ellipsoid-like surfaces given by
\begin{equation}\label{spher}
  (m_B^2)^2 + (m_F^2 + (\lambda b_4)^2)^2 = \text{constant}\ .
\end{equation}
for various different positive values of the constant. The `scale
symmetry' \eqref{untr} ensures that we lose no information by studying
our theory at any given positive value of the constant in
\eqref{spher}; in other words the phase diagram of our theory lives on
the surface \eqref{spher} at any convenient value of the
constant. Also note that the $\mc{N} = 2$ superconformal theory itself
lives at the origin of the three dimensional parameter space
$(m_B^2, \lambda b_4, m_F \sgn(\lambda)) = (0,0,0)$.

We will use the following terminology in the rest of this paper.  We
call the part of the `ellipsoid' \eqref{spher} that lies in the region
$m_F\sgn(\lambda)>0$ as the northern hemisphere of the ellipsoid. In a
similar manner, we will call the part of the ellipsoid that lies in the
region $m_F\sgn(\lambda)<0$ the southern hemisphere. Finally, the
intersection of the plane $m_F=0$ and the ellipsoid is a curve that we
call the equator of the ellipsoid.

All information about the phase diagram of the theory in its northern
hemisphere can be obtained (using the scaling \eqref{untr}) from the
study of the theory at any fixed positive value of
$m_F\sgn(\lambda)$. Similarly all information about the theory on its
southern hemisphere can be obtained by studying the theory at any
fixed negative value of $m_F\sgn(\lambda)$. Finally, the information
about the theory along the equator can clearly be obtained by studying
the theory at $m_F=0$. In the rest of this section we will separately
study the theory on the `horizontal' sections $m_F=0$, at
$m_F = |\mu| {\rm sgn}(\lambda)$ and at
$m_F = -|\mu| {\rm sgn}(\lambda)$ and then finally put together the
phase diagram by patching these results together (in order to achieve
this patching in a smooth way we also separately study a neighbourhood
of the equator).  We begin our analysis with the simplest case, namely
the study of the equator.

\subsection{The equator: \texorpdfstring{$m_F = 0$}{mF=0}}\label{mfzero}
In this case the sign
$\varepsilon = \sgn(m_F + 2\lambda \sigma) \sgn(\lambda)$ that
decides the fermionic phase simplifies:
\begin{equation}\label{fersign}
  \varepsilon =  \sgn(\lambda \sigma) \sgn(\lambda) = \sgn(\sigma)\ .
\end{equation}
This implies that the fermionic phase is tied to the bosonic phase:
when the boson is in the Higgsed / unHiggsed ($\mp$) phase, the
fermion is in the $\pm$ phase. In other words, when the boson
undergoes a phase transition, the fermion undergoes a phase transition
as well. Thus, the effective potential \eqref{LGpotrel} at $m_F = 0$
accesses only two phases viz.~$(-,+)$ and $(+,-)$. The explicit
expression for the Landau-Ginzburg potential \eqref{LGpotrel}
collapses to a form very similar to that of the regular boson theory
(Equation \eqref{RBlgpot} in Appendix \ref{critferapp}):
\begin{equation}\label{Umfzero}
  U(\sigma) =  \frac{N}{2 \pi}\left\{\arraycolsep=1.4pt\def\arraystretch{2.2}\begin{array}{cl} 
                                                                               (\psi_- - \phi_+) \lambda^2 { \sigma}^3 + 2 \lambda b_4 { \sigma}^2  +  m_B^2 { \sigma} & \quad\text{for $\sigma < 0$ i.e.~the $(-,+)$ phase}\ , \\ ( \psi_+ - \phi_-)\lambda^2 { \sigma}^3 +  2 \lambda{ b}_4  { \sigma}^2 + { m}_B^2 { \sigma}  & \quad\text{for $\sigma > 0$ i.e.~the $(+,-)$ phase}\ . \end{array}\right.
\end{equation}
\begin{figure}
  \centering
  \scalebox{0.7}{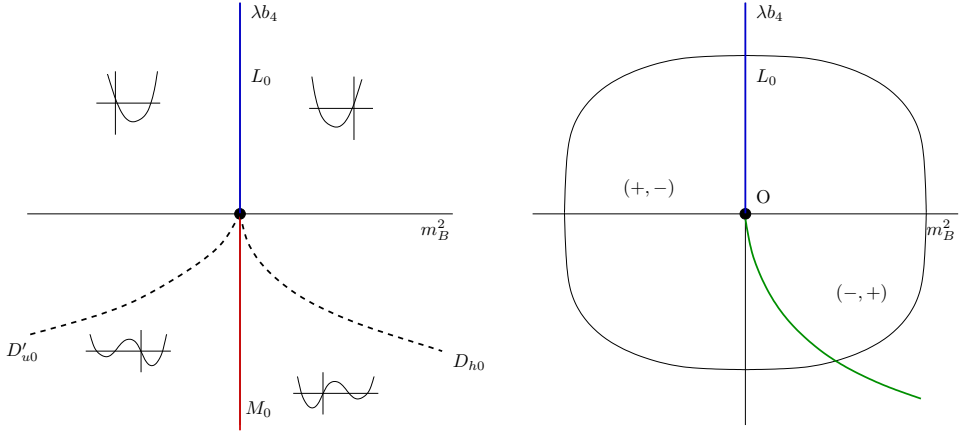}
  \caption{The $m_F =0$ section of the paraboloids in \eqref{parazero}
    is given in Figure (a). The inset plots are the profiles of the LG
    potential \eqref{Umfzero} with the horizontal axis being the
    $\sigma$ axis and the vertical axis being the value of the
    potential. Across the blue line the minimum of the potential
    passes through $\sigma=0$. Along the red line a local maximum of
    the LG potential passes through $\sigma=0$. The dashed lines
    correspond to either the appearance or disappearance of new
    extrema in the potential. The phase structure is given in Figure
    (b) with the blue line denoting a second order Higgsing phase
    transition and the green line denoting a numerically-determined
    first order phase transition. The ellipse (eq.~\eqref{mf0ellipse})
    is the intersection of the two dimensional plane with the
    ellipsoid \eqref{spher}. The first order line lies to the right of
    the diagram - as depicted in the figure above - when
    $|\lambda|<\frac{1}{2}$ but moves to the left of the figure for
    $|\lambda|>\frac{1}{2}$. It runs exactly down the negative
    $\lambda b_4$ axis at $|\lambda|=\frac{1}{2}$.}
  \label{susymfzero}
\end{figure}
We see that the above potential is stable (following the inequalities
in \eqref{phipsiordn}) and hence the analysis can be borrowed from the
regular boson analysis for the case $\phi_- < x_6^B < \phi_+$ in
\cite{Dey:2018ykx} or in Section \ref{RBphase} of Appendix
\ref{critferapp} in this paper. Note that the value of $x_6^B$ under
which \eqref{Umfzero} reduces to the regular boson theory is $\psi_-$
for $\sigma < 0$ and $\psi_+$ for $\sigma > 0$, as opposed to a single
value of $x_6^B$ for both branches in the actual regular boson
theory. For this reason the analysis in the present case has minor
quantitative differences compared to that of the regular boson. We
present the analytic features of the potential in Figure
\ref{susymfzero}(a). The main features are the following
half-parabolas:
\begin{align}\label{parazero}
  L_0:&\quad m_B^2 = 0\ ,\quad \lambda b_4 > 0\ ,\nonumber\\
  M_0:&\quad m_B^2 = 0\ ,\quad \lambda b_4 < 0\ , \nonumber\\
  D_{h0}:&\quad 16(\lambda b_4)^2 - 12\lambda^2 (\psi_+ - \phi_-) m_B^2 = 0\ ,\quad \lambda b_4 < 0\ ,\nonumber\\
  D'_{u0}:&\quad 16(\lambda b_4)^2 - 12\lambda^2 (\psi_- - \phi_+) m_B^2 = 0\ ,\quad \lambda b_4 < 0\ .
\end{align}
The half-parabola $L_0$ is significant because a minimum of the
potential goes from the range $\sigma > 0$ to the range $\sigma < 0$
and hence goes from the $(+,-)$ phase to the $(-,+)$ phase (and vice
versa). Thus, $L_0$ corresponds to a second order phase transition
between the above two phases. Similarly, the half-line $M_0$
corresponds to a maximum of the potential crossing $\sigma = 0$. The
half-parabolas $D'_{u0}$ and $D_{h0}$ correspond to the
(dis)appearance of new extrema of the potential as one crosses
them. For instance, above the $D'_{u0}$ half-parabola, the potential
has a single minimum, while below it there are two minima and one
maximum.

The phase structure is obtained by determining the global minimum in
different regions of parameter space $(m_B^2, \lambda b_4)$. There are
two competing minima in different phases in the region below $D'_{u0}$
and $D_{h0}$ in Figure \ref{susymfzero}. The dominant minimum has to
be numerically determined in this region. The line across which the
minimum in one phase becomes dominant over the other defines a
first-order phase boundary between the two phases. We present the
second order and the first order phase transition lines in Figure
\ref{susymfzero}(b).

Above, we determined the phase structure for $m_F = 0$ as a function
of two dimensionful parameters $m_B^2$ and $\lambda b_4$. As discussed
at the beginning of this section around equation \eqref{spher}, the
actual phase diagram is two dimensional and lives on the ellipsoid
\eqref{spher}. The intersection of the $m_F = 0$ slice with this
ellipsoid is given by the equatorial `ellipse'
\begin{equation}\label{mf0ellipse}
  (\lambda b_4)^4 + (m_B^2)^2 = \text{constant}\ ,
\end{equation}
as depicted in Figure \ref{susymfzero}(b). The blue line and the green
line in Figure \ref{susymfzero} intersect this ellipse at one point
each, corresponding to a second order and a first order phase
transition respectively on the equator of the phase diagram ellipsoid
\eqref{spher}.

\subsubsection{The CB-RF conformal theory}\label{mfzerocbrf}

We commented earlier around equation \eqref{fersign} in this
subsection that whenever a boson undergoes a phase transition, the
fermion undergoes a phase transition as well. Recall that the minimum
of the potential goes to zero on the half-line
\begin{equation}
  \lambda b_4 > 0\ ,\quad m_B^2 = 0\ ,
\end{equation}
and hence the gaps $c_B$, $c_F$ in \eqref{cbcf} go to zero as
well. Thus, this corresponds to a phase transition for both the boson
and the fermion. The above half-line intersects the ellipse
\eqref{mf0ellipse} at one point. The conformal dynamics at this point
is then presumably governed by a theory of critical bosons and regular
fermions coupled to a $SU(N)_k$ Chern-Simons theory.

We give more evidence for this statement in the sections below when we
study the $\mc{N} = 2$ theory at non-zero $m_F$ which, at small values
of $m_F$, should correspond to particular massive deformations of this
CB-RF conformal theory. We also obtain precisely this CB-RF point as a
scaling limit of the general theory \eqref{generalaction} in Section
\eqref{cbrflim} of Appendix \ref{scaling}.

\subsection{The northern hemisphere: \texorpdfstring{$\textnormal{sgn}(m_F) = \textnormal{sgn}(\lambda)$}{sgn(mF)=sgn(lambda)}}\label{mfplus}
\begin{figure}
  \centering
  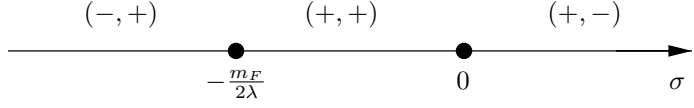
  \caption{Regions of validity for different branches of the potential
    for $\sgn(m_F)\sgn(\lambda) = +1$.}
  \label{phaseplus}
\end{figure}

We next consider the case $\sgn(m_F) \sgn(\lambda) = +1$. The regions
of validity for the various branches of the potential are in Figure
\ref{phaseplus}.
The appropriate expressions for the potential are
\begin{equation}\label{Uplus}
 U(\sigma) = \frac{N}{2\pi}\left\{\def\arraystretch{2.2}\begin{array}{lr} \def\arraystretch{1.4} \begin{array}{r} \left({ m}_B^2 + \frac{3}{4}\psi_-m _F^2\right) \sigma + \left( { b}_4 + \tfrac{3}{4} \psi_- m_F\right) 2\lambda\sigma^2\qquad \\  + (\psi_- - \phi_+) \lambda^2\sigma^3 +  \frac{\psi_- - \psi_+}{8\lambda}m_F^3 \end{array}\ , & \sigma < -\frac{m_F}{2\lambda}\ ,\\
    \def\arraystretch{1.4} \begin{array}{r} \left({ m}_B^2 + \frac{3}{4}\psi_+ m_F^2\right) \sigma + \left({ b}_4 + \frac{3}{4}\psi_+ m_F\right) 2\lambda\sigma^2\qquad \\ + (\psi_+ - \phi_+) \lambda^2\sigma^3\end{array}\ , & -\frac{m_F}{2\lambda} < \sigma < 0\ ,\\
    \def\arraystretch{1.4} \begin{array}{r}    \left({ m}_B^2 + \frac{3}{4}\psi_+ m_F^2\right) \sigma + \left({ b}_4 + \frac{3}{4}\psi_+ m_F\right) 2\lambda\sigma^2\qquad \\ + (\psi_+ - \phi_-) \lambda^2\sigma^3\end{array}\ , & \sigma > 0\ .\end{array}\right.
\end{equation}
We have subtracted an overall constant
$-\frac{m_F(b_4 m_F - m_B^2)}{4\lambda} +
\frac{\psi_+}{8\lambda}m_F^3$ from all the expressions above.

It helps to consider the interfaces $\sigma = 0$ and
$\sigma = -\frac{m_F}{2\lambda}$ separately. We write the potential in
a coordinate that is zero at the corresponding interface:
\begin{equation}\label{Usigma}
  {U}(\sigma) = \frac{N}{2\pi}\left\{\def\arraystretch{2.2}\begin{array}{lr}  \def\arraystretch{1.4} \begin{array}{r} \left({ m}_B^2 + \frac{3}{4}\psi_+ m_F^2\right) \sigma + \left({ b}_4 + \frac{3}{4}\psi_+ m_F\right)2\lambda\sigma^2\qquad\\ + (\psi_+ - \phi_+) \lambda^2\sigma^3\end{array}\ , & -\frac{m_F}{2\lambda} < \sigma < 0\ ,\\
  \def\arraystretch{1.4} \begin{array}{r} \left({ m}_B^2 + \frac{3}{4}\psi_+ m_F^2\right) \sigma + \left({ b}_4 + \frac{3}{4}\psi_+m_F\right)2\lambda\sigma^2\qquad \\ + (\psi_+ - \phi_-) \lambda^2\sigma^3\end{array}\ , & \sigma > 0\ .\end{array}\right.
\end{equation}
\begin{equation}\label{Usigmapr}
  {U}'(\sigma') = \frac{N}{2\pi}\left\{\def\arraystretch{2.2}\begin{array}{lr}  \def\arraystretch{1.5}\begin{array}{r} \left({ m}_B^2 -2 { b}_4 m_F - \frac{3}{4}\phi_+ m_F^2\right) \sigma'   + \left( { b}_4 + \tfrac{3}{4} \phi_+ m_F\right)2\lambda \sigma'{}^2  \qquad\\ + (\psi_- - \phi_+) \lambda^2\sigma'{}^3\end{array} \ , & \sigma' < 0\ ,\\
\def\arraystretch{1.5}\begin{array}{r} \left({ m}_B^2 -2 { b}_4 m_F - \frac{3}{4}\phi_+ m_F^2\right) \sigma'  + \left( { b}_4 + \tfrac{3}{4} \phi_+ m_F\right) 2\lambda\sigma'{}^2  \qquad \\ + (\psi_+ - \phi_+) \lambda^2\sigma'{}^3\end{array}\ , & 0 < \sigma' < \frac{m_F}{2\lambda}\ .\end{array}\right.
\end{equation}
where ${ \sigma'} = { \sigma} + \frac{m_F}{2\lambda}$. There is an
additional common constant
$$\tfrac{1}{8\lambda}(4 { b}_4 m_F^2 - 4 { m}_B^2 m_F + m_F^3(\phi_+ - \psi_+))$$
in both the lines above in \eqref{Usigmapr}. This is the relative
shift compared to the potential in terms of $\sigma$ above in
\eqref{Usigma}. To re-emphasize, in the region where $U(\sigma)$ and
$U'(\sigma')$ are both valid (i.e. for
$-\frac{m_F}{2 \lambda}< \sigma < 0$) the two potentials are related
via
\begin{equation} \label{reluup}
U(\sigma)=U'(\sigma + \tfrac{m_F}{2 \lambda}) + 
\tfrac{1}{8\lambda}(4 { b}_4 m_F^2 - 4 { m}_B^2 m_F + m_F^3(\phi_+ - \psi_+))
\end{equation} 
Our strategy for the rest of this subsection is the following.  We
will work out the phase diagram separately for the potential
$U(\sigma)$ (pretending that it was valid at all values of $\sigma$)
and for $U'(\sigma')$ (again pretending it was valid for all values of
$\sigma'$) and then finally patch these two results together in order
to get the actual phase diagram of the system - this time taking care
to use the results for $U$ and $U'$ only within their domains of
validity i.e.~$-\tfrac{m_F}{2\lambda} < \sigma$ and
$\sigma' < \tfrac{m_F}{2\lambda}$ respectively.

The analysis for the two sets of potentials in \eqref{Usigma} and
\eqref{Usigmapr} proceeds in the same way as that of the regular boson
theory (see Appendix \ref{critferapp} of this paper). The potential
$U(\sigma)$ in \eqref{Usigma} is stable when
$\phi_- < \psi_+ < \phi_+$. From the ordering given in
\eqref{phipsiordn}, it is clear that this is case when
$|\lambda| < \frac{1}{2}$.\footnote{When $|\lambda|>\frac{1}{2}$ the
  potential $U(\sigma)$ is unbounded from below at negative
  $\sigma$. Of course this is physically insignificant, as $U(\sigma)$
  correctly captures the quantum effective potential only for
  $\sigma> - \frac{m_F}{2 \lambda}. $}

Thus, in this case, the analysis of the regular boson for
$\phi_- < x_6^B < \phi_+$ applies (See Section \ref{RBphase} of
Appendix \ref{critferapp}). There are four semi-infinite surfaces
(half-paraboloids) that are important in describing the profile of the
potential around $\sigma = 0$. These are given by the following
conditions:
\begin{align}\label{Upara}
  L:\ & m_B^2 + \tfrac{3}{4}\psi_+ m_F^2 = 0\ ,\quad \lambda(b_4 + \tfrac{3}{4}\psi_+m_F) > 0\ ,\nonumber\\
  M:\ & m_B^2 + \tfrac{3}{4}\psi_+ m_F^2 = 0\ ,\quad \lambda(b_4 + \tfrac{3}{4}\psi_+m_F) < 0\ ,\nonumber\\
  D_u:\ & 16(\lambda b_4 + \tfrac{3}{4}\psi_+ \lambda m_F)^2 -12 \lambda^2(\psi_+-\phi_+)(m_B^2 + \tfrac{3}{4}\psi_+ m_F^2) = 0\ ,\quad \lambda(b_4 + \tfrac{3}{4}\psi_+m_F) < 0\ ,\nonumber\\
  D_h:\ & 16(\lambda b_4 + \tfrac{3}{4}\psi_+\lambda m_F)^2 -12 \lambda^2(\psi_+-\phi_-)(m_B^2 + \tfrac{3}{4}\psi_+ m_F^2) = 0\ ,\quad\lambda(b_4 + \tfrac{3}{4}\psi_+m_F) < 0\ .
\end{align}
Our notation for these curves parallels that used in the regular boson
analysis in Section \ref{RBphase} in Appendix \ref{critferapp}. The
significance of these surfaces is the following. Across the surface
$L$ the minimum of the effective potential \eqref{Usigma} goes from
the range $\sigma<0$ to the range $\sigma > 0$ or vice-versa. This
signals a second order phase transition. Similarly, as one crosses the
surface $M$ a local maximum of the effective potential \eqref{Usigma}
crosses $\sigma=0$. The surfaces $D_u$ and $D_h$ are such that when
one crosses them the effective potential \eqref{Usigma} develops or
loses a maximum-minimum pair.

In this same range of $|\lambda| < \tfrac{1}{2}$, the potential
$U'(\sigma')$ in \eqref{Usigmapr} is unbounded from below at large
positive $\sigma'$ for $\sigma' > 0$.\footnote{This is physically
  inconsequential since the range of validity of the unbounded branch
  of the potential is finite ($0 < \sigma' < \frac{m_F}{2\lambda}$)
  and hence the instability is not actually encountered.} Thus, the
$x_6^B < \phi_-$ analysis of the regular boson theory is applicable
here. Again, there are four half-paraboloids that are important for
the profile of the potential around $\sigma' = 0$
i.e.~$\sigma = -\frac{m_F}{2\lambda}$:
\begin{align}\label{Uprpara}
  L':\ & m_B^2 - 2 \lambda b_4 \frac{m_F}{\lambda} -\tfrac{3}{4}\phi_+ m_F^2 = 0\ , \quad \lambda(b_4 + \tfrac{3}{4}\phi_+ m_F) > 0\ ,\nonumber\\
  M':\ & m_B^2 - 2 \lambda b_4 \frac{m_F}{\lambda} -\tfrac{3}{4}\phi_+ m_F^2 = 0\ , \quad \lambda(b_4 + \tfrac{3}{4}\phi_+ m_F) < 0\ ,\nonumber\\
  D'_u:\ & \left\{\begin{array}{l} 16(\lambda b_4 + \tfrac{3}{4}\phi_+ \lambda m_F)^2 -12 \lambda^2 (\psi_--\phi_+)(m_B^2 -2b_4 m_F - \tfrac{3}{4}\phi_+ m_F^2) = 0\ , \\ \lambda(b_4 + \tfrac{3}{4}\phi_+m_F) < 0\ ,\end{array}\right.\nonumber\\
  D'_h:\ & \left\{\begin{array}{l} 16(\lambda b_4 + \tfrac{3}{4}\phi_+ \lambda m_F)^2 -12\lambda^2 (\psi_+-\phi_+)(m_B^2 - 2b_4 m_F - \tfrac{3}{4}\phi_+ m_F^2) = 0\, \\ \lambda(b_4 + \tfrac{3}{4}\phi_+m_F) > 0\ .\end{array}\right.
\end{align}
As above, the surfaces $L'$ and $M'$ denote the lines across which the
minimum or a local maximum, respectively, of $U'(\sigma')$ crosses
$\sigma'=0$. The surfaces $D_u'$ and $D'_h$ denote locations of
nucleation of new extrema of $U'(\sigma')$.  \footnote{In this case
  these two lines separate the region in which the potential
  $U'(\sigma')$ has no extrema from the region in which it has one
  local maximum and one local minimum; see Section \ref{RBphase} of
  Appendix \ref{critferapp} for details.}
 
\begin{figure}
  \centering
  \scalebox{0.7}{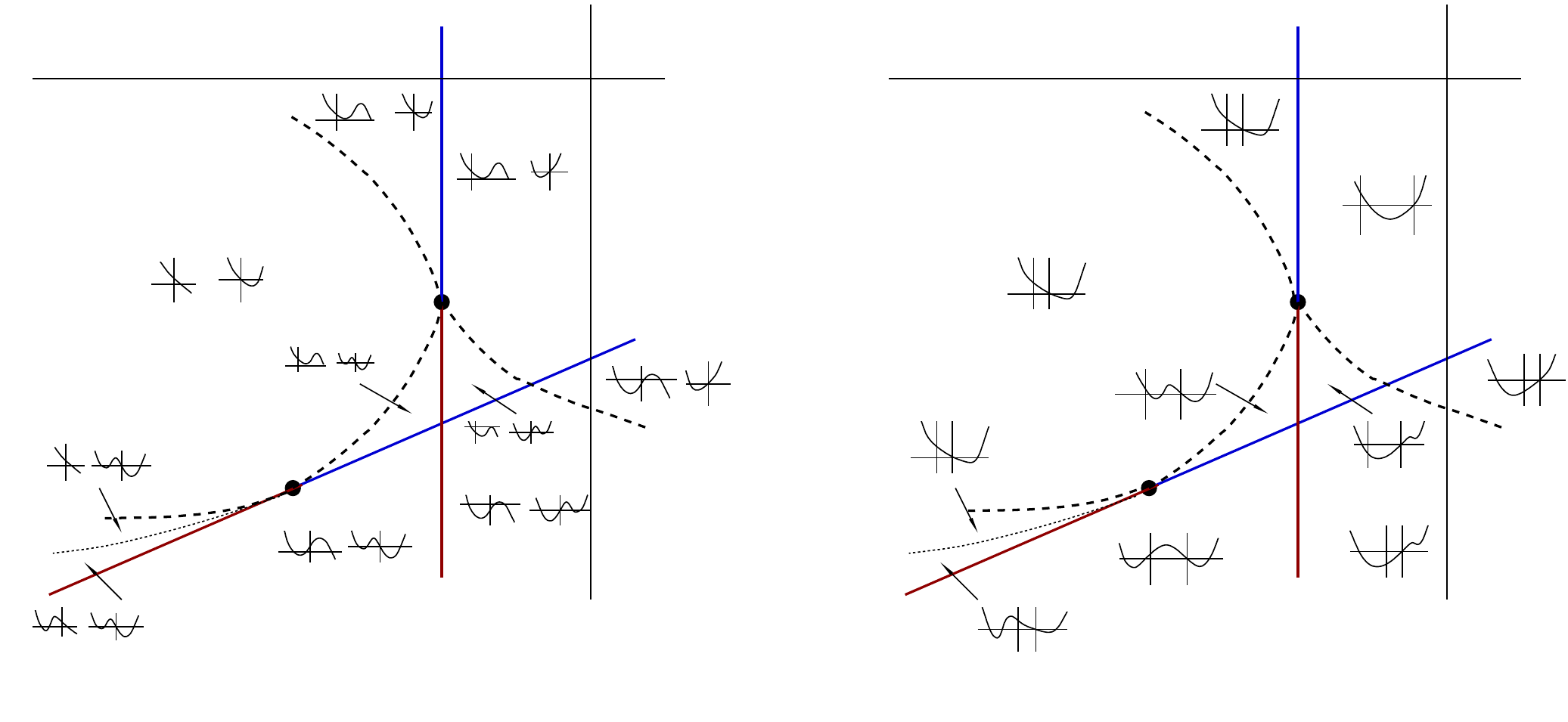}
  \caption{The $m_F =|\mu|\, \sgn(\lambda)$ section of the paraboloids
    in \eqref{Upara} and \eqref{Uprpara} for
    $|\lambda| < \tfrac{1}{2}$. Each of the inset pair-of-plots in
    Figure (a) correspond to the profiles of the potential
    $U'(\sigma')$ (the plot on the left) and $U(\sigma)$ (the plot on
    the right). The vertical lines in each of these plots correspond
    to the points $\sigma'=0$ and $\sigma = 0$ respectively. The inset
    plots in Figure (b) are the profiles of the potential
    \eqref{Uplus} and the two vertical lines in these plots correspond
    to $\sigma = -m_F / 2\lambda$ and $\sigma = 0$ respectively. The
    blue lines correspond to a minimum of the local potential crossing
    the vertical axis while the red lines correspond to a maximum
    crossing the vertical axis. The dashed and dotted lines correspond
    to either the appearance or disappearance of new extrema in the
    potential. }
  \label{phaseplotI}
\end{figure}

Recall that $D_u$ and $D'_h$ both correspond to surfaces across which
the potentials $U(\sigma)$ and $U'(\sigma')$ develop new extrema. It
turns out that the special value of $\sigma$ at which these new
extrema are nucleated lies in the region
$-\frac{m_F}{2 \lambda} < \sigma< 0$ where the potentials $U(\sigma)$
and $U'(\sigma')$ are both valid (see Figure \ref{phaseplotI}(a)).
For this reason $D_u$ and $D'_h$ define the exact same paraboloid.

\begin{figure}
  \centering
  \scalebox{0.7}{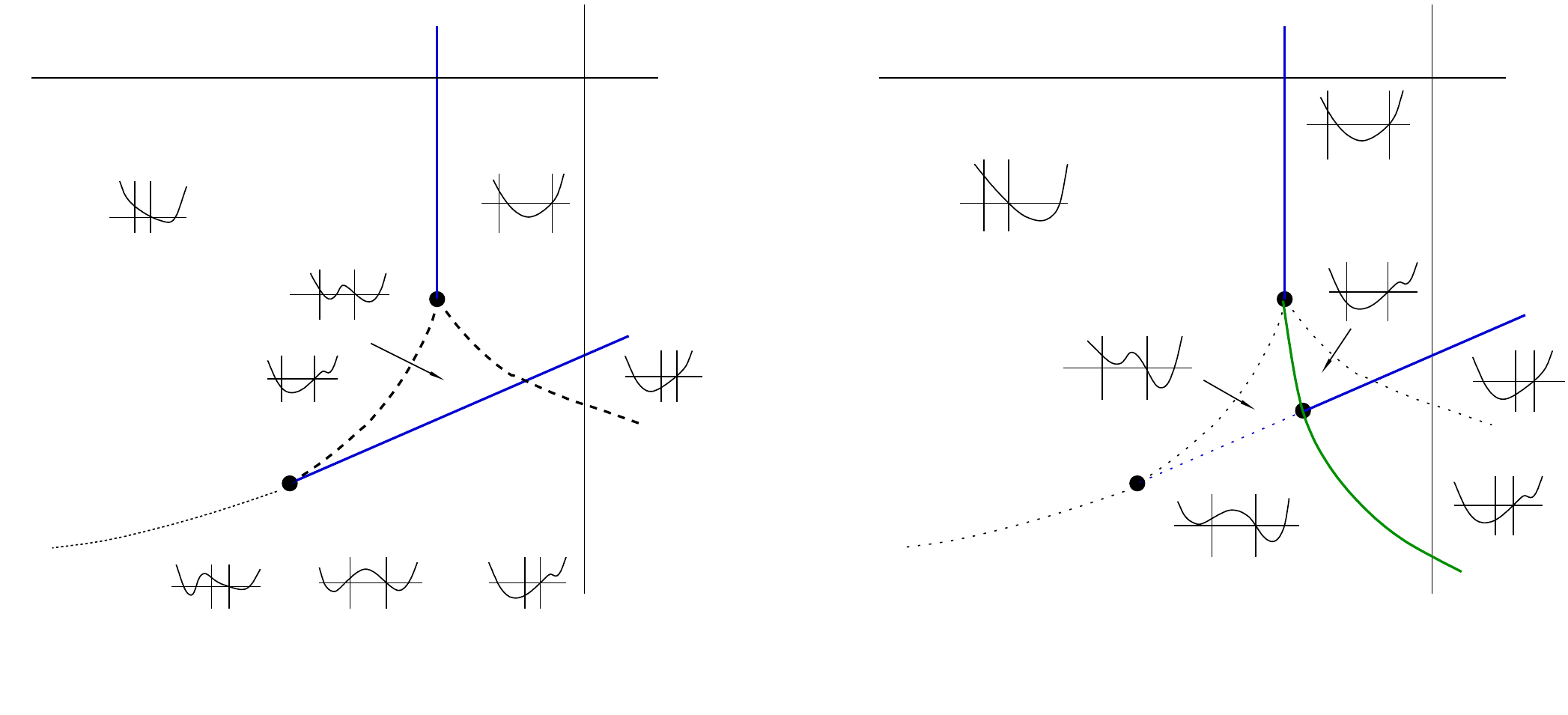}
  \caption{Deducing the phase structure for the
    $m_F = |\mu|\, \sgn(\lambda)$ section for
    $|\lambda| < \tfrac{1}{2}$. The inset plots are the profiles of
    the potential \eqref{Uplus}. The two vertical lines in each plot
    correspond to $\sigma = -m_F / 2\lambda$ and $\sigma = 0$ and
    demarcate the regions of validity of the $(-,+)$, $(+,+)$ and
    $(+,-)$ branches of the potential \eqref{Uplus} (cf.~Figure
    \ref{phaseplus}). The green line in Figure (b) corresponds to a
    numerically-determined first order transition line. It meets the
    second order transition line $L'$ emanating from the point $\CF+$
    at the point $\Op+$.}
  \label{phasediagI}
\end{figure}

As we have explained in the introduction to this section, all
information of the phase diagram in the upper hemisphere (i.e. when
$m_F \lambda>0$) can be extracted (by scaling) from the free energy at
a fixed positive value of $m_F \lambda$, let us say on the `horizontal
section'
\begin{equation}\label{mfsec}
  m_F = |\mu|\, \sgn(\lambda)\ .
\end{equation}
We make this choice in what follows. All our final results can in fact
be recast as functions of the two variables
\begin{equation}\label{homog}
  \frac{\lambda b_4}{|\mu|} \equiv \frac{\lambda b_4}{ m_F \sgn(\lambda)}\ ,\quad \frac{m_B^2}{\mu^2} \equiv \frac{m_B^2}{m_F^2}\ ,
\end{equation}
which can be thought of as a set of `coordinates' on the upper
hemisphere of the phase diagram. We next give a detailed description
of the analytic features and the phase structure of the potential
$U(\sigma)$ for $\sgn(m_F) = \sgn(\lambda)$ in Figures
\ref{phaseplotI} and \ref{phasediagI}.
We first give a brief explanation of the features of Figure
\ref{phaseplotI}. The intersections of the paraboloids in
\eqref{Upara} , \eqref{Uprpara} with the $m_F = |\mu| \sgn(\lambda)$
slice yield parabolas and we use the same notation to describe these
as their parent paraboloids.
\begin{enumerate}
\item We depict the profiles of the potentials $U(\sigma)$
  \eqref{Usigma} and $U'(\sigma')$ \eqref{Usigmapr} in the
  $(m_B^2, \lambda b_4)$ space for each of the regions demarcated by
  the curves $L$, $M$, $L'$, $M'$, $D_u$, $D_h$, $D'_u$, $D'_h$ in
  Figure \ref{phaseplotI}(a). In each region we display a pair of
  small plots: the plot on the left is the potential $U'(\sigma')$ in
  the neighbourhood of $\sigma' = 0$
  ($\sigma = -\tfrac{m_F}{2\lambda}$) and the plot on the right is the
  potential $U(\sigma)$ in the neighbourhood of $\sigma = 0$. All
  plots are based on the regular boson analysis in \cite{Dey:2018ykx}
  (or in Appendix \ref{critferapp} of this paper). Note that as one
  crosses the various lines ($L$, $M$, etc.), the plots of the
  potential \eqref{Usigma} and \eqref{Usigmapr} qualitatively change.
  The points $\RB+$ and $\CF+$ at which the second order lines $L$ and
  $L'$ end are precisely the location of the RB and CF scaling limits
  given in \eqref{rbpoint} and \eqref{limita1g1} respectively.
  
\item Since the pair of graphs in Figure \ref{phaseplotI}(a) belong to
  the same continuous potential\footnote{In fact, as can be seen from
    \eqref{Uplus}, the potential is twice differentiable, with the
    third derivative being discontinuous. Technically, the potential
    is a $C^{(2)}$ function which is piecewise smooth.} given in
  \eqref{Uplus}, we have to patch the two graphs in the region between
  $\sigma = -\frac{m_F}{2\lambda}$ and $\sigma = 0$ such that the
  continuity is maintained. The final profile of the potential
  \eqref{Uplus} in each region is given in Figure \ref{phaseplotI}(b).
  As we can see from Figure \ref{phaseplotI}(b), the curves $D_h'$
  (above RB$+$) and $D_u$ (below CF$+$) are superfluous since the
  nature of the potential remains the same when one crosses
  them. Figure \ref{phaseplotI}(b) contains complete analytic
  information about the Landau-Ginzburg potential for
  $\sgn(m_F) \sgn(\lambda) = +1$.
\end{enumerate}

We are also interested in the phase structure of the potential
\eqref{Uplus}. In other words, we focus on the minima of the potential
in various regions of the parameter space $(m_B^2, \lambda b_4)$ at
fixed $m_F = |\mu|\sgn(\lambda)$. This is summarised in Figure
\ref{phasediagI}.
\begin{enumerate}
\item One needs to study the profiles of the potential in Figure
  \ref{phaseplotI}(b) in more detail, possibly making use of numerical
  methods. For instance, in the region below the dotted and dashed
  curves in Figure \ref{phasediagI}(a), there are two competing
  minima. There is a first order phase boundary that separates the
  regions where one minimum dominates over the other. This boundary
  has to be determined numerically by computing the two local minimum
  values of the potential and the boundary is located at the parameter
  values where these two minimum values are equal. We give a schematic
  depiction of this first order line as the green line in Figure
  \ref{phasediagI}(b). In Appendix \ref{firstorder}, we give a
  detailed description of the numerics along with the required plots.

\item The dynamics on the second order transition line $L$ in Figure
  \ref{phasediagI} is governed by the conformal critical boson theory
  $\CB+$ and terminates at the point $\RB+$ which is governed by a
  theory of conformal regular bosons i.e.~free bosons (the $+$
  indicates that fermions are gapped in the neighbourhood of the line
  and are in the $+$ phase). This second order phase transition
  separates the $(+,+)$ and the $(+,-)$ phases. The phase transition
  continues beyond $\RB+$ but now switches to being a first order
  transition between the phases $(+,+)$ and $(+,-)$ till the point
  $\Op+$. Beyond the point $\Op+$, the phase transition is still first
  order but now separates the $(-,+)$ and the $(+,-)$ phases.
  
  Similarly, the line $L'$ in Figure \ref{phasediagI} corresponds to
  the regular fermion theory $\RF+$ and it ends at $\Op+$ on the first
  order line that emanates out of $\RB+$ (the $+$ indicates that the
  gapped boson is in the $+$ phase in the neighbourhood of this
  line). This second order phase transition separates the $(+,+)$ and
  the $(-,+)$ phases. Till the point $\Op+$ the dominant minimum
  undergoes the second order phase transition. At this point $\Op+$,
  it switches to being a subdominant minimum and hence becomes
  unimportant as far as the phase structure is concerned.

\item Note that if one sets $m_F = 0$ in the surfaces $L$ and $L'$ in
  \eqref{Upara} and \eqref{Uprpara}, one gets back the straight line
  $L_0$ in the case $m_F = 0$ in \eqref{parazero} and in Figure
  \ref{susymfzero}. Thus, the $\CB+$ and $\RF+$ paraboloids intersect
  with each other along the line $L_0$ on the $m_F = 0$
  section. Recall that this line $L_0$ describes the CB-RF conformal
  theory. In Section \ref{mfall}, we map the lines $L$ and $L'$ onto
  the ellipsoid \eqref{spher} and we shall see that they converge to
  the CB-RF point on equator of the ellipsoid that was described in
  Section \ref{mfzerocbrf}.
\end{enumerate}

\subsection{The southern hemisphere:
  \texorpdfstring{$\textnormal{sgn}(m_F) = -\textnormal{sgn}(\lambda)$}{sgn(mF)=-sgn(lambda)}}\label{mfminus}

\begin{figure}
  \centering
  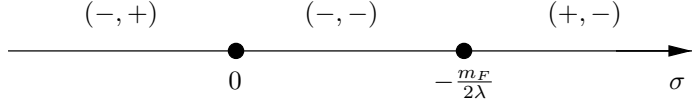
  \caption{Regions of validity for different branches of the potential
    for $\sgn(m_F)\sgn(\lambda) = -1$.}
  \label{phaseminus}
\end{figure}

This choice of $\sgn(m_F)$ accesses the $(-,+)$, $(-,-)$ and the
$(+,-)$ branches of the Landau-Ginzburg potential. The regions of
validity for the various branches on the $\sigma$ axis are displayed
in Figure \ref{phaseminus}.
The corresponding expressions for the potential in the various
branches are
\begin{equation}\label{Uminus}
 {\tl U}(\sigma) = \frac{N}{2\pi}\left\{\def\arraystretch{2.2}\begin{array}{lr}  \def\arraystretch{1.4} \begin{array}{r}  \left({ m}_B^2 + \frac{3}{4}\psi_- m_F^2\right) \sigma + \left( { b}_4 + \tfrac{3}{4} \psi_- m_F\right) 2\lambda\sigma^2\qquad \\ + (\psi_- - \phi_+) \lambda^2\sigma^3\end{array} , & \sigma < 0\ ,\\
     \def\arraystretch{1.4} \begin{array}{r} \left({ m}_B^2 + \frac{3}{4}\psi_- m_F^2\right) \sigma + \left({ b}_4 + \frac{3}{4}\psi_- m_F\right) 2\lambda\sigma^2\qquad \\ + (\psi_- - \phi_-) \lambda^2\sigma^3\end{array}\ ,\quad & 0 < \sigma < -\frac{m_F}{2\lambda}\ ,\\
        \def\arraystretch{1.4} \begin{array}{r} \left({ m}_B^2 + \frac{3}{4}\psi_+ m_F^2\right) \sigma + \left({ b}_4 + \frac{3}{4}\psi_+ m_F\right) 2\lambda\sigma^2\qquad  \\  + (\psi_+ - \phi_-) \lambda^2\sigma^3 +  \frac{\psi_+ - \psi_-}{8\lambda}m_F^3 \end{array} \ , & \sigma > -\frac{m_F}{2\lambda}\ .\end{array}\right.
\end{equation}
We have subtracted an overall constant
$-\frac{m_F(b_4 m_F - m_B^2)}{4\lambda} +
\frac{\psi_-}{8\lambda}m_F^3$ from all the expressions above for
brevity.

Again, it helps to consider the interfaces $\sigma = 0$ and
$\sigma = -\frac{m_F}{2\lambda}$ separately and write the potential in
an coordinate that is zero at the corresponding interface:
\begin{equation}\label{Usigmam}
  {\tl U}(\sigma) = \frac{N}{2\pi}\left\{\def\arraystretch{2.2}\begin{array}{lr}  \def\arraystretch{1.4} \begin{array}{r} \left({ m}_B^2 + \frac{3}{4}\psi_- m_F^2\right) \sigma + \left({ b}_4 + \frac{3}{4}\psi_- m_F\right)2\lambda\sigma^2\qquad \\ + (\psi_- - \phi_+) \lambda^2\sigma^3\end{array}\ , &  \sigma < 0\ ,\\
   \def\arraystretch{1.4} \begin{array}{r} \left({ m}_B^2 + \frac{3}{4}\psi_- m_F^2\right) \sigma + \left({ b}_4 + \frac{3}{4}\psi_- m_F\right)2\lambda\sigma^2\qquad \\ + (\psi_- - \phi_-) \lambda^2\sigma^3\end{array}\ ,\quad & 0 < \sigma < -\frac{m_F}{2\lambda}\ .\end{array}\right.
\end{equation}
\begin{equation}\label{Usigmaprm}
  {\tl U}'(\sigma') = \frac{N}{2\pi}\left\{\def\arraystretch{2.2}\begin{array}{lr} \def\arraystretch{1.5}\begin{array}{r} \left({ m}_B^2 -2 { b}_4 m_F - \frac{3}{4}\phi_- m_F^2\right) \sigma' + \left( { b}_4 + \tfrac{3}{4} \phi_- m_F\right)2\lambda \sigma'{}^2\qquad  \\ + (\psi_- - \phi_-) \lambda^2\sigma'{}^3 \end{array} \ ,\ & \frac{m_F}{2\lambda} < \sigma' < 0\ ,\\
\def\arraystretch{1.5}\begin{array}{r} \left({ m}_B^2 -2 { b}_4 m_F - \frac{3}{4}\phi_- m_F^2\right) \sigma' + \left( { b}_4 + \tfrac{3}{4} \phi_- m_F\right) 2\lambda\sigma'{}^2\qquad \\ + (\psi_+ - \phi_-) \lambda^2\sigma'{}^3\end{array}\ , & \sigma' > 0\ .\end{array}\right.
\end{equation}
where ${ \sigma'} = { \sigma} + \frac{m_F}{2\lambda}$. There is an
additional common constant
$$\tfrac{1}{8\lambda}(4 { b}_4 m_F^2 - 4 { m}_B^2 m_F + m_F^3(\phi_- - \psi_-))$$
in both the expressions above in \eqref{Usigmaprm}.  This is the
relative shift compared to the potential in terms of $\sigma$ above in
\eqref{Usigmam}.

Again, we can apply the regular boson analysis as in the previous
subsection. We stick to the range $|\lambda| < \tfrac{1}{2}$ as in the
previous cases. In this range of $|\lambda|$, the potential
${\tl U}(\sigma)$ in \eqref{Usigmam} is stable (i.e.~bounded below for
both $\sigma > 0$ and $\sigma < 0$). However, the potential
${\tl U}'(\sigma')$ in \eqref{Usigmaprm} is unbounded below for
$\sigma' < 0$ and bounded below for $\sigma' > 0$, making it
unstable. First, we consider the potential ${\tl U}(\sigma)$ in
\eqref{Usigmam}. As earlier, there are four semi-infinite surfaces
that are important for our analysis:
\begin{align}\label{Uparam}
  {\tl L}:\ & m_B^2 + \tfrac{3}{4}\psi_- m_F^2 = 0\ ,\quad \lambda(b_4 + \tfrac{3}{4}\psi_- m_F) > 0\ ,\nonumber\\
  {\tl M}:\ & m_B^2 + \tfrac{3}{4}\psi_- m_F^2 = 0\ ,\quad \lambda(b_4 + \tfrac{3}{4}\psi_- m_F) < 0\ ,\nonumber\\
  {\tl D}_u:\ & 16(\lambda b_4 + \tfrac{3}{4}\psi_- \lambda m_F)^2 -12 \lambda^2(\psi_- - \phi_+)(m_B^2 + \tfrac{3}{4}\psi_- m_F^2) = 0\ ,\quad \lambda(b_4 + \tfrac{3}{4}\psi_- m_F) < 0\ ,\nonumber\\
  {\tl D}_h:\ & 16(\lambda b_4 + \tfrac{3}{4}\psi_-\lambda m_F)^2 -12 \lambda^2(\psi_- - \phi_-)(m_B^2 + \tfrac{3}{4}\psi_- m_F^2) = 0\ ,\quad\lambda(b_4 + \tfrac{3}{4}\psi_- m_F) < 0\ .
\end{align}
Next, we focus on the potential ${\tl U}'(\sigma')$ in
\eqref{Usigmaprm} which is unbounded from below for $\sigma' <
0$. Thus, the case $x_6^B > \phi_+$ of the regular boson theory
applies (see Section \ref{RBphase} of Appendix
\ref{critferapp}). Again, there are four semi-infinite surfaces that
are important for the profile of the potential around $\sigma' = 0$
i.e.~$\sigma = -\frac{m_F}{2\lambda}$:
\begin{align}\label{Uprparam}
  {\tl L}':\ & m_B^2 - 2 \lambda b_4 \frac{m_F}{\lambda} -\tfrac{3}{4}\phi_- m_F^2 = 0\ , \quad \lambda(b_4 + \tfrac{3}{4}\phi_- m_F) > 0\ ,\nonumber\\
  {\tl M}':\ & m_B^2 - 2 \lambda b_4 \frac{m_F}{\lambda} -\tfrac{3}{4}\phi_- m_F^2 = 0\ , \quad \lambda(b_4 + \tfrac{3}{4}\phi_- m_F) < 0\ ,\nonumber\\
  {\tl D}'_u:\ &\left\{\begin{array}{l} 16(\lambda b_4 + \tfrac{3}{4}\phi_- \lambda m_F)^2 -12 \lambda^2 (\psi_--\phi_-)(m_B^2 -2b_4 m_F - \tfrac{3}{4}\phi_- m_F^2) = 0\ ,\\ \lambda(b_4 + \tfrac{3}{4}\phi_- m_F) < 0\ ,\end{array}\right.\nonumber\\
  {\tl D}'_h:\ &\left\{\begin{array}{l} 16(\lambda b_4 + \tfrac{3}{4}\phi_- \lambda m_F)^2 -12\lambda^2 (\psi_+ - \phi_-)(m_B^2 - 2b_4 m_F - \tfrac{3}{4}\phi_- m_F^2) = 0\ ,\\ \lambda(b_4 + \tfrac{3}{4}\phi_- m_F) > 0\ .\end{array}\right.
\end{align}
Again, note that the equalities in the conditions ${\tl D}_h$ and
${\tl D}'_u$ define the exact same paraboloid (see the two paragraphs
after equation \eqref{Uprpara} in the previous subsection).
\begin{figure}
  \centering
  \scalebox{0.65}{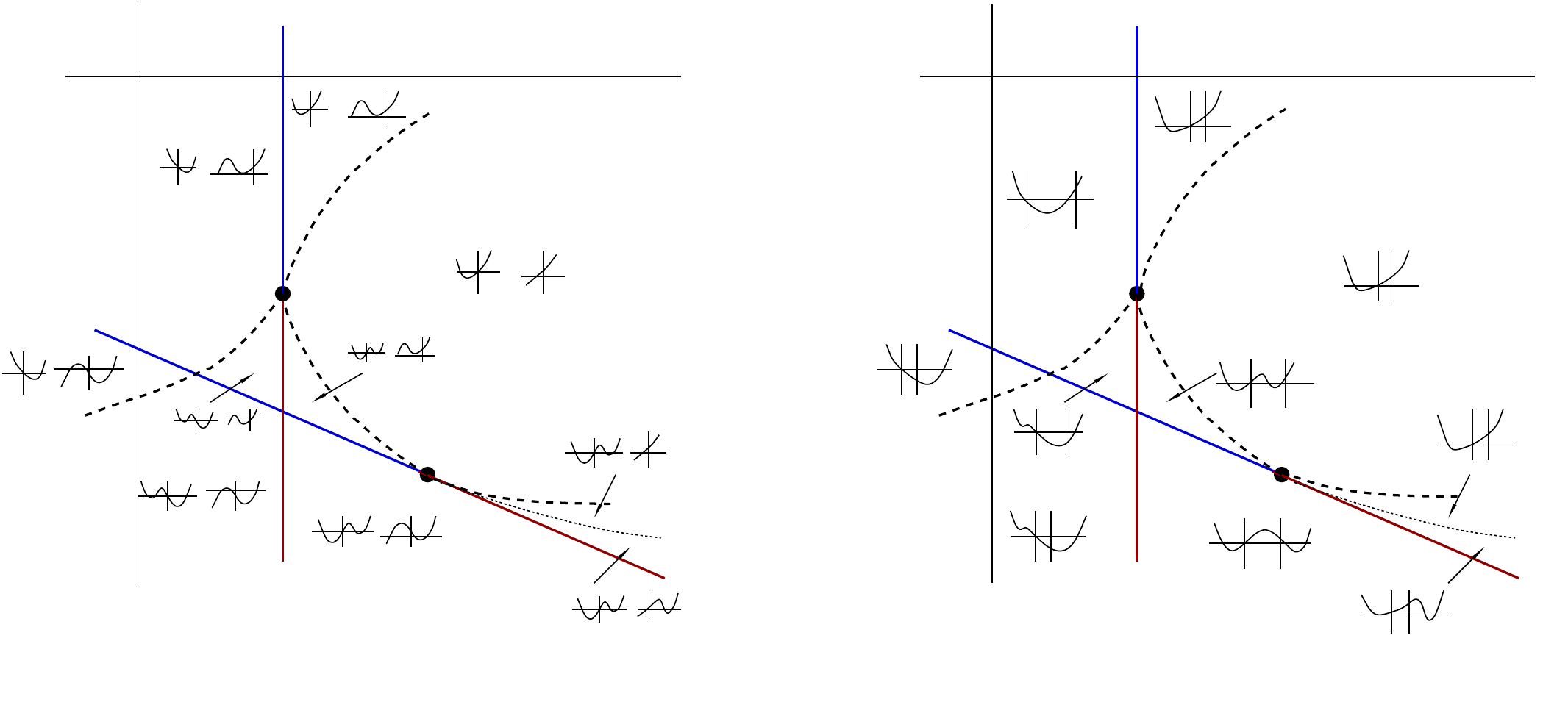}
  \caption{The $m_F =-|\mu|\, \sgn(\lambda)$ section of the surfaces
    in \eqref{Upara} and \eqref{Uprpara} for
    $|\lambda| < \tfrac{1}{2}$. Each of the inset pair-of-plots in
    Figure (a) correspond to the profiles of the potential
    ${\tl U}(\sigma)$ (the plot on the left) and ${\tl U}'(\sigma')$
    (the plot on the right). The vertical lines in each of these plots
    correspond to the points $\sigma=0$ and $\sigma' = 0$
    respectively. The inset plots in Figure (b) are the profiles of
    the potential \eqref{Uminus} and the two vertical lines in these
    plots correspond to $\sigma = 0$ and $\sigma = -m_F / 2\lambda$
    respectively. Blue lines correspond to a minimum of the local
    potential crossing the vertical axis while the red lines
    correspond to a maximum crossing the vertical axis. The dashed and
    dotted lines correspond to either the appearance or disappearance
    of new extrema in the potential. }
  \label{phaseplotII}
\end{figure}
\begin{figure}
  \centering
  \scalebox{0.7}{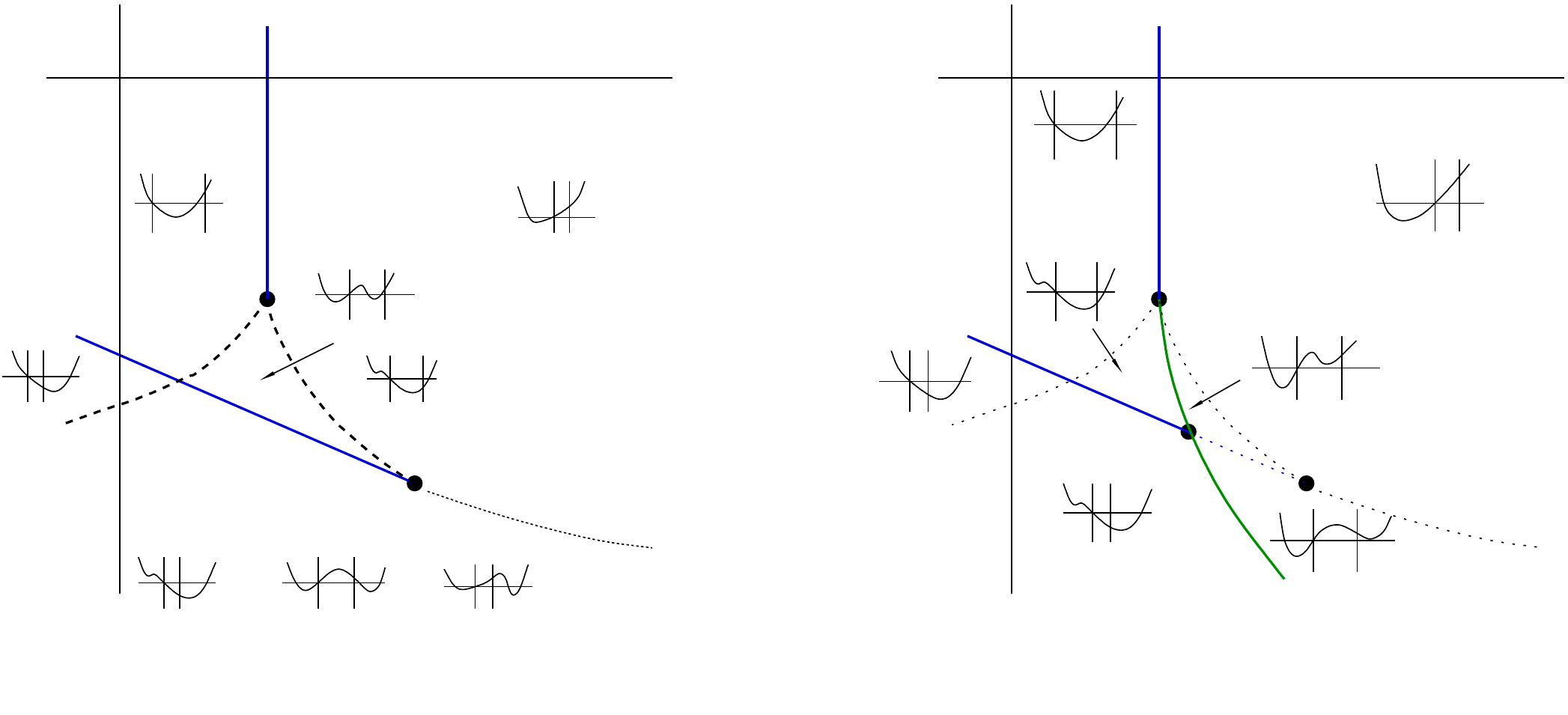}
  \caption{Deducing the phase structure for the
    $m_F =-|\mu|\, \sgn(\lambda)$ section for
    $|\lambda| < \tfrac{1}{2}$.  The inset plots are the profiles of
    the potential \eqref{Uminus}. The two vertical lines in each plot
    correspond to $\sigma = 0$ and $\sigma = -m_F / 2\lambda$
    respectively from left to right and demarcate the regions of
    validity of the $(-,+)$, $(-,-)$ and $(+,-)$ branches of the
    potential \eqref{Uminus} (cf.~Figure \ref{phaseminus}).  The green
    line in Figure (b) corresponds to a numerically-determined first
    order transition line.}
  \label{phasediagII}
\end{figure}
As in the previous subsection, we first analyse the potentials
${\tl U}$ and ${\tl U}'$ separately and then stitch them together to
obtain the profile of the Landau-Ginzburg potential
\eqref{Uminus}. The features of the $m_F = -|\mu|\sgn(\lambda)$
section of the three dimensional parameter space are shown in Figures
\ref{phaseplotII} and \ref{phasediagII}. Briefly, Figure
\ref{phaseplotII}(a) contains the profiles of the potentials
${\tl U}(\sigma)$ and ${\tl U}'(\sigma')$ separately and also displays
the intersection of the half-paraboloids in \eqref{Uparam} and
\eqref{Uprparam} with the $m_F = -|\mu|\sgn(\lambda)$ section. Figure
\ref{phaseplotII}(b) contains the plots of the full potential in all
ranges of $\sigma$ obtained by appropriately stitching together
${\tl U}$ and ${\tl U}'$ as in the previous subsection.

Figure \ref{phasediagII} contains information about the phase
structure for $\sgn(m_F) = -\sgn(\lambda)$. In obtaining Figure
\ref{phasediagII}(a), we focus only on the details of the various
dominant minima in Figure \ref{phaseplotII} and ignore other analytic
features of the potential. The final figure \ref{phasediagII}(b)
contains a schematic depiction of the first-order phase transition
line (the green line) between the relevant phases. The lines $\tl L$
and $\tl L'$ in Figure \ref{phasediagII}(b) correspond to the critical
boson theory $\CB-$ and the regular fermion theory $\RF-$
respectively. These lines terminate on the $\RB-$ and the $\CF-$
theories respectively (the $-$ sign indicates that the gapped fermion
/ boson is the $-$ phase). When mapped to the ellipsoid \eqref{spher},
these lines are expected to converge to the CB-RF point on the equator
as we shall demonstrate in the next subsection.

\subsection{Putting it together}\label{mfall}

\begin{figure}
  \centering
  \scalebox{0.6}{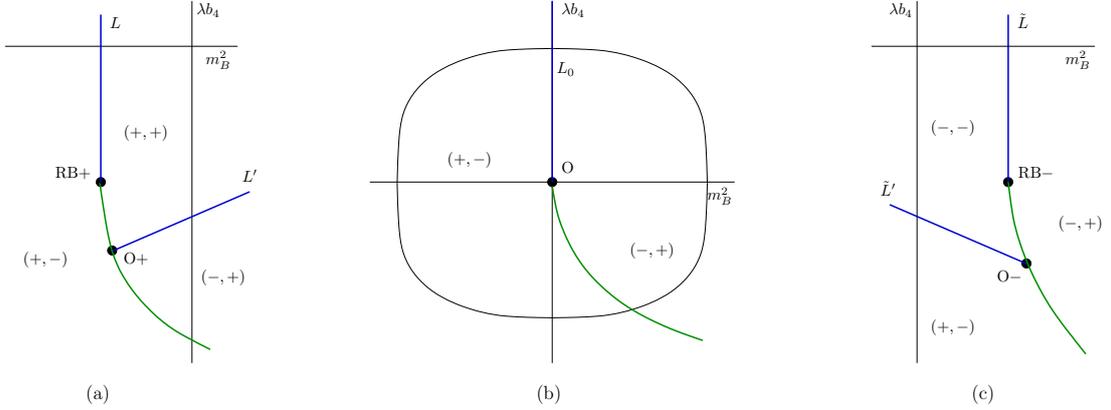}
  \caption{The phase structure for the (a)
    $m_F =|\mu|\, \sgn(\lambda)$, (b) $m_F = 0$, (c)
    $m_F = -|\mu| \sgn(\lambda)$ sections for
    $|\lambda| < \tfrac{1}{2}$. The blue lines are second order phase
    transitions while the green lines are first order transitions. The
    lines $L$, ${\tl L}$ correspond to the critical boson theory
    $\CB\pm$ and $L'$, ${\tl L}'$ correspond to the regular fermion
    theory $\RF\pm$. The phase diagram for $m_F = 0$ is the equatorial
    ellipse shown in Figure (b) and given by \eqref{mf0ellipse}. The
    intersection of the line $L_0$ with this section is a second order
    point described the CB-RF theory. The point $\Op$ in Figure (b) is
    the $\mc{N} = 2$ superconformal theory.}
  \label{phaseall}
\end{figure}

In the previous three subsections we have worked out the phase diagram
of the ${\cal N}=2$ theory separately for $m_F = 0$,
$m_F \sgn(\lambda) > 0$ and $m_F \sgn(\lambda) < 0$. We present the
phase diagrams in each of these three cases in Figure \ref{phaseall}.
However, we would
like to map back these results to the ellipsoid \eqref{spher}, whose
equation we reproduce below:
\begin{equation}
  (m_B^2)^2 + ((\lambda b_4)^2 + m_F^2)^2 = \text{constant}\ .
\end{equation}
The $m_F = 0$ section simply maps to the equator of the ellipsoid as
discussed in Section \ref{mfzero}. The $m_F \sgn(\lambda) > 0$ section
maps to the northern hemisphere via stereographic projection. In other
words, the origin of the $m_F \sgn(\lambda) > 0$ section maps to the
north pole of the ellipsoid and the equator of the ellipsoid sits at
infinity of this section. Similarly, the $m_F \sgn(\lambda) < 0$
section is mapped to the southern hemisphere of the ellipsoid. In
order to obtain the full phase diagram, all we then have to do is to
join the $m_F \sgn(\lambda) > 0$ and $m_F \sgn(\lambda) < 0$ sections
at their respective infinities with the equatorial section $m_F =
0$. We give a few more details of this `sewing procedure' in Section
\ref{sew} which can be skipped on a first reading.

\subsubsection{The sewing procedure}\label{sew}

As explained above, the phase diagram denoted in
Figure \ref{phaseall}(a) and Figure \ref{phaseall}(c) respectively are
stereographic projections, respectively, of the northern and southern
hemispheres of the ellipsoid \eqref{spher}. It follows that the points
at infinity of these phase diagrams match onto the equator of
\eqref{spher}, depicted as the `ellipse' in Figure
\ref{phaseall}(b). Note that these points (i.e. points on the equator
of \eqref{spher}) are labelled by the parabolas\footnote{We need to
  choose the leaves of the foliation such that they are scale
  invariant. Given the dimensions of $m_B^2$ and $b_4$ are $2$ and $1$
  respectively, this naturally gives us the parabolas in
  \eqref{parao}.}
\begin{equation}\label{parao}
m_B^2 = a (\lambda b_4)^2 \ ,
\end{equation}
(and the choice of branch of the parabola) in
Figure \ref{phaseall}(b). In order to complete our global construction
of the phase diagram we need a rule for assigning points at infinity
in Figure \ref{phaseall}(a) and Figure \ref{phaseall}(c) to parabolas
\eqref{parao} in Figure \ref{phaseall}(b).  Such a rule is needed in
order to provide an unambiguous sewing of the northern hemisphere to
the southern hemisphere through the equator.

In order to obtain the sewing rule one foliates the phase diagrams
Figure \ref{phaseall}(a) and Figure \ref{phaseall}(c) with the parabolas
\eqref{parao}. Unlike in the case of the phase diagram of
Figure \ref{phaseall}(b), in this case (branches of) parabolas do not
uniquely label points on the phase diagram. However (branches of)
these parabolas do uniquely label points at infinity in the diagrams
Figure \ref{phaseall}(a) and Figure \ref{phaseall}(c). A moments thought
will convince the reader that the sewing rule is simply that, any
point at infinity labelled by a parabola (and choice of branch) in the
northern/southern hemisphere phase diagram simply maps to the point on
the $m_F=0$ phase diagram labelled by the same parabola and same
choice of branch in Figure \ref{phaseall}(b).  This rule follows because
taking $m_B^2$ and $b_4$ to infinity at fixed $m_F$ is the same (upto
a scaling) as taking $m_F$ to zero at fixed $m_B^2$ and $b_4$.
 
Implementing this sewing rule yields our final result for the `phase
diagram ellipsoid'. In Figures \ref{3dplot14}, \ref{3dplot12} and
\ref{3dplot34} we present a Mathematica generated 3D plot of this
ellipsoid for $|\lambda|=\frac{1}{4}$ , $|\lambda|=\frac{1}{2}$ and
$\lambda=\frac{3}{4}$ respectively.

\subsubsection{First order phase transitions}\label{fomain}

The first order phase transition line in the phase diagram starts at
the point $\RB+$ in Figure \ref{phaseall}(a) and proceeds in a smooth
manner until it meets the point $\Op+$. Upto this point the first
order transition line separates the $(+,+)$ and $(+,-)$ phases. At
$\Op+$ this phase transition line is non-analytic. As the line
proceeds beyond $\Op+$ it now separates the $(-,+)$ and $(+,-)$
phases. This line then proceeds to infinity in the diagram of Figure
\ref{phaseall}(a) along the parabola
\begin{equation}\label{parafo}
  m_B^2 = \nu(\lambda)\ (\lambda b_4)^2\ ,
\end{equation}
where the function $\nu(\lambda)$ is determined by equating the
potential energies at the competing minima of the potential in the
$(+,-)$ and the $(-,+)$ phases (cf.~Appendix \ref{firstorder}). The
first order line for $m_F = 0$ in Figure \ref{phaseall}(b) also has
the exact same form given in \eqref{parafo} since it is obtained by
comparing the same potential energies at the competing minima for the
same phases as in Figure \ref{phaseall}(a). Thus, the first order line
in Figure \ref{phaseall}(a) (i.e.~in the northern hemisphere) smoothly
continues across the equator ($m_F = 0$) and becomes the part of the
first order phase transition line in Figure \ref{phaseall}(c)
(i.e.~the southern hemisphere) between the point $\Op-$ and infinity
always separating the same two phases $(-,+)$ and $(+,-)$. At the
point $\Op-$ this line is non-analytic. As it continues beyond $\Op-$
it now separates the $(-,-)$ and $(-,+)$ phases. This line finally
terminates at the point $\RB-$.

\subsubsection{Second order phase transitions}

The situation with the second order phase transition lines $L$, $L'$,
${\tl L}$ and ${\tl L}'$ in Figure \ref{phaseall} is more
interesting. These lines correspond to the $\CB+$, $\RF+$, $\CB-$ and
$\RF-$ conformal theories respectively. We describe the mapping of
these lines onto the ellipsoid below.

The lines $L$ and ${\tl L}$ in Figures \ref{phaseall}(a)
and \ref{phaseall}(c) respectively are straight lines and correspond
to the ($\lambda b_4 > 0$ branch of the) parabola $a=0$ in
\eqref{parao}. Hence, they are sewn to the point at infinity of the
straight line $L_0$ in Figure \ref{phaseall}(b) which also corresponds
to the parabola \eqref{parao} with $a=0$. Similarly, the reader can
easily convince herself that both the second order lines $L'$ and
${\tl L}'$ in Figures \ref{phaseall}(a) and \ref{phaseall}(c)
respectively eventually approach the ($\lambda b_4 > 0$ branch of the)
parabola $a = 0$ and hence are also sewn together with the same point
at infinity of the line $L_0$ in Figure \ref{phaseall}(b).

It follows that the phase diagram ellipsoid has an extremely
interesting point; one at which four different phases $(\pm, \pm)$ and
four different second order phase transition lines ($L$, $L'$,
${\tl L}$, ${\tl L}'$) meet. The dynamics at this special point is
conformal and is given by CS theory simultaneously coupled to both a
critical boson and a regular fermion. This precise theory is obtained
in the CB-RF scaling limit that is developed in detail in Section
\ref{cbrflim} of Appendix \ref{scaling}.

\begin{figure}
	\centering
	{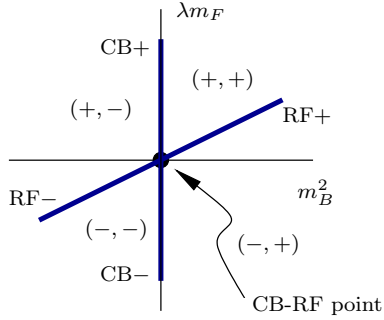}
	\caption{The features of the phase diagram of the $\mc{N} = 2$
          theory in the neighbourhood of the CB-RF conformal point
          depicted as a thick dot at the origin. The blue lines
          correspond to the $\CB\pm$, $\RF\pm$ theories that emanate /
          terminate at the origin. The various topological phases are
          also shown.}
	\label{cbrflimit}
\end{figure}

The phase diagram in the neighbourhood of this very special point is
best viewed in terms of the dimensionless coordinates
\begin{equation}
  \left(\frac{m_B^2}{(\lambda b_4)^2}\, , \frac{\lambda m_F}{\lambda b_4}\right)\ ,
\end{equation}
and is plotted in Figure \ref{cbrflimit} (for a representative value
of $\lambda b_4 = 1$).
The information in this figure is obtained by using the sewing rules
just described and the equations governing the four second order lines
$L$, $L'$, ${\tl L}$ and ${\tl L}'$ given in \eqref{Upara},
\eqref{Uprpara}, \eqref{Uparam} and \eqref{Uprparam} respectively. The
information can also be obtained by considering a small neighbourhood
of the CB-RF scaling limit in Section \ref{cbrflim} of Appendix
\ref{scaling}. There we reproduce the above schematic diagram in
Figure \ref{cbrflimit} for values of $x_4$ and $x_6$ other than those
for the $\mc{N} = 2$ theory.

\subsection{The phase diagram}\label{phasedN2}
We have already presented a schematic version of this phase diagram in
Figure \ref{3dplot} which we reproduce here for convenience.
\begin{figure}
	\centering
	\scalebox{0.43}{\includegraphics{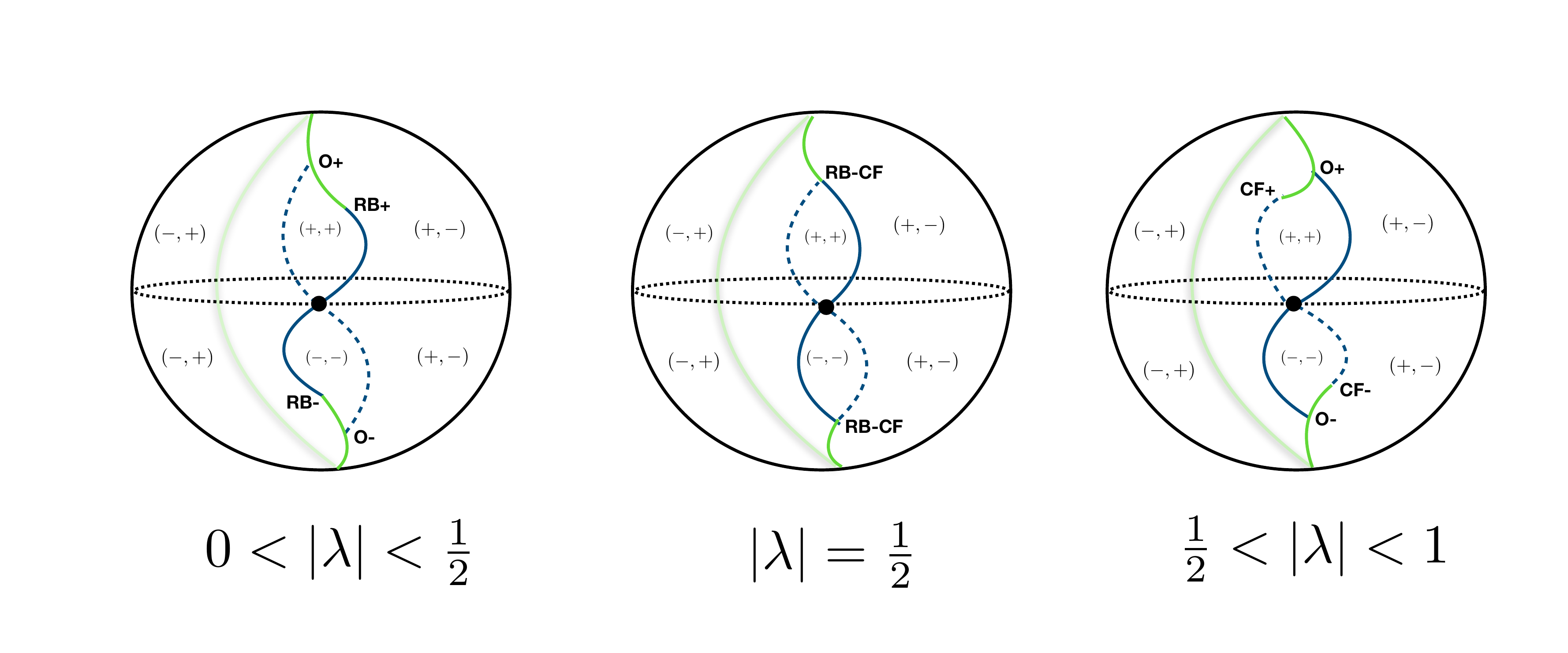}}
	\caption{The schematic phase diagram for the $\mc{N} = 2$
          theory presented in the introduction for three
          representative values of $|\lambda|$. The black dot on the
          equator is the CB-RF CFT. The remaining notation is
          explained in this section (Section \ref{N2phase}) or in the
          footnote to Figure \ref{3dplot} in the introduction.}
	\label{3dplotmain}
\end{figure}
We have plotted the full phase diagram ellipsoid at $|\lambda|=1/4$ in
Figure \ref{3dplot14}.
\begin{figure}
	\centering
	\includegraphics[width=4in,height=3in]{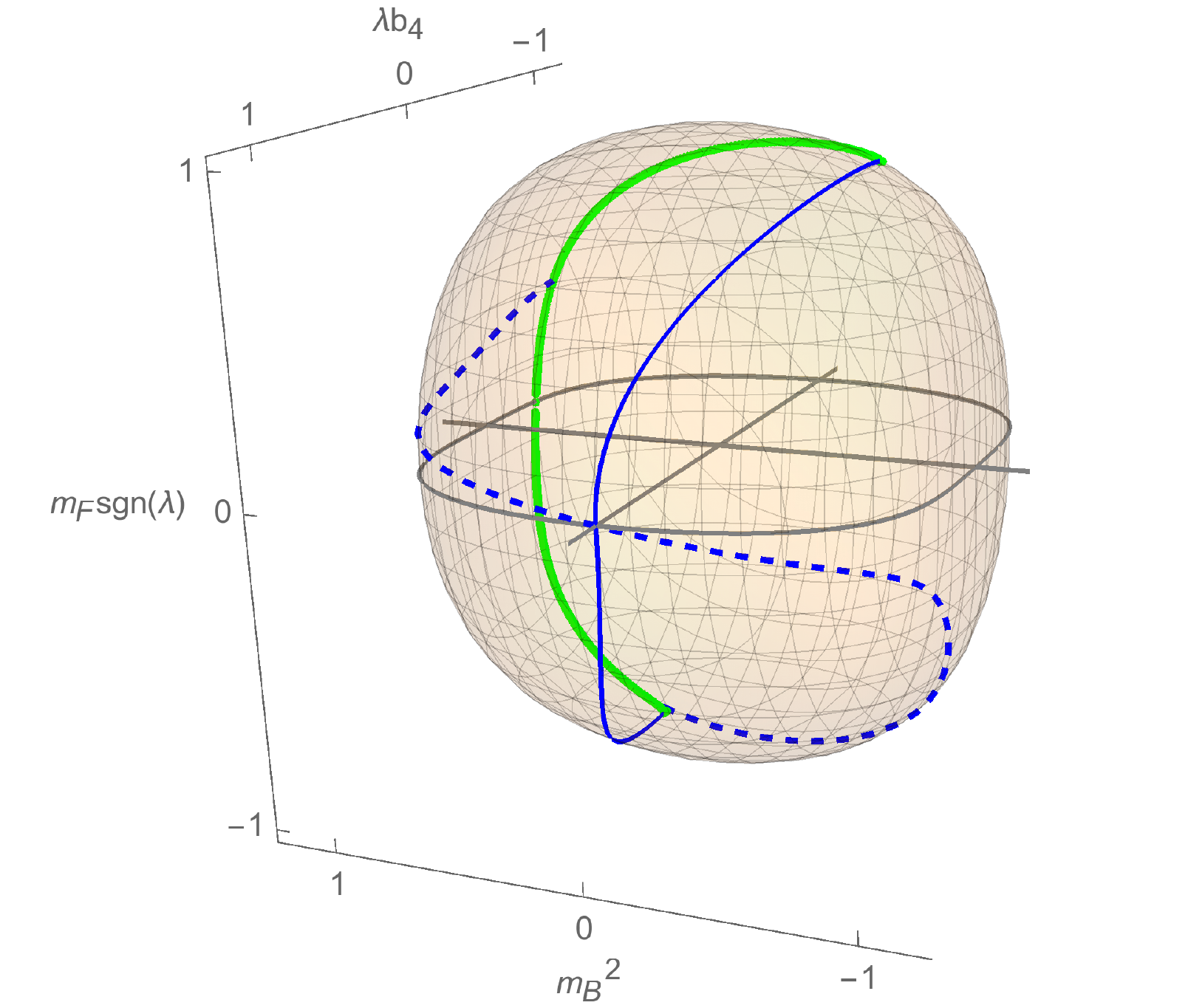}
	\caption{The phase diagram ellipsoid of the $\mc{N} = 2$
          theory for the value $|\lambda| = \tfrac{1}{4}$. The equator
          $m_F = 0$ is displayed in gray. There are two solid and two
          dashed blue lines (one each in the northern hemisphere and
          one each in the southern hemisphere). The solid blue lines
          are described by CB conformal theories while the dashed blue
          lines are RF conformal theories -- both correspond to second
          order phase transitions. The green line is a first order
          transition line and lies on the far side of the
          ellipsoid. The solid blue lines meet the green line at two
          points (one each on the two hemispheres) - these points are
          described by the RB conformal field theory. The solid blue
          (CB) and dashed blue (RF) lines meet at the point
          $(m_B^2, \lambda b_4, m_F\sgn(\lambda)) = (0,1,0)$ on the
          equator which is described by the CB-RF conformal field
          theory.}
	\label{3dplot14}
\end{figure}
\begin{figure}
	\centering
	\includegraphics[width=4in,height=3in]{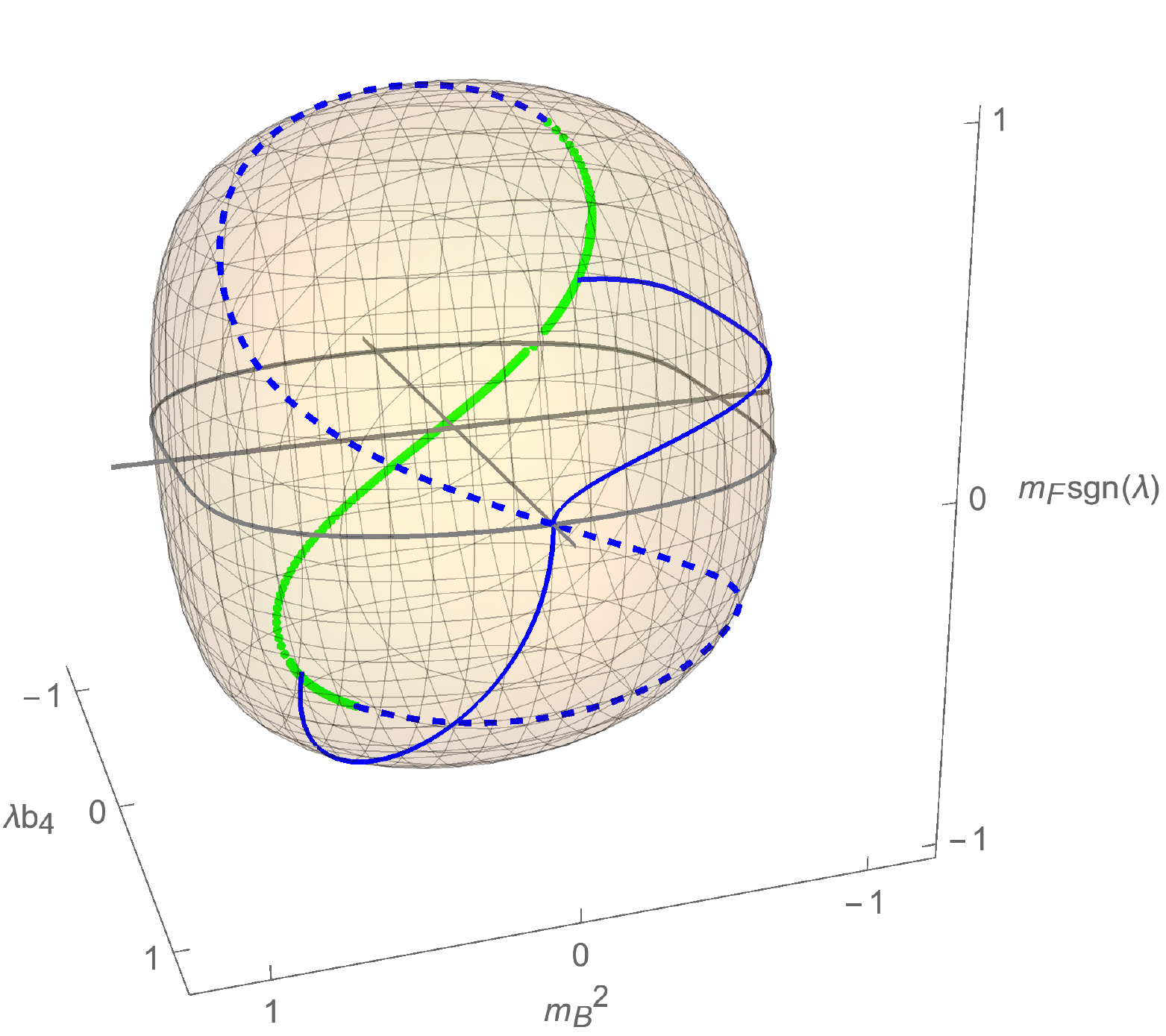}
	\caption{The phase diagram ellipsoid of the $\mc{N} = 2$
          theory for the value $|\lambda| = \tfrac{3}{4}$.  The first
          order phase transition line in green is on the far side of
          the ellipsoid. Note that the first order line meets the
          dashed blue lines described by regular fermion CFTs. These
          meeting points themselves are described by critical fermion
          CFTs. The solid blue lines are described by critical boson
          CFTs and the intersection point of the two solid and two
          dashed blue lines is described by the CB-RF CFT. The
          features of this phase diagram can be obtained by applying
          the duality map \eqref{dualitymap} to the case
          $|\lambda| = \tfrac{1}{4}$.}
	\label{3dplot34}
\end{figure}
\begin{figure}
	\centering
	\includegraphics[width=4in,height=3in]{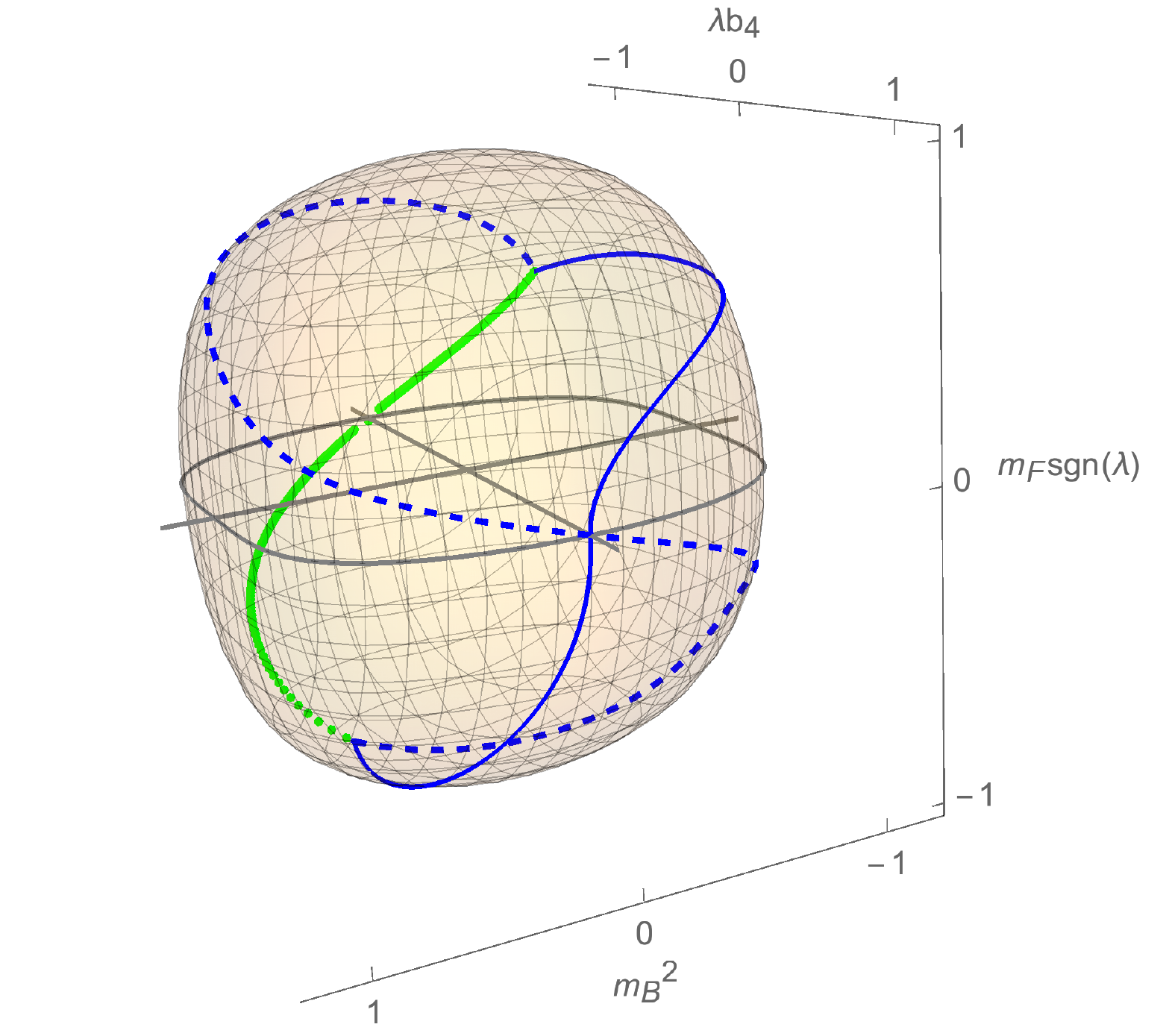}
	\caption{The phase diagram ellipsoid of the $\mc{N} = 2$
          theory for the value $|\lambda| = \tfrac{1}{2}$.  The dashed
          blue lines are described by regular fermion CFTs and the
          solid blue lines are critical boson CFTs. The first order
          phase transition line in green is on the far side of the
          ellipsoid. Note that the first order line meets the dashed
          blue line and solid blue line at one point in the northern
          hemisphere and at one point in the southern
          hemisphere. These meeting points are presumably governed by
          a combination of critical fermion and regular boson CFTs
          i.e.~an RB-CF CFT. The intersection point of the two solid
          and two dashed blue lines is described by the CB-RF CFT. The
          three special points, one on the equator, one in the
          northern hemisphere and one in the southern hemisphere are
          all expected to be described by CFTs that are self-dual
          under the duality \eqref{dualitymap}.}
	\label{3dplot12}
\end{figure}
Upto this point in this paper we have worked out the phase diagram of
our theory only in the special case $|\lambda|<\frac{1}{2}$.  The
analysis for the case $|\lambda|>\frac{1}{2}$ proceeds in an entirely
similar manner; the final results for this analysis could be
anticipated from duality. We again give a three dimensional plot of
the phase diagram for the representative value
$|\lambda| = \tfrac{3}{4}$ in Figure \ref{3dplot34}.

Finally, we present the phase diagram for the case
$|\lambda| = \tfrac{1}{2}$ in Figure \ref{3dplot12} which is expected
to be self-dual under the duality map \eqref{dualitymap}.  Note that
the phase diagram in this case is very special. In particular, the
line of second order CB phase transitions and the line of second order
RF phase transitions end at a single critical point in each of the
northern and southern hemisphere of phase diagram. The low energy
dynamics of this critical point is presumably a theory of regular
bosons and Gross-Neveu (or critical) fermions simultaneously
interacting with a single $SU(N)$ Chern Simons gauge field at
$|\lambda|=\frac{1}{2}$. This theory has never been studied before in
the literature to the best of our knowledge, and there are several
questions about it that would be interesting to investigate. In
particular, this theory has four naively marginal operators. At first
subleading order the coupling behind each of these terms will develop
a $\beta$ function. It would be very interesting to study these
$\beta$ functions and investigate the resultant structure of fixed
points.

As we have mentioned above, moreover, $|\lambda|=\frac{1}{2}$ is
special because this value is duality invariant. At precisely this
value of $|\lambda|$, therefore, it should be possible to orbifold
both the original UV theory \eqref{generalaction} and the very special
IR theory described in the previous paragraph by the duality operation
\eqref{dualitymap}\footnote{A similar procedure is used to construct
  ${\cal N}=3$ gauge theories in $d=4$ starting from ${\cal N}=4$ Yang
  Mills theory.}. The resultant theory sounds particularly interesting
to us because it is obtained by orbifolding with a Bose-Fermi duality:
the excitations that survive this orbifolding are therefore maximally
anyonic.  The resultant theory should be simpler than its parent
unorbifolded theory in some ways (for instance it will have only two
rather than 4 naively marginal operators whose $\beta$ function we
would have to control). It is conceivable that this orbifolded fixed
point continues to exist at small finite values of $N$ and $k$ and
could turn out to have applications in condensed matter physics. We
feel that the detailed study of this theory is an interesting
direction for future study.

\section{The \texorpdfstring{$\mc{N}=1$}{N=1} locus}\label{N1phase}
The class of theories \eqref{generalaction} has a subset consisting of
a two parameter set of ${\cal N}=1$ theories. The coupling constants
in the Lagrangian are given in terms of the two parameters $\mu$ and
$w$ as
\begin{equation}\label{susyN1}
  m_F=\mu\ ,\ \ m_B^2=\mu^2\ ,\ \ b_4= \mu w\ ,\quad x_4=\frac{1+w}{2}\ ,\quad x_6=w^2-1\ ,\quad y_4' = 3 + w\ ,\quad y_4'' = w - 1\ .
\end{equation}
\subsection{Classical analysis} \label{class}
From \eqref{generalaction} one can read the classical potential for the field $\bar{\phi}\phi$ and the effective mass term for $\bar\psi\psi$ as
\begin{align}
U_{\text{cl}}(\bar\phi \phi) &= \kappa \left(m_B^2 \frac{\bar\phi \phi}{\kappa} + 4\pi b_4  \left(\frac{\bar\phi \phi}{ \kappa}\right)^2 
+ 4 \pi^2 (x_6+1) \left(\frac{\bar\phi\phi}{\kappa}\right)^3\right) + \left(4 \pi x_4 \frac{\bar\phi \phi}{ \kappa} + m_F\right) \bar{\psi}\psi\ ,
\end{align}
which in the $\mathcal{N}=1$ theory becomes
\begin{align}\label{Uclass}
U_{\text{cl}} &=  {\bar\phi \phi} \left(\mu+2\pi w  \frac{\bar\phi \phi}{\kappa} \right)^2 + \left(2 \pi (1+w) \frac{\bar\phi \phi}{\kappa} + \mu\right) \bar{\psi}\psi\ .
\end{align}
The vacua of the above theory are characterised by $\phi = 0$ and the
following two possibilities for the scalar $\phi$:
\begin{equation}\label{bosvac}
  \text{unHiggsed,} +:\quad \bar\phi \phi = 0\ ,\qquad \text{Higgsed,}-:\quad 2\pi\bar\phi \phi = -\frac{\mu \kappa}{w}\ .
\end{equation}
The condition $\sgn(m_F) \sgn(\kappa) = \pm 1$ which decides the level
of the low-energy Chern-Simons theories becomes
\begin{equation}
  \pm:\qquad \sgn(\kappa)\sgn\left(\mu + 2\pi (1 + w) \frac{\bar\phi \phi}{\kappa}\right) = \pm 1\ , 
\end{equation}
where $\bar\phi\phi$ has to be evaluated at one of the bosonic vacua
in \eqref{bosvac}. The four possible phases $(\pm, \pm)$ are then
characterised by
\begin{align}
  &(+,+):\quad \bar\phi \phi = 0\ ,\ \sgn(\mu \kappa) = +1\ ,\qquad   (+,-):\quad 2\pi\bar\phi \phi = -\frac{\mu \kappa}{w}\ ,\ \sgn\left(-\frac{\mu \kappa}{w}\right) = +1\ ,\nonumber\\&(-,+):\quad \bar\phi \phi = 0\ ,\ \sgn(\mu \kappa) = -1\ ,\qquad   (-,-):\quad 2\pi\bar\phi \phi = -\frac{\mu \kappa}{w}\ ,\ \sgn\left(-\frac{\mu \kappa}{w}\right) = -1\ .
\end{align}
Classically, we need to satisfy ${\bar\phi}\phi \geq 0$ and we see
that this is possible in the $(+,+)$, $(-,+)$, $(+,-)$ phases for
appropriate choices of signs of $\mu$, $w$ and $k$. However
${\bar \phi} \phi$ is required to be negative in the $(-,-)$ phase
which is impossible classically. Thus, the $\mc{N} = 1$ theory never
has a classical vacuum in this phase.

\subsection{At finite \texorpdfstring{$\lambda$}{lambda}}

\begin{figure}
	\centering
	\scalebox{0.7}{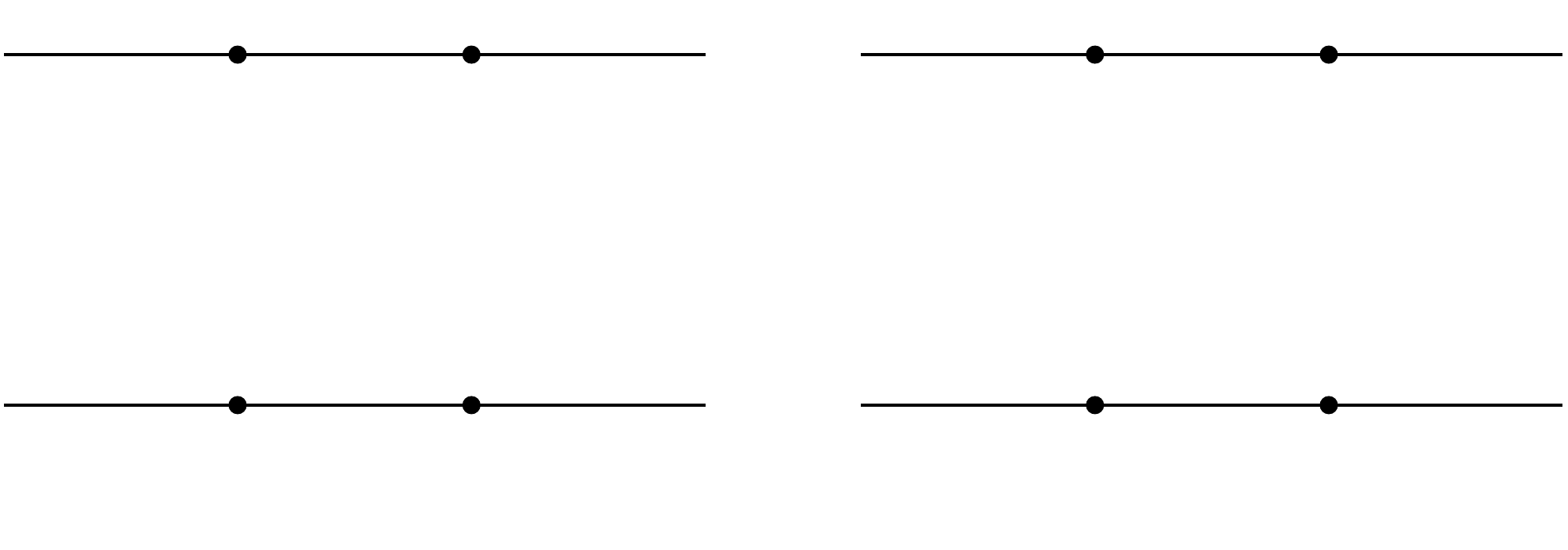}
	\caption{Regions of validity for different branches of the potential
		for various choices of $\sgn(\mu)\sgn(\lambda)$ and
		$\sgn(1+w)$.}
              \label{phaseseqsusy}
\end{figure}

Recall that the large $N$ analysis provides us with an exact (i.e.~to
all orders in $\lambda$) Landau-Ginzburg effective potential for the
theory for the variable $\sigma$ (in fact, for the variable
$(\bar\phi\phi)_{\rm cl}$ which is related to $\sigma$ via
\eqref{sigph}). The phases are determined by the minima of the
Landau-Ginzburg potential. The four phases are characterised in terms
of the minimum value of $\sigma$ as (cf. \eqref{fermcond} and
\eqref{bosecond})
\begin{alignat}{2}\label{phasevalid}
  &(+,+):\quad \sgn(1+w) \sigma > -\frac{\mu}{\lambda}\frac{1}{|1+w|}\ ,\qquad &&\sigma < 0\ ,\nonumber\\
  &(+,-):\quad \sgn(1+w) \sigma > -\frac{\mu}{\lambda}\frac{1}{|1+w|}\ ,\qquad &&\sigma > 0\ ,\nonumber\\
  &(-,+):\quad \sgn(1+w) \sigma < -\frac{\mu}{\lambda}\frac{1}{|1+w|}\ ,\qquad &&\sigma < 0\ ,\nonumber\\
  &(-,-):\quad \sgn(1+w) \sigma < -\frac{\mu}{\lambda}\frac{1}{|1+w|}\ ,\qquad &&\sigma > 0\ .
\end{alignat}
For any given set of values for the signs $\sgn(1+w)$ and
$\sgn(\mu) \sgn(\lambda)$, the potential explores three among the four
phases above as given in Figure \ref{phaseseq}. We reproduce this
figure in terms of the variables $\mu$, $\lambda$ and $w$ in Figure
\ref{phaseseqsusy} for easy reference.

Plugging in the special locus of values \eqref{susyN1} in the general
expression for the Landau-Ginzburg potential in \eqref{Ustab}, we get
\begin{multline}\label{UstabN1}
U^{(\varepsilon,\pm)}(\sigma) = \frac{N}{2\pi}\Big[ \left(w^2 - 1 - \phi_\pm + \tfrac{1}{8}(1+w)^3\psi_\varepsilon\right)  \lambda^2 { \sigma}^3  +  \\ \left( w  + \tfrac{3}{16}(1+w)^2 \psi_\varepsilon \right) 2 \mu \lambda  { \sigma}^2
+ \left(1 + \tfrac{3}{8} (1+w) \psi_\varepsilon \right) \mu^2 { \sigma} + \tfrac{1}{8\lambda} \psi_\varepsilon \mu^3 \Big]\ .
\end{multline}
where as earlier, $\varepsilon=\pm$ denotes the fermionic phases and the explicit $\pm$ denotes the bosonic phases.

\begin{figure}
	\centering
	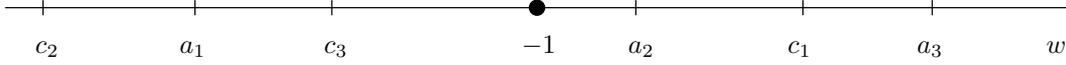
	\caption{Locations of the various functions of $|\lambda|$ in
          \eqref{roots} on the $w$-line for all values of
          $|\lambda|$.}
              \label{wphase}
\end{figure}

It turns out that, on the $\mc{N} = 1$ locus of parameters, the
Landau-Ginzburg potential in each of the four branches can be written
in a very specific form:
\begin{align}\label{UeffN1factor}
	U_{\text{eff}}^{({\rm +,+})}(\sigma) &= \frac{N}{2\pi} \frac{\lambda^2\psi_+}{8} (w - a_1)^2 (w - c_1) \times  \left(\sigma +\frac{\mu}{\lambda}\frac{1}{  w - a_1}\right)^2 \left(\sigma + \frac{\mu}{\lambda}\frac{1 }{w -c_1}\right)\ , \nonumber\\
	U_{\text{eff}}^{({\rm +,-})}(\sigma) &= \frac{N}{2\pi}  \frac{\lambda^2\psi_+}{8} (w - a_2)^2 (w - c_2)\times\left(\sigma +\frac{ \mu }{\lambda}\frac{1}{w - a_2}\right)^2 \left(\sigma + \frac{\mu}{\lambda}\frac{1 }{w - c_2}\right)\ , \nonumber \\
	U_{\text{eff}}^{({\rm -,+})}(\sigma) &= \frac{N}{2\pi} \frac{\lambda^2\psi_-}{8} ( w - a_3)^2 (w - c_3) \times  \left(\sigma + \frac{\mu}{\lambda}\frac{1}{w - a_3}\right)^2\left(\sigma + \frac{\mu}{\lambda}\frac{1}{w - c_3}\right)\ ,\nonumber\\
  U_{\text{eff}}^{({\rm -,-})}(\sigma) &= \frac{N}{2\pi} \frac{\lambda^2\psi_-}{8} (w - a_4)^2 (w - c_4) \times\left(\sigma + \frac{ \mu }{\lambda}\frac{1}{w - a_4}\right)^2 \left(\sigma + \frac{\mu}{\lambda}\frac{1}{w - c_4}\right) \nonumber\\ &\quad -\frac{N}{2\pi} \frac{32 |\lambda|^3 }{3(4-|\lambda|^2)^2}\sigma^3\ ,
\end{align}
where $\psi_\pm$ are defined in \eqref{psipm} and $a_i$ and $c_i$ are the following functions of $|\lambda|$:
\begin{align} \label{roots}
a_1 &= -\frac{2-|\lambda| }{|\lambda| }\ ,\quad a_2 = \frac{1}{a_1}\ ,\quad c_1 = \frac{2 - 2|\lambda| - |\lambda|^2}{ |\lambda|(2-|\lambda|)}\ ,\quad c_2 = -\frac{6 - 6|\lambda| + |\lambda|^2}{ |\lambda|(2-|\lambda|)}\ ,\nonumber \\
a_3 &= \frac{2+|\lambda| }{|\lambda|}\ ,\quad a_4 = \frac{1}{a_3}\ ,\quad c_3 = -\frac{2 + 2|\lambda| - |\lambda|^2}{ |\lambda|(2 + |\lambda|)}\ ,\quad c_4 = \frac{|\lambda|^2 + 6|\lambda| + 6}{|\lambda|(2+|\lambda|)}\ .
\end{align}
We give the locations of the quantities $a_1$, $a_2$, $a_3$, $c_1$,
$c_2$, $c_3$ on the $w$-line in Figure \ref{wphase} since they clearly
signify special points for the potential above. The ordering of these
quantities shown in Figure \ref{wphase} holds for any value of
$|\lambda|$ between $0$ and $1$.

The first three potentials are of the form
\begin{equation}\label{specform}
  U(\sigma) = \frac{N}{2\pi} \frac{\lambda^2 \psi}{8} (w - a)^2 (w - c) \left(\sigma + \frac{\mu}{\lambda}\frac{1}{w-a}\right)^2\left(\sigma + \frac{\mu}{\lambda}\frac{1}{w-c}\right)\ ,
\end{equation}
whereas the potential for the $(-,-)$ branch fails to be of the above
form due to the additional term proportional to $\sigma^3$. We analyse
the $(+,+)$, $(+,-)$ and $(-,+)$ branches first by studying the
function in \eqref{specform} above and then study the $(-,-)$ branch
separately.

\begin{figure}
	\centering
	\scalebox{0.9}{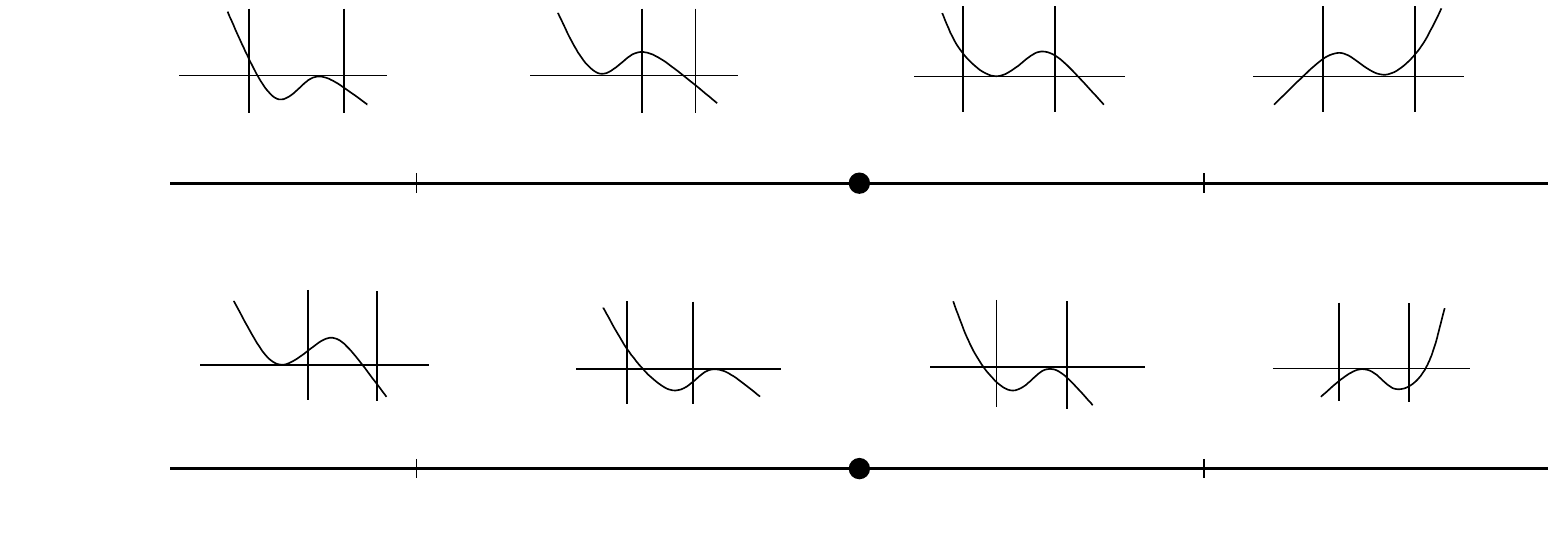}
	\caption{Profile of the function $U_{\rm eff}^{(+,+)}(\sigma)$
          as $w$ (horizontal line in this graph) runs from $-\infty$
          to $\infty$. The vertical line in each potential plot with
          label $0$ denotes the point $\sigma = 0$ and the unlabelled
          vertical line is the point
          $\sigma = -\frac{\mu}{\lambda}\frac{1}{1+w}$.}
              \label{wphasepp}
\end{figure}

\begin{figure}
	\centering
	\scalebox{0.9}{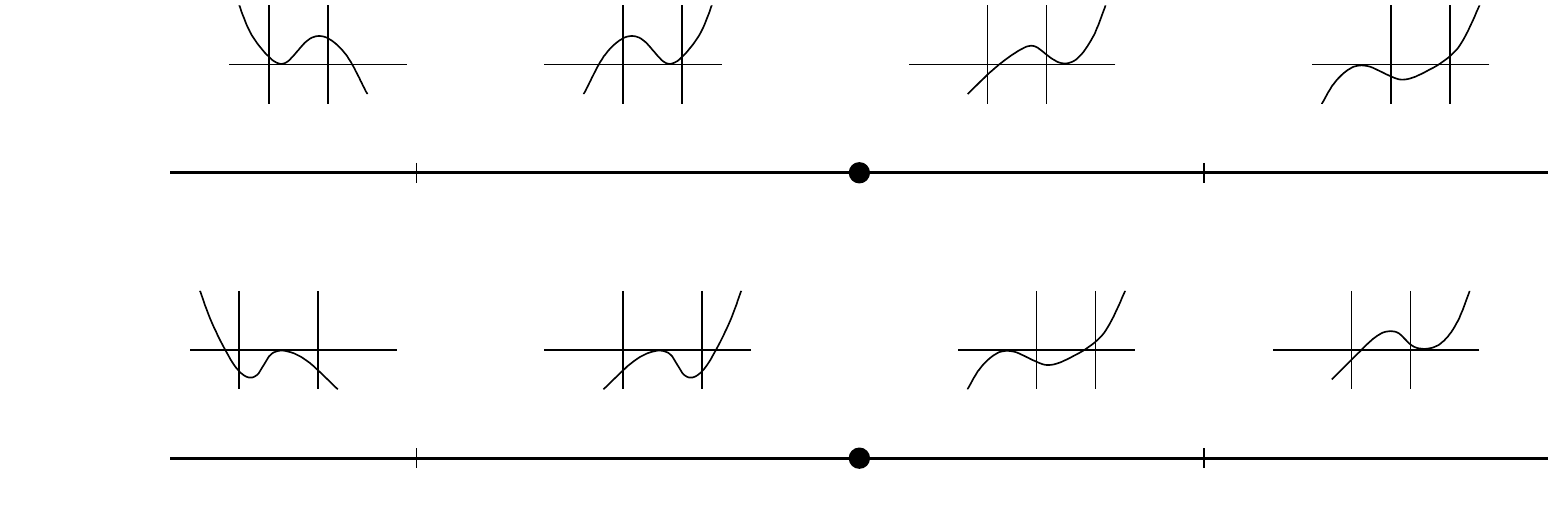}
	\caption{Profile of the function $U_{\rm eff}^{(+,-)}(\sigma)$ as $w$ (horizontal line
		in this graph) runs from $-\infty$ to $\infty$. The vertical line in each potential plot with
          label $0$ denotes the point $\sigma = 0$ and the unlabelled
          vertical line is the point
          $\sigma = -\frac{\mu}{\lambda}\frac{1}{1+w}$.}
              \label{wphasepm}
\end{figure}

To begin with, we analyse the function \eqref{specform} when
$w \neq a, c$. The function \eqref{specform} has zeroes at
\begin{equation}\label{Uzeroes}
  \sigma = -\frac{\mu}{\lambda} \frac{1}{w-a}\ ,-\frac{\mu}{\lambda} \frac{1}{w-a}\ , -\frac{\mu}{\lambda} \frac{1}{w-c}\ .
\end{equation}
Since there is a double zero at
$\sigma = -\frac{\mu}{\lambda}\frac{1}{w-a}$, this zero is also an
extremum of the potential. It is a minimum when
\begin{equation}\label{Umindef}
  \psi  \frac{\mu}{\lambda} \left(\frac{c-a}{w-a}\right) > 0\ .
\end{equation}
The other extremum of the potential lies between the two zeroes at
\begin{equation}\label{Uext2}
\sigma = - \frac{\mu}{3\lambda}\left(\frac{2}{w-c} + \frac{1}{w-a}\right)\ .
\end{equation}
Based on these facts and the ordering of the $a_i$ and $c_i$ in Figure
\ref{wphase}, the profile of the potential can be deduced in the
$(+,+)$, $(+,-)$ and $(-,+)$ branches listed in
\eqref{UeffN1factor}. We have displayed the result of such an analysis
in Figure \ref{wphasepp}, \ref{wphasepm} and \ref{wphasepm} for the
$(+,+)$, $(+,-)$ and $(-,+)$ branches respectively.

Note that there are special values of $w$ viz.~$w = a,c$ where the
form of the function \eqref{specform} changes from being cubic to
being either quadratic or linear in $\sigma$. When $w = a$, we have
\begin{equation}\label{speclin}
U(\sigma)\Big|_{w=a} =  \frac{N}{2\pi} \frac{\mu^2 \psi}{8} (a - c) \left(\sigma + \frac{\mu}{\lambda}\frac{1}{a-c}\right)\ ,
\end{equation}
It is easy to see that when $w$ approaches $a$, the double zero (hence
also an extremum) at $\sigma = -\frac{\mu}{\lambda}\frac{1}{w-a}$ and
the other extremum \eqref{Uext2} both go away to $\sigma = \pm\infty$
and the potential becomes linear. More specifically, let us start with
a value of $w$ just below $a$. As $w$ is increased towards $a$ the
extremum listed in the first of \eqref{Uzeroes} and the extremum
listed in \eqref{Uext2} both move towards $\sgn (\mu \lambda)
\infty$. As $w$ is increased above $a$, these two extrema reappear,
but this time at $-\sgn (\mu \lambda) \infty$.

When $w = c$, we have
\begin{equation}\label{specquad}
  U(\sigma)\Big|_{w=c} =  \frac{N}{2\pi} \frac{\mu\lambda \psi}{8} (a - c)^2 \left(\sigma + \frac{\mu}{\lambda}\frac{1}{c-a}\right)^2\ ,
\end{equation}
Again, it is easy to see that as one approaches $w = c$ from below the
single zero $\sigma = -\frac{\mu}{\lambda}\frac{1}{w-c}$ and the
extremum \eqref{Uext2} go away to $\sgn(\mu\lambda) \infty$ (the
double zero in \eqref{Uzeroes} survives and becomes the extremum of
the quadratic potential \eqref{specquad} above). When $w$ is just
larger than $c$ both the zero and the extremum which disappeared at
$w=c$ reappaer at $\sigma = -\sgn(\mu\lambda) \infty$.
\begin{figure}
	\centering
	\scalebox{0.9}{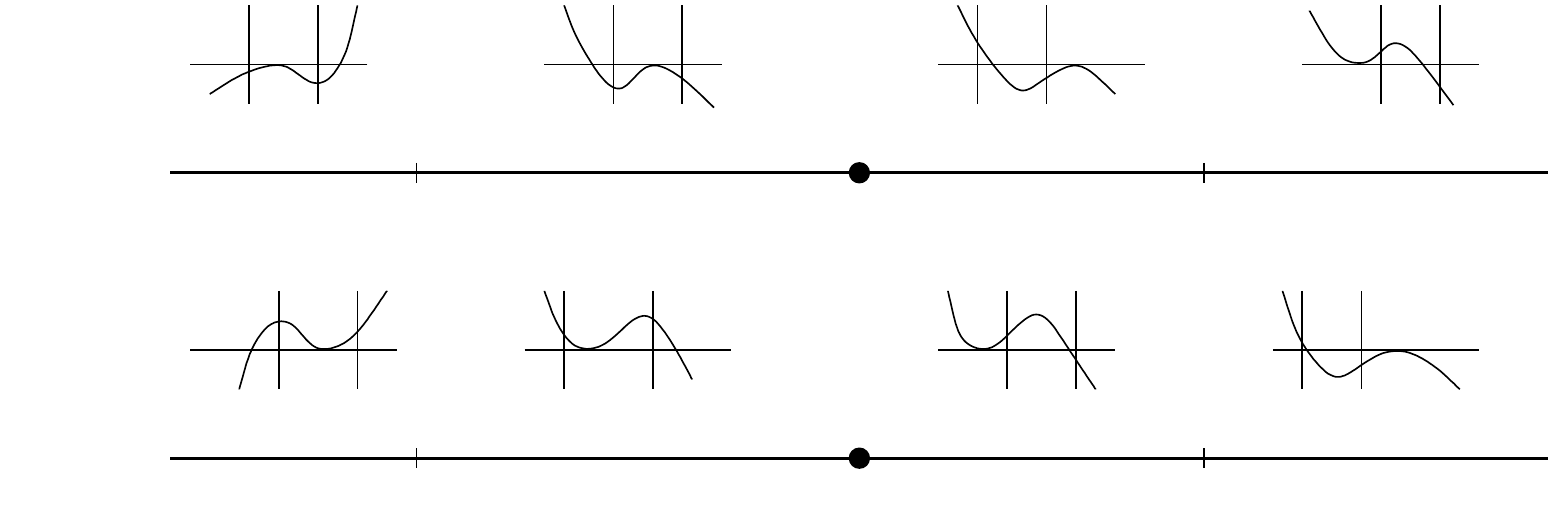}
	\caption{Profile of the function $U_{\rm eff}^{(-,+)}(\sigma)$
          as $w$ (horizontal line in this graph) runs from $-\infty$
          to $\infty$.  The vertical line in each potential plot with
          label $0$ denotes the point $\sigma = 0$ and the unlabelled
          vertical line is the point
          $\sigma = -\frac{\mu}{\lambda}\frac{1}{1+w}$.}
              \label{wphasemp}
\end{figure}

The $(-,-)$ branch potential is not completely of
the form \eqref{specform}. There is an additional cubic term in
$\sigma$ which changes the behaviour quite drastically. In fact, it
can be easily verified that the cubic potential on the $(-,-)$ branch
always has three real zeroes and (hence) none of these zeroes are
extrema of the potential. We do not give all the cumbersome details of
the potential for different values of $w$ and only plot the
appropriate part of the potential on the $(-,-)$ branch wherever it is
required for the analysis of the vacuum structure below.

Note that in the first three branches in \eqref{UeffN1factor}, the
form of the potential \eqref{specform} closely mirrors that of the
tree-level potential \eqref{Uclass}. We reproduce below the tree-level
potential in terms of the variable $\sigma$
\begin{align}
U_{\text{cl}}(\sigma) &=  \frac{N}{2\pi}  \sigma \times \lambda^2 w^2 \left(\sigma + \frac{\mu}{\lambda} \frac{1}{w} \right)^2 \ .
\end{align}
The specific form of the classical potential is not an accident and is
in fact due to $\mc{N} = 1$ supersymmetry. The potential energy
density for an $\mc{N} = 1$ supersymmetric theory with a scalar
superfield $\phi$ with the usual superspace kinetic term for $\phi$
and a superspace potential term $W(\phi,\bar\phi)$ is of the form
\begin{equation}
  \bar\phi\phi \left|\frac{\partial W}{\partial \phi}\right|^2\ .
\end{equation}
In particular, note that the potential above is the product of a
positive definite term $\bar\phi\phi$ that comes from the kinetic term
and another positive definite term $|W'|^2$ that comes from the
superspace potential term. The Landau-Ginzburg potential for the
branches $(+,+)$, $(+,-)$ and $(-,+)$ is also precisely of this form:
\begin{align}
    U_{\text{qu}}(\sigma) &= \frac{N}{2\pi} \frac{\psi_\pm}{8}  (w - c) \left(\sigma + \frac{\mu}{\lambda}\frac{1}{w-c}\right)\times \lambda^2 (w - a)^2 \left(\sigma + \frac{\mu}{\lambda}\frac{1}{w-a}\right)^2\ .
\end{align}
The first factor above is linear in $\sigma$ and is a
$\lambda$-dependent deformation of the $\sigma$ factor in the
classical potential. It can be easily checked that this factor is
positive definite in the domain of validity in $\sigma$ of the
appropriate branch (one of $(+,+)$, $(+,-)$, $(-,+)$) of the
potential. Given the comparison with the classical potential it is
tempting to conjecture that this factor arises from the superspace
kinetic term of appropriate scalar superfield whose bosonic component
involves $\sigma$. The second factor is the square of a term linear in
$\sigma$ and is again a $\lambda$-dependent deformation of the
$|W'|^2$ term in the classical potential. Presumably, this complete
square term arises from an appropriate superspace potential for the
same superfield whose bosonic component involves $\sigma$. Note that
the potential on the $(-,-)$ branch is not of this form and this may
be tied to the fact that a classical supersymmetric vacuum does not
exist in the $(-,-)$ branch.

\subsection{The vacuum structure}
We study the exact Landau-Ginzburg potential for the $\mc{N} = 1$
theory at different values of $w$.  The regions of validity of the
different branches of the potential are given in Figure
\ref{phaseseqsusy} for different values of $\sgn(\mu\lambda)$ and
$\sgn(1+w)$. We borrow the appropriate parts of the potential for the
$(+,+)$, $(+,-)$ and $(-,+)$ branches from Figures \ref{wphasepp},
\ref{wphasepm}, \ref{wphasemp} and separately plot the $(-,-)$ branch
wherever required, and patch them together to obtain the
Landau-Ginzburg potential. We provide a representative plot of the
potential for each interval in the $w$-line displayed in Figure
\ref{wphase}. We only indicate those special values of $w$ where the
phase structure of the potential changes and hide the remaining values
in order to avoid cluttering.

\begin{figure}
	\centering
	\scalebox{0.88}{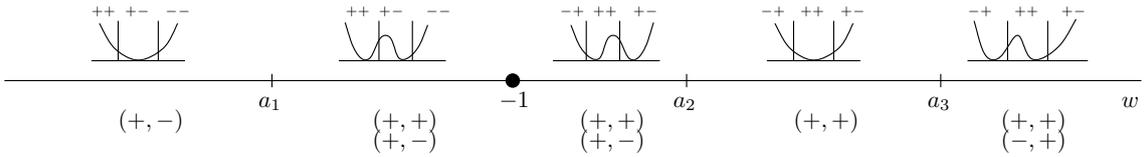}
	\caption{The Landau-Ginzburg potential $U_{\rm eff}(\sigma)$
          plotted for $w$ (horizontal line in this graph) ranging from
          $-\infty$ to $\infty$, in the case that
          $\sgn(\mu) = \sgn(\lambda)$. One of the two vertical lines
          in each plot of $U_{\rm eff}(\sigma)$ - the line that
          separates a bosonic (second index) $+$ phase from a bosonic
          $-$ phase - lies at $\sigma=0$. The second vertical line
          separates the fermionic (first index) $+$ phase from the
          fermionic $-$ phase and lies at
          $\sigma = -\frac{\mu}{\lambda}\frac{1}{1+w}$ and occurs at
          either positive or negative values of $\sigma$ depending on
          the value of $w$. The brackets below the horizontal line
          denote the phase(s) of the theory for the appropriate
          ranges of $w$. The quantities $a_1$, $a_2$ and $a_3$ are
          defined in \eqref{roots} and are given by
          $a_1 = a_2^{-1} = -\frac{2-|\lambda}{|\lambda|}$,
          $a_3 = \frac{2 + |\lambda|}{|\lambda|}$.}
              \label{wphaseLGpotp}
\end{figure}

\begin{figure}
	\centering
	\scalebox{0.88}{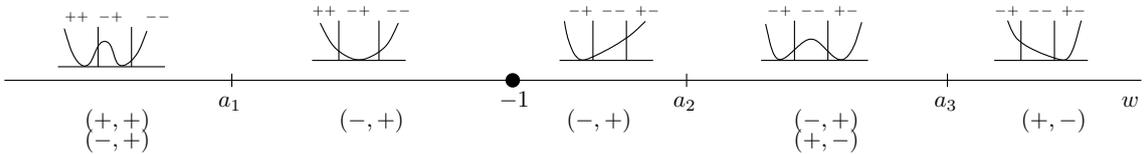}
	\caption{The Landau-Ginzburg potential $U_{\rm eff}(\sigma)$
          plotted for values of the parameter $w$ (horizontal line in
          this graph) ranging from $-\infty$ to $\infty$ in the case
          that $\sgn(\mu) = -\sgn(\lambda)$. We follow the same
          notation as in Figure \ref{wphaseLGpotp}.}
              \label{wphaseLGpotm}
\end{figure}

As is apparent from Figure \ref{phaseseqsusy}, it is useful to
separate the two cases $\sgn(\mu)\sgn(\lambda) = \pm 1$. We display
the potential for different values of $w$ in Figure \ref{wphaseLGpotp}
for $\sgn(\mu) = \sgn(\lambda)$ and in Figure \ref{wphaseLGpotm} for
$\sgn(\mu) = -\sgn(\lambda)$.  We explain briefly the features of
Figures \ref{wphaseLGpotp} and \ref{wphaseLGpotm}. Firstly, note that
as $w$ crosses the special value $-1$ the various branches among
$(\pm,\pm)$ that the Landau-Ginzburg potential accesses changes. Next,
note that the $(-,-)$ that appears for $\mu\lambda < 0$ never has a
minimum in it. As $w$ crosses one of $a_1$, $a_2$ or $a_3$, new vacua
appear from infinity or already-existing vacua disappear to infinity
in the $(+,+)$, $(+,-)$ or $(-,+)$ branches respectively.  This is due
to the potential becoming linear from being cubic in the branch in
which the vacuum structure changes, as explained in detail around
\eqref{speclin}.


\subsection{Quantum versus Classical} 

In the classical limit $|\lambda| \to 0$ the vacuum structure of the
${\cal N}=1$ theory simplifies. In this limit $a_1$ and $a_3$ in Fig
\ref{wphaseLGpotp} and Figure \ref{wphaseLGpotm} respectively tend to
$-\infty$ and $+\infty$. It follows that in this limit we have either
one or two vacua depending on whether $\mu \lambda w>0$ or
$\mu \lambda w<0$, in agreement with the result of the classical
analysis in Section \ref{class}.

Let us now consider the case $\lambda \neq 0$ but $\lambda$ small. In
this case the quantum vacuum (or phase) structure agrees with the
classical vacuum structure at values of $w$ that are order unity, but
differs from the classical result when $w$ is of order
$\frac{1}{ \lambda}$. In more detail, let us first consider the case
$\lambda \mu>0$, depicted in Figure \ref{wphaseLGpotp}. In this case
we have two classical vacua - one at $\sigma=0$ and one at $\sigma >0$
- when $w<0$, but only one such vacuum (at $\sigma=0$) when $w>0$. As
mentioned above, this result continues to hold in the quantum case
when $w$ is of order unity. Let us now follow the fate of these
`classical' vacua (i.e vacua that have a clear classical counterpart)
as $|w|$ is increased to order $1/\lambda$. When $w$ is taken large
and positive, for $w > a_3$ we find that the single classical vacuum
at $\sigma < 0$ splits up into two vacua, both at $\sigma < 0$ (see
Fig \ref{wphaseLGpotp}). The extra vacuum comes in from infinity when
$w$ crosses $a_3$. On the other hand when $w$ is taken so large and
negative that $w < a_1$ we find that one of the vacua (the unHiggsed
vacuum, namely the continuation of the classical vacuum at $\phi=0$)
goes away to infinity. In this range the quantum theory has only one
vacuum - the Higgsed vacuum. Thus we see that quantum effects, when
sufficiently strong, lead to new vacua coming in from infinity as well
as vacua going away to infinity, including those that exist
classically. Clearly this phenomenon is non-perturbative (as it
happens at values of $w$ of order $1/\lambda$).

The situation is `reversed' when $\mu \lambda <0$.  In this case the
single classical vacuum (the unHiggsed vacuum) that existed for $w<0$
splits into two unHiggsed vacua when $w<a_1$ (i.e.~one extra vacuum
comes in from infinity in the unHiggsed phase of the boson), while the
unHiggsed vacuum (one of the two vacua that exist for $w>0$ in the
classical limit) goes away to infinity when $w> a_3$.

We find the fact that we can reliably track the non-perturbative
appearance and disappearance of supersymmetric vacua as a function of
$w$ quite remarkable. It is important, however, to remember that we
only have the ability to vary $w$ continuously in the strict large $N$
limit. At any finite value of $N$, the parameter $w$ will be forced to
lie at a one of discrete set of points (an issue investigated in some
detail in the soon-to-appear paper \cite{ofer}), and so cannot be
varied continuously.

\section{Acknowledgements}
We would like to thank O.~Aharony, F.~Benini, L.~Janagal, A.~Mishra,
D.~Radicevic and A.~Sharon for useful discussions. We would also like
to thank O.~Aharony and A.~Sharon for sharing an advance version of
their draft \cite{ofer} with us.  The work of A.~D., I.~H., S.~M., and
N.~P. was supported by the Infosys Endowment for the study of the
Quantum Structure of Spacetime. S.~J. would like to thank TIFR, Mumbai
for hospitality during the completion of this work. The work of
S.~J. is supported by the Ramanujan Fellowship. Finally we would all
like to acknowledge our debt to the steady support of the people of
India for research in the basic sciences.

\appendix

\section{A review of the critical fermion and regular boson theories
  and their zero-temperature phase diagrams}\label{critferapp}

The regular boson (RB) theory is defined by the action
\begin{align}
  S_{\rm RB}  &= \int d^3 x  \biggl[\frac{i\kappa_B}{4\pi} \varepsilon^{\mu\nu\rho}
         \Tr( X_\mu\partial_\nu X_\rho -\tfrac{2 i}{3}  X_\mu X_\nu X_\rho)
         + D_\mu \bar \phi D^\mu\phi  \nonumber\\
       &\qquad\qquad\quad
         +m_B^2 \bar\phi \phi +  \frac{4\pi { b}_4}{ \kappa_B}(\bar\phi \phi)^2
         + \frac{4\pi^2}{\kappa_B^2} \left( x^B_6 + 1\right)  (\bar\phi \phi)^3\biggl]\ ,
\label{rst}
\end{align}
while the following action defines the critical fermion (CF) theory
\begin{align}\label{csfnonlinear}
S_{\rm CF}  &= \int d^3 x \bigg[\frac{i\kappa_F}{4\pi} \varepsilon^{\mu\nu\rho}  \Tr( X_\mu\partial_\nu X_\rho -\tfrac{2 i}{3} X_\mu X_\nu
X_\rho) + \bar{\psi} \gamma_\mu D^{\mu} \psi \nonumber\\
&\qquad\qquad\quad
-\frac{4\pi}{\kappa_F}\zeta_F \left(\bar\psi \psi - \frac{\kappa_F
	y_2^2}{4\pi}\right) - \frac{ 4\pi y_4}{\kappa_F} \zeta_F^2 +
\frac{4\pi^2}{\kappa_F^2} x_6^F \zeta_F^3\bigg]\ .
\end{align}
where
\begin{equation}\label{kkmt}
\kappa_B={\sgn}(k_B) \left( |k_B|+N_B \right)\ , \quad \kappa_F
= {\sgn}(k_F) \left( |k_F|+N_F \right)\ ,
\end{equation}
are the renormalized Chern-Simons levels and $k_F$ and $k_B$ are the
levels of the WZW theory dual to the pure Chern-Simons theory. These
two theories are conjectured to be dual to each other under the
following map between the various parameters appearing in the
Lagrangians \eqref{rst}, \eqref{csfnonlinear}:
\begin{equation}
\kappa_B = -\kappa_F\ , \quad N_B=|k_F|\ ,\quad x_6^F =x_6^B\ , \quad y_4 = { b}_4\ , \quad y_2^2 = m_B^2\ .
\label{dualitytransformRBCF}
\end{equation}
The theories can be solved exactly in the large $N$ limit by a
saddle-point computation. For instance, the thermal free energies have
been computed to all orders in the 't Hooft coupling
$\lambda = N / k$. Further, the same computation yields equations for
the pole masses $c_B$ and $c_F$ of the RB and CF theories, as well as
the equations governing the vacuum expectation value of the gauge
invariant operator $\bar\phi \phi$ in the RB theory and the operator
$\zeta_F$ in the CF theory. The thermal free energies and the
equations for the pole masses and vev's map to each other under the
duality map \eqref{dualitytransformRBCF}.

In a previous paper \cite{Dey:2018ykx}, a three variable off-shell
generalisation of the thermal free energy of the RB (CF) theory was
presented, which upon extremization with respect to its three
variables, yielded the saddle-point equations for the pole mass $c_B$
($c_F$) and the vev of $\bar\phi\phi$ ($\zeta_F$). We present this
off-shell generalisation below, and the corresponding saddle-point
equations for the regular boson theory first.

\subsection{The regular boson theory}

\subsubsection{The thermal free energy}

We define the variable $\sigma_B$ as short-hand for the vacuum
expectation value of the operator $\bar\phi\phi$:
\begin{equation}\label{sigmadef}
  \sigma_B = \frac{2\pi}{N_B} \langle \bar\phi \phi\rangle \ .
\end{equation}
The three-variable off-shell free energy for the regular boson theory
is then given by (see equation (4.2) of \cite{Dey:2018ykx}),
\begin{align}\label{FoffRB}
  &F(c_B, \sigma_B,  \tl\cS)\nonumber\\
  &= \frac{N_B }{6\pi} \Bigg[-3 {\hat c}_B^2 {\hat \sigma}_B + \lambda_B^2 {\hat \sigma}_B^3 +  3\left( {\hat m}_B^2 {\hat \sigma}_B  + 2\lambda_B {\hat b}_4 {\hat \sigma}_B^2 +  (x_6 + 1)  \lambda_B^2 {\hat \sigma}_B^3\right) \nonumber\\
  &\qquad\qquad - 4\lambda_B^2 ({\tl \cS} + {\hat \sigma}_B)^3 + 6|\lambda_B|{\hat c}_B ({\tl \cS} + {\hat \sigma}_B)^2 \nonumber\\
  &\qquad\qquad - {\hat c}_B^3 + 3 \int_{-\pi}^{\pi}d\alpha\, \rho_{B}(\alpha)\,\int_{{\hat c}_B}^{\infty} dy ~y~\left( \ln\left(1-e^{-y-i\alpha}\right)  + \ln\left(1-e^{-y+i\alpha}\right)\right)\Bigg]\ ,
\end{align}
where $\rho_B(\alpha)$ is the distribution function for the holonomy
$\alpha$ of the $SU(N)$ gauge field around the thermal circle. The
last line in \eqref{FoffRB} arises from the one-loop determinant of
the fundamental bosonic field.

Note that \eqref{FoffRB} is a function of three `field' variables,
namely $c_B$, $\sigma_B$ and ${\tilde\cS}$ (the hats on some of the
variables indicate that they have been scaled by appropriate powers of
the temperature to make them dimensionless). We use the same notation
for the off-shell variables $c_B$ and $\sigma_B$ and their
saddle-point values which are the pole mass of the fundamental bosonic
excitation and the vacuum expectation value \eqref{sigmadef}
respectively. We do not have a physical interpretation of the variable
${\tl \cS}$ but, as we shall see below, it becomes the following
function of the pole mass $c_B$ on-shell:
\begin{equation}
\cS(c_B) = \frac{1}{2} \int_{-\pi}^\pi d\alpha\,\rho_B(\alpha)   \( \log(2 \sinh (\tfrac{\hat{c}_B + i\alpha}{2})) + \log(2 \sinh (\tfrac{\hat{c}_B - i\alpha}{2})) \)\ ,
\end{equation}
Extremizing \eqref{FoffRB} w.r.t. ${\tilde\cS}$, $c_B$ and $\sigma_B$
respectively yields the equations (see equation (4.3) of \cite{Dey:2018ykx})
\begin{align} \label{foll}
&({\tilde\cS}+{\hat \sigma}_B) (-{\hat c}_B + |\lambda_B| ({\tilde\cS}+{\hat \sigma}_B))=0\ ,\nonumber\\
&{\hat c}_B (\cS(c_B)+{\hat \sigma}_B)-|\lambda_B| ({\tilde\cS}+{\hat \sigma}_B)^2=0\ ,\nonumber\\
&{\hat c}_B^2 - {\hat m}_B^2-4 {\hat c}_B |\lambda_B|  ({\tilde\cS}+{\hat \sigma}_B) +\lambda_B \left(4 {\tilde\cS}^2 \lambda_B- 4 {\hat b}_4 {\hat \sigma}_B+ 8 \lambda_B {\hat \sigma}_B {\tilde\cS}-3 \lambda_B {\hat \sigma}_B^2 x_6^B \right)=0\ .
\end{align}
The first equation offers us a choice between two solutions:
\begin{align}\label{foll2}
  \text{unHiggsed}(\text{uH}):&\quad  {\tl \cS} = -{\hat\sigma}_B\ ,\nonumber\\
  \text{Higgsed}(\text{H}):&\quad {\tl \cS} =  \frac{{\hat c}_B}{|\lambda_B|} - {\hat \sigma}_B\ .
\end{align}
Substituting either of the two solutions for ${\tl \cS}$ above into
the second equation of \eqref{foll} gives ${\tl \cS} = \cS(c_B)$
on-shell as claimed earlier. More importantly, we get two expressions
for the thermal free energy as a function of two variables $c_B$ and
$\sigma_B$:
\begin{align}\label{feuhnh}
  F^{\rm (uH)}(c_B, \sigma_B) &= \frac{N_B}{2\pi}\Bigg[-\frac{1}{3}{\hat c}_B^3 -  {\hat c}_B^2 {\hat \sigma}_B +  {\hat m}_B^2 {\hat \sigma}_B + 2 \lambda_B {\hat b}_4 {\hat \sigma}_B^2 + (\tfrac{4}{3} + x_6^B) \lambda_B^2 {\hat \sigma}_B^3\nonumber\\
                              &\qquad\qquad  +  \int_{-\pi}^{\pi}d\alpha\, \rho_{B}(\alpha)\,\int_{{\hat c}_B}^{\infty} dy\,y\,\left( \ln\left(1-e^{-y-i\alpha}\right)  + \ln\left(1-e^{-y+i\alpha}\right)\right)\Bigg]\ ,\nonumber\\
  F^{\rm (H)}(c_B, \sigma_B) &= \frac{N_B}{2\pi}\Bigg[-\frac{|\lambda_B|-2}{3|\lambda_B|}{\hat c}_B^3 - {\hat c}_B^2 {\hat \sigma}_B + {\hat m}_B^2 {\hat \sigma}_B + 2 \lambda_B {\hat b}_4 {\hat \sigma}_B^2 + (\tfrac{4}{3} + x_6^B) \lambda_B^2 {\hat \sigma}_B^3\nonumber\\
                              &\qquad\qquad  + \int_{-\pi}^{\pi}d\alpha\, \rho_{B}(\alpha)\,\int_{{\hat c}_B}^{\infty} dy\,y\,\left( \ln\left(1-e^{-y-i\alpha}\right)  + \ln\left(1-e^{-y+i\alpha}\right)\right)\Bigg]\ .
\end{align}
The interpretation of the two branches of solution in \eqref{foll2}
(and consequently the two branches of free energy) is that one of them
corresponds to the unHiggsed phase and the other corresponds to the
Higgsed phase. By comparing with direct computations of the free
energies in both the phases (\cite{Aharony:2012ns,Jain:2013py} for the
unHiggsed phase and \cite{Dey:2018ykx} for the Higgsed phase), we
observe that the first equation in \eqref{foll2} corresponds to the
unHiggsed phase and the second to the Higgsed phase.

Extremizing the expressions in \eqref{feuhnh} w.r.t.~$c_B$ gives the
equations for $\sigma_B$ in \eqref{foll2} with ${\tl \cS}$ replaced by
its on-shell value $\cS(c_B)$:
\begin{align}\label{foll2pr}
  \text{unHiggsed}:&\quad  {\hat\sigma}_B = -\cS(c_B)\ ,\nonumber\\
  \text{Higgsed}:&\quad {\hat \sigma}_B = \frac{{\hat c}_B}{|\lambda_B|} - {\cS}(c_B)\ .
\end{align}
Extremizing \eqref{feuhnh} w.r.t.~$\sigma_B$ gives the following
equations for $c_B$ which are the equations for the pole mass of the
fundamental bosonic excitation in the respective phases:
\begin{align}\label{foll3}
  \text{unHiggsed}:&\quad {\hat c}_B^2 - {\hat m}_B^2  + 4  {\hat b}_4 \lambda_B \cS - (4 + 3x_6^B) \lambda_B^2 \cS^2 = 0\ ,\nonumber\\
\text{Higgsed}:&\quad {\hat c}_B^2 - {\hat m}_B^2 + 4  {\hat b}_4 ( \lambda_B \cS - \sgn(\lambda_B) {\hat c}_B) - (4 + 3 x_6^B)( \lambda_B \cS - \sgn(\lambda_B) {\hat c}_B)^2  = 0\ .
\end{align} 
Solving the theory then amounts to solving the above equations for
$c_B$ in the respective phases and plugging back into \eqref{foll2pr}
to obtain the corresponding values of $\sigma_B$. Once we have the
values of $c_B$ and $\sigma_B$, we plug them into the free energy
expressions in \eqref{feuhnh} to obtain the on-shell thermal free
energies in the two phases.

\subsubsection{The zero temperature LG potential}

At zero temperature, the free energies and saddle-point equations
becomes very simple. The simplifications are as follows:
\begin{enumerate}
\item The term with the integral over $y$ in the second line of the
  free energy expressions vanishes.
\item The quantity $\cS(c_B)$ simplifies to
  \begin{equation}\label{cbzero}
    \cS(c_B) = \frac{{\hat c}_B}{2}\ ,
  \end{equation}
  which in turn simplifies the equations \eqref{foll2pr}:
  \begin{align}\label{foll2przero}
  \text{unHiggsed}:&\quad  {\sigma}_B = -\frac{c_B}{2}\ ,\nonumber\\
  \text{Higgsed}:&\quad { \sigma}_B = \frac{{ c}_B}{|\lambda_B|} - \frac{c_B}{2}\ ,
  \end{align}
  and equations \eqref{foll3}:
  \begin{align}\label{foll3zero}
    \text{unHiggsed}:&\quad \left(1 - \lambda_B^2(1 + \tfrac{3}{4}x_6^B)\right){ c}_B^2 - 2  \lambda_B { b}_4 c_B - { m}_B^2  = 0\ ,\nonumber\\
    \text{Higgsed}:&\quad \left(1 - \left(\tfrac{2-|\lambda_B|}{2|\lambda_B|}\right)^2\lambda_B^2(1 + \tfrac{3}{4}x_6^B)\right) { c}_B^2 - 2 \lambda_B { b}_4 \left( \tfrac{2-|\lambda_B|}{|\lambda_B|}\right) c_B  - { m}_B^2 = 0\ .
  \end{align} 
  Thus, the quadratic equation \eqref{foll3zero} determines the pole
  mass $c_B$ and the linear equation \eqref{foll2przero} determines
  the vacuum expectation value $\sigma_B$.
  
\item As a consequence of \eqref{foll2przero} we see that $\sigma_B$
  can assume only negative values in the unHiggsed phase and only
  positive values in the Higgsed phase since $c_B$ is by definition a
  positive quantity.
\end{enumerate}

It is useful and conceptually attractive to rewrite the zero
temperature limit of the two-variable free energy \eqref{feuhnh} in
terms of $\sigma_B$ alone using the simple relations
\eqref{foll2przero} between $c_B$ and $\sigma_B$. As one can see from
the expressions in \eqref{feuhnh}, this changes the coefficient of the
$\sigma_B^3$ term in the free energy.

The resulting function of $\sigma_B$ is precisely the quantum
effective potential for the field $(\bar\phi\phi)_{\rm cl}$
corresponding to the expectation value of the gauge invariant operator
$\bar\phi\phi$ (cf. Section \ref{qea} in the main text). The effective
potential turns out to be (see equation (5.10) in \cite{Dey:2018ykx})
\begin{align}\label{RBlgpot}
&U_{\rm RB}(\sigma_B) = \left\{\arraycolsep=1.4pt\def\arraystretch{2.2}\begin{array}{cl} \displaystyle \frac{N_B}{2 \pi} 
\left[ (x_6^B - \phi_+) \lambda_B^2 { \sigma}_B^3 + 2 \lambda_B b_4 { \sigma}_B^2  +  m_B^2 { \sigma}_B \right] & \quad\text{for }\sigma_B < 0\ , \\  \displaystyle \frac{N_B}{2\pi}\left[( x_6^B - \phi_-) \lambda_B^2 { \sigma}_B^3 +  2 \lambda_B{ b}_4  { \sigma}_B^2 + { m}_B^2 { \sigma}_B\right]  & \quad\text{for }\sigma_B > 0\ , \end{array}\right.
\end{align}
where $\phi_\pm$ are the following functions of $|\lambda_B|$:
\begin{equation}\label{phipmapp}
  \phi_+ = \frac{4}{3}\left(\frac{1}{|\lambda_B|^2}-1\right)\ ,\quad \phi_- = \frac{4}{3}\left(\frac{1}{(2-|\lambda_B|)^2}-1\right)\ ,
\end{equation}
with the ordering $\phi_- < 0 < \phi_+$ for all values of
$|\lambda_B|$. These functions are referred to as $\phi_1$ and
$\phi_2$ respectively in the earlier work
\cite{Aharony:2018pjn,Dey:2018ykx}. The solutions of the saddle-point
equations \eqref{foll3zero} and \eqref{foll2przero} can then be
phrased as the problem of minimisation of the quantum effective
potential $U_{\rm RB}(\sigma_B)$.

\subsubsection{Stability of the regular boson theory}

The stability of the regular boson theory can be studied very easily
using the effective potential in \eqref{RBlgpot}. The potential is
stable (meaning, it increases without bound as $|\sigma_B|$ increases)
when the marginal (at large $N_B$) parameter $x_6^B$ satisfies the
following bound
\begin{equation}\label{RBstab}
  \phi_- < x_6^B < \phi_+\ .
\end{equation}
When the above bound is not satisfied, the potential either has an
instability for large and negative $\sigma_B$ (when $x_6^B > \phi_+$)
or for large and positive $\sigma_B$ (when $x_6^B < \phi_-$).

\subsubsection{The zero temperature phase diagram}\label{RBphase}

\begin{figure}
	\begin{subfigure}{0.31\textwidth}
		\scalebox{0.65}{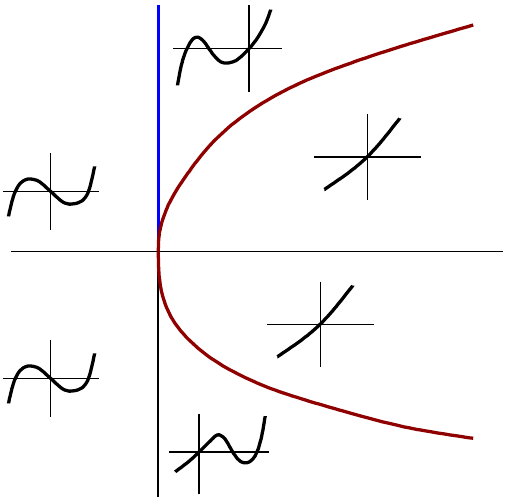}
		\caption{$x_6^B < \phi_-$}
	\end{subfigure}\hspace{10pt}
	\begin{subfigure}{0.31\textwidth}
		\scalebox{0.65}{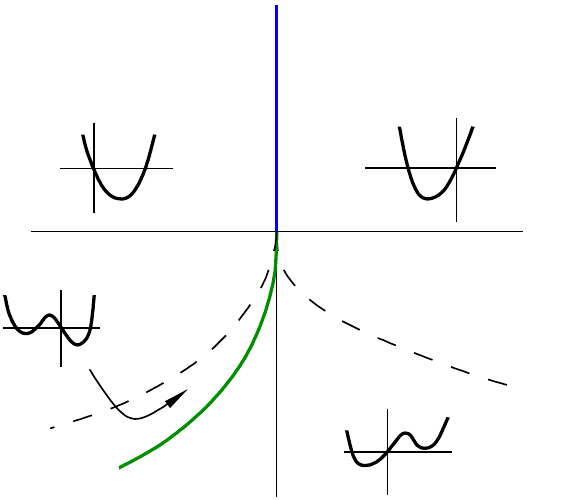}
		\caption{$\phi_- < x_6^B < \phi_+$}
              \end{subfigure}\hspace{10pt}
	\begin{subfigure}{0.31\textwidth}
		\scalebox{0.65}{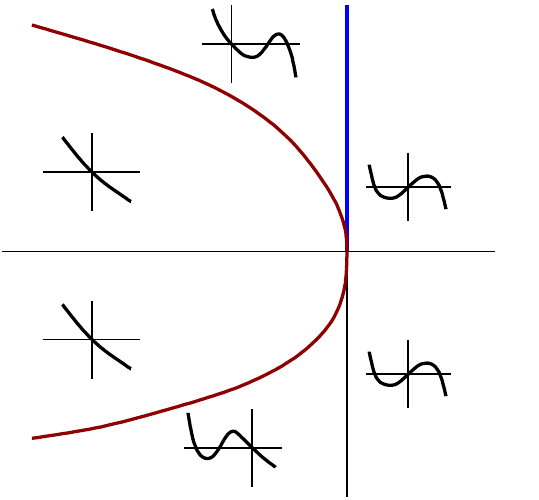}
		\caption{$\phi_+ < x_6^B$}
	\end{subfigure}
	\caption{The phase diagram of the RB theory for various
          choices of the parameter $x_6^B$. The LG potential
          \eqref{RBlgpot} is plotted as a small inlay in each region
          of parameter space. In each inlay, the range $\sigma_B < 0$
          corresponds to the unHiggsed branch of the potential while
          the range $\sigma_B > 0$ corresponds to the Higgsed
          branch. The curves demarcating different regions of
          parameter space are explained in the text.}
        \label{phaseappRB}
\end{figure}

In the large $N_B$ limit, the classically marginal parameter $x_6^B$
in the regular boson action \eqref{rst} remains marginal at all orders
in the 't Hooft coupling. Hence, one can study the theory by tuning
the value of $x_6^B$ to any value that one wishes. The theory also
possesses two relevant parameters $\lambda_B b_4$ and $m_B^2$. Thus,
given a value of $x_6^B$, one can study the phase structure of the
theory by turning on arbitrary initial values of the relevant
parameters $\lambda_B b_4$ and $m_B^2$. The Landau-Ginzburg potential
\eqref{RBlgpot} depends on these parameters in a simple way and one
can qualitatively sketch these potentials for different choices of the
initial values of $\lambda_B b_4$ and $m_B^2$. Local extrema of the
potential then correspond to the 'phases'\footnote{We use the word
  phase to mean local minima as well as local maxima in this paper.}
of the theory. We summarize our findings pictorially in Figure
\ref{phaseappRB} for the three different ranges of $x_6^B$ given by
$x_6^B < \phi_-$, $\phi_- < x_6^B < \phi_+$ and $\phi_+ < x_6^B$ (For detailed analysis see \cite{Dey:2018ykx}, Figure 6, 8 and 11 in \cite{Dey:2018ykx} correspond to the three different ranges of $x_6^B$).

We next explain the different curves that appear in the plots in
Figure \ref{phaseappRB}. The blue curve $L$ corresponds to a minimum
crossing over from the unHiggsed branch to the Higgsed branch or vice
versa. From the LG potential \eqref{RBlgpot}, it clear that this
happens when to $m_B^2$ changes sign, and further, when
$\lambda_B b_4$ is positive, a minimum crosses $\sigma_B = 0$. Thus, the equation of the blue curve $L$ is
\begin{equation}
  L:\quad m_B^2 = 0\ ,\quad \lambda_B b_4 > 0\ .
\end{equation}
Similarly, the negative $\lambda_B b_4$ axis corresponds to a maximum
crossing $\sigma_B = 0$. Though not shown in Figure \ref{phaseappRB}, it
is an important analytic feature of the LG potential and we label it
\begin{equation}
  M:\quad m_B^2 = 0\ ,\quad \lambda_B b_4 < 0\ .
\end{equation}
The green curve $D_\nu$ is a parabola which separates regions of
parameter space where there are two competing minima for the
potential. Across the green curve, the dominant minimum changes from
the unHiggsed branch to the Higgsed branch or vice versa. The
curvature of the green parabola is computed numerically by comparing
the values of the LG potential at the two competing minima and
checking where they are equal.

The red curves and the dashed curves are parabolas across which one
pair of minimum and maximum vanishes or a new pair is created. The
curve $D_u$ corresponds to the creation of a new minimum and
maximum in the unHiggsed phase and its equation is given by studying
the discriminant of the extremization equation of $U_{\rm RB}$ in the
unHiggsed phase:
\begin{equation}
  D_u:\quad 16 (\lambda_B b_4)^2 - 12 \lambda_B^2  (x_6^B - \phi_+) m_B^2 = 0\ .
\end{equation}
Similarly, the parabola $D_h$ corresponds to the case where the new
set of extrema develop in the Higgsed branch of the potential. Its
equation is similarly given by
\begin{equation}
  D_h:\quad 16 (\lambda_B b_4)^2 - 12 \lambda_B^2  (x_6^B - \phi_-) m_B^2 = 0\ .
\end{equation}

\begin{figure}
	\centering
	\scalebox{0.8}{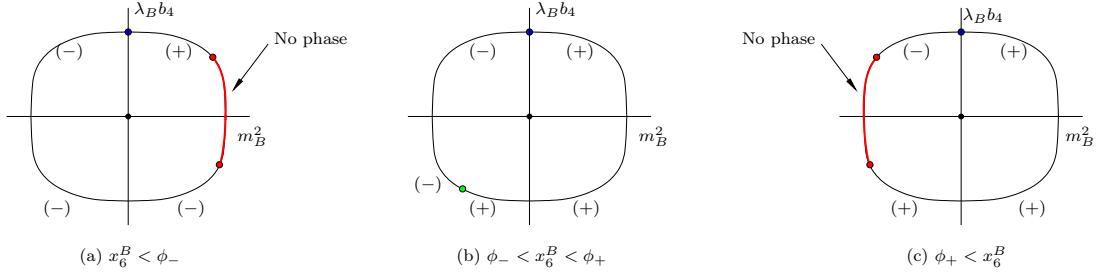}
	\caption{The phase diagram of the regular boson theory. The
          unHiggsed phase is labelled by $(+)$ and the Higgsed phase
          is labelled by $(-)$. The blue dot on the positive
          $\lambda_B b_4$ axis corresponds to a second order phase
          transition governed by the CB conformal theory. The red
          regions in figures (a) and (c) are regions where there e is
          no phase and the potential is monotonic. The green dot in
          figure (b) is a first order phase transition.}
              \label{ellipsephase}
\end{figure}

Finally, the true phase diagram is in fact only a function of the
dimensionless ratio
\begin{equation}
  \frac{m_B^2}{(\lambda_B b_4)^2}\ ,
\end{equation}
since one of the two relevant parameters can be exchanged for the
energy scale of the theory. Thus, the true phase diagram, for example,
can be obtained by taking an ellipsoidal section of the diagrams in
Figure \ref{phaseappRB}:
\begin{equation}\label{2del}
  (\lambda_B b_4)^4 + (m_B^2)^2 = \text{a large positive constant}.
\end{equation}
We give this phase diagram as a function of $x_6^B$ in Figure
\ref{ellipsephase}.

\subsection{The critical fermion theory}
The thermal free energy for the critical fermion theory was computed
in \cite{Minwalla:2015sca} and an off-shell version was presented in
\cite{Dey:2018ykx} which we reproduce here:
\begin{align}\label{offshcf}
&F_F(c_F, \zeta_F, \tl\cC)\nonumber\\
  &=\frac{N_F }{6\pi} \Bigg[ -8 \lambda_F^2 {\tl \cC}^3 + 6\lambda_F {\tl \cC}^2 \Big(\frac{4\pi{\hat \zeta}_F}{\kappa_F}\Big) - 3{\tl \cC} \bigg({\hat c}_F^2 - \Big( 2 \lambda_F {\tl \cC} -\frac{4\pi\hat\zeta_F}{\kappa_F} \Big)^2\bigg)\nonumber\\
&\qquad\qquad + 3 \bigg(\frac{{\hat y}_2^2}{2\lambda_F}\frac{4\pi {\hat \zeta}_F }{\kappa_F}-\frac{{\hat y}_4}{2\lambda_F} \Big(\frac{4 \pi {\hat \zeta}_F }{\kappa_F}\Big)^2+\frac{x_6^F}{8\lambda_F}\Big(\frac{4 \pi {\hat \zeta}_F}{\kappa_F}\Big)^3\bigg)  \nonumber \\ 
&\qquad\qquad + {\hat c}_F^3 -3 \int_{-\pi}^{\pi}d\alpha\, \rho_{F}(\alpha)\int_{{\hat c}_F}^{\infty} dy\,y\,\left( \ln\left(1+e^{-y-i\alpha}\right)  + \ln\left(1+e^{-y+i\alpha}\right)\right)\Bigg].
\end{align}
It is straightforward to verify that the above expression can be
obtained by applying the duality map \eqref{dualitytransformRBCF} and
making the following field redefinitions:
\begin{align}\label{fieldredefRBCF}
  &{\hat c}_B = {\hat c}_F\ ,\quad \lambda_B {\tilde \cS} = -\frac{ \sgn(\lambda_F)}{ 2 } \hat{c}_F +\lambda_F {\tilde \cC}\ ,\quad 2 \lambda_B \sigma_B = -\frac{4\pi \zeta_F}{\kappa_F}\ ,
\end{align}
along with the following redefinition of the holonomy distribution function
\begin{equation}
  |\lambda_B|\rho_B(\alpha) = \frac{1}{2\pi} - |\lambda_F|\rho_F(\pi - \alpha)\ .
\end{equation}
The saddle point value of the variable $c_F$ is the pole mass of the
fundamental fermionic excitation which we also denote by the same
symbol. The variable $\zeta_F$ is precisely the (constant part of the)
field $\zeta_F$ that appears in the critical fermion Lagrangian
\eqref{csfnonlinear}. We do not have an interpretation for the
variable ${\tilde \cC}$ (as in the case of the regular boson theory's
${\tilde \cS}$). However, as one shall see below, it becomes the following function of $c_F$ at the saddle point:
\begin{equation}
  \cC(c_F) = \frac{1}{2} \int_{-\pi}^\pi d\alpha\,\rho_F(\alpha)   \( \log(2 \cosh (\tfrac{\hat{c}_F +i\alpha}{2}))+ \log(2 \cosh (\tfrac{\hat{c}_F - i\alpha}{2})) \)\ .
\end{equation}
Extremizing the off-shell expression \eqref{offshcf} w.r.t.~$c_F$,
${\tilde \cC}$ and $\zeta_F$ gives the following equations:
\begin{align}\label{CFsaddle}
  &{\tilde \cC} = \cC(c_F)\ ,\nonumber\\
  &{\hat c}_F^2 = \bigg(2\lambda_F {\tilde \cC} - \frac{4\pi {\hat \zeta}_F}{\kappa_F}\bigg)^2\ ,\nonumber\\
  &4\lambda_F^2 {\tilde \cC}^2 - 4\lambda_F {\tilde \cC} \frac{4\pi {\hat \zeta}_F}{\kappa_F} - {\hat y}_2^2 + 2{\hat y}_4 \frac{4\pi{\hat \zeta}_F}{\kappa_F} - \frac{3 x_6^F}{4} \Big(\frac{4\pi{\hat \zeta}_F}{\kappa_F}\Big)^2 = 0\ .
\end{align}
Since $c_F$ is positive by definition, the second equation above can
be rewritten as
\begin{align}
  {\hat c}_F = |X_F|\ ,\quad X_F=  2\lambda_F \cC(c_F) - \frac{4\pi{\hat \zeta}_F}{\kappa_F}\ ,
\end{align}
The quantity $X_F$ is to be thought of as the thermal effective mass
for the fundamental fermion since it appears in the thermal two-point
function in the correct place for an effective mass \footnote{In fact,
  the second term in
  $X_F = 2\lambda_F \cC(c_F) - \frac{4\pi{\hat \zeta}_F}{\kappa_F}$
  appears in the tree-level action as the coefficient of
  $\bar\psi\psi$.}. The phase of a Chern-Simons gauged fermion with
effective mass $X_F$ is determined by the sign
\begin{equation}\label{ferphase}
\varepsilon_F = \sgn(X_F) \sgn(\lambda_F) = \pm 1\ .
\end{equation}
The sign is correlated with the shift in the level of the low-energy
pure Chern-Simons theory w.r.t.~the level in the ultraviolet. In terms
of the sign $\varepsilon_F$, the second saddle-point equation in
\eqref{CFsaddle} becomes
\begin{align}\label{cfgapapp}
\varepsilon_F = \pm\ \text{phase}:\quad  {\hat c}_F = \varepsilon_F \left(2|\lambda_F| {\tl \cC} - \frac{4\pi {\hat \zeta}_F}{|\kappa_F|}\right)\ .
\end{align}
We next solve for ${\tl \cC}$ in terms of $c_F$ and $\zeta_F$ to
substitute back in the off-shell free energy \eqref{offshcf}
\begin{align}
  2|\lambda_F|{\tl \cC} = \varepsilon_F {\hat c}_F + \frac{4\pi {\hat \zeta}_F}{|\kappa_F|}\ .
\end{align}
Plugging this into \eqref{offshcf}, we get the free energies in the
$\varepsilon_F = \pm$ phases as functions of two variables $c_F$ and
$\zeta_F$:
\begin{align}\label{2varcfos}
&F_F^{(\varepsilon_F)}(c_F, \zeta_F)\nonumber\\
  &=\frac{N_F }{6\pi} \Bigg[-\frac{\varepsilon_F}{|\lambda_F|} {\hat c}_F^3 - \frac{3}{2\lambda_F}{\hat c}_F^2\Big(\frac{4\pi{\hat \zeta}_F}{\kappa_F}\Big) + \frac{1}{2\lambda_F}\Big(\frac{4\pi{\hat \zeta}_F}{\kappa_F}\Big)^3\nonumber\\
&\qquad\qquad + 3 \bigg(\frac{{\hat y}_2^2}{2\lambda_F}\frac{4\pi {\hat \zeta}_F }{\kappa_F}-\frac{{\hat y}_4}{2\lambda_F} \Big(\frac{4 \pi {\hat \zeta}_F }{\kappa_F}\Big)^2+\frac{x_6^F}{8\lambda_F}\Big(\frac{4 \pi {\hat \zeta}_F}{\kappa_F}\Big)^3\bigg)  \nonumber \\ 
&\qquad\qquad + {\hat c}_F^3 -3 \int_{-\pi}^{\pi}d\alpha\, \rho_{F}(\alpha)\int_{{\hat c}_F}^{\infty} dy\,y\,\left( \ln\left(1+e^{-y-i\alpha}\right)  + \ln\left(1+e^{-y+i\alpha}\right)\right)\Bigg].
\end{align}
where $c_F$ and $\zeta_F$ satisfy the saddle-point equations in
\eqref{CFsaddle}.

In the zero temperature limit, the sign $\varepsilon_F$ simplifies to
\begin{equation}
  \varepsilon_F = -\sgn(\zeta_F)\ ,
\end{equation}
and the equation for the gap $c_F$ in \eqref{cfgapapp} simplifies to
\begin{equation}
  c_F = \frac{|4\pi \zeta_F|}{|\kappa_F|}\frac{1}{1 - \varepsilon_F |\lambda_F|}\ .
\end{equation}
Plugging the expression for $c_F$ above into the two-variable free
energy \eqref{2varcfos} we obtain an explicit Landau-Ginzburg quantum
effective potential as a function of $\zeta_F$ for the two phases as
\begin{equation}\label{critpotapp}
U_{\rm CF}(\zeta_F)
=\left\{\def\arraystretch{2.2}\begin{array}{ll}\frac{N_F}{2\pi}\left[\frac{(x_6^{F} - \psi_+)}{\lambda_F} \left(\frac{2\pi \zeta_F}{\kappa_F}\right)^3 - \frac{2 y_4}{\lambda_F} \left(\frac{2\pi \zeta_F}{\kappa_F}\right)^2 + \frac{y_2^2}{\lambda_F} \left(\frac{2\pi \zeta_F}{\kappa_F}\right) \right]\ ,\quad & \zeta_F < 0\ , \\ \frac{N_F}{2\pi}\left[\frac{(x_6^{F} - \psi_-)}{\lambda_F} \left(\frac{2\pi \zeta_F}{\kappa_F}\right)^3 - \frac{2 y_4}{\lambda_F} \left(\frac{2\pi \zeta_F}{\kappa_F}\right)^2 + \frac{y_2^2}{\lambda_F} \left(\frac{2\pi \zeta_F}{\kappa_F}\right) \right]\ ,\quad & \zeta_F > 0\ ,\end{array}\right.
\end{equation}
where the quantities $\psi_\pm$ are defined as
\begin{align}\label{psipmapp}
  &\psi_- = \frac{4}{3}\left(\frac{1}{(1+|\lambda|)^2} - 1\right)\ ,\quad \psi_+ = \frac{4}{3} \left(\frac{1}{(1 - |\lambda|)^2} - 1\right)\ .
\end{align}
Note that these map to the functions $\phi_\pm$ defined in
\eqref{phipm} under the duality
\eqref{dualitytransformRBCF}. Moreover, under the field redefinitions
\eqref{fieldredefRBCF} and the duality map
\eqref{dualitytransformRBCF}, the above LG potential maps exactly to
the LG potential for the regular boson theory given in
\eqref{RBlgpot}. 

The stability analysis can be done in the same way as for the regular
boson and we come to the conclusion that the critical fermion theory
is stable when $x_6^F$ satisfies
\begin{equation}\label{CFstab}
  \psi_- < x_6^F < \psi_+\ .
\end{equation}
Note that, as expected, this maps to the stability condition for the
regular boson theory given in \eqref{RBstab}.

Similarly, The phase diagram of the critical fermion theory looks
exactly like that of the regular boson theory under the usual change
of variables given by \eqref{dualitytransformRBCF}.

In Section \ref{scaling} in the main body of the paper, we use the
alternate notation $\zeta = \frac{4\pi \zeta_F}{\kappa_F}$ to make
expressions more compact. In terms of this variable, the effective
potential becomes
\begin{equation}
  U_{\rm CF}(\zeta)
  =\left\{\def\arraystretch{2.2}\begin{array}{ll}\displaystyle\frac{N_F}{2\pi}\left[(x_6^{F} - \psi_+) \frac{\zeta^3}{8 \lambda_F} - \frac{ y_4}{2 \lambda_F} \zeta^2 + \frac{ y_2^2}{2\lambda_F}\zeta \right]\ ,\quad & \sgn(\zeta)\sgn(\lambda_F) < 0\\ \displaystyle \frac{N_F}{2\pi}\left[(x_6^{F} - \psi_-) \frac{\zeta^3}{8 \lambda_F} - \frac{ y_4}{2 \lambda_F} \zeta^2 + \frac{ y_2^2}{2\lambda_F}\zeta \right]\ ,\quad & \sgn(\zeta)\sgn(\lambda_F) > 0\ . \end{array}\right.
\end{equation}

\section{The thermal free energy of the Chern-Simons matter theory
  with one boson and one fermion}\label{federapp}
We start with the action \eqref{generalaction} which we reproduce here
for convenience:
\begin{align}
S  &= \int d^3 x  \biggl[ \frac{i\kappa}{4 \pi}\epsilon^{\mu\nu\rho}
\Tr( X_\mu\partial_\nu X_\rho -\tfrac{2 i}{3}  X_\mu X_\nu X_\rho) \nonumber\\
&+  \overline{D_\mu  \phi}  D^\mu\phi + \bar\psi \gamma^\mu D_\mu \psi 
+m_B^2 \bar\phi \phi + m_F \bar\psi \psi + \frac{4\pi b_4}{ \kappa} (\bar\phi \phi)^2 
+ \frac{4 \pi^2 (x_6+1)}{\kappa^2} (\bar\phi\phi)^3 \nonumber\\
&
+ \frac{4 \pi x_4}{\kappa} (\bar\psi \psi) (\bar\phi\phi)
+ \frac{2 \pi (y_4'-3)}{\kappa} (\bar\psi\phi)( \bar\phi \psi)
+ \frac{2 \pi y_4''}{\kappa} \left((\bar\psi \phi)( \bar \psi \phi ) 
+(\bar \phi \psi)( \bar \phi \psi )\right) \biggl].
\label{generalactionapp}
\end{align}
We shall be interested in the Higgsed phase of the boson $\phi$ and
work in the unitary gauge
\begin{equation}
  \phi = \begin{pmatrix} 0 \\ \vdots \\ 0 \\ \sqrt{|\kappa|} V \end{pmatrix}\ ,
\end{equation}
which explicitly breaks the $SU(N)$ gauge invariance to $SU(N-1)$. It
is helpful to reorganise the fermion and the gauge field to suit this
choice of gauge:
\begin{equation}
  \psi=\begin{pmatrix} \psi_a \\ \sqrt{\kappa} \ \psi_N \end{pmatrix} \ ,\quad X = \begin{pmatrix} A_a{}^b & \frac{1}{\sqrt{\kappa}}W_a \\ \frac{1}{\sqrt{\kappa}}\bar{W}{}^b & Z \end{pmatrix}\ ,
\end{equation}
where the $a,b$ indices runs over $1$ to $N-1$ (Eventually, we choose
to work in the lightcone gauge $A_- = 0$ for the $SU(N-1)$ gauge
field).

The above action action can be written in terms of the new field
variables $A$, $Z$, $W$, $\psi$, $\psi_N$, $V$ as
\begin{align}\label{asb}
  S_{\text{E}}
  &= \frac{i \kappa}{4 \pi}\int \Tr\left(A d A - \tfrac{2i}{3} A^3\right) + \frac{i }{4 \pi}\int  \Big[2 \bar{W}\! D W+\kappa ZdZ - 2iZ\bar{W}W\Big] \nonumber\\ 
  &\quad + \int d^3x (|\kappa| V^2 Z_\mu Z^\mu + {\sgn}(\kappa) V^2 \bar{W}_\mu W^\mu)\nonumber\\
  &\quad+|\kappa| \int d^3x \Big[  \partial_\mu V \partial^\mu V + m_B^2 V^2 + 
    4\pi \sgn(\kappa) b_4 V^4 + 4\pi^2 (x_6+1) V^6 \Big]\nonumber\\ 
  &\quad+\int d^3x \Big[\bar\psi^a \gamma^\mu D_\mu \psi_a + (m_F + 4\pi  {\sgn}(\kappa) x_4 V^2)\bar{\psi}^a\psi_a\Big]\nonumber\\
  &\quad + \int d^3 x \Big[\kappa \bar\psi^N \gamma^\mu\partial_\mu \psi_N + \bar\psi^N \bar{W}{}^a \psi_a + \bar\psi^a W_a \psi_N\Big]\nonumber\\
  &\quad +  \int d^3x \Big[|\kappa|(4\pi x_4  + 2\pi (y'_4-3))V^2 + \kappa m_F + \kappa Z \Big]\bar\psi^N \psi_N \nonumber\\
  &\quad + \int d^3x \Big[ 2\pi y''_4 |\kappa| V^2\Big]  (\bar\psi^N \bar\psi^N  + \psi_N \psi_N)\ .
\end{align}
where $D_\mu = \partial_\mu -i A_\mu$ and the exterior product $ABC$
stands for $ d^3 x \epsilon^{\mu\nu \rho} A_\mu B_\nu C_\rho$. In the
large $N$ limit, the singlet fermionic fields $\bar\psi^a W_a$,
$\psi_N$ and their complex conjugates can be set to zero since the
path integral over them is dominated by their classical values
viz.~zero. When $N$ is finite but large, these singlet fermions (and
the signs of the couplings $y_4'$ and $y_4''$) are important for the
duality to work as, for instance, explained in \cite{Jensen:2017bjo,
  Benini:2017aed}.

We break up the resulting action $S_{\text{E}}[A,W,Z,\psi,V]$ into two
parts
\begin{equation}
\label{betp}
S_{\rm E}[A,W,Z,\psi,V]= S_1[A,W,Z,\psi,V] + S_2[V]\ ,
\end{equation}
where 
\begin{align}\label{asb1}
  S_1[A,W,Z,\psi,V]&=\frac{i \kappa}{4 \pi}\int d^3x\, \epsilon^{\mu\nu\rho}\,\Tr\left(A_\mu\partial_\nu A_\rho - \frac{2i}{3}A_\mu A_\nu A_\rho\right)\nonumber\\
  &\quad + \frac{i }{4 \pi}\int d^3x\, \epsilon^{\mu\nu\rho} \left(2 \bar{W}^a_\mu (D_\nu W_\rho)_a + \kappa Z_\mu \partial_\nu Z_\rho - 2iZ_\mu \bar{W}^a_\nu W_{a\rho}\right) \nonumber\\ 
&\quad +\int d^3x\, (|\kappa| V^2 Z_\mu Z^\mu + {\sgn}(\kappa) V^2\bar{W}^{a\mu} W_{a\mu}) \nonumber\\
&\quad +\int d^3x \Big[\bar\psi^a \gamma^\mu (D_\mu \psi)_a + (m_F+4\pi {\sgn}(\kappa) x_4 V^2 )\bar{\psi}^a\psi_a\Big]\ ,
\end{align}
and 
\begin{align}\label{vpot}
S_2[V] &= \int d^3x \Big(|\kappa| \partial_\mu V \partial^\mu V +  U_{\rm cl}(V)\Big)\ ,\nonumber\\
U_{\rm cl}(V) &= |\kappa| m_B^2 V^2  + 4\pi b_4\kappa V^4 + 4\pi^2 |\kappa| (x_6 + 1) V^6\ .
\end{align}
We shall often denote the effective mass
$m_F + 4\pi \sgn(\kappa) x_4 V^2$ of the fermion $\psi_a$ by
${\tl m}_F$:
\begin{equation}\label{tlmfdefapp}
  {\tl m}_F \equiv m_F + 4\pi \sgn(\kappa) V^2\ .
\end{equation}
We now outline the salient steps in the derivation of the large
$N$ saddle-point equations, the thermal free energy and the exact
two-point functions.

\begin{enumerate}
\item We follow the fairly standard procedure of first choosing the
  lightcone gauge $A_- = 0$ for the $SU(N-1)$ gauge boson and
  integrating out the gauge boson and the $Z$ boson
  \cite{Choudhury:2018iwf}.

\item Following this, we introduce one pair of $SU(N-1)$-singlet
  bilocal fields $(\alpha_{\mu\nu}(q,p)$, $\Sigma^{\mu\nu}(q,p))$ and
  $(\alpha_F(q,p), \Sigma_F(q,p))$ corresponding to the singlet
  combinations $\bar{W}^a_\mu W_{a\nu}$ and $\bar\psi^a \psi_a$
  respectively using the Hubbard-Stratonovich transform. The fields in
  the path integral are then the bilocal gauge-singlet fields
  $\alpha$, $\Sigma$ and the Higgs field $V$.

\item The action governing the bilocal gauge-singlet fields and the
  Higgs field has an explicit prefactor of $N$ and hence can be
  evaluated by saddle point approximation in the large $N$
  limit. Finally, as explained in detail in Section 3.2 of
  \cite{Dey:2018ykx}, the large $N$ saddle point occurs at constant
  values of the Higgs field $V$. This enormously simplifies the task
  of obtaining the saddle-point equations and the large $N$ thermal
  free energy. The saddle-point values of the $\alpha$'s, $\Sigma$'s
  and $V$ respectively give the exact two-point functions of the
  fundamental excitations ($W$ and $\psi$), their self-energies and
  the vev of the Higgs field at the large $N$ saddle point.

\item Again, following Section 3.2 of \cite{Dey:2018ykx}, one can
  compute the path integral of our current theory in two steps: First,
  we integrate out the gauge-singlet bilocal fields for a fixed
  \emph{a priori} undetermined constant Higgs field $V$ and obtain an
  effective action for the Higgs field $V$. In the second step, we
  carry out the saddle point approximation for the effective action
  as a function of $V$. We describe each of these steps briefly below.

\item We first note that the action $S_1$ in \eqref{asb1} is simply
  the sum of the actions for a critical boson in the Higgsed phase and
  for a regular fermion (in either of its phases) provided we make the
  identifications
  \begin{equation}\label{crireg}
    m_B^{\rm cri} = -\frac{4\pi}{|\lambda|}V^2\ ,\quad m_F^{\rm reg} = {\tl m}_F = m_F + 4\pi \sgn(\lambda) x_4 V^2\ .
  \end{equation}
  Thus, the result of the path integral over the bilocal gauge-singlet
  fields is simply the sum of the critical boson and regular fermion
  free energies (evaluated at their large $N$ saddle points) along
  with apriori unknown cosmological constant counterterms:
  \begin{align}\label{S1pr}
    &S'_1[V] = F_{\rm CB}[m_B^{\rm cri}(V)] + F_{\rm RF}[m_F^{\rm reg}(V)]\ ,\nonumber\\
    &= \frac{N}{6\pi}\bigg[\frac{2}{|\lambda|} {\hat c}_B^3 + \frac{3}{2} {\hat c}_B^2 {\hat m}_B^{\rm cri} + \Lambda_B ({\hat m}_B^{\rm cri})^3 \nonumber\\
    &\qquad\qquad  -{\hat c}_B^3 + 3 \int_{-\pi}^{\pi}d\alpha\, \rho(\alpha)\,\int_{{\hat c}_B}^{\infty} dy\,y\,\left( \log\left(1-e^{-y-i\alpha}\right) + \log\left(1-e^{-y+i\alpha}\right)\right)  \nonumber\\
    &\qquad\qquad - \frac{\sgn(X_F)}{\lambda} {\hat c}_F^3 + \frac{3}{2\lambda} {\hat c}_F^2  {\hat m}_F^{\rm reg} + \Lambda_F ({\hat m}_F^{\rm reg})^3 \nonumber\\
    & \qquad\qquad +{\hat c}_F^3  - 3 \int_{-\pi}^{\pi}d\alpha\, \rho(\alpha)\,\int_{{\hat c}_F}^{\infty} dy\,y\,\left( \log\left(1+e^{-y-i\alpha}\right) + \log\left(1+e^{-y+i\alpha}\right)\right)\Bigg]\ ,
  \end{align}
  where ${\hat X}_F = 2\lambda \cC(c_F) + {\hat m}_F^{\rm reg}$ and ${c}_B$,
  ${ c}_F$ are the pole masses of the critical boson and the regular
  fermion and are given by the solutions of the following implicit
  equations
  \begin{align}
    {\hat c}_B &= |\lambda| (\cS(c_B) - \tfrac{1}{2} {\hat m}_B^{\rm cri})\ ,\quad {\hat c}_F = |2\lambda \cC(c_F) + {\hat m}_F^{\rm reg}|\ ,\label{gapeq}\\
    \cC(c_F) &= \frac{1}{2} \int_{-\pi}^\pi d\alpha\,\rho(\alpha)   \( \log(2 \cosh (\tfrac{\hat{c}_F +i\alpha}{2}))+ \log(2 \cosh (\tfrac{\hat{c}_F - i\alpha}{2})) \)\ , \label{ccdef}\\
    \cS(c_B) &= \frac{1}{2} \int_{-\pi}^\pi d\alpha\,\rho(\alpha)   \( \log(2 \sinh (\tfrac{\hat{c}_B + i\alpha}{2}))+ \log(2 \sinh (\tfrac{\hat{c}_B - i\alpha}{2})) \)\ .\label{csdef}
  \end{align}
\item We thus get an effective action for the constant Higgs field $V$
  which is the sum of the actions $S_1'[V]$ in \eqref{S1pr} and
  $S_2[V]$ in \eqref{vpot}:
  \begin{equation}\label{Seffapp}
    S_{\rm eff}[V] = S_1'[V] + S_2[V]\ .
  \end{equation}
  For future purposes, we introduce
  the variable $\sigma$ which is the following function of $V^2$:
  \begin{equation}\label{sigmadefapp}
    \sigma = \frac{2\pi V^2}{|\lambda|}\ .
  \end{equation}
  In terms of this variable, the effective action for $V$ becomes
  \begin{align}\label{SeffV}
    &S_{\rm eff}[\sigma]\nonumber\\
    &= \frac{N}{6\pi}\bigg[\frac{2}{|\lambda|} {\hat c}_B^3 - 3 {\hat c}_B^2 {\hat \sigma} - 8 \Lambda_B {\hat \sigma}^3 + 3 ({\hat m}_B^2 {\hat \sigma} + 2\lambda {\hat b}_4 {\hat \sigma}^2 + (1+x_6)\lambda^2 {\hat \sigma}^3) \nonumber\\
    &\qquad\qquad  -{\hat c}_B^3 + 3 \int_{-\pi}^{\pi}d\alpha\, \rho(\alpha)\,\int_{{\hat c}_B}^{\infty} dy\,y\,\left( \log\left(1-e^{-y-i\alpha}\right) + \log\left(1-e^{-y+i\alpha}\right)\right)  \nonumber\\
    &\qquad\qquad  - \frac{\sgn(X_F)}{\lambda} {\hat c}_F^3 + \frac{3}{2\lambda} {\hat c}_F^2 ({\hat m}_F + 2\lambda x_4 {\hat \sigma}) + \Lambda_F ({\hat m}_F + 2\lambda x_4 {\hat \sigma})^3 \nonumber\\
    & \qquad\qquad +{\hat c}_F^3  - 3 \int_{-\pi}^{\pi}d\alpha\, \rho(\alpha)\,\int_{{\hat c}_F}^{\infty} dy\,y\,\left( \log\left(1+e^{-y-i\alpha}\right) + \log\left(1+e^{-y+i\alpha}\right)\right)\Bigg]\ .
  \end{align}
  
\item The so-far unknown `cosmological constants' $\Lambda_B$ and
  $\Lambda_F$ multiply terms which depend on $\sigma$ and are
  important parts of the effective potential of the one boson one
  fermion theory. An honest evaluation of the free energy of the one
  boson one fermion theory would of course leave no such ambiguous
  terms that are proportional to $\sigma$ in the final answer. The
  values of $\Lambda_B$ and $\Lambda_F$ can indeed be fixed by setting
  the one-point function of the fluctuations of the Higgs field $V$
  about its vev $v$ to zero (also known as the \emph{tadpole
    cancellation condition}). This equation should coincide with the
  equation obtained by extremization of the effective action
  $S_{\rm eff}[\sigma]$ in \eqref{SeffV} if one chooses appropriate
  values for $\Lambda_B$ and $\Lambda_F$. An exactly analogous step
  occurs in the computation of the regular boson free energy in
  \cite{Dey:2018ykx} (see for example equation (3.19) of that
  reference). We look at this step next.

\item We next extremize \eqref{SeffV} with respect to $\sigma$. It it
  sufficient to differentiate only explicit occurrences of $\sigma$
  because the terms involving derivatives of $c_B$ and $c_F$
  w.r.t.~$\sigma$ are always multiplied by the gap equations
  \eqref{gapeq}. We thus get
  \begin{multline}\label{cosmotad}
    -({\hat c}_B^2 + 8 \Lambda_B {\hat \sigma}^2) + x_4 ({\hat c}_F^2 + 2\lambda \Lambda_F ({\hat m}_F + 2 \lambda x_4 {\hat \sigma} )^2) \\ +  ({\hat m}_B^2 + 4\lambda {\hat b}_4 {\hat \sigma} + 3(x_6 + 1) \lambda^2 {\hat \sigma}^2) = 0\ .
  \end{multline}    
  The tadpole cancellation condition will be derived in the next
  subsection (Section \ref{tadpole}) and we give the result here:
  \begin{multline}\label{mhat}
    -(\hat{c}_B^2 - \lambda^2\hat{\sigma}^2) + x_4({\hat c}_F^2 - ({\hat m}_F + 2\lambda x_4 {\hat \sigma})^2) \\ + (\hat{m}_B^2 + 4 \hat{b}_4 \lambda \hat{\sigma}+3(x_6+1) \lambda^2 \hat{\sigma}^2)=0\ .
  \end{multline}

\item Clearly, comparing \eqref{cosmotad} and \eqref{mhat}, we see
  that the unknown constants $\Lambda_B$ and $\Lambda_F$ are fixed to be
  \begin{equation}\label{ccval}
    \Lambda_B = -\frac{\lambda^2}{8}\ ,\quad \Lambda_F = -\frac{1}{2\lambda} \ .
  \end{equation}

\item The free energy in the Higgsed phase of the boson and either
  phase of the fermion is then given by $S_{\rm eff}[V]$ in
  \eqref{SeffV} with the cosmological constants assuming the values in
  \eqref{ccval}:
  \begin{align}\label{finalfe}
    F[\sigma]  &= \frac{N}{6\pi}\bigg[\frac{2}{|\lambda|} {\hat c}_B^3 - 3 {\hat c}_B^2 {\hat \sigma} + \lambda^2 {\hat \sigma}^3 +  3\left( {\hat m}_B^2 {\hat \sigma}  + 2\lambda {\hat b}_4 {\hat \sigma}^2 +  (x_6 + 1)  \lambda^2 {\hat \sigma}^3\right) \nonumber\\ 
    &\qquad\quad -{\hat c}_B^3 + 3 \int_{-\pi}^{\pi}d\alpha\, \rho(\alpha)\,\int_{{\hat c}_B}^{\infty} dy\,y\,\left( \log\left(1-e^{-y-i\alpha}\right) + \log\left(1-e^{-y+i\alpha}\right)\right)  \nonumber\\
    &\qquad\quad - \frac{\sgn(X_F)}{\lambda} {\hat c}_F^3 + \frac{3}{2\lambda}  {\hat c}_F^2 ({\hat m}_F + 2\lambda x_4 {\hat \sigma}) -\frac{({\hat m}_F + 2\lambda x_4 {\hat \sigma})^3}{2\lambda} \nonumber\\
    & \qquad\quad + {\hat c}_F^3  - 3 \int_{-\pi}^{\pi}d\alpha\, \rho(\alpha)\,\int_{{\hat c}_F}^{\infty} dy\,y\,\left( \log\left(1+e^{-y-i\alpha}\right) + \log\left(1+e^{-y+i\alpha}\right)\right)\Bigg]\ .   
  \end{align}
\end{enumerate}

\subsection{Tadpole cancellation for \texorpdfstring{$V$}{V}}\label{tadpole}

The scalar field $V$ (the only non-zero component of $\phi$ in the
unitary gauge) has a non-zero vacuum expectation value $v$ in the
Higgsed phase. We write $V(x)$ as the sum of its vev $v$ and the
fluctuation $H$:
\begin{equation} \label{deffh} V(x)= v + H(x)\ .
\end{equation}
Then, the fluctuation $H(x)$ should have vanishing expectation value
(i.e. one point function, i.e. tadpole) about the true vacuum $v$:
\begin{equation}\label{nseffH}
\int_{\mbb{R}^2 \times S^1} 
[dH dW dZ d\psi dA]\  H(x)\ e^{-S_{\text{E}}[A,W,Z,\psi,V=v+H]} =0\ .
\end{equation}
Recall that we are working in the constant $V$ subspace of the
configuration space of the field $V$ in the large $N$ limit. In this
case, the above one-point function can be perturbatively evaluated to
all orders in the 't Hooft coupling $\lambda$. The tadpole
cancellation condition \eqref{nseffH} becomes
\begin{equation} \label{exptc}
  {\rm sgn}(\kappa) \langle
  \bar{W}^{a\mu}(x) W_{a\mu}(x) \rangle + |\kappa| \langle Z_\mu(x)
  Z^\mu(x) \rangle \ +4\pi x_4 \sgn(\kappa) \langle
  \bar{\psi}^{a}(x) \psi_{a}(x) \rangle + \frac{\partial } {\partial
    (V^2)}U_{\rm cl}(V^2) =0\ .
\end{equation} 
where $U_{\text{cl}}(V)$ is the potential for the Higgs field as $V$
given in \eqref{vpot}. In \eqref{exptc} all the expectation values are
to be computed about the `vacuum' where $V(x)=v$.

Note that the first, third and fourth term in \eqref{exptc} are of
order $N$, while the second term - proportional to
$\langle Z_\mu (x)^2 \rangle $ - is of order unity (this is because
the kinetic term for $Z$ in the action \eqref{asb1} scales like
$\kappa$ and hence the propagator scales like $1/\kappa$).  So while
working in the large $N$ limit, we can safely drop the second term. We
now write the integrated version of the tadpole cancellation condition
\eqref{nseffH} after substituting $\sigma$ in place of $V$ from
\eqref{sigmadefapp}:
\begin{equation} \label{exptcn} 
\frac{\lambda}{2 \pi \cV_3} \int d^3 x \langle \bar{W}^{a\mu}(x)  W_{a\mu}(x) \rangle + \frac{\lambda}{2 \pi \cV_3} \int d^3 x \langle \bar{\psi}^a(x)  \psi_a(x) \rangle + \frac{\partial U_{\rm cl}(\sigma)} {\partial \sigma} =0\ ,
\end{equation} 
In going from \eqref{exptc} to \eqref{exptcn}, we integrated
\eqref{exptc} over spacetime and divided the resulting expression by
the volume of spacetime $\cV_3$.  Writing \eqref{exptcn} in momentum
space, we get
\begin{equation} \label{vgap}
  \frac{\lambda}{2 \pi} \int \frac{\td^3
    p}{(2\pi)^3} g^{\mu\nu}\,G^a_{a\mu\nu}(p) +\frac{\lambda}{2 \pi}
  \int \frac{\td^3 p}{(2\pi)^3}\, \tr\,K^a_{a}(p) +
  \frac{\partial U_{\rm cl}(\sigma)} {\partial \sigma} =0\ ,
\end{equation}
where the trace in the second term is in the space of $2 \times 2$
gamma matrices and the two-point functions $G^a_b$ and $K^a_b$ are
given by
\begin{align} \label{propmomn}
& \langle \bar{W}^{a\mu}(-p) W_{b\nu} (p')\rangle = 
G^a_{b\mu\nu}(p')\,  (2 \pi)^{3} \delta^{(3)}(p'-p) = \delta^a_b\, G_{\mu\nu}(p')\,  (2 \pi)^{3} \delta^{(3)}(p'-p)\ ,\nonumber\\
& \langle \bar{\psi}^a(-p)  \psi_a(p') \rangle = K^a_b(p') (2 \pi)^{3} \delta^{(3)}(p'-p) = \delta^a_b K(p') (2 \pi)^{3} \delta^{(3)}(p'-p)\ .
\end{align}
The explicit expressions for $G_{\mu\nu}(p)$ can be found in
\cite{Dey:2018ykx} and for $K(p)$ in \cite{Jain:2013gza} (following
\cite{Giombi:2011kc}, \cite{Aharony:2012ns}, \cite{Jain:2013py}). The
expression for the classical potential $U_{\rm cl}$ is given in
\eqref{vpot}. Plugging these expressions into \eqref{vgap}, we get the
equation
\begin{multline}\label{vgapsim}
-\frac{N}{2\pi} \left({ c}_B^2-\lambda^2{ \sigma}^2 \right) + \frac{N}{2\pi} 4\lambda x_4(\lambda \cC^2+(m_F+2x_4 \lambda \sigma) \cC) \\ + \frac{N}{2\pi} \left({\hat m}_B^2 + 4 {\hat b}_4 \lambda {\hat \sigma} + 3 (x_6+1) \lambda^2 \sigma^2\right) = 0\ .
\end{multline}
Using the gap equation for $c_F$ in \eqref{gapeq}, we can rewrite the above equation in the form
\begin{multline}\label{mhatder}
- \left({\hat c}_B^2-\lambda^2{\hat \sigma}^2 \right) +   x_4({\hat c}_F^2 - ({\hat m}_F + 2\lambda x_4 {\hat \sigma})^2) \\ +  \left({\hat m}_B^2 + 4 {\hat b}_4 \lambda {\hat \sigma} + 3 (x_6+1) \lambda^2 {\hat \sigma}^2\right) = 0\ .
\end{multline}

\subsection{The five variable off-shell free energy}\label{offsh5}

In this subsection, we derive the five variable off-shell free energy
that is presented in \eqref{Feffofsh}. In Section \ref{off-shell}, we
start with the five variable off-shell free energy and extremize it
with respect to its variables. Plugging in the solutions of the
extremization equations into the off-shell free energy results in
different free energy expressions for different phases of the
theory. Here, we start with the free energies computed in the previous
subsection for the Higgsed phase of the boson and either phase of the
fermion and in \cite{Jain:2013gza} for the unHiggsed phase of the
boson and either phase of the fermion, and build up the expression for
the five variable off-shell free energy by combining the two
computations appropriately.

The expression for the free energy \eqref{finalfe} in the Higgsed
phase of the boson and either phase of the fermion was a function of
the single variable $\sigma$. It is not hard to promote the pole
masses $c_B$ and $c_F$ to variables since the variation of
\eqref{finalfe} w.r.t.~them is proportional to their equations
\eqref{gapeq}. Thus, we have a three variable off-shell version of the
free energy in the Higgsed phase of the boson:
\begin{align}\label{fehig}
  F[c_B, c_F, \sigma]  &= \frac{N}{6\pi}\bigg[\frac{2}{|\lambda|} {\hat c}_B^3 - 3 {\hat c}_B^2 {\hat \sigma} + \lambda^2 {\hat \sigma}^3 +  3\left( {\hat m}_B^2 {\hat \sigma}  + 2\lambda {\hat b}_4 {\hat \sigma}^2 +  (x_6 + 1)  \lambda^2 {\hat \sigma}^3\right) \nonumber\\ 
                       &\qquad\quad -{\hat c}_B^3 + 3 \int_{-\pi}^{\pi}d\alpha\, \rho(\alpha)\,\int_{{\hat c}_B}^{\infty} dy\,y\,\left( \log\left(1-e^{-y-i\alpha}\right) + \log\left(1-e^{-y+i\alpha}\right)\right)  \nonumber\\
                       &\qquad\quad - \frac{\sgn(X_F)}{\lambda} {\hat c}_F^3 + \frac{3}{2\lambda}  {\hat c}_F^2 ({\hat m}_F + 2\lambda x_4 {\hat \sigma}) -\frac{({\hat m}_F + 2\lambda x_4 {\hat \sigma})^3}{2\lambda} \nonumber\\
                       & \qquad\quad + {\hat c}_F^3  - 3 \int_{-\pi}^{\pi}d\alpha\, \rho(\alpha)\,\int_{{\hat c}_F}^{\infty} dy\,y\,\left( \log\left(1+e^{-y-i\alpha}\right) + \log\left(1+e^{-y+i\alpha}\right)\right)\Bigg]\ .   
\end{align}
Extremizing this w.r.t.~$c_B$, $c_F$ and $\sigma$ yields the equations
\eqref{gapeq} and \eqref{mhat}.

A similar expression was obtained in \cite{Jain:2013gza} for the free
energy in the unHiggsed phase of the boson and either phase of the
fermion:
\begin{align}\label{feunhig}
  F[c_B, c_F, {\tl \cS}]  &= \frac{N}{6\pi}\bigg[3{\hat c}_B^2 {\tl \cS}  - \lambda^2 {\tl \cS}^3 +  3\left( -{\hat m}_B^2 {\tl \cS}  + 2\lambda {\hat b}_4 {\tl \cS}^2 - (x_6 + 1)  \lambda^2 {\tl \cS}^3\right) \nonumber\\ 
                       &\qquad\quad -{\hat c}_B^3 + 3 \int_{-\pi}^{\pi}d\alpha\, \rho(\alpha)\,\int_{{\hat c}_B}^{\infty} dy\,y\,\left( \log\left(1-e^{-y-i\alpha}\right) + \log\left(1-e^{-y+i\alpha}\right)\right)  \nonumber\\
                       &\qquad\quad - \frac{\sgn(X_F)}{\lambda} {\hat c}_F^3 + \frac{3}{2\lambda}  {\hat c}_F^2 ({\hat m}_F - 2\lambda x_4 {\tl \cS}) -\frac{({\hat m}_F - 2\lambda x_4 {\tl \cS})^3}{2\lambda} \nonumber\\
                       & \qquad\quad + {\hat c}_F^3  - 3 \int_{-\pi}^{\pi}d\alpha\, \rho(\alpha)\,\int_{{\hat c}_F}^{\infty} dy\,y\,\left( \log\left(1+e^{-y-i\alpha}\right) + \log\left(1+e^{-y+i\alpha}\right)\right)\Bigg]\ .   
\end{align}
Extremizing w.r.t.~$c_B$, $c_F$ and ${\tl \cS}$ gives
\begin{align}\label{unhiggap}
  c_B:&\quad  6 {\hat c}_B ({\tl \cS} - \cS(c_B)) = 0\ ,\nonumber\\
  c_F:&\quad \frac{3{\hat c}_F}{\lambda}\left(- \sgn(X_F){\hat c}_F + ({\hat m}_F - 2\lambda x_4 {\tl \cS}) + 2\lambda  \cC(c_F)\right) = 0\ ,\nonumber\\
  {\tl \cS}:&\quad {\hat c}_B^2 - {\hat m}_B^2 - \lambda^2 {\tl \cS}^2 + 4\lambda {\hat b}_4 {\tl \cS} - 3(x_6 + 1)\lambda^2 {\tl \cS}^2 -3 x_4 ({\hat c}_F^2 - ({\hat m}_F - 2\lambda x_4 {\tl \cS})^2) = 0\ ,
\end{align}
which are the large $N$ saddle point equations obtained in
\cite{Jain:2013gza}.

The above `off-shell' free energies are not satisfactorily off-shell
for the following reasons:
\begin{enumerate}
\item The expressions for the off-shell free energies are different in the different phases of the boson.
\item Though both phases of the fermion are incorporated in each of
  the two expressions \eqref{fehig} and \eqref{feunhig}, they are
  non-analytic in the variable $c_F$ since
  $\sgn(X_F) = \pm \sgn(\lambda)$ in the $\pm$ phase of the fermion.
\end{enumerate}
In order for a candidate free energy to be called truly off-shell, it
must be analytic in all its variables and its extremization
w.r.t.~said variables yields the individual non-analytic free energies
in the different phases of the boson and fermion. This can be achieved
at the cost of introducing new variables. We address each of the
points above in turn.

Firstly, we look at the issue with the bosonic phases. Note that
\eqref{fehig} agrees with \eqref{feunhig} sans the ${\hat c}_B^3$ term
present in the first line of \eqref{fehig} if one makes the
replacement ${\hat \sigma} \to -\tl \cS$. Hence, a candidate off-shell
free energy would be generated as follows: Retain the expression
\eqref{fehig} with the ${\hat c}_B^3$ term in the first line deleted,
and introduce additional terms depending on a new variable such that
when this new variable in integrated out, it either generates the
${\hat c}_B^3$ term in the first line \eqref{fehig} or performs the
replacement ${\hat \sigma} \to -{\tl \cS}$ to get \eqref{feunhig}.

Indeed, consider the expression depending on the two variables
${\hat \sigma}$ and ${\tl \cS}$ (hence, one extra variable from either
the Higgsed or the unHiggsed point of view):
\begin{equation}\label{oscand}
  -3{\hat c}_B^2 {\hat \sigma} + \lambda^2 {\hat \sigma}^3 + 3\left({\hat m}_B^2 {\hat \sigma} +  2\lambda {\hat b}_4 {\hat \sigma}^2 +  (x_6 + 1)  \lambda^2 {\hat \sigma}^3\right) - \alpha \left(2\lambda^2({\tl \cS} + {\hat \sigma})^3 - 3 |\lambda| {\hat c}_B ({\tl\cS} + {\hat \sigma})^2\right)\ ,
\end{equation}
where $\alpha$ is a constant to be determined. Extremizing
w.r.t.~${\tl \cS}$, we get the equation
\begin{equation}
  6|\lambda| ({\tl \cS} + {\hat \sigma}) \left(|\lambda|({\tl \cS} + {\hat \sigma}) - {\hat c}_B\right) = 0\ .
\end{equation}
The significance of the above equation is as follows. From the first
of \eqref{unhiggap}, we see that the variable ${\tl \cS}$ in the
unHiggsed phase is set to the function $\cS(c_B)$. However, in the
Higgsed phase, this exact same function of $c_B$ appears in the gap
equation for $c_B$ in \eqref{gapeq}. One thus engineers the extra
terms in \eqref{oscand} to obtain either the replacement
${\hat \sigma} \to -{\tl \cS}$ in the unHiggsed phase or the gap
equation in the Higgsed phase which is guaranteed to generate the
${\hat c}_B^3$ present in the first line of \eqref{fehig}.

Indeed, the solution ${\tl \cS} + {\hat \sigma} = 0$ sets the new
additional terms in \eqref{oscand} to zero and enforces the
replacement ${\hat \sigma} = -{\tl \cS}$ in the original set of
terms. Moreover, the second solution
${\tl \cS} + {\hat \sigma} = {\hat c}_B / |\lambda|$ generates the
${\hat c}_B^3$ term present in the first line of \eqref{fehig} if we
choose $\alpha = 2$.

Next, we look at the non-analyticity in the ${\hat c}_F^3$ term in the
third line of either \eqref{fehig}. This can be cured as follows. Upon
extremization of \eqref{fehig} w.r.t.~$c_F$, we get the equation
\begin{equation}\label{cfgapeq}
  - \sgn(X_F)  {\hat c}_F + ({\hat m}_F + 2\lambda x_4 {\hat \sigma}) + 2 \lambda \cC(c_F) = 0\ ,
\end{equation}
We make the function $\cC$ a variable now and enforce the relation \eqref{cfgapeq} using a Lagrange multiplier ${\tl \cC}$:
\begin{align}
  & - \frac{\sgn(X_F)}{\lambda} {\hat c}_F^3 + \frac{3}{2\lambda}  {\hat c}_F^2 ({\hat m}_F + 2\lambda x_4 {\hat \sigma}) -\frac{({\hat m}_F + 2\lambda x_4 {\hat \sigma})^3}{2\lambda}\nonumber\\ & - 3 {\tl \cC} \left({\hat c}_F^2 - ({\hat m}_F + 2\lambda x_4 {\hat \sigma} + 2 \lambda \cC)^2\right)\nonumber\\
  & + {\hat c}_F^3  - 3 \int_{-\pi}^{\pi}d\alpha\, \rho(\alpha)\,\int_{{\hat c}_F}^{\infty} dy\,y\,\left( \log\left(1+e^{-y-i\alpha}\right) + \log\left(1+e^{-y+i\alpha}\right)\right)\ ,
\end{align}
which using the ${\tl \cC}$ equation of motion can be written as
\begin{align}
=& -8 \lambda^2 \cC^3 - 6\lambda \cC^2 ({\hat m}_F + 2\lambda x_4 {\hat \sigma}) - 3{\tl \cC} \left({\hat c}_F^2 - ({\hat m}_F + 2\lambda x_4 {\hat \sigma} + 2 \lambda \cC)^2\right)\nonumber\\
  & + {\hat c}_F^3  - 3 \int_{-\pi}^{\pi}d\alpha\, \rho(\alpha)\,\int_{{\hat c}_F}^{\infty} dy\,y\,\left( \log\left(1+e^{-y-i\alpha}\right) + \log\left(1+e^{-y+i\alpha}\right)\right)\ ,
\end{align}
which is indeed an analytic in $c_F$, $\cC$ and ${\tl
  \cC}$. Extremizing w.r.t.~$c_F$, $\cC$ and ${\tl \cC}$ yields the
equations
\begin{align}\label{feroffeq}
  c_F:&\quad - 6 {\hat c}_F ({\tl \cC} - \cC(c_F)) = 0\ ,\nonumber\\
  \cC:&\quad  12\lambda  ({\tl \cC} - \cC) ({\hat {\tl m}}_F + 2\lambda \cC) = 0\ ,\nonumber\\
  {\tl \cC}:&\quad {\hat c}_F^2 - ({\hat {\tl m}}_F + 2 \lambda \cC)^2 = 0\ .
\end{align}
The above three equations together enforce the original gap equation
\eqref{cfgapeq}. We seem to have introduced two additional variables
$\cC$ and ${\tl \cC}$ in order to get rid of the non-analyticity in
the ${\hat c}_F^3$ term in the free energy. However, one can easily
drop one of the variables, say $\cC$, by using the second equation in
\eqref{feroffeq} which replaces all instances of $\cC$ by ${\tl
  \cC}$. We are now ready to write down an expression for the
off-shell free energy which addresses the non-analyticity in both the
bosonic and fermionic variables in \eqref{fehig} and \eqref{feunhig}.

The final off-shell free energy depends on five variables $c_B$,
$c_F$, ${\tl \cS}$, ${\tl \cC}$ and $\sigma$ and is given by
\begin{align}\label{fosapp}
  &F[c_B, c_F, {\tl \cS}, {\tl \cC}, \sigma]\nonumber\\
  &= \frac{N}{6\pi}\bigg[-3 {\hat c}_B^2 {\hat \sigma} + \lambda^2 {\hat \sigma}^3 +  3\left( {\hat m}_B^2 {\hat \sigma}  + 2\lambda {\hat b}_4 {\hat \sigma}^2 +  (x_6 + 1)  \lambda^2 {\hat \sigma}^3\right) \nonumber\\
  &\qquad\quad - 4\lambda^2 ({\tl \cS} + {\hat \sigma})^3 + 6|\lambda|{\hat c}_B ({\tl \cS} + {\hat \sigma})^2\nonumber\\
  &\qquad\quad -{\hat c}_B^3 + 3 \int_{-\pi}^{\pi}d\alpha\, \rho(\alpha)\,\int_{{\hat c}_B}^{\infty} dy\,y\,\left( \log\left(1-e^{-y-i\alpha}\right) + \log\left(1-e^{-y+i\alpha}\right)\right)  \nonumber\\
  &\qquad\quad -8 \lambda^2 {\tl \cC}^3 - 6\lambda {\tl \cC}^2 ({\hat m}_F + 2\lambda x_4 {\hat \sigma}) - 3{\tl \cC} \left({\hat c}_F^2 - ({\hat m}_F + 2\lambda x_4 {\hat \sigma} + 2 \lambda {\tl \cC})^2\right)\nonumber\\
  & \qquad\quad + {\hat c}_F^3  - 3 \int_{-\pi}^{\pi}d\alpha\, \rho(\alpha)\,\int_{{\hat c}_F}^{\infty} dy\,y\,\left( \log\left(1+e^{-y-i\alpha}\right) + \log\left(1+e^{-y+i\alpha}\right)\right)\Bigg]\ .   
\end{align}

\section{Numerical analysis of the first order transition curves}\label{firstorder}

This section is meant to read in conjunction with Section \ref{fomain}
in the main body of this paper. In this section, we explain the
procedure to obtain the first order transition lines in Section
\ref{N2phase}. In Figure \ref{phaseall} in Section \ref{N2phase}, the
green lines are first order transition lines. The green line in Figure
\ref{phaseall}(a) segment between $\RB+$ and $\Op+$ separates the
$(+,+)$ and the $(+,-)$ phases while the segment from $\Op+$ onwards
separates the $(-,+)$ and the $(+,-)$ phases. Similarly, in Figure
\ref{phaseall}(c), there are two segments separating the
$(-,-)$-$(-,+)$ and the $(-,+)$-$(+,-)$ phases respectively. In Figure
\ref{phaseall}(b) corresponding to $m_F = 0$, there is a continuous
first order line which separates the $(-,+)$ and the $(+,-)$ phases.

These segments are determined as follows. Take for instance the
segment between $\RB+$ and $\Op+$ which separates the $(+,+)$ and the
$(+,-)$ phases. There is one minimum each in each of the phases and
one of these minima is dominant on one side of the first order
line. Thus, on the first order line, the values of the potential at
these two minima must be equal to each other. In other words, we need
to solve the equation
\begin{equation}\label{1stordseg1}
U^{(+,+)}(\sigma)|_{\sigma_{\text{min}}^{(+,+)}} \ - \ U^{(+,-)}(\sigma)|_{\sigma_{\text{min}}^{(+,-)}} = 0\ ,
\end{equation}
where $U^{(+,\pm)}(\sigma)$ are the expressions for the
Landau-Ginzburg potential in the $(+, \pm)$ phases (see
\eqref{LGpotrel}) and $\sigma_{\text{min}}^{(+,\pm)}$ are the minima
of the potential in the $(+,\pm)$ phases respectively. These values of
$\sigma$ are functions of the parameters $m_B^2$, $\lambda b_4$ and
$m_F$ present in the potential. The equation in \eqref{1stordseg1} is
a then a constraint between the three parameters and describes a
surface in three dimensional space. The intersection of this surface
with the section $m_F = |\mu| \sgn(\lambda)$ gives rise to a curve on
this section which is the green line in Figure
\ref{phaseall}(a). Similarly, the intersection of this surface with
the ellipsoid \eqref{spher} is again a curve on the ellipsoid
(e.g.~the green line in Figure \ref{3dplot14}). This procedure is then
repeated for all such segments in Figure \ref{phaseall}. Note that in
the case of Figure \ref{phaseall}(b) corresponding to $m_F = 0$, the
phase diagram is one-dimensional and is given by an elliptical section
\ref{mf0ellipse} of the two dimensional plane. The numerically
determined green line intersects this ellipse at one point and this is
the first order transition point on the ellipse (see Figure
\ref{ellipsephase}(b) for a depiction of this in the context of the
regular boson theory.)

\begin{figure}
	\begin{subfigure}{0.3\textwidth}
	\includegraphics[width=1.7in,height=1.5in]{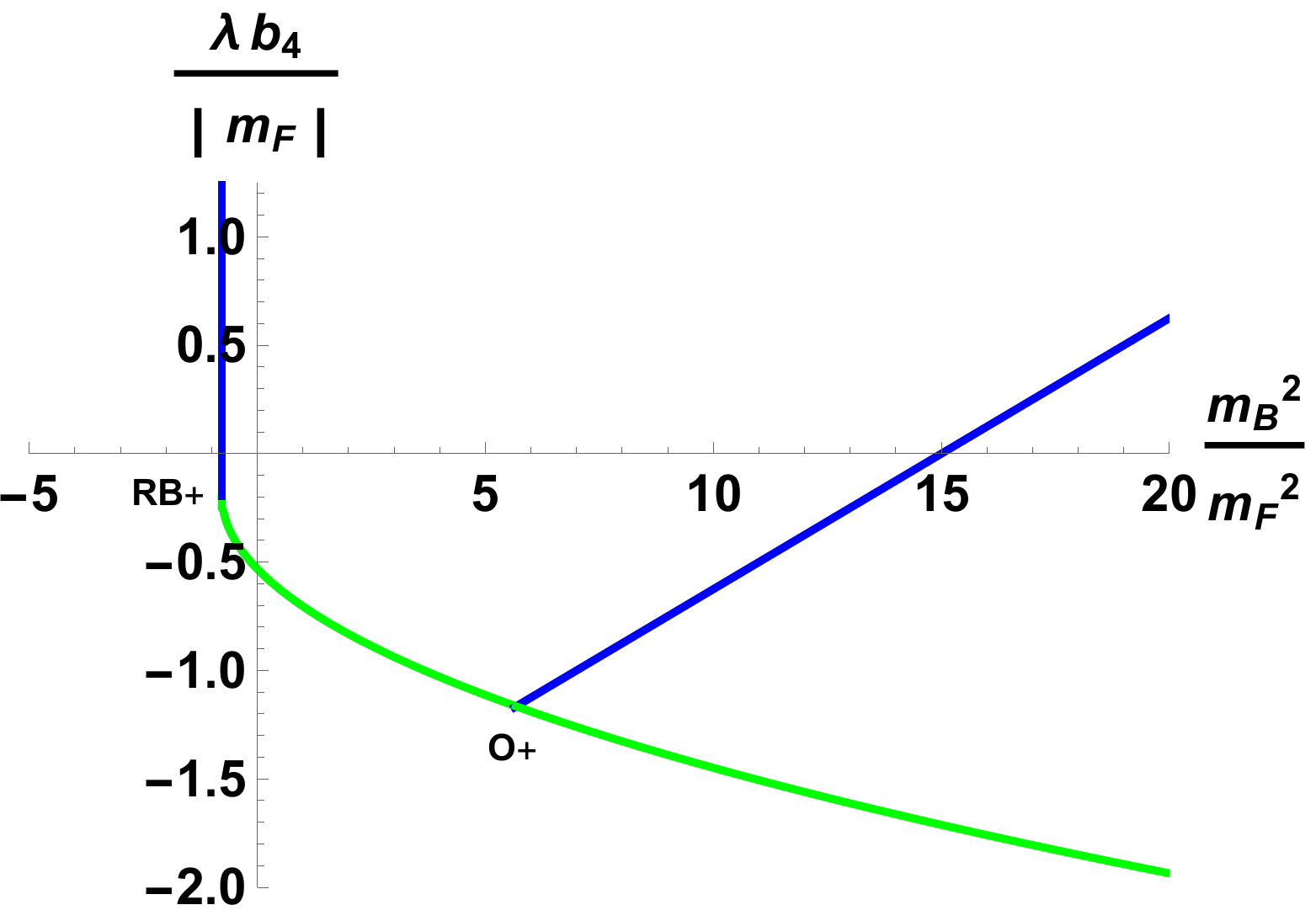}
		\caption{$|\lambda| =0.25$}
		\label{1stOrda}
	\end{subfigure}\hspace{10pt}
\begin{subfigure}{0.3\textwidth}
	\includegraphics[width=1.7in,height=1.5in]{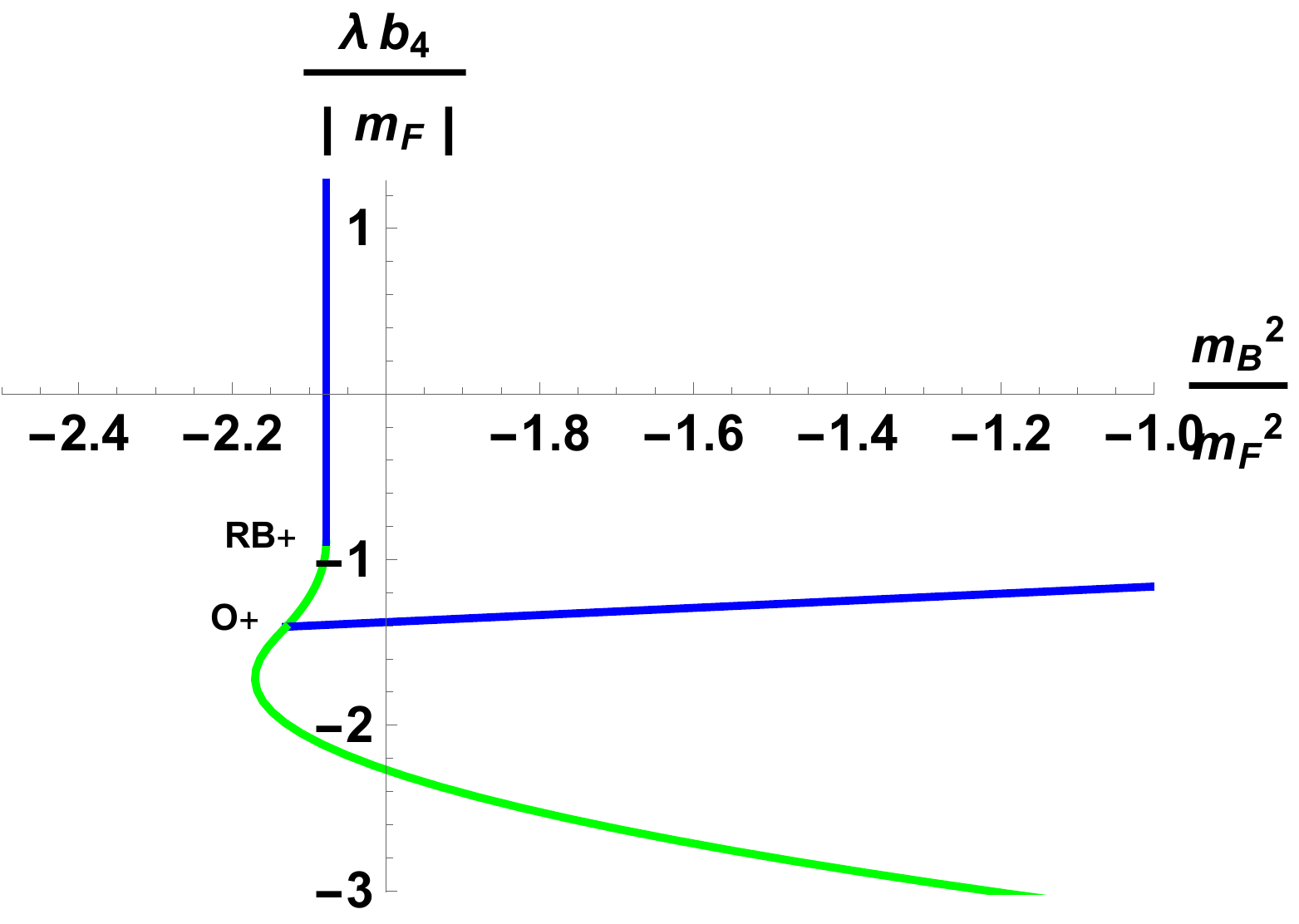}
	\caption{$|\lambda| =0.43$}
	\label{1stOrdb}
\end{subfigure}\hspace{10pt}
	\begin{subfigure}{0.3\textwidth}
	\includegraphics[width=1.7in,height=1.5in]{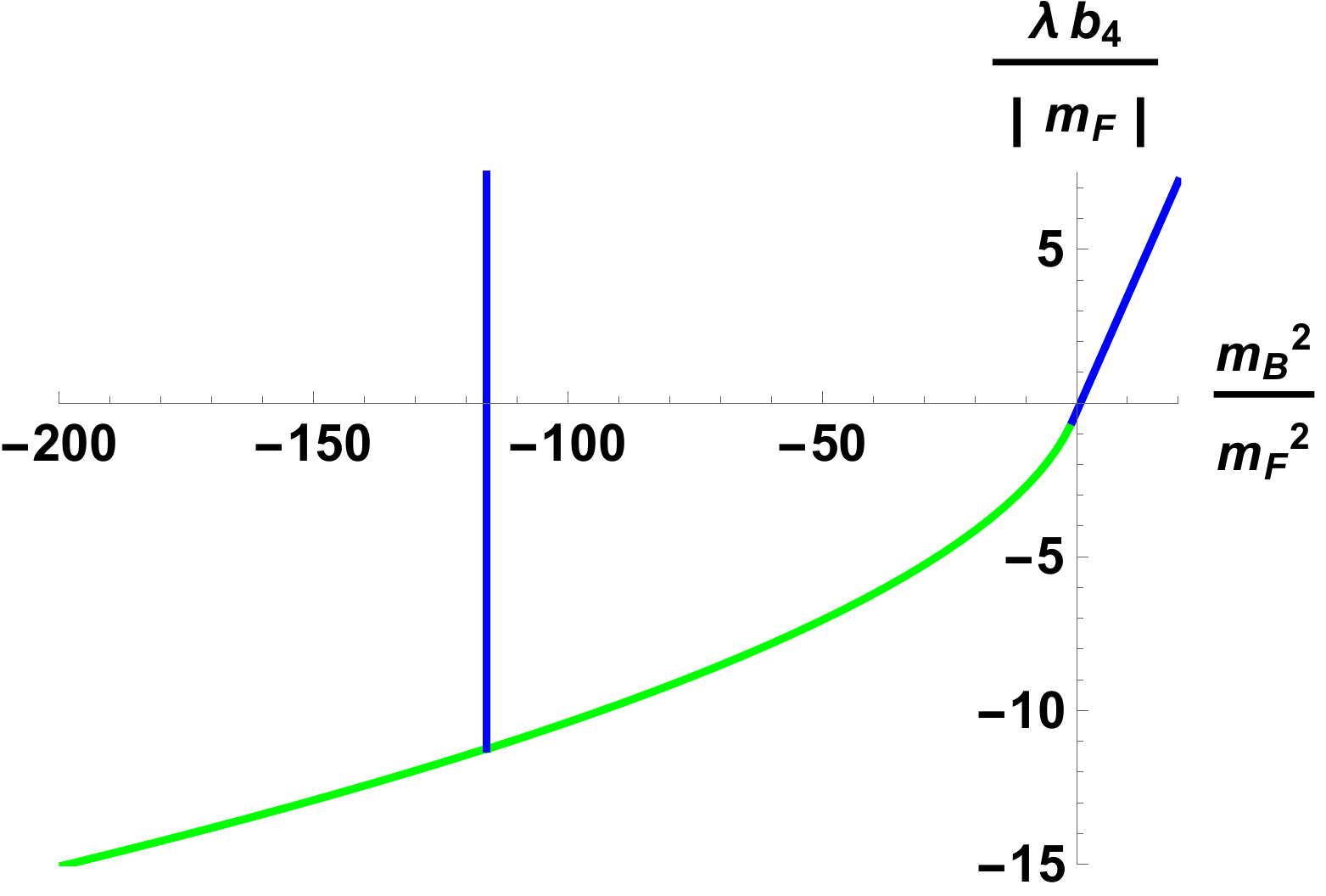}
		\caption{$|\lambda|=0.75$}
		\label{1stOrdc}
	\end{subfigure}
        \caption{Numerically determined first order transition lines
          with $m_F \sgn(\lambda) = 1$ for different values of
          $|\lambda|$.}
\label{1stOrd}
\end{figure}

We provide a few numerical plots of the first order transition lines
in Figure (\ref{1stOrd}) for the $|\lambda| = 0.25, 0.43, 0.75$ for
the section $m_F\sgn(\lambda) = 1$. The corresponding schematic
diagram is in Figure \ref{phaseall}(a).
In each plot, the blue lines are second order transition lines while
the green lines are the numerically determined first order transition
lines. See equations \eqref{Upara}, \eqref{Uprpara} in Section
\ref{N2phase} for the equations describing the second order transition
lines.

We should also mention that we can also \emph{analytically} determine
the first order line segment separating the $(+,+)$ and the $(+,-)$
phases. For $|\lambda|=0.25$, the analytic expression of this
segment is given by,
\begin{equation}
\frac{m_B^2}{m_F^2}=\frac{9}{80} \left(26+9 \sqrt{15}\right) \Big(\frac{\lambda \  b_4}{|m_F|}\Big)^2+\frac{7}{160} \left(26+9 \sqrt{15}\right) \frac{\lambda \ b_4}{|m_F|} +\frac{7 \left(7 \sqrt{15}-122\right)}{1280}\ .
\label{1stordseganal}
\end{equation}
We have checked that our numerical solution perfectly matches with the
analytically expression above.

\begin{figure}
	\centering
	\includegraphics[width=4in,height=3in]{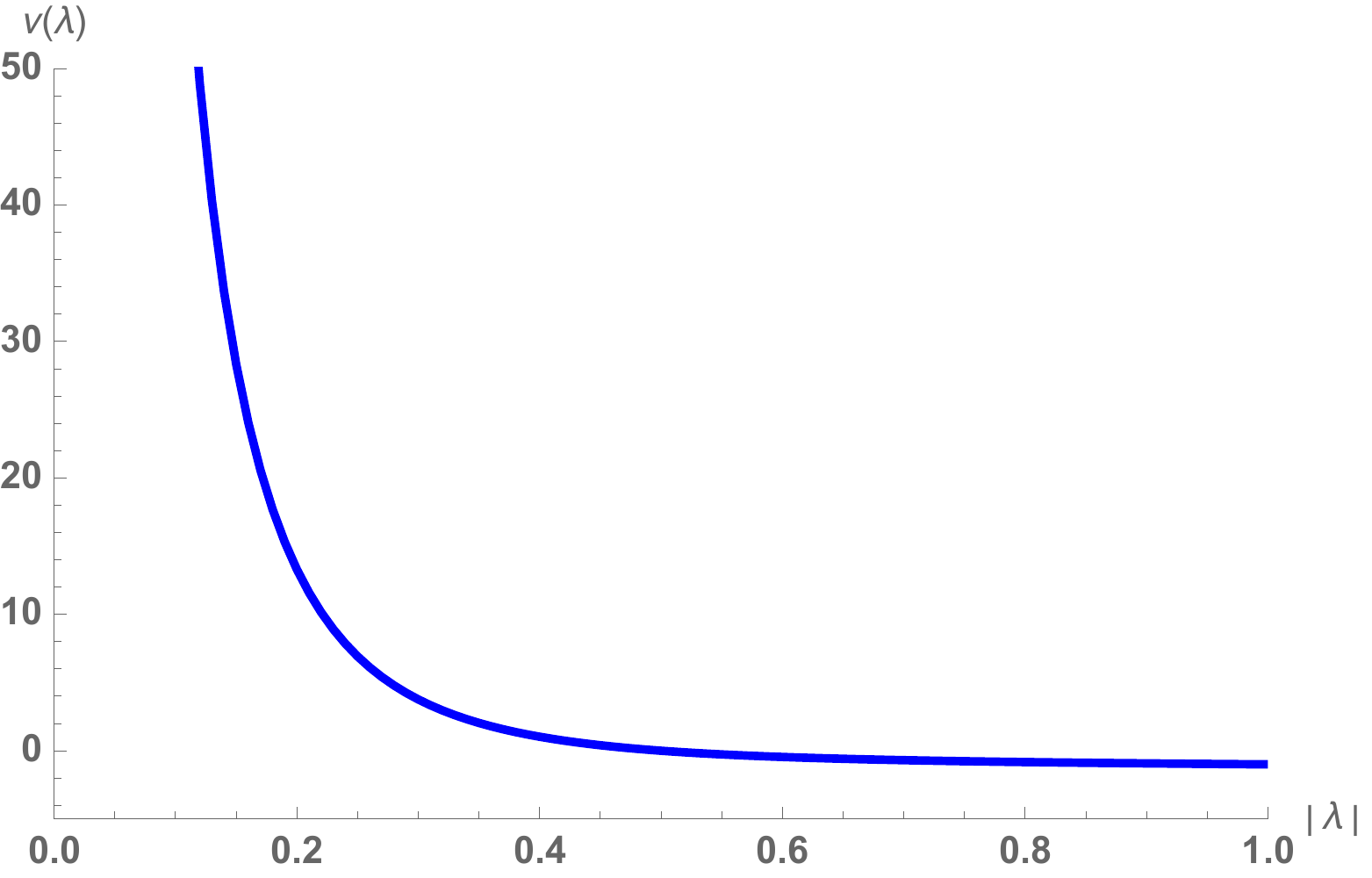}
	\caption{$\nu(\lambda)$ as a function of $|\lambda|$. The
          function $\nu(\lambda)$ appears in the equation for the
          first order line \eqref{mf0fo} in the $m_F = 0$ section of
          the $\mc{N} = 2$ phase diagram.}
	\label{nuvslambdamF0}
\end{figure}

We can also obtain an analytic expression for the first order line in
the case of the $m_F = 0$ section of the three dimensional parameter
space, analysed in Section \ref{mfzero}. The equation for the first
order line is given by
\begin{equation}\label{mf0fo}
  m_B^2 = \nu(\lambda)\ (\lambda b_4)^2\ ,
\end{equation}
where $\nu(\lambda)$ is a numerically determined function of
$|\lambda|$. We plot the profile of $\nu(\lambda)$ vs.~$|\lambda|$ in
Figure \ref{nuvslambdamF0}.
In the case of the first order lines in the northern and southern
hemisphere sections $m_F = \pm |\mu| \sgn(\lambda)$, the behaviour for
large $m_B^2$ and $\lambda b_4$ is very well approximated
by \eqref{mf0fo} since keeping $m_F$ fixed and taking $m_B^2$ and
$\lambda b_4$ to $\infty$ is equivalent via the scalings \eqref{untr}
to keeping $m_B^2$ and $\lambda b_4$ fixed and taking $m_F$ to zero.

\section{Scaling limits}\label{scaling}

In this section we study scaling limits of the theory
\eqref{generalaction} in which the physics - or more precisely the
dynamics about a particular saddle point of the theory
\eqref{generalaction} - simplifies. We will find scaling limits in
which the dynamics of the corresponding saddle points of our theory
reduces, in turn, to the critical fermion (CF) theory, the regular
fermion (RF) theory, the regular boson (RB) theory, the critical boson
(CB) theory. We also describe a scaling limit in which the resultant
theory is a CFT describing the interaction of an RF plus CB theory.

The scaling limits we study in this section are all (generalisations
of) scaling limits that have already been studied in
\cite{Jain:2013gza} (in the case of the RF, CB and mixed limits) and
\cite{Aharony:2018pjn} (in the case of the RB and CF limits). Our
discussion in this section generalises the discussions of
\cite{Jain:2013gza} and \cite{Aharony:2018pjn} in the following
directions. First the discussions in the earlier literature focused
exclusively on scaling limits in the $(+,+)$ and $(-,+)$ phases
(i.e. the unHiggsed phase of the boson with either phase for the
fermion). The new results for the Higgsed phase of the boson reported
earlier in this paper allow us to study also the $(+,-)$ and $(-,-)$
phases and thereby discover new scaling limits. Second, the
discussions in the earlier literature were always `local' in the
following sense: while the authors of \cite{Jain:2013gza} and
\cite{Aharony:2018pjn} established that the dynamics of particular
saddle points of the theory \eqref{generalaction} reduces to the
dynamics of simpler theories, the authors of those works were agnostic
about whether the saddle point being studied is in fact the dominant
saddle point of the theory under consideration. It is certainly
possible (and we will see below that it happens many times) for the
dynamics of a particular saddle point to simplify dramatically in a
particular scaling limit, without implying anything significant for
the physical behaviour of the full theory in the same scaling
limit. This happens simply because the saddle point under consideration is
subdominant in the scaling limit. In the current section we will shed
light on this issue by studying the quantum effective potential
(defined and computed earlier in this paper) in the scaling limits
mentioned above.  As the quantum effective potential has `global'
information of all phases of the theory, it allows us to definitively
determine when (and whether) the `simplified' scaled saddle points
that we study below actually dominate the physical dynamics of our
system.

Taking our lead from \cite{Jain:2013gza} and \cite{Aharony:2018pjn},
we first study scaling limits in which one of $c_B$ or $c_F$ is scaled
to $\infty$ with the other fixed. This is conveniently done by setting
\begin{equation}\label{scalingparam}
m_F = \mu\ ,\quad m_B^2 = a_1 \mu^2 + a_2 \mu + a_3\ ,\quad b_4 = g_1 \mu + g_2\ ,
\end{equation}
and taking the limit $|\mu| \to \infty$ with both the dimensionless
parameters ($a_1$ and $g_1$) and the dimensionful parameters ($a_2$,
$a_3$ and $g_2$) held fixed.

Of course, only dimensionless ratios of dimensionful parameters are
truly physical. The parametrization \eqref{scalingparam} is thus more
invariantly stated in terms of ratios as
\begin{equation}\label{scalingparamn}
  \frac{m_B^2}{m_F^2} = a_1 + \frac{a_2}{\mu}  + \frac{a_3}{\mu^2}\ ,\quad  \frac{b_4}{m_F} = g_1 + 
  \frac{g_2}{\mu}\ ,
\end{equation}
The parametrization \eqref{scalingparamn} (or \eqref{scalingparam}) is
extremely redundant:~we have chosen to parametrize the two physical
dimensionless ratios using two dimensionless parameters ($a_1$ and
$g_1$) together with 3 dimensionful parameters ($a_2$, $a_3$ and
$g_2$) and one additional auxiliary scale $\mu$.  This highly
redundant parametrization is useful under some circumstances as we now
describe.

\textbf{Quasi-bosonic scaling limits:} As we will see below, there are
special points in the space of ratios
$(\frac{m_B^2}{m_F^2}, \frac{b_4}{m_F})$ corresponding to the discrete values
$$\frac{m_B^2}{m_F^2} =\alpha_i\ ,\quad \frac{b_4}{m_F}=\gamma_i\ ,$$ where the
discrete parameter $i$ labels the various special points at which our
theory behaves in a very special way (see the CF and RB scaling limits
below for the values of $\alpha_i$ and $\gamma_i$). It is also
interesting to study our theory in the neighbourhood of these special
points corresponding to deforming the values of $\frac{m_B^2}{m_F^2}$
and $\frac{b_4}{m_F}$ away from their special values. For this purpose
we use the parametrization \eqref{scalingparamn} together with the
following natural convention:
$$a_1 =\alpha_i\ ,\quad g_1=\gamma_i\quad\text{throughout the
	procedure of taking the scaling limit.}$$ With these conventions, we
are now effectively using three dimensionful quantities and one
auxiliary scale $\mu$ (so a total of three dimensionless ratios) to
parametrize the two physically important dimensionless ratios. Clearly
our parametrization still has a one parameter redundancy; the
superfluity is simply the fact that the redefinitions
\begin{equation}\label{redef}
a_2 = a_2' + \frac{\xi}{\mu}\ ,\quad a_3'=a_3 - {\xi}\ ,
\end{equation}
leave the ratios \eqref{scalingparamn} unchanged and so define the
same theory. In order to obtain a unique parametrization of our ratio
space, we need a further convention to fix the ambiguity
\eqref{redef}. We address this in each of the individual subsections
below.

\textbf{Quasi-fermionic scaling limits:} We will also use the
parametrization \eqref{scalingparamn} to study the neighbourhood of
special lines (rather than special points) in the space of ratios
\eqref{scalingparamn} (see the CB and RF limits described below). In
this situation we simply choose the parameters $a_1$ and $g_1$ to lie
on the special lines. In addition to the ambiguity \eqref{redef},
there is now a new potential ambiguity in this context. This is
associated with motion \emph{along} the special line -- that is
changing $a_1$ and $g_1$ such that they still lie on the special line
and reabsorbing the changes in appropriate redefinitions of $a_2$,
$a_3$ and $g_2$. We need an additional convention to fix this
ambiguity.

To end the introductory part of this section we present the formulae
for the fermionic and bosonic gaps in the $(\varepsilon, \pm)$ phases
for convenient reference:
\begin{equation}\label{scalinggaps}
c_F = \frac{|{\tl m}_F|}{1 -\varepsilon |\lambda|}\quad\text{with}\quad {\tl m}_F = m_F + 2 x_4 \lambda \sigma\ ,\qquad c_B = \frac{2 - 2 |\lambda| \pm 2}{-(2 - |\lambda|)}\sigma\ ,
\end{equation}
where $\sigma$ is determined by extremizing the quantum effective
potential. 

\subsection{Fermionic scaling limits}
In order to scale $c_B$ to $\infty$ with $c_F$ fixed, we need to scale
$\sigma$ to $\infty$ such that the combination ${\tl m}_F$ is
fixed. We thus perform the following field redefinition from $\sigma$
to $\zeta$:
\begin{equation}\label{sigmazeta}
\sigma = -\frac{m_F}{2 x_4 \lambda} - \frac{\zeta}{2 x_4 \lambda}\ ,
\end{equation}
with $\zeta$ held fixed in the limit $\mu \to \infty$. 
In the scaling limits under study, in other words, we 
are focusing our attention on the part of the $\sigma$ axis 
that lies around the `phase boundary' at 
$\sigma=-\frac{m_F}{2 x_4 \lambda}$ in Figure \ref{phaseseq}.

\textbf{Note:} Our variable $\zeta$ is related to the variable
$\zeta_F$ (defined in Appendix \ref{critferapp}) that usually appears
in discussions of the critical fermion theory as
\begin{equation}
\zeta = \frac{4\pi \zeta_F}{\kappa_F}\ .
\end{equation}
In the main text of this paper we will continue to use the variable $\zeta$ instead of $\zeta_F$  to avoid cluttering of notation.

When $x_4 \lambda m_F > 0$, the fermionic scaling limit focuses on the
transition between $(-,+)$ and $(+,+)$ phases in Figure \ref{phaseseq}
and loses all information about the third phase\footnote{This third
	phase is $(+,-)$ when $x_4>0$ and $\lambda m_F>0$ while it is the
	$(-,-)$ phase when $x_4<0$ and $\lambda m_F<0$.} in Figures
\ref{phaseseq}(a) and \ref{phaseseq}(d) (since $m_F \to \infty$ with
$x_4$ and $\lambda$ fixed). Note that when $x_4 \lambda m_F>0$ the
scaling limit lies entirely in the unHiggsed phase of the boson.

Similarly, when $x_4 \lambda m_F < 0$ the fermionic scaling limit
focuses on the transition between the $(-,-)$ and $(+,-)$ phases and
loses all information about the third phase in Figures
\ref{phaseseq}(b) and \ref{phaseseq}(c). It follows that when
$x_4 \lambda m_F<0$, the scaling limit lies entirely in Higgsed
phase of the boson.

Plugging \eqref{sigmazeta} into the effective potential
$U^{(\varepsilon,\pm)}$ \eqref{LGpot}, we get
\begin{align}\label{Uredef}
\frac{2\pi}{N}  U^{(\varepsilon,\pm)}(\zeta)
&=  -\left(\psi_\varepsilon + \frac{x_6 - \phi_\pm}{x_4^3}\right)\frac{\zeta^3}{8\lambda} -\left(-\frac{b_4}{2 x_4^2 \lambda} + \frac{3 m_F}{8 x_4^3 \lambda}(x_6 - \phi_\pm)\right) \zeta^2 \nonumber\\
&+ \left(\frac{b_4 m_F}{x_4^2 \lambda} - \frac{m_B^2}{2 x_4 \lambda} - \frac{3 m_F^2}{8 x_4^3 \lambda}(x_6 - \phi_\pm)\right)\zeta\ ,
\end{align}
where we have ignored the terms independent of $\zeta$\footnote{The
	constant term in the potential has a piece proportional to $\mu^3$
	which just corrects the cosmological constant counterterm, and is
	not required to vanish.}. We also plug in the parametrization
\eqref{scalingparam} in the above formula to get
\begin{align}\label{Uredefn}
\frac{2\pi}{N}  U^{(\varepsilon,\pm)}(\zeta)
&=  -\left(\psi_\varepsilon + \frac{x_6 - \phi_\pm}{x_4^3}\right)\frac{\zeta^3}{8\lambda} +\left(\frac{\mu}{2x_4^2\lambda} \left(g_1 - \tfrac{3}{4 x_4}(x_6 - \phi_\pm)\right) + \frac{g_2}{2x_4^2 \lambda}\right) \zeta^2 \nonumber\\
&+ \left(\frac{\mu^2}{2 x_4^2\lambda}\left(2 g_1 - x_4 a_1 - \tfrac{3}{4 x_4}(x_6-\phi_\pm)\right) + \frac{\mu}{2x_4^2\lambda} (2g_2 - x_4 a_2) - \frac{a_3}{2x_4\lambda}\right)\zeta\ ,
\end{align}
In order for the assumptions of this subsection to be self-consistent,
the extremum value of $\zeta$, obtained by extremizing \eqref{Uredef}
w.r.t $\zeta$, should be of order unity (as we have assumed in
\eqref{sigmazeta}) rather than of order $\mu$. This requirement puts
some constraints on the parameters $a_1$, $a_2$, $a_3$, $g_1$ and
$g_2$ described above; solutions to this constraint generically yield
the RF limit.  If we also demand that not just the value of $\zeta$ at
its extremum but that its whole quantum effective potential has a good
$\mu \to \infty$ limit, we get additional constraints and land on the
CF scaling limit. We take these up in turn, starting with the CF
scaling limit.

\subsubsection{The critical fermion limit} \label{cflim}

We wish to find a scaling limit in which the potential for the field
$\zeta$ is finite, i.e. a limit in which the field $\zeta$ can
fluctuate with a finite cost (as opposed to an infinite cost, which would be the case in an 
infinitely steep potential) in
potential. As noted above, when $x_4 \lambda m_F>0$, our scaling
limit lies in the bosonic unHiggsed ($+$) phase; in this case the CF
limit has previously been studied in \cite{Aharony:2018pjn}.  We will
call this the $\CF+$ limit. On the other hand when $x_4 \lambda m_F<0$
the CF scaling limit lies in the Higgsed ($-$) phase of the boson;
this limit has not been studied before in the previous literature
(because it involves working in the bosonic Higgsed phase). We will
call this the $\CF-$ limit.

Note that while the coefficient of $\zeta^3$ in \eqref{Uredefn} is
independent of $\mu$, the coefficients of $\zeta^2$ and $\zeta$ have
terms proportional to $\mu$ and $\mu^2$. On physical grounds we would
like this potential to be independent of $\mu$. One way to ensure this
is to set the coefficient of every $\mu$-dependent term in the
potential separately to zero. \footnote{This strategy is not
	completely forced on us but involves a choice as we explain in much
	more detail at the end of this subsubsection.}  This gives the
following constraints on the parameters in \eqref{scalingparam}:
\begin{align}\label{limita1g1}
g_1 = \gamma^F_{\pm}\ ,\quad a_1 = \alpha^F_\pm\ ,\quad a_2 = \frac{2g_2}{x_4}\ .
\end{align}
where the special values $\gamma^F_\pm$ and $\alpha^F_\pm$ correspond
to the $\CF+$ and $\CF-$ limits and are given by
\begin{align}\label{critvalF}
\gamma^F_{\pm} = \frac{3 (x_6 - \phi_\pm)}{4 x_4}\ ,\quad \alpha^F_{\pm} = \frac{3 (x_6 - \phi_\pm)}{4 x_4^2}\ .
\end{align}
(Here, of course, $x_6$ and $x_4$ refer to the dimensionless coupling
constants of the original UV theory \eqref{Feffofsh}). In these two
scaling limits the effective potential reduces to the quantum
effective potential for the CF theory
\begin{align}\label{actforcf}
S_{\rm CF}  &= \int d^3 x \bigg[ i \varepsilon^{\mu\nu\rho} \frac{\kappa}{ 4
	\pi} \Tr( A_\mu\partial_\nu A_\rho -\frac{2 i}{3} A_\mu A_\nu
A_\rho) + \bar{\psi} \gamma_\mu D^{\mu} \psi \nn 
&\qquad\qquad\quad
-\zeta (\bar\psi \psi - \frac{\kappa
	y_2^2}{4\pi} ) - \frac{\kappa y_4}{ 4\pi} \zeta^2 +
\frac{\kappa}{16\pi} x_6^F \zeta^3\bigg]\ ,
\end{align}
whose effective potential is given by (see Appendix \ref{critferapp})
\begin{equation}\label{critpot}
U_{\rm CF}(\zeta)
=\left\{\def\arraystretch{2.2}\begin{array}{ll}\displaystyle\frac{N}{2\pi}\left[(x_6^{F} - \psi_+) \frac{\zeta^3}{8 \lambda} - \frac{ y_4}{2 \lambda} \zeta^2 + \frac{ y_2^2}{2\lambda}\zeta \right]\ ,\quad & \sgn(\zeta)\sgn(\lambda) < 0 \\ \displaystyle\frac{N}{2\pi}\left[(x_6^{F} - \psi_-) \frac{\zeta^3}{8 \lambda} - \frac{ y_4}{2 \lambda} \zeta^2 + \frac{ y_2^2}{2\lambda}\zeta \right]\ ,\quad & \sgn(\zeta)\sgn(\lambda) > 0\ .\end{array}\right.
\end{equation}
where $\psi_{\varepsilon}$ was defined in \eqref{psipm}. 

The parameters $x_6^{F}$, $y_2$ and $y_4$ are given in terms of
$g_2$, $a_3$ by
\begin{equation}\label{critferparam}
x_6^F =  x_6^{F\pm} = -\frac{x_6 - \phi_\pm}{x_4^3}\ ,\quad y_4 = -\frac{g_2}{ x_4^2}\ ,\quad y_2^2 =  - \frac{a_3}{x_4}\ . 
\end{equation}
From \eqref{sigmazeta}, it is apparent that we reach the $\CF\pm$
limit when $\sgn(m_F / x_4 \lambda) = \pm 1$.

The effective potential \eqref{critpot} is bounded from below if and
only if $x_6^F$ lies in the interval $(\psi_-, \psi_+)$. When this is
not the case, \eqref{critpot} unbounded from below, and one of the
following things must be true of the UV theory:
\begin{enumerate}
	\item Either the potential of the UV theory is also unbounded from
	below,
	\item Else the UV potential has a minimum that lies outside the
	scaling limit with a lower free energy than any configuration that
	can be accessed within the scaling limit. In this case the dominant
	saddle point of UV theory is not captured by our scaling limit.
\end{enumerate}
In either of the two cases above the scaling limit is not really
interesting (in the first case because the UV theory itself is
unstable, and in the second case because the scaling limit focuses on
a subdominant saddle point).

If, on the other hand, $x_6^F$ does lie in the interval
$(\psi_-, \psi_+)$ it is at least possible that the scaling limit
focuses on a region of $\sigma$ that includes the dominant saddle
point of a well-defined UV theory. Whether that is actually the case
or not requires a more detailed study of saddle points outside the
scaling limit; we will perform such an analysis in the special case of
the ${\cal N}=2$ theory below.

We now address the ambiguities in our description referred to around
equation \eqref{scalingparamn}. In our scaling ansatz, the ultraviolet
theory is parametrized by two dimensionless parameters ($a_1$ and
$g_1$), three dimensionful parameters ($a_2$, $a_3$ and $g_2$) and one
mass scale $\mu$. In our computation above, we have chosen the
convention that each $\mu$-dependent term in the potential is set to
zero separately. This ensures that the dimensionless parameters $a_1$
and $g_1$ are fixed to their limiting values (cf.~\eqref{limita1g1})
even when $\mu$ is finite but large. Similarly, of the three
dimensionful parameters, two are related to each other by the last
equality $2g_2 = x_4 a_2$ in \eqref{limita1g1}, again even when $\mu$
is finite but large. Thus, at the end of the scaling limit (where we
have lost the mass scale $\mu$), we have two dimensionful parameters
(conveniently chosen to be $g_2$ and $a_3$, cf.~\eqref{critferparam})
which matches the physics of the critical fermion theory deformed by
its \emph{two} relevant parameters.

In the just-concluded computation, we adopted the convention that each
$\mu$-dependent term is set to zero separately. However, we emphasize
that this is a choice that fixes the ambiguity of parametrization
(discussed around \eqref{scalingparam}) and not a physical requirement
as we now pause to explain.

Let us focus on the coefficient of $\zeta$ in \eqref{Uredefn} (the
discussion for the coefficient of $\zeta^2$ proceeds analogously). If
we denote the coefficient of $\zeta$ by $A \mu^2 +B \mu + C$ it is not
strictly really necessary to set $A$ and $B$ separately to
zero. Instead, we could set $A = \frac{p_1}{\mu} + \frac{p_2}{\mu^2}$
and $B= -p_1 + \frac{q_2}{\mu}$ so that the $\mu$-dependent pieces
vanish. In order to maintain the $\mu$-independent part at its
previous value, we need to redefine it by $C' = C - p_2 - q_2$. In
terms of the parametrization \eqref{scalingparamn}, this corresponds
to the redefinition
\begin{align}\label{paramredef}
&2g_1' - x_4 a_1' = 2g_1 - x_4 a_1 + \frac{p_1}{\mu} + \frac{p_2}{\mu^2}\ ,\nonumber\\
&2g_2' - x_4 a_2' = 2g_2 - x_4 a_2 - p_1 + \frac{q_2}{\mu}\ ,\nonumber\\
&-x_4 a_3' = -x_4 a_3 - p_2 - q_2\ . 
\end{align}
Suppose we first fix the convention alluded to in the discussion
around \eqref{scalingparamn} that $a_1$ and $g_1$ are fixed to the
special values $\alpha^F_\pm$ and $\gamma^F_\pm$ (this is ensured in
our choice of setting each $\mu$-dependent term to zero
separately). This kills the arbitrary constants $p_1$ and $p_2$ in
\eqref{paramredef}. The remaining ambiguity resides in the constant
$q_2$ and is the ambiguity referred to \eqref{redef}. This ambiguity
is also killed by our choice of setting each $\mu$-dependent term to
zero separately.

\subsubsection{The regular fermion limit}\label{rflim}

In the previous subsection we studied a scaling limit (or RG flow)
under which the theory \eqref{Feffofsh} reduces to the CF theory. As
we have reviewed in Appendix \ref{critferapp}, the quantum effective action for
the CF theory is nontrivial (i.e. allows for a fluctuating $\zeta$
field). This is the reason we determined our scaling limit in the last
subsection by demanding that the quantum effective action for $\zeta$
is nontrivial.

In this subsection we search for a scaling limit in which our theory
reduces instead to the (less special) RF theory. As we review in
Appendix \ref{critferapp}, the quantum effective action for the RF is
trivial - or more precisely infinitely deep around an extremum which,
however, occurs at a finite value of $\zeta$. For this reason in this
section we search for a scaling limit in which the effective potential
takes the appropriate form; i.e. is infinitely deep at a finite value
of $\zeta$.

One way of achieving this goal is simply to set the coefficient of
$\mu^2$ in coefficient of $\zeta$ in the quantum effective potential
\eqref{Uredefn} to zero. This gives
\begin{equation} \label{rflo}
2 g_1 =  x_4 a_1 + \frac{3}{4 x_4}(x_6 - \phi_\pm)\ ,
\end{equation}
which can be rewritten as
\begin{equation} \label{rflt}
2 \left( g_1-\gamma_{\pm}^F \right) =
x_4 \left( a_1 - \alpha_{\pm}^F \right)\ .
\end{equation}
where $\gamma_{\pm}^F$ and $\alpha_{\pm}^F$, defined in
\eqref{critvalF}, are the values of $g_1$ and $a_1$ at which we obtain
the CF theory. The potential in \eqref{Uredefn} becomes
\begin{align}\label{Uregfer}
\frac{2\pi}{N}  U(\zeta)
&=  -\left(\psi_\varepsilon + \frac{x_6 - \phi_\pm}{ x_4^3}\right)\frac{\zeta^3}{8\lambda}  +  \frac{\zeta^2}{2 x_4^2 \lambda} \left(\mu(g_1 - \gamma^F_{\pm}) + g_2 \right)\nonumber\\
& +  \frac{\zeta}{2 x_4^2 \lambda} \left(\mu( 2g_2 -  x_4 a_2) - x_4 a_3 \right)\ ,
\end{align}
The terms in the potential \eqref{Uregfer} that are proportional to
$\mu$ can be rewritten by completing squares as
\begin{align}
\frac{\mu(g_1 - \gamma^F_{\pm})}{2x_4^2 \lambda}\left(  \left(\zeta + \frac{(2 g_2 - x_4 a_2)}{2(g_1 - \gamma^F_{\pm})}\right)^2 - \frac{(2g_2 - x_4 a_2)^2}{4(g_1 - \gamma^F_{\pm})^2}\right)\ .
\end{align}
In the $\mu \to \infty$ limit, the potential is dominated by the above
term. Suppose the prefactor
$ \frac{\mu(g_1 - g^F_{1\pm})}{2x_4^2 \lambda}$ is positive:
\begin{equation}\label{condmin}
\frac{\mu(g_1 - \gamma^F_{\pm})}{2x_4^2 \lambda} \geq 0\ ,\quad{\rm i.e.}\quad {\sgn}(g_1 - \gamma^F_{\pm}) ={\sgn}(\lambda \mu)\ .
\end{equation}
In this case the following value of $\zeta$ is a minimum of the above
term and the variable $\zeta$ freezes to this value in the
$\mu \to \infty$ limit:
\begin{equation} \label{zerf} \zeta = - \frac{2g_2 - x_4 a_2 }{2(g_1 -
	\gamma^F_{\pm})} +{\cal O}(1/\mu)\ ,
\end{equation}
where we have indicated possible $1/\mu$ corrections to the extremum
which vanish in the $\mu \to \infty$ limit. In this section, we adopt
the convention that the ${\cal O}(1/\mu)$ correction to \eqref{zerf}
vanishes so that $\zeta$ takes the following precise value at the
minimum of the potential:
\begin{equation}\label{mfreg}
\zeta_{\rm min.} =  - \frac{2g_2 - x_4 a_2 }{2(g_1 - \gamma^F_{\pm})}\ .
\end{equation}
It is natural to label the above minimum of $\zeta$ as
\begin{equation} \label{mfregn}
m_F^{\rm reg} = -\zeta_{\rm min.} =  \frac{2g_2 - x_4 a_2 }{2(g_1 - \gamma^F_{\pm})}\ ,
\end{equation}
since this minimum value of $\zeta$ will serve as the mass parameter
of the regular fermion theory.

The requirement that we have already enunciated - that the ${\cal O}(1/\mu)$ correction to 
\eqref{zerf} vanish - holds only if 
the special value of $\zeta$ described in
\eqref{mfreg} also separately extremize the $\mu$-independent part of
the potential listed in \eqref{Uregfer}. This occurs provided that
\begin{align}\label{secondcond}
& \left(\psi_\varepsilon + \frac{x_6 - \phi_\pm}{ x_4^3}\right)\frac{3(m_F^{\rm reg})^2}{8\lambda} + \frac{g_2 m_F^{\rm reg}}{x_4^2 \lambda} + \frac{a_3}{2 x_4 \lambda} = 0\ .
\end{align}
The solution \eqref{zerf} is a local minimum - rather than a local
maximum - of the potential $U(\zeta)$ in \eqref{Uregfer} provided that
\eqref{condmin} holds. Note that the LHS of \eqref{condmin} changes
sign as $g_1$ crosses $\gamma^F_{\pm}$, i.e. passes the critical
fermion scaling limit. It follows that the regular fermion fixed line
is locally stable only on one side of the CF scaling limit, as
expected on general grounds.

\eqref{rflt} and \eqref{secondcond} and 
\eqref{mfregn} are three conditions on the 5 parameters that appear in 
\eqref{scalingparamn}. In more detail 
\eqref{rflt} gives one relation between the two dimensionless parameters $a_1$ and $g_1$  
in \eqref{scalingparamn} while \eqref{secondcond}
and \eqref{mfregn} gives two relationships between the three dimensionful parameters $g_2$, $a_2$ and $a_3$ and the one physical dimensionful parameter of the IR theory, namely $m_F^{\rm reg}$. 
Even with all the conventions adopted in this 
subsection, in other words, we still have  
landed on a two parameter set of scaling limits 
rather than a unique scaling limit as in the 
previous subsection. The reason for this, of course, is that the RF limit occurs on a line 
in the phase diagram of our theory- rather than on a point in the phase diagram of our theory
as was the case in the previous section. 
The shift in dimensionless parameters that was 
unfixed by \eqref{rflt} simply moves us along 
the line of RF theories in our phase diagram. 
And the one unfixed massive parameter also actually
does the same. To see the last point note that 
that $m_F^{\rm reg}$ depends only on the combination $2g_2 - x_4 a_2$. Consequently, 
the unfixed massive parameter is parametrized 
by the simultaneous shift
\begin{equation}\label{a2g2}
2 g_2 = 2g'_2 - p\ ,\quad x_4 a_2 = x_4 a'_2 - p\ ,
\end{equation}
that leaves $m_F^{\rm reg}$ invariant. A glance at 
\eqref{scalingparamn} will convince the reader 
that this shift also moves us along the line of 
RF theories. 

As both the moves above, i.e. shifts in $a_1$ and 
$g_1$ that continue to obey \eqref{rflt} and the shifts in $a_2$, $g_2$ in 
\eqref{a2g2} move us along the same line in phase 
space, there is a linear combination of these 
shifts that do not move us in phase space; this 
linear combination is an unfixed ambiguity of 
our parametrization of this scaling limit. This 
ambiguity is given by \eqref{a2g2} together 
with 
\begin{equation}\label{a1g1}
2g_1 = 2g'_1 + \frac{p}{\mu}\ ,\quad x_4 a_1 = x_4 a'_1 + \frac{p}{\mu}\ .
\end{equation}
(as a check one can verify that above two shifts \eqref{a2g2}, \eqref{a1g1} leave
the potential in \eqref{Uregfer} unchanged). 

In order to reach the scaling limit of this 
section, on physical grounds we needed only to 
impose a single condition, namely that 
\eqref{rflt} hold upto corrections of order 
$1/\mu$. The extra conditions we have imposed, 
namely that this equation hold exactly rather 
than upto corrections of order $1/\mu$ and 
also that the terms order $1/\mu$ in \eqref{zerf} 
vanish - were imposed for convenience rather 
than necessity, and served to fix some of the
ambiguities of our parametrization (discussed
under \eqref{scalingparamn}) in a manner similar 
to the discussion of the previous subsection.

\subsection{Bosonic scaling limits}

By definition, a bosonic scaling limit is one in which the bosonic
pole mass $c_B$ stays finite while $c_F$ is scaled to $\infty$ so that
the resultant theory is purely bosonic. From the second of
\eqref{scalinggaps} we see that $c_B$ is proportional to $\sigma$,
which implies that $\sigma$ must stay finite in the scaling limit
$\mu \to \infty$.  As a consequence the effective potential in the
bosonic scaling limit is a function of $\sigma$ (rather than of
$\zeta$ as in the fermionic limit), and the bosonic analogue of
\eqref{Uredefn} is obtained by plugging the definition of
${\tl m}_F = m_F + 2\lambda x_4 \sigma$ in terms of $\sigma$ into the
original expression for the Landau-Ginzburg potential
\eqref{LGpot}. We find
\begin{multline}\label{Uregbosp}
U^{(\varepsilon,\pm)}(\sigma) = \frac{N}{2\pi}\Big[ \left(x_6 - \phi_\pm + x_4^3\psi_\varepsilon\right)  \lambda^2 { \sigma}^3  +  \left( { b}_4 + \tfrac{3}{4}  x_4^2\psi_\varepsilon m_F \right) 2 \lambda  { \sigma}^2\\
+ \left({ m}_B^2 + \tfrac{3}{4} x_4 \psi_\varepsilon m_F^2\right){
	\sigma}\Big]\ ,
\end{multline}
where we have ignored a constant term proportional to
$\mu^3$. Plugging in the scalings in \eqref{scalingparam}, the above
expression becomes
\begin{multline}\label{Uregbos}
U^{(\varepsilon,\pm)}(\sigma) = \frac{N}{2\pi}\Big[ \left(x_6 - \phi_\pm + x_4^3\psi_\varepsilon\right)  \lambda^2 { \sigma}^3  +  \left( \mu (g_1  + \tfrac{3}{4}  x_4^2\psi_\varepsilon) + g_2 \right) 2 \lambda  { \sigma}^2\\
+ \left( \mu^2 (a_1 + \tfrac{3}{4} x_4 \psi_\varepsilon)  + a_2 \mu + a_3\right){ \sigma} \Big]\ ,
\end{multline}

\subsubsection{Regular boson limit}\label{rblim}

As in the CF limit, we obtain the regular boson limit by substituting
in the scalings \eqref{scalingparam} into \eqref{Uregbos}. The
coefficient of $\sigma$ in \eqref{Uregbos} has a term proportional to
$\mu^2$ and a term proportional to $\mu$ (in addition to the constant
piece), while the coefficient of $\sigma^2$ has a term proportional to
$\mu$ (in addition to a constant). As in Section \ref{cflim} we tune
to the RB limit by setting all three coefficients mentioned above to
zero\footnote{As in the CF limit two of these conditions are forced on
	us on physical grounds while the third is a choice that fixes the
	ambiguity \eqref{redef}.}. This requirement gives
\begin{equation} \label{rbpoint}
g_1 = \gamma^B_{\pm}\ ,\quad
a_1 = \alpha^B_{\pm}\ ,\quad a_2 = 0\ ,
\end{equation}
where $\gamma_\pm^B$ and $\alpha_\pm^B$ are given by
\begin{equation} \label{gammaalpha}
\gamma^B_{\pm} = - \frac{3 x_4^2}{4} \psi_{\pm}\ ,\quad   \alpha^B_{\pm} = - \frac{3 x_4}{4} \psi_{\pm}\ ,
\end{equation}
where $\psi_\pm$ was defined in \eqref{psipm}. The effective
potential \eqref{Uregbos} reduces to that of the regular boson:
\begin{equation} \label{bosep}
U_{\rm RB}(\sigma) = \left\{\def\arraystretch{2.2}\begin{array}{lr} \frac{N}{2\pi} \left[(x_6^{B} - \phi_+) \lambda^2 \sigma^3 + 2 {\tl b}_4 \lambda \sigma^2 + {\tl m}_B^2 \sigma\right]\quad & \sigma < 0\ ,\\ \frac{N}{2\pi} \left[(x_6^{B} - \phi_-) \lambda^2 \sigma^3 + 2 {\tl b}_4 \lambda \sigma^2 + {\tl m}_B^2 \sigma\right]\quad & \sigma > 0\ ,\end{array}\right.
\end{equation}
with 
\begin{equation} \label{xsixval}
x_6^B = x_6^{B\pm} = x_6 + x_4^3 \psi_{\pm}\ ,\quad {\tl b}_4 = g_2\ ,\quad {\tl m}_B^2 = a_3\ .
\end{equation}
where the sign $\pm$ applies when the infinitely massive fermion which
is being integrated out is in its $+$ or the $-$ phase respectively.
In the scaling limit where we keep $\sigma$ to be $\mc{O}(1)$, we see
that $\sgn({\tl m}_F) = \sgn(m_F)$; it follows that the fermion is in
the $\pm$ phase when $\sgn(m_F) \sgn(\lambda) = \pm 1$ respectively.
We refer to these two cases as the ${\rm RB\pm}$ scaling limits
respectively.

As the RB limit focuses on finite values of $\sigma$, it zeroes in on
the phases around $\sigma=0$ in each of the four cases listed in Fig
\ref{phaseseq}. In particular when $\lambda m_F>0$ our limit (the
$\RB+$ limit) focuses on the `transition region' between the $(+,+)$
and $(+,-)$ phases (discarding the $(-,+)$ phase in the case the
$x_4>0$ and discarding the $(-,-)$ phase in the case that $x_4<0$). On
the other hand in the case that $\lambda m_F<0$ our limit focuses on
the `transition between' $(-,+)$ and $(-,-)$ (discarding the $(+,-)$
phase when $x_4>0$ and discarding the $(-,-)$ phase when $x_4<0$.

When $x_6^B$ lies outside the interval $(\phi_-, \phi_+)$, the
effective potential \eqref{bosep} is unbounded from below (as
described in Appendix \ref{critferapp}). When this happens (as in
Section \ref{cflim}) the saddle point we are focusing on in this
subsubsection cannot represent the global minimum of the UV effective
potential. In this situation, as in Section \ref{cflim}, the UV
potential is either unbounded from below or has another saddle point
(other than the one focused on in this subsubsection) which
represents the dominant phase of the theory. When, on the other hand,
$x_6^B$ lies in the $(\phi_-, \phi_+)$, the saddle point focused on
in this subsubsection may (or may not) represent the global minimum of
the UV effective potential depending on details.

\subsubsection{The critical boson scaling limit}\label{cblim}

The discussion here is very similar to the regular fermion scaling
limit. In order to go close to the critical boson fixed point, one
needs to have a competition between the $\sigma^2$ and $\sigma$
deformations. This allows at most a linear dependence on the mass
scale $\mu$ in the effective potential \eqref{Uregbos}. One way to
achieve this - the way we adopt in this paper - is to set the
coefficient of $\mu^2$ to zero \footnote{Another way to do this is to
	set the coefficient of $\mu^2$ to be of $\frac{1}{\mu}$. By choosing
	the value zero in this subsubsection we are making a choice that
	effectively fixes the ambiguity \eqref{redef}. This discussion is
	exactly the same as for the regular fermion limit}.

Demanding that the term proportional to $\mu^2$ in the coefficient of
$\sigma$ is zero gives
\begin{equation}
a_1 = - \frac{3}{4}x_4 \psi_\pm\ ,
\end{equation}
or 
\begin{equation}\label{cblimm}
a_1 -\alpha_{\pm}^B= 0\ ,
\end{equation}
and the effective potential \eqref{Uregbos} becomes
\begin{align}
\frac{2\pi}{N} U^{(\varepsilon,\pm)}(\sigma) & =  \left(x_6 - \phi_\pm + x_4^3\psi_\varepsilon\right)  \lambda^2 { \sigma}^3  +   \left( \left(g_1 + \tfrac{3}{4}x_4^2\psi_\varepsilon\right)\mu + g_2 \right) 2\lambda{ \sigma}^2 + \left(a_2 \mu + a_3\right){ \sigma}\ ,\label{Ucritbos}
\end{align}
up to a constant that scales as $\mu^3$. In the $\mu \to \infty$
limit, the part of the potential that is dominant is
\begin{equation}
\mu \left(2\lambda (g_1 - \gamma^B_{\pm}) \sigma^2 + a_2 \sigma\right) = 2\lambda\mu(g_1 - \gamma_{\pm}^B) \left(\left(\sigma + \frac{a_2}{4\lambda(g_1 - \gamma_{\pm}^B)}\right)^2 - \frac{a_2^2}{16\lambda^2(g_1 - \gamma_{\pm}^B)^2}\right)\ .
\end{equation}
The above potential has an extremum at
\begin{equation}
\sigma = -\frac{a_2}{4\lambda(g_1 - \gamma_{\pm}^B)}\ ,
\end{equation}
which is
a local minimum provided that
\begin{equation}\label{minmax}
2\lambda\mu(g_1 - \gamma_{\pm}^B) \geq 0.
\end{equation}
When this is satisfied, $\sigma$ freezes at the extremum in the
$\mu \to \infty$ limit:
\begin{equation} \label{exval}
\sigma  \equiv - \tfrac{1}{2}m_B^{\rm cri} = -\frac{a_2}{4\lambda(g_1 - g_{1\pm}^B)} \ .
\end{equation}
The minimum value of $\sigma$ also has to separately minimise the
$\mu$-independent part of the potential, giving rise to the equation
\begin{equation}
a_3 = 2\lambda m_B^{\rm cri} g_2 - \frac{3}{4}(x_6 - \phi_\pm + \psi_\varepsilon)(m_B^{\rm cri})^2\ .
\end{equation}
When the converse of \eqref{minmax} is true, the extremum
\eqref{exval} is a local maximum rather than a minimum, and cannot
represent the true phase of the UV theory. As in the case of the RF
scaling limit, \eqref{cblimm} is a line in parameter space that runs
through the RB point \eqref{rbpoint}. The condition \eqref{minmax} is
obeyed on one side of the RB point and not the other. In the
neighbourhood of the point \eqref{rbpoint}, the condition
\eqref{minmax} is obeyed along the part of the RB phase diagram that
corresponds to the second order phase transition line that emanates
out of the RB conformal point (see the phase diagrams listed in Figure
\ref{phaseappRB} and also Figures 7,8,9 in \cite{Aharony:2018pjn} and
Figures 6,8,11 in \cite{Dey:2018ykx}).

\subsection{The CB-RF scaling limit}\label{cbrflim}

\subsubsection{Anticipating the scaling limit}
We have explained above that the $\RF\pm$ limit occurs in the
neighbourhood of the line
\begin{equation} \label{rfltrev}
2 \left( \frac{b_4}{m_F}-\gamma_{\pm}^F \right) = x_4 
\left( \frac{m_B^2}{m_F^2} - \alpha_{\pm}^F \right)\quad\text{with}\quad\frac{m_F}{\lambda} 
\left( \frac{b_4}{m_F} - \gamma_{\pm}^F \right) \geq 0\ .
\end{equation}
where $\gamma^F_\pm$ and $\alpha^F_\pm$ were 
defined in \eqref{critvalF}. 
The $+$ sign in the equation above applies when $x_4 \lambda m_F>0$,
while the $-$ sign applies when $x_4 \lambda m_F<0$. To remind the
reader, the $\RF+$ limit corresponds to regular fermions coupled to
$SU(N)_k$ Chern-Simons gauge fields while the $\RF-$ limit corresponds
to regular fermions coupled to $SU(N-1)_k$ Chern-Simons gauge fields.

On the other hand the $\CB\pm$ limit occurs in the neighbourhood of
the line
\begin{equation}\label{nowp}
\frac{m_B^2}{m_F^2} - \alpha_{\pm}^B= 0\quad\text{with}\quad \frac{m_F}{\lambda} \left(\frac{b_4}{m_F} - \gamma_{\pm}^B\right) \geq 0\ ,
\end{equation}
where where $\gamma^B_\pm$ and $\alpha^B_\pm$ were 
defined in \eqref{gammaalpha} and 
the $\pm$ sign applies when $\lambda m_F$ is correspondingly
positive or negative. Again, the $\CB+$ limit corresponds to critical
bosons coupled to $SU(N)_k$ Chern-Simons gauge fields whereas the
$\CB-$ limit corresponds to critical bosons coupled to $SU(N)_{k-1}$
Chern-Simons gauge fields.

Let us, for a moment, view the equations 
\eqref{rfltrev} and \eqref{nowp} each as 
defining two dimensional manifolds in  the 
three dimensional space parametrized by 
the three variables $b_4$, $m_F$ and 
$m_B^2$. For any given values of $x_4$ and $x_6$, all four half-paraboloids in
\eqref{rfltrev} and \eqref{nowp} (two for $\lambda m_F > 0$ and two
for $\lambda m_F < 0$) occur in the three dimensional relevant
parameter space. It is easy to see that these four half-paraboloids
have a common intersection at the following half-line in
three-dimensional parameter space:
\begin{equation}\label{rfcbline}
\lambda m_F = 0\ ,\quad m_B^2 = 0\ ,\quad \lambda b_4 \geq 0\ .
\end{equation}

A point in the phase diagram of our theory 
is associated with an equivalence class of 
points in $(m_F, b_4, m_B^2)$ space; points 
in the same equivalence classes are related by scalings which act as $(m_F, b_4, m_B^2) 
\rightarrow (\Lambda m_F, \Lambda b_4, \Lambda^2 m_B^2) $ for any positive $ \Lambda$. Clearly the 
points on the ray \eqref{rfcbline} all lie in 
the same equivalence class and so correspond to a single point in the phase diagram
of the theory, which we call the CB-RF conformal point. The
neighbourhood of this point corresponds to massive critical bosons and
regular fermions and should also be obtained as an appropriate scaling
limit of the theory \eqref{generalaction}.

\subsubsection{The scaling limit}
Thus, in order to study the CB-RF point, one has to keep both $c_B$
and $c_F$ finite in the scaling limit. The equations for $c_B$ and
$c_F$ in \eqref{scalinggaps} tell us that ${\tl m}_F$ and $\sigma$
have to be kept fixed in the scaling limit, and consequently that
$m_F$ has to be kept fixed. The scaling behaviour in
\eqref{scalingparam} is not well-suited to this end; hence, we adopt a
different scaling limit first discussed in \cite{Jain:2013gza}: we
take the limit $\mu \to \infty$ with
\begin{equation} \label{oho}
m_B^2 = a_1 \mu^2\ ,\quad  b_4 = g_0 \mu^2 + g_1 \mu + g_2\ ,\quad m_F\ \text{fixed}\ .
\end{equation}
\footnote{The term proportional to $\mu^2$ in  $b_4$ will turn out to be multiplied by the inverse of one of the mass parameters, namely  
	$m_B^{\rm cri}$, of the IR theory and so is dimensionally 
	allowed.}
Some features of the scaling \eqref{oho} are 
easy to understand. Recall that the Landau 
Ginzburg potential four branches whose 
ranges of validity are given in \eqref{cbcf}.
The transition between a Higgsed and unHiggsed 
phase occurs at $\sigma= 0$ while transition 
between a $\pm$ fermionic phase occurs at 
$\sigma= -\frac{m_F}{2 x_4 \lambda}$. In the 
CB-RF limit, our IR theory is able to undergo 
both kinds of phase transition. This can only happen
if $ -\frac{m_F}{2 x_4 \lambda}$ stays 
fixed in the limit $\mu \to \infty$, explaining
why $m_F$ is held fixed in the limit. In order
to reproduce CB-RF physics, the quantum 
effective potential must also have a minimum
at a value of $\sigma$ that is held fixed 
as $\mu \to \infty$; this requirement essentially 
fixes the relative scalings of $b_4$ and $m_B^2$, 
as we will see in more detail below. 

Let us now
look at the potential of the theory:
\begin{align}
\frac{2\pi}{N} U^{(\varepsilon,\pm)}(\sigma) & =  \left(x_6 - \phi_\pm + x_4^3\psi_\varepsilon\right)  \lambda^2 { \sigma}^3  +  \left( { b}_4 + \tfrac{3}{4}x_4^2 m_F\psi_{\varepsilon}\right)  2\lambda{ \sigma}^2 + \left({ m}_B^2 + \tfrac{3}{4} x_4 m_F^2\psi_\varepsilon\right){ \sigma}\ ,\label{Unewcrit}
\end{align}
where we have again dropped the constant term which scales as
$\mu^3$. In the scaling limit, the terms proportional to $\mu^2$
dominate:
\begin{equation}\label{rbcfpot}
\mu^2( 2\lambda g_0 \sigma^2 + a_1 \sigma) = 2\lambda \mu^2 g_0\left(\left(\sigma+ \frac{a_1}{4\lambda g_0}\right)^2 - \frac{a_1^2}{16\lambda^2 g_0^2}\right)\ .
\end{equation}
Suppose the prefactor $\lambda g_0$ in the above expression is
positive
\begin{equation}\label{rbcfineq}
\lambda g_0 \geq 0\ .
\end{equation}
Then the extremum of the potential \eqref{rbcfpot} is a minimum and
$\sigma$ freezes at the minimum given by
\begin{equation}\label{sigmafreeze}
\sigma = - \frac{a_1}{4\lambda g_0} = \lim_{\mu \to \infty} -\frac{m_B^2}{4\lambda b_4} \ .
\end{equation}
Note that this looks exactly like the scaling limit that produces the
critical boson theory from the regular boson theory with critical
boson mass parameter
$m_B^{\rm cri} = \lim \frac{m_B^2}{2\lambda b_4}$. Thus, it is natural
to call the extremum value of $\sigma$ in \eqref{sigmafreeze} as
\begin{equation}\label{mbcridef}
\sigma =  -\frac{a_1}{4\lambda g_0} \equiv -\tfrac{1}{2}m_B^{\rm cri}\ .
\end{equation}
This also freezes the effective fermion mass parameter ${\tl m}_F$ at
the value
\begin{equation}\label{mftldef}
{\tl m}_F = m_F - \lambda x_4 m_B^{\rm cri} \equiv m_F^{\rm reg}\ ,
\end{equation}
which is exactly of the form of a mass for the regular fermion
theory.

Moreover, as in previous subsections, we fix some of the 
ambiguities of our redundant parametrization of our scaling 
limit by demanding that \eqref{sigmafreeze} hold exactly 
rather than only upto correction terms of order $1/\mu^2$. 
This precisely determined value of 
$\sigma$ must extremize the quantum effective potential 
\begin{equation}
3\left(x_6 - \phi_\pm + x_4^3\psi_\varepsilon\right)  \lambda^2 { \sigma}^2  +  \left( { b}_4 + \tfrac{3}{4}x_4^2 m_F\psi_{\varepsilon}\right) 4\lambda { \sigma} + \left({ m}_B^2 + \tfrac{3}{4} x_4 m_F^2\psi_\varepsilon\right) = 0\ .
\end{equation}
The terms proportional to $\mu^2$ already cancel by virtue of
\eqref{sigmafreeze}. The term proportional to $\mu$ is zero if and only if we set
$g_1 = 0$; we make this choice. The $\mu$-independent term is zero if the following
relation is satisfied:
\begin{equation}
- 2\lambda g_2 m_B^{\rm cri} + \tfrac{3}{4}(x_6 - \phi_\pm) \lambda^2 (m_B^{\rm cri})^2 + \tfrac{3}{4}x_4  \psi_{\varepsilon}\left(- 2 \lambda m_B^{\rm cri} x_4 m_F  +  m_F^2 + x_4^2\lambda^2 (m_B^{\rm cri})^2\right) = 0\ .
\end{equation}
That is
\begin{equation}\label{g2val}
g_2 =  \frac{\tfrac{3}{4}\left(x_6  -\phi_\pm\right)\lambda^2 (m_B^{\rm cri})^2 + \frac{3}{4}x_4 \psi_\varepsilon (m_F^{\rm reg})^2}{2\lambda m_B^{\rm cri}}\ .
\end{equation}
Thus, we seem to obtain a theory of gauged critical bosons and regular
fermions with masses $m_B^{\rm cri}$ and $m_F^{\rm reg}$ with the
ultraviolet parameters given in terms of these by\footnote{Here, we
	have set $a_1 = \sgn(m_B^2)$ without loss of generality}
\begin{equation}
m_B^2 = \sgn(m_B^2)\mu^2\ ,\quad m_F = m_F^{\rm reg} + x_4 \lambda m_B^{\rm cri}\ ,\quad b_4 = \frac{1}{2\lambda m_B^{\rm cri}} \mu^2 + g_2\ ,
\end{equation}
with the value of $g_2$ given in \eqref{g2val}.

\subsubsection{Matching with previous scaling limits}

The condition \eqref{rbcfineq} can be rewritten in terms of $b_4$ as
\begin{equation}\label{rfcbcond}
\lambda b_4 \geq 0\quad\text{in the limit}\quad \mu \to \infty\ .
\end{equation}
This is exactly the condition on $\lambda b_4$ that one has in the
intersection of the RF and CB half-paraboloids in \eqref{rfcbline}.

The conformal RF-CB point corresponds to the point where
$c_B = c_F = 0$, which occurs when
$m_B^{\rm cri} = m_F^{\rm reg} = 0$. In terms of the ultraviolet
parameters this point is given by
\begin{equation}\label{rfcblinen}
m_F = 0\ ,\quad m_B^2 = 0\ ,
\end{equation}
where we have used \eqref{mftldef} and the second equality of
\eqref{sigmafreeze}. The equations in \eqref{rfcblinen} and the
condition in \eqref{rfcbcond} precisely describe the intersection of
the $\RF\pm$ and $\CB\pm$ half-paraboloids in \eqref{rfcbline}.

Suppose one turns on the mass parameter $m_F^{\rm reg}$ for the
fermion while keeping the boson mass at zero. This is expected to
further flow down to the critical boson conformal theories $\CB\pm$
where the sign $\pm$ corresponds to
$\sgn(\lambda m_F^{\rm reg}) = \pm 1$. This sign coincides with the
sign of $\lambda m_F$ when $m_B^{\rm cri} = 0$ as is obvious from the
formula \eqref{mftldef}. The condition $\sgn(\lambda m_F) = \pm 1$ for
the occurrence of $\CB\pm$ limits is precisely the condition we
obtained in the critical boson scaling limit (cf.~Section
\ref{cblim}).

Similarly, if one turns on $m_B^{\rm cri}$ and keeps $m_F^{\rm reg}$
fixed at zero, the theory further flows down to the $\RF\pm$ theories,
where the sign is given by the sign of $m_B^{\rm cri}$. The condition
$m_F^{\rm reg} = 0$ gives (from \eqref{mftldef})
\begin{equation}
m_F = \lambda x_4 m_B^{\rm cri}\ ,
\end{equation}
which relates the sign of $m_B^{\rm cri}$ to the sign of
$\lambda m_F x_4$ as $\sgn(m_B^{\rm cri}) = \sgn(\lambda m_F
x_4)$. Thus, we get the $\RF\pm$ theories when
$\sgn(\lambda m_F x_4) = \pm 1$ which is precisely the condition we
obtained in the regular fermion scaling limit (cf.~Section
\ref{rflim}).

\begin{figure}
	\centering
	{\input{cbrflimit.pdf_t}}
	\caption{The features of the CB-RF theory in the neighbourhood of
		the CB-RF conformal point depicted as a thick dot at the
		origin. The blue lines correspond to the various quasi-fermionic
		conformal theories that emanate / terminate at the origin. The
		various topological phases are also shown. The figure is plotted
		for $x_4 > 0$.}
	\label{cbrflimitapp}
\end{figure}
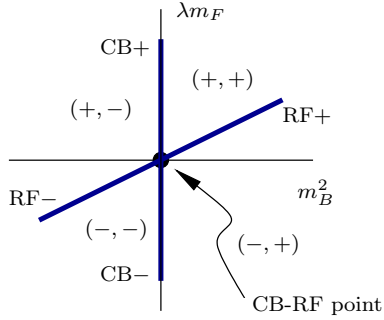

The $m_B^{\rm cri} = 0$ and $m_F^{\rm reg} = 0$ loci correspond to
\begin{alignat}{2}
\text{Critical boson},\ m_B^{\rm cri} = 0:&\qquad\ \ m_B^2 &&= 0\ ,\nonumber\\
\text{Regular fermion},\ m_F^{\rm reg} = 0:&\qquad \frac{\lambda m_F}{\lambda b_4} &&= \frac{\lambda^2 x_4}{2} \frac{m_B^2}{(\lambda b_4)^2}\ .
\end{alignat}
These are precisely the same equations that define the CB and RF
conformal scaling limits in \eqref{nowp} and \eqref{rfltrev} in the
neighbourhood of $m_F = 0$, as can be easily seen from those
equations. For a fixed large (positive) value of $\lambda b_4$ we plot
these lines in $(m_B^2, \lambda m_F)$ space in Figure \ref{cbrflimitapp}.
Clearly, these lines coincide with the CB and RF
lines in \eqref{rfltrev} and \eqref{nowp} at the same value of
$\lambda b_4$. A generic deformation with masses for both the bosons
and fermions would lead to the appropriate one of the four topological
phases $(\pm, \pm)$. These are separated by the four transition lines.

The previous discussion further demonstrates that the regular fermion
and the critical boson lines terminate at the CB-RF conformal point.

\bigskip \providecommand{\href}[2]{#2}\begingroup\raggedright
 \endgroup
  
\end{document}

%% file: N1susyp.pdf_t
\begin{picture}(0,0)%
\includegraphics{N1susyp.pdf}%
\end{picture}%
\setlength{\unitlength}{4144sp}%
\begingroup\makeatletter\ifx\SetFigFont\undefined%
\gdef\SetFigFont#1#2#3#4#5{%
  \reset@font\fontsize{#1}{#2pt}%
  \fontfamily{#3}\fontseries{#4}\fontshape{#5}%
  \selectfont}%
\fi\endgroup%
\begin{picture}(7767,841)(346,-1610)
\put(1171,-1546){\makebox(0,0)[lb]{\smash{{\SetFigFont{10}{12.0}{\familydefault}{\mddefault}{\updefault}{\color[rgb]{0,0,0}$(-,+)$}%
}}}}
\put(1171,-1411){\makebox(0,0)[lb]{\smash{{\SetFigFont{10}{12.0}{\familydefault}{\mddefault}{\updefault}{\color[rgb]{0,0,0}$(+,+)$}%
}}}}
\put(2791,-916){\makebox(0,0)[lb]{\smash{{\SetFigFont{10}{12.0}{\familydefault}{\mddefault}{\updefault}{\color[rgb]{0,0,0}$(+,+)$}%
}}}}
\put(2791,-1051){\makebox(0,0)[lb]{\smash{{\SetFigFont{10}{12.0}{\familydefault}{\mddefault}{\updefault}{\color[rgb]{0,0,0}$(+,-)$}%
}}}}
\put(4951,-1411){\makebox(0,0)[lb]{\smash{{\SetFigFont{10}{12.0}{\familydefault}{\mddefault}{\updefault}{\color[rgb]{0,0,0}$(-,+)$}%
}}}}
\put(4951,-1546){\makebox(0,0)[lb]{\smash{{\SetFigFont{10}{12.0}{\familydefault}{\mddefault}{\updefault}{\color[rgb]{0,0,0}$(+,-)$}%
}}}}
\put(6931,-916){\makebox(0,0)[lb]{\smash{{\SetFigFont{10}{12.0}{\familydefault}{\mddefault}{\updefault}{\color[rgb]{0,0,0}$(+,+)$}%
}}}}
\put(6931,-1051){\makebox(0,0)[lb]{\smash{{\SetFigFont{10}{12.0}{\familydefault}{\mddefault}{\updefault}{\color[rgb]{0,0,0}$(-,+)$}%
}}}}
\put(7966,-1366){\makebox(0,0)[lb]{\smash{{\SetFigFont{10}{12.0}{\familydefault}{\mddefault}{\updefault}{\color[rgb]{0,0,0}$w$}%
}}}}
\put(361,-916){\makebox(0,0)[lb]{\smash{{\SetFigFont{10}{12.0}{\familydefault}{\mddefault}{\updefault}{\color[rgb]{0,0,0}$\mu\lambda>0$}%
}}}}
\put(361,-1546){\makebox(0,0)[lb]{\smash{{\SetFigFont{10}{12.0}{\familydefault}{\mddefault}{\updefault}{\color[rgb]{0,0,0}$\mu\lambda<0$}%
}}}}
\put(1171,-1051){\makebox(0,0)[lb]{\smash{{\SetFigFont{10}{12.0}{\familydefault}{\mddefault}{\updefault}{\color[rgb]{0,0,0}$(+,-)$}%
}}}}
\put(2791,-1411){\makebox(0,0)[lb]{\smash{{\SetFigFont{10}{12.0}{\familydefault}{\mddefault}{\updefault}{\color[rgb]{0,0,0}$(-,+)$}%
}}}}
\put(4951,-1051){\makebox(0,0)[lb]{\smash{{\SetFigFont{10}{12.0}{\familydefault}{\mddefault}{\updefault}{\color[rgb]{0,0,0}$(+,+)$}%
}}}}
\put(1891,-1456){\makebox(0,0)[lb]{\smash{{\SetFigFont{10}{12.0}{\familydefault}{\mddefault}{\updefault}{\color[rgb]{0,0,0}$\frac{|\lambda|-2}{|\lambda|}$}%
}}}}
\put(4051,-1456){\makebox(0,0)[lb]{\smash{{\SetFigFont{10}{12.0}{\familydefault}{\mddefault}{\updefault}{\color[rgb]{0,0,0}$\frac{|\lambda|}{|\lambda|-2}$}%
}}}}
\put(6121,-1456){\makebox(0,0)[lb]{\smash{{\SetFigFont{10}{12.0}{\familydefault}{\mddefault}{\updefault}{\color[rgb]{0,0,0}$\frac{2+|\lambda|}{|\lambda|}$}%
}}}}
\put(6931,-1411){\makebox(0,0)[lb]{\smash{{\SetFigFont{10}{12.0}{\familydefault}{\mddefault}{\updefault}{\color[rgb]{0,0,0}$(+,-)$}%
}}}}
\put(5536,-1456){\makebox(0,0)[lb]{\smash{{\SetFigFont{10}{12.0}{\familydefault}{\mddefault}{\updefault}{\color[rgb]{0,0,0}$1$}%
}}}}
\end{picture}%

%% file: phaseseq.pdf_t
\begin{picture}(0,0)%
\includegraphics{phaseseq.pdf}%
\end{picture}%
\setlength{\unitlength}{4144sp}%
\begingroup\makeatletter\ifx\SetFigFont\undefined%
\gdef\SetFigFont#1#2#3#4#5{%
  \reset@font\fontsize{#1}{#2pt}%
  \fontfamily{#3}\fontseries{#4}\fontshape{#5}%
  \selectfont}%
\fi\endgroup%
\begin{picture}(9044,3177)(1779,-3811)
\put(8056,-1231){\makebox(0,0)[lb]{\smash{{\SetFigFont{10}{12.0}{\familydefault}{\mddefault}{\updefault}{\color[rgb]{0,0,0}$0$}%
}}}}
\put(4456,-1231){\makebox(0,0)[lb]{\smash{{\SetFigFont{10}{12.0}{\familydefault}{\mddefault}{\updefault}{\color[rgb]{0,0,0}$0$}%
}}}}
\put(3106,-3256){\makebox(0,0)[lb]{\smash{{\SetFigFont{10}{12.0}{\familydefault}{\mddefault}{\updefault}{\color[rgb]{0,0,0}$0$}%
}}}}
\put(9406,-3256){\makebox(0,0)[lb]{\smash{{\SetFigFont{10}{12.0}{\familydefault}{\mddefault}{\updefault}{\color[rgb]{0,0,0}$0$}%
}}}}
\put(2251,-781){\makebox(0,0)[lb]{\smash{{\SetFigFont{10}{12.0}{\familydefault}{\mddefault}{\updefault}{\color[rgb]{0,0,0}$(-,+)$}%
}}}}
\put(4951,-781){\makebox(0,0)[lb]{\smash{{\SetFigFont{10}{12.0}{\familydefault}{\mddefault}{\updefault}{\color[rgb]{0,0,0}$(+,-)$}%
}}}}
\put(7201,-781){\makebox(0,0)[lb]{\smash{{\SetFigFont{10}{12.0}{\familydefault}{\mddefault}{\updefault}{\color[rgb]{0,0,0}$(-,+)$}%
}}}}
\put(9901,-781){\makebox(0,0)[lb]{\smash{{\SetFigFont{10}{12.0}{\familydefault}{\mddefault}{\updefault}{\color[rgb]{0,0,0}$(+,-)$}%
}}}}
\put(2251,-2806){\makebox(0,0)[lb]{\smash{{\SetFigFont{10}{12.0}{\familydefault}{\mddefault}{\updefault}{\color[rgb]{0,0,0}$(+,+)$}%
}}}}
\put(4951,-2806){\makebox(0,0)[lb]{\smash{{\SetFigFont{10}{12.0}{\familydefault}{\mddefault}{\updefault}{\color[rgb]{0,0,0}$(-,-)$}%
}}}}
\put(9901,-2806){\makebox(0,0)[lb]{\smash{{\SetFigFont{10}{12.0}{\familydefault}{\mddefault}{\updefault}{\color[rgb]{0,0,0}$(-,-)$}%
}}}}
\put(7201,-2806){\makebox(0,0)[lb]{\smash{{\SetFigFont{10}{12.0}{\familydefault}{\mddefault}{\updefault}{\color[rgb]{0,0,0}$(+,+)$}%
}}}}
\put(3601,-781){\makebox(0,0)[lb]{\smash{{\SetFigFont{10}{12.0}{\familydefault}{\mddefault}{\updefault}{\color[rgb]{0,0,0}$(+,+)$}%
}}}}
\put(8551,-781){\makebox(0,0)[lb]{\smash{{\SetFigFont{10}{12.0}{\familydefault}{\mddefault}{\updefault}{\color[rgb]{0,0,0}$(-,-)$}%
}}}}
\put(8551,-2806){\makebox(0,0)[lb]{\smash{{\SetFigFont{10}{12.0}{\familydefault}{\mddefault}{\updefault}{\color[rgb]{0,0,0}$(-,+)$}%
}}}}
\put(3601,-2806){\makebox(0,0)[lb]{\smash{{\SetFigFont{10}{12.0}{\familydefault}{\mddefault}{\updefault}{\color[rgb]{0,0,0}$(+,-)$}%
}}}}
\put(2836,-1231){\makebox(0,0)[lb]{\smash{{\SetFigFont{10}{12.0}{\familydefault}{\mddefault}{\updefault}{\color[rgb]{0,0,0}$\displaystyle-\frac{m_F}{2x_4\lambda}$}%
}}}}
\put(5626,-1096){\makebox(0,0)[lb]{\smash{{\SetFigFont{10}{12.0}{\familydefault}{\mddefault}{\updefault}{\color[rgb]{0,0,0}$\sigma$}%
}}}}
\put(10576,-1096){\makebox(0,0)[lb]{\smash{{\SetFigFont{10}{12.0}{\familydefault}{\mddefault}{\updefault}{\color[rgb]{0,0,0}$\sigma$}%
}}}}
\put(9136,-1231){\makebox(0,0)[lb]{\smash{{\SetFigFont{10}{12.0}{\familydefault}{\mddefault}{\updefault}{\color[rgb]{0,0,0}$\displaystyle-\frac{m_F}{2x_4\lambda}$}%
}}}}
\put(5626,-3121){\makebox(0,0)[lb]{\smash{{\SetFigFont{10}{12.0}{\familydefault}{\mddefault}{\updefault}{\color[rgb]{0,0,0}$\sigma$}%
}}}}
\put(4186,-3256){\makebox(0,0)[lb]{\smash{{\SetFigFont{10}{12.0}{\familydefault}{\mddefault}{\updefault}{\color[rgb]{0,0,0}$\displaystyle-\frac{m_F}{2x_4\lambda}$}%
}}}}
\put(7786,-3256){\makebox(0,0)[lb]{\smash{{\SetFigFont{10}{12.0}{\familydefault}{\mddefault}{\updefault}{\color[rgb]{0,0,0}$\displaystyle-\frac{m_F}{2x_4\lambda}$}%
}}}}
\put(10576,-3121){\makebox(0,0)[lb]{\smash{{\SetFigFont{10}{12.0}{\familydefault}{\mddefault}{\updefault}{\color[rgb]{0,0,0}$\sigma$}%
}}}}
\put(2926,-1726){\makebox(0,0)[lb]{\smash{{\SetFigFont{12}{14.4}{\familydefault}{\mddefault}{\updefault}{\color[rgb]{0,0,0}$\lambda m_F>0$, $x_4 > 0$}%
}}}}
\put(7876,-1726){\makebox(0,0)[lb]{\smash{{\SetFigFont{12}{14.4}{\familydefault}{\mddefault}{\updefault}{\color[rgb]{0,0,0}$\lambda m_F < 0$, $x_4>0$}%
}}}}
\put(2926,-3751){\makebox(0,0)[lb]{\smash{{\SetFigFont{12}{14.4}{\familydefault}{\mddefault}{\updefault}{\color[rgb]{0,0,0}$\lambda m_F > 0$, $x_4<0$}%
}}}}
\put(7876,-3751){\makebox(0,0)[lb]{\smash{{\SetFigFont{12}{14.4}{\familydefault}{\mddefault}{\updefault}{\color[rgb]{0,0,0}$\lambda m_F < 0$, $x_4<0$}%
}}}}
\end{picture}%

%% file: susymfzero.pdf_t
\begin{picture}(0,0)%
\includegraphics{susymfzero.pdf}%
\end{picture}%
\setlength{\unitlength}{4144sp}%
\begingroup\makeatletter\ifx\SetFigFont\undefined%
\gdef\SetFigFont#1#2#3#4#5{%
  \reset@font\fontsize{#1}{#2pt}%
  \fontfamily{#3}\fontseries{#4}\fontshape{#5}%
  \selectfont}%
\fi\endgroup%
\begin{picture}(8082,3689)(256,-8228)
\put(2341,-4696){\makebox(0,0)[lb]{\smash{{\SetFigFont{10}{12.0}{\familydefault}{\mddefault}{\updefault}{\color[rgb]{0,0,0}$\lambda b_4$}%
}}}}
\put(3781,-6541){\makebox(0,0)[lb]{\smash{{\SetFigFont{10}{12.0}{\familydefault}{\mddefault}{\updefault}{\color[rgb]{0,0,0}$m_B^2$}%
}}}}
\put(5491,-6181){\makebox(0,0)[lb]{\smash{{\SetFigFont{10}{12.0}{\familydefault}{\mddefault}{\updefault}{\color[rgb]{0,0,0}$(+,-)$}%
}}}}
\put(6616,-4696){\makebox(0,0)[lb]{\smash{{\SetFigFont{10}{12.0}{\familydefault}{\mddefault}{\updefault}{\color[rgb]{0,0,0}$\lambda b_4$}%
}}}}
\put(8056,-6541){\makebox(0,0)[lb]{\smash{{\SetFigFont{10}{12.0}{\familydefault}{\mddefault}{\updefault}{\color[rgb]{0,0,0}$m_B^2$}%
}}}}
\put(7291,-7081){\makebox(0,0)[lb]{\smash{{\SetFigFont{10}{12.0}{\familydefault}{\mddefault}{\updefault}{\color[rgb]{0,0,0}$(-,+)$}%
}}}}
\put(271,-7576){\makebox(0,0)[lb]{\smash{{\SetFigFont{10}{12.0}{\familydefault}{\mddefault}{\updefault}{\color[rgb]{0,0,0}$D_{u0}'$}%
}}}}
\put(4051,-7666){\makebox(0,0)[lb]{\smash{{\SetFigFont{10}{12.0}{\familydefault}{\mddefault}{\updefault}{\color[rgb]{0,0,0}$D_{h0}$}%
}}}}
\put(2341,-5236){\makebox(0,0)[lb]{\smash{{\SetFigFont{10}{12.0}{\familydefault}{\mddefault}{\updefault}{\color[rgb]{0,0,0}$L_0$}%
}}}}
\put(2296,-8071){\makebox(0,0)[lb]{\smash{{\SetFigFont{10}{12.0}{\familydefault}{\mddefault}{\updefault}{\color[rgb]{0,0,0}$M_0$}%
}}}}
\put(6616,-6271){\makebox(0,0)[lb]{\smash{{\SetFigFont{10}{12.0}{\familydefault}{\mddefault}{\updefault}{\color[rgb]{0,0,0}O}%
}}}}
\put(6616,-5236){\makebox(0,0)[lb]{\smash{{\SetFigFont{10}{12.0}{\familydefault}{\mddefault}{\updefault}{\color[rgb]{0,0,0}$L_0$}%
}}}}
\end{picture}%

%% file: phaseplus.pdf_t
\begin{picture}(0,0)%
\includegraphics{phaseplus.pdf}%
\end{picture}%
\setlength{\unitlength}{4144sp}%
\begingroup\makeatletter\ifx\SetFigFont\undefined%
\gdef\SetFigFont#1#2#3#4#5{%
  \reset@font\fontsize{#1}{#2pt}%
  \fontfamily{#3}\fontseries{#4}\fontshape{#5}%
  \selectfont}%
\fi\endgroup%
\begin{picture}(4074,616)(2689,-1700)
\put(6616,-1636){\makebox(0,0)[lb]{\smash{{\SetFigFont{10}{12.0}{\familydefault}{\mddefault}{\updefault}{\color[rgb]{0,0,0}$\sigma$}%
}}}}
\put(3151,-1231){\makebox(0,0)[lb]{\smash{{\SetFigFont{10}{12.0}{\familydefault}{\mddefault}{\updefault}{\color[rgb]{0,0,0}$(-,+)$}%
}}}}
\put(4456,-1231){\makebox(0,0)[lb]{\smash{{\SetFigFont{10}{12.0}{\familydefault}{\mddefault}{\updefault}{\color[rgb]{0,0,0}$(+,+)$}%
}}}}
\put(5896,-1231){\makebox(0,0)[lb]{\smash{{\SetFigFont{10}{12.0}{\familydefault}{\mddefault}{\updefault}{\color[rgb]{0,0,0}$(+,-)$}%
}}}}
\put(5356,-1636){\makebox(0,0)[lb]{\smash{{\SetFigFont{10}{12.0}{\familydefault}{\mddefault}{\updefault}{\color[rgb]{0,0,0}$0$}%
}}}}
\put(3871,-1636){\makebox(0,0)[lb]{\smash{{\SetFigFont{10}{12.0}{\familydefault}{\mddefault}{\updefault}{\color[rgb]{0,0,0}$-\frac{m_F}{2\lambda}$}%
}}}}
\end{picture}%

%% file: phaseplotI.pdf_t
\begin{picture}(0,0)%
\includegraphics{phaseplotI.pdf}%
\end{picture}%
\setlength{\unitlength}{4144sp}%
\begingroup\makeatletter\ifx\SetFigFont\undefined%
\gdef\SetFigFont#1#2#3#4#5{%
  \reset@font\fontsize{#1}{#2pt}%
  \fontfamily{#3}\fontseries{#4}\fontshape{#5}%
  \selectfont}%
\fi\endgroup%
\begin{picture}(9477,4351)(1381,-4400)
\put(6571,-3436){\makebox(0,0)[lb]{\smash{{\SetFigFont{10}{12.0}{\familydefault}{\mddefault}{\updefault}{\color[rgb]{0,0,0}$D_u'$}%
}}}}
\put(10486,-2716){\makebox(0,0)[lb]{\smash{{\SetFigFont{10}{12.0}{\familydefault}{\mddefault}{\updefault}{\color[rgb]{0,0,0}$D_h$}%
}}}}
\put(10441,-2131){\makebox(0,0)[lb]{\smash{{\SetFigFont{10}{12.0}{\familydefault}{\mddefault}{\updefault}{\color[rgb]{0,0,0}$L'$}%
}}}}
\put(9271,-646){\makebox(0,0)[lb]{\smash{{\SetFigFont{10}{12.0}{\familydefault}{\mddefault}{\updefault}{\color[rgb]{0,0,0}$L$}%
}}}}
\put(8416,-3121){\makebox(0,0)[lb]{\smash{{\SetFigFont{10}{12.0}{\familydefault}{\mddefault}{\updefault}{\color[rgb]{0,0,0}CF$+$}%
}}}}
\put(9316,-1861){\makebox(0,0)[lb]{\smash{{\SetFigFont{10}{12.0}{\familydefault}{\mddefault}{\updefault}{\color[rgb]{0,0,0}RB$+$}%
}}}}
\put(7471,-3481){\makebox(0,0)[lb]{\smash{{\SetFigFont{10}{12.0}{\familydefault}{\mddefault}{\updefault}{\color[rgb]{0,0,0}$M'$}%
}}}}
\put(9271,-3526){\makebox(0,0)[lb]{\smash{{\SetFigFont{10}{12.0}{\familydefault}{\mddefault}{\updefault}{\color[rgb]{0,0,0}$M$}%
}}}}
\put(10261,-691){\makebox(0,0)[lb]{\smash{{\SetFigFont{10}{12.0}{\familydefault}{\mddefault}{\updefault}{\color[rgb]{0,0,0}$m_B^2$}%
}}}}
\put(10171,-196){\makebox(0,0)[lb]{\smash{{\SetFigFont{10}{12.0}{\familydefault}{\mddefault}{\updefault}{\color[rgb]{0,0,0}$\lambda b_4$}%
}}}}
\put(8461,-4336){\makebox(0,0)[lb]{\smash{{\SetFigFont{12}{14.4}{\familydefault}{\mddefault}{\updefault}{\color[rgb]{0,0,0}(b)}%
}}}}
\put(8011,-736){\makebox(0,0)[lb]{\smash{{\SetFigFont{10}{12.0}{\familydefault}{\mddefault}{\updefault}{\color[rgb]{0,0,0}$D_h'$}%
}}}}
\put(6931,-3166){\makebox(0,0)[lb]{\smash{{\SetFigFont{10}{12.0}{\familydefault}{\mddefault}{\updefault}{\color[rgb]{0,0,0}$D_u$}%
}}}}
\put(4996,-196){\makebox(0,0)[lb]{\smash{{\SetFigFont{10}{12.0}{\familydefault}{\mddefault}{\updefault}{\color[rgb]{0,0,0}$\lambda b_4$}%
}}}}
\put(2971,-916){\makebox(0,0)[lb]{\smash{{\SetFigFont{10}{12.0}{\familydefault}{\mddefault}{\updefault}{\color[rgb]{0,0,0}$D_h'$}%
}}}}
\put(5311,-2716){\makebox(0,0)[lb]{\smash{{\SetFigFont{10}{12.0}{\familydefault}{\mddefault}{\updefault}{\color[rgb]{0,0,0}$D_h$}%
}}}}
\put(5266,-2131){\makebox(0,0)[lb]{\smash{{\SetFigFont{10}{12.0}{\familydefault}{\mddefault}{\updefault}{\color[rgb]{0,0,0}$L'$}%
}}}}
\put(1756,-3166){\makebox(0,0)[lb]{\smash{{\SetFigFont{10}{12.0}{\familydefault}{\mddefault}{\updefault}{\color[rgb]{0,0,0}$D_u$}%
}}}}
\put(1396,-3391){\makebox(0,0)[lb]{\smash{{\SetFigFont{10}{12.0}{\familydefault}{\mddefault}{\updefault}{\color[rgb]{0,0,0}$D_u'$}%
}}}}
\put(4096,-646){\makebox(0,0)[lb]{\smash{{\SetFigFont{10}{12.0}{\familydefault}{\mddefault}{\updefault}{\color[rgb]{0,0,0}$L$}%
}}}}
\put(3241,-3121){\makebox(0,0)[lb]{\smash{{\SetFigFont{10}{12.0}{\familydefault}{\mddefault}{\updefault}{\color[rgb]{0,0,0}CF$+$}%
}}}}
\put(4141,-1861){\makebox(0,0)[lb]{\smash{{\SetFigFont{10}{12.0}{\familydefault}{\mddefault}{\updefault}{\color[rgb]{0,0,0}RB$+$}%
}}}}
\put(2296,-3481){\makebox(0,0)[lb]{\smash{{\SetFigFont{10}{12.0}{\familydefault}{\mddefault}{\updefault}{\color[rgb]{0,0,0}$M'$}%
}}}}
\put(4096,-3526){\makebox(0,0)[lb]{\smash{{\SetFigFont{10}{12.0}{\familydefault}{\mddefault}{\updefault}{\color[rgb]{0,0,0}$M$}%
}}}}
\put(5086,-691){\makebox(0,0)[lb]{\smash{{\SetFigFont{10}{12.0}{\familydefault}{\mddefault}{\updefault}{\color[rgb]{0,0,0}$m_B^2$}%
}}}}
\put(3241,-4336){\makebox(0,0)[lb]{\smash{{\SetFigFont{12}{14.4}{\familydefault}{\mddefault}{\updefault}{\color[rgb]{0,0,0}(a)}%
}}}}
\end{picture}%

%% file: phasediagI.pdf_t
\begin{picture}(0,0)%
\includegraphics{phasediagI.pdf}%
\end{picture}%
\setlength{\unitlength}{4144sp}%
\begingroup\makeatletter\ifx\SetFigFont\undefined%
\gdef\SetFigFont#1#2#3#4#5{%
  \reset@font\fontsize{#1}{#2pt}%
  \fontfamily{#3}\fontseries{#4}\fontshape{#5}%
  \selectfont}%
\fi\endgroup%
\begin{picture}(9567,4351)(1156,-4400)
\put(1171,-3436){\makebox(0,0)[lb]{\smash{{\SetFigFont{10}{12.0}{\familydefault}{\mddefault}{\updefault}{\color[rgb]{0,0,0}$D_u'$}%
}}}}
\put(4771,-196){\makebox(0,0)[lb]{\smash{{\SetFigFont{10}{12.0}{\familydefault}{\mddefault}{\updefault}{\color[rgb]{0,0,0}$\lambda b_4$}%
}}}}
\put(5086,-2716){\makebox(0,0)[lb]{\smash{{\SetFigFont{10}{12.0}{\familydefault}{\mddefault}{\updefault}{\color[rgb]{0,0,0}$D_h$}%
}}}}
\put(5041,-2131){\makebox(0,0)[lb]{\smash{{\SetFigFont{10}{12.0}{\familydefault}{\mddefault}{\updefault}{\color[rgb]{0,0,0}$L'$}%
}}}}
\put(3871,-646){\makebox(0,0)[lb]{\smash{{\SetFigFont{10}{12.0}{\familydefault}{\mddefault}{\updefault}{\color[rgb]{0,0,0}$L$}%
}}}}
\put(3016,-3121){\makebox(0,0)[lb]{\smash{{\SetFigFont{10}{12.0}{\familydefault}{\mddefault}{\updefault}{\color[rgb]{0,0,0}CF$+$}%
}}}}
\put(3916,-1861){\makebox(0,0)[lb]{\smash{{\SetFigFont{10}{12.0}{\familydefault}{\mddefault}{\updefault}{\color[rgb]{0,0,0}RB$+$}%
}}}}
\put(4861,-691){\makebox(0,0)[lb]{\smash{{\SetFigFont{10}{12.0}{\familydefault}{\mddefault}{\updefault}{\color[rgb]{0,0,0}$m_B^2$}%
}}}}
\put(2881,-3661){\makebox(0,0)[lb]{\smash{{\SetFigFont{10}{12.0}{\familydefault}{\mddefault}{\updefault}{\color[rgb]{0,0,0}or}%
}}}}
\put(3916,-3661){\makebox(0,0)[lb]{\smash{{\SetFigFont{10}{12.0}{\familydefault}{\mddefault}{\updefault}{\color[rgb]{0,0,0}or}%
}}}}
\put(3286,-2131){\makebox(0,0)[lb]{\smash{{\SetFigFont{10}{12.0}{\familydefault}{\mddefault}{\updefault}{\color[rgb]{0,0,0}or}%
}}}}
\put(3106,-4336){\makebox(0,0)[lb]{\smash{{\SetFigFont{12}{14.4}{\familydefault}{\mddefault}{\updefault}{\color[rgb]{0,0,0}(a)}%
}}}}
\put(8461,-4336){\makebox(0,0)[lb]{\smash{{\SetFigFont{12}{14.4}{\familydefault}{\mddefault}{\updefault}{\color[rgb]{0,0,0}(b)}%
}}}}
\put(9946,-196){\makebox(0,0)[lb]{\smash{{\SetFigFont{10}{12.0}{\familydefault}{\mddefault}{\updefault}{\color[rgb]{0,0,0}$\lambda b_4$}%
}}}}
\put(10261,-2716){\makebox(0,0)[lb]{\smash{{\SetFigFont{10}{12.0}{\familydefault}{\mddefault}{\updefault}{\color[rgb]{0,0,0}$D_h$}%
}}}}
\put(10036,-691){\makebox(0,0)[lb]{\smash{{\SetFigFont{10}{12.0}{\familydefault}{\mddefault}{\updefault}{\color[rgb]{0,0,0}$m_B^2$}%
}}}}
\put(7921,-2896){\makebox(0,0)[lb]{\smash{{\SetFigFont{10}{12.0}{\familydefault}{\mddefault}{\updefault}{\color[rgb]{0,0,0}CF$+$}%
}}}}
\put(9046,-331){\makebox(0,0)[lb]{\smash{{\SetFigFont{10}{12.0}{\familydefault}{\mddefault}{\updefault}{\color[rgb]{0,0,0}$L$}%
}}}}
\put(8551,-1816){\makebox(0,0)[lb]{\smash{{\SetFigFont{10}{12.0}{\familydefault}{\mddefault}{\updefault}{\color[rgb]{0,0,0}RB$+$}%
}}}}
\put(9226,-2671){\makebox(0,0)[lb]{\smash{{\SetFigFont{10}{12.0}{\familydefault}{\mddefault}{\updefault}{\color[rgb]{0,0,0}O$+$}%
}}}}
\put(10420,-1861){\makebox(0,0)[lb]{\smash{{\SetFigFont{10}{12.0}{\familydefault}{\mddefault}{\updefault}{\color[rgb]{0,0,0}$L'$}%
}}}}
\end{picture}%

%% file: phaseminus.pdf_t
\begin{picture}(0,0)%
\includegraphics{phaseminus.pdf}%
\end{picture}%
\setlength{\unitlength}{4144sp}%
\begingroup\makeatletter\ifx\SetFigFont\undefined%
\gdef\SetFigFont#1#2#3#4#5{%
  \reset@font\fontsize{#1}{#2pt}%
  \fontfamily{#3}\fontseries{#4}\fontshape{#5}%
  \selectfont}%
\fi\endgroup%
\begin{picture}(4074,616)(2689,-1700)
\put(3151,-1231){\makebox(0,0)[lb]{\smash{{\SetFigFont{10}{12.0}{\familydefault}{\mddefault}{\updefault}{\color[rgb]{0,0,0}$(-,+)$}%
}}}}
\put(5896,-1231){\makebox(0,0)[lb]{\smash{{\SetFigFont{10}{12.0}{\familydefault}{\mddefault}{\updefault}{\color[rgb]{0,0,0}$(+,-)$}%
}}}}
\put(4456,-1231){\makebox(0,0)[lb]{\smash{{\SetFigFont{10}{12.0}{\familydefault}{\mddefault}{\updefault}{\color[rgb]{0,0,0}$(-,-)$}%
}}}}
\put(4006,-1636){\makebox(0,0)[lb]{\smash{{\SetFigFont{10}{12.0}{\familydefault}{\mddefault}{\updefault}{\color[rgb]{0,0,0}$0$}%
}}}}
\put(6616,-1636){\makebox(0,0)[lb]{\smash{{\SetFigFont{10}{12.0}{\familydefault}{\mddefault}{\updefault}{\color[rgb]{0,0,0}$\sigma$}%
}}}}
\put(5221,-1636){\makebox(0,0)[lb]{\smash{{\SetFigFont{10}{12.0}{\familydefault}{\mddefault}{\updefault}{\color[rgb]{0,0,0}$-\frac{m_F}{2\lambda}$}%
}}}}
\end{picture}%

%% file: phaseplotII.pdf_t
\begin{picture}(0,0)%
\includegraphics{phaseplotII.pdf}%
\end{picture}%
\setlength{\unitlength}{4144sp}%
\begingroup\makeatletter\ifx\SetFigFont\undefined%
\gdef\SetFigFont#1#2#3#4#5{%
  \reset@font\fontsize{#1}{#2pt}%
  \fontfamily{#3}\fontseries{#4}\fontshape{#5}%
  \selectfont}%
\fi\endgroup%
\begin{picture}(9738,4351)(898,-4400)
\put(10621,-3436){\makebox(0,0)[rb]{\smash{{\SetFigFont{10}{12.0}{\familydefault}{\mddefault}{\updefault}{\color[rgb]{0,0,0}${\tl D}_h'$}%
}}}}
\put(1711,-196){\makebox(0,0)[rb]{\smash{{\SetFigFont{10}{12.0}{\familydefault}{\mddefault}{\updefault}{\color[rgb]{0,0,0}$\lambda b_4$}%
}}}}
\put(1396,-2716){\makebox(0,0)[rb]{\smash{{\SetFigFont{10}{12.0}{\familydefault}{\mddefault}{\updefault}{\color[rgb]{0,0,0}${\tl D}_u$}%
}}}}
\put(1441,-2131){\makebox(0,0)[rb]{\smash{{\SetFigFont{10}{12.0}{\familydefault}{\mddefault}{\updefault}{\color[rgb]{0,0,0}${\tl L}'$}%
}}}}
\put(3466,-3121){\makebox(0,0)[rb]{\smash{{\SetFigFont{10}{12.0}{\familydefault}{\mddefault}{\updefault}{\color[rgb]{0,0,0}CF$-$}%
}}}}
\put(2566,-1861){\makebox(0,0)[rb]{\smash{{\SetFigFont{10}{12.0}{\familydefault}{\mddefault}{\updefault}{\color[rgb]{0,0,0}RB$-$}%
}}}}
\put(2611,-3526){\makebox(0,0)[rb]{\smash{{\SetFigFont{10}{12.0}{\familydefault}{\mddefault}{\updefault}{\color[rgb]{0,0,0}${\tl M}$}%
}}}}
\put(3466,-4336){\makebox(0,0)[rb]{\smash{{\SetFigFont{12}{14.4}{\familydefault}{\mddefault}{\updefault}{\color[rgb]{0,0,0}(a)}%
}}}}
\put(6706,-2716){\makebox(0,0)[rb]{\smash{{\SetFigFont{10}{12.0}{\familydefault}{\mddefault}{\updefault}{\color[rgb]{0,0,0}${\tl D}_u$}%
}}}}
\put(8776,-3121){\makebox(0,0)[rb]{\smash{{\SetFigFont{10}{12.0}{\familydefault}{\mddefault}{\updefault}{\color[rgb]{0,0,0}CF$-$}%
}}}}
\put(7876,-1861){\makebox(0,0)[rb]{\smash{{\SetFigFont{10}{12.0}{\familydefault}{\mddefault}{\updefault}{\color[rgb]{0,0,0}RB$-$}%
}}}}
\put(7921,-3526){\makebox(0,0)[rb]{\smash{{\SetFigFont{10}{12.0}{\familydefault}{\mddefault}{\updefault}{\color[rgb]{0,0,0}${\tl M}$}%
}}}}
\put(7021,-196){\makebox(0,0)[rb]{\smash{{\SetFigFont{10}{12.0}{\familydefault}{\mddefault}{\updefault}{\color[rgb]{0,0,0}$\lambda b_4$}%
}}}}
\put(8731,-4336){\makebox(0,0)[rb]{\smash{{\SetFigFont{12}{14.4}{\familydefault}{\mddefault}{\updefault}{\color[rgb]{0,0,0}(b)}%
}}}}
\put(9181,-736){\makebox(0,0)[rb]{\smash{{\SetFigFont{10}{12.0}{\familydefault}{\mddefault}{\updefault}{\color[rgb]{0,0,0}${\tl D}_u'$}%
}}}}
\put(10261,-3166){\makebox(0,0)[rb]{\smash{{\SetFigFont{10}{12.0}{\familydefault}{\mddefault}{\updefault}{\color[rgb]{0,0,0}${\tl D}_h$}%
}}}}
\put(6751,-2131){\makebox(0,0)[rb]{\smash{{\SetFigFont{10}{12.0}{\familydefault}{\mddefault}{\updefault}{\color[rgb]{0,0,0}${\tl L}'$}%
}}}}
\put(4006,-781){\makebox(0,0)[rb]{\smash{{\SetFigFont{10}{12.0}{\familydefault}{\mddefault}{\updefault}{\color[rgb]{0,0,0}${\tl D}_u'$}%
}}}}
\put(5401,-3436){\makebox(0,0)[rb]{\smash{{\SetFigFont{10}{12.0}{\familydefault}{\mddefault}{\updefault}{\color[rgb]{0,0,0}${\tl D}_h'$}%
}}}}
\put(5086,-3211){\makebox(0,0)[rb]{\smash{{\SetFigFont{10}{12.0}{\familydefault}{\mddefault}{\updefault}{\color[rgb]{0,0,0}${\tl D}_h$}%
}}}}
\put(4366,-3526){\makebox(0,0)[rb]{\smash{{\SetFigFont{10}{12.0}{\familydefault}{\mddefault}{\updefault}{\color[rgb]{0,0,0}${\tl M}'$}%
}}}}
\put(9766,-3571){\makebox(0,0)[rb]{\smash{{\SetFigFont{10}{12.0}{\familydefault}{\mddefault}{\updefault}{\color[rgb]{0,0,0}${\tl M}'$}%
}}}}
\put(7921,-376){\makebox(0,0)[rb]{\smash{{\SetFigFont{10}{12.0}{\familydefault}{\mddefault}{\updefault}{\color[rgb]{0,0,0}${\tl L}$}%
}}}}
\put(10351,-691){\makebox(0,0)[rb]{\smash{{\SetFigFont{10}{12.0}{\familydefault}{\mddefault}{\updefault}{\color[rgb]{0,0,0}$m_B^2$}%
}}}}
\put(2611,-376){\makebox(0,0)[rb]{\smash{{\SetFigFont{10}{12.0}{\familydefault}{\mddefault}{\updefault}{\color[rgb]{0,0,0}${\tl L}$}%
}}}}
\put(5041,-691){\makebox(0,0)[rb]{\smash{{\SetFigFont{10}{12.0}{\familydefault}{\mddefault}{\updefault}{\color[rgb]{0,0,0}$m_B^2$}%
}}}}
\end{picture}%

%% file: phasediagII.pdf_t
\begin{picture}(0,0)%
\includegraphics{phasediagII.pdf}%
\end{picture}%
\setlength{\unitlength}{4144sp}%
\begingroup\makeatletter\ifx\SetFigFont\undefined%
\gdef\SetFigFont#1#2#3#4#5{%
  \reset@font\fontsize{#1}{#2pt}%
  \fontfamily{#3}\fontseries{#4}\fontshape{#5}%
  \selectfont}%
\fi\endgroup%
\begin{picture}(9564,4351)(1114,-4400)
\put(5401,-3436){\makebox(0,0)[rb]{\smash{{\SetFigFont{10}{12.0}{\familydefault}{\mddefault}{\updefault}{\color[rgb]{0,0,0}${\tl D}_h'$}%
}}}}
\put(1801,-196){\makebox(0,0)[rb]{\smash{{\SetFigFont{10}{12.0}{\familydefault}{\mddefault}{\updefault}{\color[rgb]{0,0,0}$\lambda b_4$}%
}}}}
\put(1486,-2716){\makebox(0,0)[rb]{\smash{{\SetFigFont{10}{12.0}{\familydefault}{\mddefault}{\updefault}{\color[rgb]{0,0,0}${\tl D}_u$}%
}}}}
\put(1531,-2131){\makebox(0,0)[rb]{\smash{{\SetFigFont{10}{12.0}{\familydefault}{\mddefault}{\updefault}{\color[rgb]{0,0,0}${\tl L}'$}%
}}}}
\put(3556,-3121){\makebox(0,0)[rb]{\smash{{\SetFigFont{10}{12.0}{\familydefault}{\mddefault}{\updefault}{\color[rgb]{0,0,0}CF$-$}%
}}}}
\put(2656,-1861){\makebox(0,0)[rb]{\smash{{\SetFigFont{10}{12.0}{\familydefault}{\mddefault}{\updefault}{\color[rgb]{0,0,0}RB$-$}%
}}}}
\put(3691,-3661){\makebox(0,0)[rb]{\smash{{\SetFigFont{10}{12.0}{\familydefault}{\mddefault}{\updefault}{\color[rgb]{0,0,0}or}%
}}}}
\put(2656,-3661){\makebox(0,0)[rb]{\smash{{\SetFigFont{10}{12.0}{\familydefault}{\mddefault}{\updefault}{\color[rgb]{0,0,0}or}%
}}}}
\put(3286,-2131){\makebox(0,0)[rb]{\smash{{\SetFigFont{10}{12.0}{\familydefault}{\mddefault}{\updefault}{\color[rgb]{0,0,0}or}%
}}}}
\put(3466,-4336){\makebox(0,0)[rb]{\smash{{\SetFigFont{12}{14.4}{\familydefault}{\mddefault}{\updefault}{\color[rgb]{0,0,0}(a)}%
}}}}
\put(8731,-4336){\makebox(0,0)[rb]{\smash{{\SetFigFont{12}{14.4}{\familydefault}{\mddefault}{\updefault}{\color[rgb]{0,0,0}(b)}%
}}}}
\put(7246,-196){\makebox(0,0)[rb]{\smash{{\SetFigFont{10}{12.0}{\familydefault}{\mddefault}{\updefault}{\color[rgb]{0,0,0}$\lambda b_4$}%
}}}}
\put(6931,-2716){\makebox(0,0)[rb]{\smash{{\SetFigFont{10}{12.0}{\familydefault}{\mddefault}{\updefault}{\color[rgb]{0,0,0}${\tl D}_u$}%
}}}}
\put(6976,-2131){\makebox(0,0)[rb]{\smash{{\SetFigFont{10}{12.0}{\familydefault}{\mddefault}{\updefault}{\color[rgb]{0,0,0}${\tl L}'$}%
}}}}
\put(9271,-2896){\makebox(0,0)[rb]{\smash{{\SetFigFont{10}{12.0}{\familydefault}{\mddefault}{\updefault}{\color[rgb]{0,0,0}CF$-$}%
}}}}
\put(8146,-331){\makebox(0,0)[rb]{\smash{{\SetFigFont{10}{12.0}{\familydefault}{\mddefault}{\updefault}{\color[rgb]{0,0,0}${\tl L}$}%
}}}}
\put(8641,-1816){\makebox(0,0)[rb]{\smash{{\SetFigFont{10}{12.0}{\familydefault}{\mddefault}{\updefault}{\color[rgb]{0,0,0}RB$-$}%
}}}}
\put(2701,-376){\makebox(0,0)[rb]{\smash{{\SetFigFont{10}{12.0}{\familydefault}{\mddefault}{\updefault}{\color[rgb]{0,0,0}${\tl L}$}%
}}}}
\put(8326,-2851){\makebox(0,0)[rb]{\smash{{\SetFigFont{10}{12.0}{\familydefault}{\mddefault}{\updefault}{\color[rgb]{0,0,0}O$-$}%
}}}}
\put(5176,-691){\makebox(0,0)[rb]{\smash{{\SetFigFont{10}{12.0}{\familydefault}{\mddefault}{\updefault}{\color[rgb]{0,0,0}$m_B^2$}%
}}}}
\put(10621,-691){\makebox(0,0)[rb]{\smash{{\SetFigFont{10}{12.0}{\familydefault}{\mddefault}{\updefault}{\color[rgb]{0,0,0}$m_B^2$}%
}}}}
\end{picture}%

%% file: phaseall.pdf_t
\begin{picture}(0,0)%
\includegraphics{phaseall.pdf}%
\end{picture}%
\setlength{\unitlength}{4144sp}%
\begingroup\makeatletter\ifx\SetFigFont\undefined%
\gdef\SetFigFont#1#2#3#4#5{%
  \reset@font\fontsize{#1}{#2pt}%
  \fontfamily{#3}\fontseries{#4}\fontshape{#5}%
  \selectfont}%
\fi\endgroup%
\begin{picture}(10779,4046)(1114,-8585)
\put(5491,-6181){\makebox(0,0)[lb]{\smash{{\SetFigFont{10}{12.0}{\familydefault}{\mddefault}{\updefault}{\color[rgb]{0,0,0}$(+,-)$}%
}}}}
\put(6616,-4696){\makebox(0,0)[lb]{\smash{{\SetFigFont{10}{12.0}{\familydefault}{\mddefault}{\updefault}{\color[rgb]{0,0,0}$\lambda b_4$}%
}}}}
\put(8056,-6541){\makebox(0,0)[lb]{\smash{{\SetFigFont{10}{12.0}{\familydefault}{\mddefault}{\updefault}{\color[rgb]{0,0,0}$m_B^2$}%
}}}}
\put(7291,-7081){\makebox(0,0)[lb]{\smash{{\SetFigFont{10}{12.0}{\familydefault}{\mddefault}{\updefault}{\color[rgb]{0,0,0}$(-,+)$}%
}}}}
\put(6616,-8521){\makebox(0,0)[rb]{\smash{{\SetFigFont{12}{14.4}{\familydefault}{\mddefault}{\updefault}{\color[rgb]{0,0,0}(b)}%
}}}}
\put(6616,-6271){\makebox(0,0)[lb]{\smash{{\SetFigFont{10}{12.0}{\familydefault}{\mddefault}{\updefault}{\color[rgb]{0,0,0}O}%
}}}}
\put(10081,-4696){\makebox(0,0)[rb]{\smash{{\SetFigFont{10}{12.0}{\familydefault}{\mddefault}{\updefault}{\color[rgb]{0,0,0}$\lambda b_4$}%
}}}}
\put(11161,-7351){\makebox(0,0)[rb]{\smash{{\SetFigFont{10}{12.0}{\familydefault}{\mddefault}{\updefault}{\color[rgb]{0,0,0}O$-$}%
}}}}
\put(11476,-6316){\makebox(0,0)[rb]{\smash{{\SetFigFont{10}{12.0}{\familydefault}{\mddefault}{\updefault}{\color[rgb]{0,0,0}RB$-$}%
}}}}
\put(10261,-5866){\makebox(0,0)[lb]{\smash{{\SetFigFont{10}{12.0}{\familydefault}{\mddefault}{\updefault}{\color[rgb]{0,0,0}$(-,-)$}%
}}}}
\put(10261,-7846){\makebox(0,0)[lb]{\smash{{\SetFigFont{10}{12.0}{\familydefault}{\mddefault}{\updefault}{\color[rgb]{0,0,0}$(+,-)$}%
}}}}
\put(11521,-6811){\makebox(0,0)[lb]{\smash{{\SetFigFont{10}{12.0}{\familydefault}{\mddefault}{\updefault}{\color[rgb]{0,0,0}$(-,+)$}%
}}}}
\put(11836,-5191){\makebox(0,0)[rb]{\smash{{\SetFigFont{10}{12.0}{\familydefault}{\mddefault}{\updefault}{\color[rgb]{0,0,0}$m_B^2$}%
}}}}
\put(10891,-8521){\makebox(0,0)[rb]{\smash{{\SetFigFont{12}{14.4}{\familydefault}{\mddefault}{\updefault}{\color[rgb]{0,0,0}(c)}%
}}}}
\put(3016,-4696){\makebox(0,0)[lb]{\smash{{\SetFigFont{10}{12.0}{\familydefault}{\mddefault}{\updefault}{\color[rgb]{0,0,0}$\lambda b_4$}%
}}}}
\put(3106,-5191){\makebox(0,0)[lb]{\smash{{\SetFigFont{10}{12.0}{\familydefault}{\mddefault}{\updefault}{\color[rgb]{0,0,0}$m_B^2$}%
}}}}
\put(1621,-6316){\makebox(0,0)[lb]{\smash{{\SetFigFont{10}{12.0}{\familydefault}{\mddefault}{\updefault}{\color[rgb]{0,0,0}RB$+$}%
}}}}
\put(2296,-7171){\makebox(0,0)[lb]{\smash{{\SetFigFont{10}{12.0}{\familydefault}{\mddefault}{\updefault}{\color[rgb]{0,0,0}O$+$}%
}}}}
\put(3061,-7351){\makebox(0,0)[lb]{\smash{{\SetFigFont{10}{12.0}{\familydefault}{\mddefault}{\updefault}{\color[rgb]{0,0,0}$(-,+)$}%
}}}}
\put(2296,-5911){\makebox(0,0)[lb]{\smash{{\SetFigFont{10}{12.0}{\familydefault}{\mddefault}{\updefault}{\color[rgb]{0,0,0}$(+,+)$}%
}}}}
\put(1306,-7171){\makebox(0,0)[lb]{\smash{{\SetFigFont{10}{12.0}{\familydefault}{\mddefault}{\updefault}{\color[rgb]{0,0,0}$(+,-)$}%
}}}}
\put(2161,-8521){\makebox(0,0)[rb]{\smash{{\SetFigFont{12}{14.4}{\familydefault}{\mddefault}{\updefault}{\color[rgb]{0,0,0}(a)}%
}}}}
\put(11116,-4831){\makebox(0,0)[lb]{\smash{{\SetFigFont{10}{12.0}{\familydefault}{\mddefault}{\updefault}{\color[rgb]{0,0,0}${\tl L}$}%
}}}}
\put(6571,-5281){\makebox(0,0)[lb]{\smash{{\SetFigFont{10}{12.0}{\familydefault}{\mddefault}{\updefault}{\color[rgb]{0,0,0}$L_0$}%
}}}}
\put(3466,-6361){\makebox(0,0)[lb]{\smash{{\SetFigFont{10}{12.0}{\familydefault}{\mddefault}{\updefault}{\color[rgb]{0,0,0}$L'$}%
}}}}
\put(9766,-6496){\makebox(0,0)[lb]{\smash{{\SetFigFont{10}{12.0}{\familydefault}{\mddefault}{\updefault}{\color[rgb]{0,0,0}${\tl L}'$}%
}}}}
\put(2161,-4831){\makebox(0,0)[lb]{\smash{{\SetFigFont{10}{12.0}{\familydefault}{\mddefault}{\updefault}{\color[rgb]{0,0,0}$L$}%
}}}}
\end{picture}%

%% file: cbrflimit.pdf_t
\begin{picture}(0,0)%
\includegraphics{cbrflimit.pdf}%
\end{picture}%
\setlength{\unitlength}{4144sp}%
\begingroup\makeatletter\ifx\SetFigFont\undefined%
\gdef\SetFigFont#1#2#3#4#5{%
  \reset@font\fontsize{#1}{#2pt}%
  \fontfamily{#3}\fontseries{#4}\fontshape{#5}%
  \selectfont}%
\fi\endgroup%
\begin{picture}(2223,1917)(1786,-2362)
\put(3511,-1636){\makebox(0,0)[lb]{\smash{{\SetFigFont{8}{9.6}{\familydefault}{\mddefault}{\updefault}{\color[rgb]{0,0,0}$m_B^2$}%
}}}}
\put(2791,-556){\makebox(0,0)[lb]{\smash{{\SetFigFont{8}{9.6}{\familydefault}{\mddefault}{\updefault}{\color[rgb]{0,0,0}$\lambda m_F$}%
}}}}
\put(3241,-2311){\makebox(0,0)[lb]{\smash{{\SetFigFont{8}{9.6}{\familydefault}{\mddefault}{\updefault}{\color[rgb]{0,0,0}CB-RF point}%
}}}}
\put(3151,-1951){\makebox(0,0)[lb]{\smash{{\SetFigFont{8}{9.6}{\familydefault}{\mddefault}{\updefault}{\color[rgb]{0,0,0}$(-,+)$}%
}}}}
\put(2251,-1861){\makebox(0,0)[lb]{\smash{{\SetFigFont{8}{9.6}{\familydefault}{\mddefault}{\updefault}{\color[rgb]{0,0,0}$(-,-)$}%
}}}}
\put(2161,-1141){\makebox(0,0)[lb]{\smash{{\SetFigFont{8}{9.6}{\familydefault}{\mddefault}{\updefault}{\color[rgb]{0,0,0}$(+,-)$}%
}}}}
\put(2881,-961){\makebox(0,0)[lb]{\smash{{\SetFigFont{8}{9.6}{\familydefault}{\mddefault}{\updefault}{\color[rgb]{0,0,0}$(+,+)$}%
}}}}
\put(2341,-781){\makebox(0,0)[lb]{\smash{{\SetFigFont{8}{9.6}{\familydefault}{\mddefault}{\updefault}{\color[rgb]{0,0,0}CB$+$}%
}}}}
\put(2341,-2131){\makebox(0,0)[lb]{\smash{{\SetFigFont{8}{9.6}{\familydefault}{\mddefault}{\updefault}{\color[rgb]{0,0,0}CB$-$}%
}}}}
\put(3421,-1186){\makebox(0,0)[lb]{\smash{{\SetFigFont{8}{9.6}{\familydefault}{\mddefault}{\updefault}{\color[rgb]{0,0,0}RF$+$}%
}}}}
\put(1801,-1681){\makebox(0,0)[lb]{\smash{{\SetFigFont{8}{9.6}{\familydefault}{\mddefault}{\updefault}{\color[rgb]{0,0,0}RF$-$}%
}}}}
\end{picture}%

%% file: phaseseqsusy.pdf_t
\begin{picture}(0,0)%
\includegraphics{phaseseqsusy.pdf}%
\end{picture}%
\setlength{\unitlength}{4144sp}%
\begingroup\makeatletter\ifx\SetFigFont\undefined%
\gdef\SetFigFont#1#2#3#4#5{%
  \reset@font\fontsize{#1}{#2pt}%
  \fontfamily{#3}\fontseries{#4}\fontshape{#5}%
  \selectfont}%
\fi\endgroup%
\begin{picture}(9044,3174)(1779,-3820)
\put(8056,-1231){\makebox(0,0)[lb]{\smash{{\SetFigFont{10}{12.0}{\familydefault}{\mddefault}{\updefault}{\color[rgb]{0,0,0}$0$}%
}}}}
\put(4456,-1231){\makebox(0,0)[lb]{\smash{{\SetFigFont{10}{12.0}{\familydefault}{\mddefault}{\updefault}{\color[rgb]{0,0,0}$0$}%
}}}}
\put(3106,-3256){\makebox(0,0)[lb]{\smash{{\SetFigFont{10}{12.0}{\familydefault}{\mddefault}{\updefault}{\color[rgb]{0,0,0}$0$}%
}}}}
\put(9406,-3256){\makebox(0,0)[lb]{\smash{{\SetFigFont{10}{12.0}{\familydefault}{\mddefault}{\updefault}{\color[rgb]{0,0,0}$0$}%
}}}}
\put(2251,-781){\makebox(0,0)[lb]{\smash{{\SetFigFont{10}{12.0}{\familydefault}{\mddefault}{\updefault}{\color[rgb]{0,0,0}$(-,+)$}%
}}}}
\put(4951,-781){\makebox(0,0)[lb]{\smash{{\SetFigFont{10}{12.0}{\familydefault}{\mddefault}{\updefault}{\color[rgb]{0,0,0}$(+,-)$}%
}}}}
\put(7201,-781){\makebox(0,0)[lb]{\smash{{\SetFigFont{10}{12.0}{\familydefault}{\mddefault}{\updefault}{\color[rgb]{0,0,0}$(-,+)$}%
}}}}
\put(9901,-781){\makebox(0,0)[lb]{\smash{{\SetFigFont{10}{12.0}{\familydefault}{\mddefault}{\updefault}{\color[rgb]{0,0,0}$(+,-)$}%
}}}}
\put(2251,-2806){\makebox(0,0)[lb]{\smash{{\SetFigFont{10}{12.0}{\familydefault}{\mddefault}{\updefault}{\color[rgb]{0,0,0}$(+,+)$}%
}}}}
\put(4951,-2806){\makebox(0,0)[lb]{\smash{{\SetFigFont{10}{12.0}{\familydefault}{\mddefault}{\updefault}{\color[rgb]{0,0,0}$(-,-)$}%
}}}}
\put(9901,-2806){\makebox(0,0)[lb]{\smash{{\SetFigFont{10}{12.0}{\familydefault}{\mddefault}{\updefault}{\color[rgb]{0,0,0}$(-,-)$}%
}}}}
\put(7201,-2806){\makebox(0,0)[lb]{\smash{{\SetFigFont{10}{12.0}{\familydefault}{\mddefault}{\updefault}{\color[rgb]{0,0,0}$(+,+)$}%
}}}}
\put(3601,-781){\makebox(0,0)[lb]{\smash{{\SetFigFont{10}{12.0}{\familydefault}{\mddefault}{\updefault}{\color[rgb]{0,0,0}$(+,+)$}%
}}}}
\put(8551,-781){\makebox(0,0)[lb]{\smash{{\SetFigFont{10}{12.0}{\familydefault}{\mddefault}{\updefault}{\color[rgb]{0,0,0}$(-,-)$}%
}}}}
\put(8551,-2806){\makebox(0,0)[lb]{\smash{{\SetFigFont{10}{12.0}{\familydefault}{\mddefault}{\updefault}{\color[rgb]{0,0,0}$(-,+)$}%
}}}}
\put(3601,-2806){\makebox(0,0)[lb]{\smash{{\SetFigFont{10}{12.0}{\familydefault}{\mddefault}{\updefault}{\color[rgb]{0,0,0}$(+,-)$}%
}}}}
\put(2836,-1231){\makebox(0,0)[lb]{\smash{{\SetFigFont{10}{12.0}{\familydefault}{\mddefault}{\updefault}{\color[rgb]{0,0,0}$\displaystyle-\frac{\mu}{\lambda}\frac{1}{1+w}$}%
}}}}
\put(5626,-1096){\makebox(0,0)[lb]{\smash{{\SetFigFont{10}{12.0}{\familydefault}{\mddefault}{\updefault}{\color[rgb]{0,0,0}$\sigma$}%
}}}}
\put(10576,-1096){\makebox(0,0)[lb]{\smash{{\SetFigFont{10}{12.0}{\familydefault}{\mddefault}{\updefault}{\color[rgb]{0,0,0}$\sigma$}%
}}}}
\put(9136,-1231){\makebox(0,0)[lb]{\smash{{\SetFigFont{10}{12.0}{\familydefault}{\mddefault}{\updefault}{\color[rgb]{0,0,0}$\displaystyle-\frac{\mu}{\lambda}\frac{1}{1+w}$}%
}}}}
\put(5626,-3121){\makebox(0,0)[lb]{\smash{{\SetFigFont{10}{12.0}{\familydefault}{\mddefault}{\updefault}{\color[rgb]{0,0,0}$\sigma$}%
}}}}
\put(4186,-3256){\makebox(0,0)[lb]{\smash{{\SetFigFont{10}{12.0}{\familydefault}{\mddefault}{\updefault}{\color[rgb]{0,0,0}$\displaystyle-\frac{\mu}{\lambda}\frac{1}{1+w}$}%
}}}}
\put(7786,-3256){\makebox(0,0)[lb]{\smash{{\SetFigFont{10}{12.0}{\familydefault}{\mddefault}{\updefault}{\color[rgb]{0,0,0}$\displaystyle-\frac{\mu}{\lambda}\frac{1}{1+w}$}%
}}}}
\put(10576,-3121){\makebox(0,0)[lb]{\smash{{\SetFigFont{10}{12.0}{\familydefault}{\mddefault}{\updefault}{\color[rgb]{0,0,0}$\sigma$}%
}}}}
\put(2926,-1726){\makebox(0,0)[lb]{\smash{{\SetFigFont{12}{14.4}{\familydefault}{\mddefault}{\updefault}{\color[rgb]{0,0,0}$\lambda \mu>0$, $1+w > 0$}%
}}}}
\put(7876,-1726){\makebox(0,0)[lb]{\smash{{\SetFigFont{12}{14.4}{\familydefault}{\mddefault}{\updefault}{\color[rgb]{0,0,0}$\lambda \mu < 0$, $1+w>0$}%
}}}}
\put(2926,-3751){\makebox(0,0)[lb]{\smash{{\SetFigFont{12}{14.4}{\familydefault}{\mddefault}{\updefault}{\color[rgb]{0,0,0}$\lambda \mu > 0$, $1+w<0$}%
}}}}
\put(7876,-3751){\makebox(0,0)[lb]{\smash{{\SetFigFont{12}{14.4}{\familydefault}{\mddefault}{\updefault}{\color[rgb]{0,0,0}$\lambda \mu < 0$, $1+w<0$}%
}}}}
\end{picture}%

%% file: wphase.pdf_t
\begin{picture}(0,0)%
\includegraphics{wphase.pdf}%
\end{picture}%
\setlength{\unitlength}{4144sp}%
\begingroup\makeatletter\ifx\SetFigFont\undefined%
\gdef\SetFigFont#1#2#3#4#5{%
  \reset@font\fontsize{#1}{#2pt}%
  \fontfamily{#3}\fontseries{#4}\fontshape{#5}%
  \selectfont}%
\fi\endgroup%
\begin{picture}(6324,387)(1789,-1516)
\put(6436,-1456){\makebox(0,0)[lb]{\smash{{\SetFigFont{10}{12.0}{\familydefault}{\mddefault}{\updefault}{\color[rgb]{0,0,0}$c_1$}%
}}}}
\put(4861,-1456){\makebox(0,0)[lb]{\smash{{\SetFigFont{10}{12.0}{\familydefault}{\mddefault}{\updefault}{\color[rgb]{0,0,0}$-1$}%
}}}}
\put(3691,-1456){\makebox(0,0)[lb]{\smash{{\SetFigFont{10}{12.0}{\familydefault}{\mddefault}{\updefault}{\color[rgb]{0,0,0}$c_3$}%
}}}}
\put(2836,-1456){\makebox(0,0)[lb]{\smash{{\SetFigFont{10}{12.0}{\familydefault}{\mddefault}{\updefault}{\color[rgb]{0,0,0}$a_1$}%
}}}}
\put(1981,-1456){\makebox(0,0)[lb]{\smash{{\SetFigFont{10}{12.0}{\familydefault}{\mddefault}{\updefault}{\color[rgb]{0,0,0}$c_2$}%
}}}}
\put(7201,-1456){\makebox(0,0)[lb]{\smash{{\SetFigFont{10}{12.0}{\familydefault}{\mddefault}{\updefault}{\color[rgb]{0,0,0}$a_3$}%
}}}}
\put(7966,-1456){\makebox(0,0)[lb]{\smash{{\SetFigFont{10}{12.0}{\familydefault}{\mddefault}{\updefault}{\color[rgb]{0,0,0}$w$}%
}}}}
\put(5491,-1456){\makebox(0,0)[lb]{\smash{{\SetFigFont{10}{12.0}{\familydefault}{\mddefault}{\updefault}{\color[rgb]{0,0,0}$a_2$}%
}}}}
\end{picture}%

%% file: wphasepp.pdf_t
\begin{picture}(0,0)%
\includegraphics{wphasepp.pdf}%
\end{picture}%
\setlength{\unitlength}{4144sp}%
\begingroup\makeatletter\ifx\SetFigFont\undefined%
\gdef\SetFigFont#1#2#3#4#5{%
  \reset@font\fontsize{#1}{#2pt}%
  \fontfamily{#3}\fontseries{#4}\fontshape{#5}%
  \selectfont}%
\fi\endgroup%
\begin{picture}(7102,2461)(1021,-2825)
\put(6436,-2761){\makebox(0,0)[lb]{\smash{{\SetFigFont{10}{12.0}{\familydefault}{\mddefault}{\updefault}{\color[rgb]{0,0,0}$c_1$}%
}}}}
\put(4861,-2761){\makebox(0,0)[lb]{\smash{{\SetFigFont{10}{12.0}{\familydefault}{\mddefault}{\updefault}{\color[rgb]{0,0,0}$-1$}%
}}}}
\put(2836,-2761){\makebox(0,0)[lb]{\smash{{\SetFigFont{10}{12.0}{\familydefault}{\mddefault}{\updefault}{\color[rgb]{0,0,0}$a_1$}%
}}}}
\put(7966,-2626){\makebox(0,0)[lb]{\smash{{\SetFigFont{10}{12.0}{\familydefault}{\mddefault}{\updefault}{\color[rgb]{0,0,0}$w$}%
}}}}
\put(1036,-2536){\makebox(0,0)[lb]{\smash{{\SetFigFont{10}{12.0}{\familydefault}{\mddefault}{\updefault}{\color[rgb]{0,0,0}$\mu\lambda<0$}%
}}}}
\put(6436,-1456){\makebox(0,0)[lb]{\smash{{\SetFigFont{10}{12.0}{\familydefault}{\mddefault}{\updefault}{\color[rgb]{0,0,0}$c_1$}%
}}}}
\put(4861,-1456){\makebox(0,0)[lb]{\smash{{\SetFigFont{10}{12.0}{\familydefault}{\mddefault}{\updefault}{\color[rgb]{0,0,0}$-1$}%
}}}}
\put(2836,-1456){\makebox(0,0)[lb]{\smash{{\SetFigFont{10}{12.0}{\familydefault}{\mddefault}{\updefault}{\color[rgb]{0,0,0}$a_1$}%
}}}}
\put(7966,-1321){\makebox(0,0)[lb]{\smash{{\SetFigFont{10}{12.0}{\familydefault}{\mddefault}{\updefault}{\color[rgb]{0,0,0}$w$}%
}}}}
\put(1036,-1231){\makebox(0,0)[lb]{\smash{{\SetFigFont{10}{12.0}{\familydefault}{\mddefault}{\updefault}{\color[rgb]{0,0,0}$\mu\lambda>0$}%
}}}}
\put(3916,-1006){\makebox(0,0)[lb]{\smash{{\SetFigFont{10}{12.0}{\familydefault}{\mddefault}{\updefault}{\color[rgb]{0,0,0}$0$}%
}}}}
\put(5806,-1006){\makebox(0,0)[lb]{\smash{{\SetFigFont{10}{12.0}{\familydefault}{\mddefault}{\updefault}{\color[rgb]{0,0,0}$0$}%
}}}}
\put(7471,-1006){\makebox(0,0)[lb]{\smash{{\SetFigFont{10}{12.0}{\familydefault}{\mddefault}{\updefault}{\color[rgb]{0,0,0}$0$}%
}}}}
\put(2701,-2356){\makebox(0,0)[lb]{\smash{{\SetFigFont{10}{12.0}{\familydefault}{\mddefault}{\updefault}{\color[rgb]{0,0,0}$0$}%
}}}}
\put(4141,-2356){\makebox(0,0)[lb]{\smash{{\SetFigFont{10}{12.0}{\familydefault}{\mddefault}{\updefault}{\color[rgb]{0,0,0}$0$}%
}}}}
\put(5536,-2356){\makebox(0,0)[lb]{\smash{{\SetFigFont{10}{12.0}{\familydefault}{\mddefault}{\updefault}{\color[rgb]{0,0,0}$0$}%
}}}}
\put(7111,-2356){\makebox(0,0)[lb]{\smash{{\SetFigFont{10}{12.0}{\familydefault}{\mddefault}{\updefault}{\color[rgb]{0,0,0}$0$}%
}}}}
\put(2116,-1006){\makebox(0,0)[lb]{\smash{{\SetFigFont{10}{12.0}{\familydefault}{\mddefault}{\updefault}{\color[rgb]{0,0,0}$0$}%
}}}}
\end{picture}%

%% file: wphasepm.pdf_t
\begin{picture}(0,0)%
\includegraphics{wphasepm.pdf}%
\end{picture}%
\setlength{\unitlength}{4144sp}%
\begingroup\makeatletter\ifx\SetFigFont\undefined%
\gdef\SetFigFont#1#2#3#4#5{%
  \reset@font\fontsize{#1}{#2pt}%
  \fontfamily{#3}\fontseries{#4}\fontshape{#5}%
  \selectfont}%
\fi\endgroup%
\begin{picture}(7102,2416)(1021,-2825)
\put(6436,-2761){\makebox(0,0)[lb]{\smash{{\SetFigFont{10}{12.0}{\familydefault}{\mddefault}{\updefault}{\color[rgb]{0,0,0}$a_2$}%
}}}}
\put(4861,-2761){\makebox(0,0)[lb]{\smash{{\SetFigFont{10}{12.0}{\familydefault}{\mddefault}{\updefault}{\color[rgb]{0,0,0}$-1$}%
}}}}
\put(2836,-2761){\makebox(0,0)[lb]{\smash{{\SetFigFont{10}{12.0}{\familydefault}{\mddefault}{\updefault}{\color[rgb]{0,0,0}$c_2$}%
}}}}
\put(7966,-2626){\makebox(0,0)[lb]{\smash{{\SetFigFont{10}{12.0}{\familydefault}{\mddefault}{\updefault}{\color[rgb]{0,0,0}$w$}%
}}}}
\put(1036,-2536){\makebox(0,0)[lb]{\smash{{\SetFigFont{10}{12.0}{\familydefault}{\mddefault}{\updefault}{\color[rgb]{0,0,0}$\mu\lambda<0$}%
}}}}
\put(6436,-1456){\makebox(0,0)[lb]{\smash{{\SetFigFont{10}{12.0}{\familydefault}{\mddefault}{\updefault}{\color[rgb]{0,0,0}$a_2$}%
}}}}
\put(4861,-1456){\makebox(0,0)[lb]{\smash{{\SetFigFont{10}{12.0}{\familydefault}{\mddefault}{\updefault}{\color[rgb]{0,0,0}$-1$}%
}}}}
\put(2836,-1456){\makebox(0,0)[lb]{\smash{{\SetFigFont{10}{12.0}{\familydefault}{\mddefault}{\updefault}{\color[rgb]{0,0,0}$c_2$}%
}}}}
\put(7966,-1321){\makebox(0,0)[lb]{\smash{{\SetFigFont{10}{12.0}{\familydefault}{\mddefault}{\updefault}{\color[rgb]{0,0,0}$w$}%
}}}}
\put(1036,-1231){\makebox(0,0)[lb]{\smash{{\SetFigFont{10}{12.0}{\familydefault}{\mddefault}{\updefault}{\color[rgb]{0,0,0}$\mu\lambda>0$}%
}}}}
\put(2296,-1006){\makebox(0,0)[rb]{\smash{{\SetFigFont{10}{12.0}{\familydefault}{\mddefault}{\updefault}{\color[rgb]{0,0,0}$0$}%
}}}}
\put(3826,-1006){\makebox(0,0)[lb]{\smash{{\SetFigFont{10}{12.0}{\familydefault}{\mddefault}{\updefault}{\color[rgb]{0,0,0}$0$}%
}}}}
\put(7606,-1006){\makebox(0,0)[lb]{\smash{{\SetFigFont{10}{12.0}{\familydefault}{\mddefault}{\updefault}{\color[rgb]{0,0,0}$0$}%
}}}}
\put(4186,-2311){\makebox(0,0)[lb]{\smash{{\SetFigFont{10}{12.0}{\familydefault}{\mddefault}{\updefault}{\color[rgb]{0,0,0}$0$}%
}}}}
\put(5716,-2311){\makebox(0,0)[lb]{\smash{{\SetFigFont{10}{12.0}{\familydefault}{\mddefault}{\updefault}{\color[rgb]{0,0,0}$0$}%
}}}}
\put(7156,-2311){\makebox(0,0)[lb]{\smash{{\SetFigFont{10}{12.0}{\familydefault}{\mddefault}{\updefault}{\color[rgb]{0,0,0}$0$}%
}}}}
\put(2521,-2311){\makebox(0,0)[rb]{\smash{{\SetFigFont{10}{12.0}{\familydefault}{\mddefault}{\updefault}{\color[rgb]{0,0,0}$0$}%
}}}}
\put(5761,-1006){\makebox(0,0)[lb]{\smash{{\SetFigFont{10}{12.0}{\familydefault}{\mddefault}{\updefault}{\color[rgb]{0,0,0}$0$}%
}}}}
\end{picture}%

%% file: wphasemp.pdf_t
\begin{picture}(0,0)%
\includegraphics{wphasemp.pdf}%
\end{picture}%
\setlength{\unitlength}{4144sp}%
\begingroup\makeatletter\ifx\SetFigFont\undefined%
\gdef\SetFigFont#1#2#3#4#5{%
  \reset@font\fontsize{#1}{#2pt}%
  \fontfamily{#3}\fontseries{#4}\fontshape{#5}%
  \selectfont}%
\fi\endgroup%
\begin{picture}(7102,2416)(1021,-2825)
\put(6436,-2761){\makebox(0,0)[lb]{\smash{{\SetFigFont{10}{12.0}{\familydefault}{\mddefault}{\updefault}{\color[rgb]{0,0,0}$a_3$}%
}}}}
\put(4861,-2761){\makebox(0,0)[lb]{\smash{{\SetFigFont{10}{12.0}{\familydefault}{\mddefault}{\updefault}{\color[rgb]{0,0,0}$-1$}%
}}}}
\put(2836,-2761){\makebox(0,0)[lb]{\smash{{\SetFigFont{10}{12.0}{\familydefault}{\mddefault}{\updefault}{\color[rgb]{0,0,0}$c_3$}%
}}}}
\put(7966,-2626){\makebox(0,0)[lb]{\smash{{\SetFigFont{10}{12.0}{\familydefault}{\mddefault}{\updefault}{\color[rgb]{0,0,0}$w$}%
}}}}
\put(1036,-2536){\makebox(0,0)[lb]{\smash{{\SetFigFont{10}{12.0}{\familydefault}{\mddefault}{\updefault}{\color[rgb]{0,0,0}$\mu\lambda<0$}%
}}}}
\put(6436,-1456){\makebox(0,0)[lb]{\smash{{\SetFigFont{10}{12.0}{\familydefault}{\mddefault}{\updefault}{\color[rgb]{0,0,0}$a_3$}%
}}}}
\put(4861,-1456){\makebox(0,0)[lb]{\smash{{\SetFigFont{10}{12.0}{\familydefault}{\mddefault}{\updefault}{\color[rgb]{0,0,0}$-1$}%
}}}}
\put(2836,-1456){\makebox(0,0)[lb]{\smash{{\SetFigFont{10}{12.0}{\familydefault}{\mddefault}{\updefault}{\color[rgb]{0,0,0}$c_3$}%
}}}}
\put(7966,-1321){\makebox(0,0)[lb]{\smash{{\SetFigFont{10}{12.0}{\familydefault}{\mddefault}{\updefault}{\color[rgb]{0,0,0}$w$}%
}}}}
\put(1036,-1231){\makebox(0,0)[lb]{\smash{{\SetFigFont{10}{12.0}{\familydefault}{\mddefault}{\updefault}{\color[rgb]{0,0,0}$\mu\lambda>0$}%
}}}}
\put(7066,-2311){\makebox(0,0)[rb]{\smash{{\SetFigFont{10}{12.0}{\familydefault}{\mddefault}{\updefault}{\color[rgb]{0,0,0}$0$}%
}}}}
\put(4096,-2311){\makebox(0,0)[rb]{\smash{{\SetFigFont{10}{12.0}{\familydefault}{\mddefault}{\updefault}{\color[rgb]{0,0,0}$0$}%
}}}}
\put(5716,-2311){\makebox(0,0)[rb]{\smash{{\SetFigFont{10}{12.0}{\familydefault}{\mddefault}{\updefault}{\color[rgb]{0,0,0}$0$}%
}}}}
\put(2611,-2311){\makebox(0,0)[lb]{\smash{{\SetFigFont{10}{12.0}{\familydefault}{\mddefault}{\updefault}{\color[rgb]{0,0,0}$0$}%
}}}}
\put(7651,-1006){\makebox(0,0)[rb]{\smash{{\SetFigFont{10}{12.0}{\familydefault}{\mddefault}{\updefault}{\color[rgb]{0,0,0}$0$}%
}}}}
\put(5851,-1006){\makebox(0,0)[rb]{\smash{{\SetFigFont{10}{12.0}{\familydefault}{\mddefault}{\updefault}{\color[rgb]{0,0,0}$0$}%
}}}}
\put(2116,-1006){\makebox(0,0)[lb]{\smash{{\SetFigFont{10}{12.0}{\familydefault}{\mddefault}{\updefault}{\color[rgb]{0,0,0}$0$}%
}}}}
\put(3871,-1006){\makebox(0,0)[rb]{\smash{{\SetFigFont{10}{12.0}{\familydefault}{\mddefault}{\updefault}{\color[rgb]{0,0,0}$0$}%
}}}}
\end{picture}%

%% file: wphaseLGpotp.pdf_t
\begin{picture}(0,0)%
\includegraphics{wphaseLGpotp.pdf}%
\end{picture}%
\setlength{\unitlength}{4144sp}%
\begingroup\makeatletter\ifx\SetFigFont\undefined%
\gdef\SetFigFont#1#2#3#4#5{%
  \reset@font\fontsize{#1}{#2pt}%
  \fontfamily{#3}\fontseries{#4}\fontshape{#5}%
  \selectfont}%
\fi\endgroup%
\begin{picture}(7674,1087)(439,-1700)
\put(1036,-736){\makebox(0,0)[lb]{\smash{{\SetFigFont{5}{6.0}{\familydefault}{\mddefault}{\updefault}{\color[rgb]{0,0,0}$++$}%
}}}}
\put(1261,-736){\makebox(0,0)[lb]{\smash{{\SetFigFont{5}{6.0}{\familydefault}{\mddefault}{\updefault}{\color[rgb]{0,0,0}$+-$}%
}}}}
\put(1531,-736){\makebox(0,0)[lb]{\smash{{\SetFigFont{5}{6.0}{\familydefault}{\mddefault}{\updefault}{\color[rgb]{0,0,0}$--$}%
}}}}
\put(5806,-736){\makebox(0,0)[lb]{\smash{{\SetFigFont{5}{6.0}{\familydefault}{\mddefault}{\updefault}{\color[rgb]{0,0,0}$++$}%
}}}}
\put(6076,-736){\makebox(0,0)[lb]{\smash{{\SetFigFont{5}{6.0}{\familydefault}{\mddefault}{\updefault}{\color[rgb]{0,0,0}$+-$}%
}}}}
\put(5536,-736){\makebox(0,0)[lb]{\smash{{\SetFigFont{5}{6.0}{\familydefault}{\mddefault}{\updefault}{\color[rgb]{0,0,0}$-+$}%
}}}}
\put(6931,-736){\makebox(0,0)[lb]{\smash{{\SetFigFont{5}{6.0}{\familydefault}{\mddefault}{\updefault}{\color[rgb]{0,0,0}$-+$}%
}}}}
\put(7246,-736){\makebox(0,0)[lb]{\smash{{\SetFigFont{5}{6.0}{\familydefault}{\mddefault}{\updefault}{\color[rgb]{0,0,0}$++$}%
}}}}
\put(7561,-736){\makebox(0,0)[lb]{\smash{{\SetFigFont{5}{6.0}{\familydefault}{\mddefault}{\updefault}{\color[rgb]{0,0,0}$+-$}%
}}}}
\put(7966,-1366){\makebox(0,0)[lb]{\smash{{\SetFigFont{10}{12.0}{\familydefault}{\mddefault}{\updefault}{\color[rgb]{0,0,0}$w$}%
}}}}
\put(4411,-736){\makebox(0,0)[lb]{\smash{{\SetFigFont{5}{6.0}{\familydefault}{\mddefault}{\updefault}{\color[rgb]{0,0,0}$++$}%
}}}}
\put(4681,-736){\makebox(0,0)[lb]{\smash{{\SetFigFont{5}{6.0}{\familydefault}{\mddefault}{\updefault}{\color[rgb]{0,0,0}$+-$}%
}}}}
\put(4186,-736){\makebox(0,0)[lb]{\smash{{\SetFigFont{5}{6.0}{\familydefault}{\mddefault}{\updefault}{\color[rgb]{0,0,0}$-+$}%
}}}}
\put(3286,-736){\makebox(0,0)[lb]{\smash{{\SetFigFont{5}{6.0}{\familydefault}{\mddefault}{\updefault}{\color[rgb]{0,0,0}$--$}%
}}}}
\put(2746,-736){\makebox(0,0)[lb]{\smash{{\SetFigFont{5}{6.0}{\familydefault}{\mddefault}{\updefault}{\color[rgb]{0,0,0}$++$}%
}}}}
\put(2971,-736){\makebox(0,0)[lb]{\smash{{\SetFigFont{5}{6.0}{\familydefault}{\mddefault}{\updefault}{\color[rgb]{0,0,0}$+-$}%
}}}}
\put(4321,-1501){\makebox(0,0)[lb]{\smash{{\SetFigFont{10}{12.0}{\familydefault}{\mddefault}{\updefault}{\color[rgb]{0,0,0}$(+,+)$}%
}}}}
\put(4321,-1636){\makebox(0,0)[lb]{\smash{{\SetFigFont{10}{12.0}{\familydefault}{\mddefault}{\updefault}{\color[rgb]{0,0,0}$(+,-)$}%
}}}}
\put(2926,-1501){\makebox(0,0)[lb]{\smash{{\SetFigFont{10}{12.0}{\familydefault}{\mddefault}{\updefault}{\color[rgb]{0,0,0}$(+,+)$}%
}}}}
\put(2926,-1636){\makebox(0,0)[lb]{\smash{{\SetFigFont{10}{12.0}{\familydefault}{\mddefault}{\updefault}{\color[rgb]{0,0,0}$(+,-)$}%
}}}}
\put(3781,-1366){\makebox(0,0)[lb]{\smash{{\SetFigFont{10}{12.0}{\familydefault}{\mddefault}{\updefault}{\color[rgb]{0,0,0}$-1$}%
}}}}
\put(2161,-1366){\makebox(0,0)[lb]{\smash{{\SetFigFont{10}{12.0}{\familydefault}{\mddefault}{\updefault}{\color[rgb]{0,0,0}$a_1$}%
}}}}
\put(1216,-1501){\makebox(0,0)[lb]{\smash{{\SetFigFont{10}{12.0}{\familydefault}{\mddefault}{\updefault}{\color[rgb]{0,0,0}$(+,-)$}%
}}}}
\put(5761,-1501){\makebox(0,0)[lb]{\smash{{\SetFigFont{10}{12.0}{\familydefault}{\mddefault}{\updefault}{\color[rgb]{0,0,0}$(+,+)$}%
}}}}
\put(6661,-1366){\makebox(0,0)[lb]{\smash{{\SetFigFont{10}{12.0}{\familydefault}{\mddefault}{\updefault}{\color[rgb]{0,0,0}$a_3$}%
}}}}
\put(7156,-1501){\makebox(0,0)[lb]{\smash{{\SetFigFont{10}{12.0}{\familydefault}{\mddefault}{\updefault}{\color[rgb]{0,0,0}$(+,+)$}%
}}}}
\put(7156,-1636){\makebox(0,0)[lb]{\smash{{\SetFigFont{10}{12.0}{\familydefault}{\mddefault}{\updefault}{\color[rgb]{0,0,0}$(-,+)$}%
}}}}
\put(4951,-1366){\makebox(0,0)[lb]{\smash{{\SetFigFont{10}{12.0}{\familydefault}{\mddefault}{\updefault}{\color[rgb]{0,0,0}$a_2$}%
}}}}
\end{picture}%

%% file: wphaseLGpotm.pdf_t
\begin{picture}(0,0)%
\includegraphics{wphaseLGpotm.pdf}%
\end{picture}%
\setlength{\unitlength}{4144sp}%
\begingroup\makeatletter\ifx\SetFigFont\undefined%
\gdef\SetFigFont#1#2#3#4#5{%
  \reset@font\fontsize{#1}{#2pt}%
  \fontfamily{#3}\fontseries{#4}\fontshape{#5}%
  \selectfont}%
\fi\endgroup%
\begin{picture}(7674,1087)(439,-1700)
\put(7111,-736){\makebox(0,0)[lb]{\smash{{\SetFigFont{5}{6.0}{\familydefault}{\mddefault}{\updefault}{\color[rgb]{0,0,0}$-+$}%
}}}}
\put(7336,-736){\makebox(0,0)[lb]{\smash{{\SetFigFont{5}{6.0}{\familydefault}{\mddefault}{\updefault}{\color[rgb]{0,0,0}$--$}%
}}}}
\put(7561,-736){\makebox(0,0)[lb]{\smash{{\SetFigFont{5}{6.0}{\familydefault}{\mddefault}{\updefault}{\color[rgb]{0,0,0}$+-$}%
}}}}
\put(5581,-736){\makebox(0,0)[lb]{\smash{{\SetFigFont{5}{6.0}{\familydefault}{\mddefault}{\updefault}{\color[rgb]{0,0,0}$-+$}%
}}}}
\put(5806,-736){\makebox(0,0)[lb]{\smash{{\SetFigFont{5}{6.0}{\familydefault}{\mddefault}{\updefault}{\color[rgb]{0,0,0}$--$}%
}}}}
\put(6031,-736){\makebox(0,0)[lb]{\smash{{\SetFigFont{5}{6.0}{\familydefault}{\mddefault}{\updefault}{\color[rgb]{0,0,0}$+-$}%
}}}}
\put(4861,-736){\makebox(0,0)[rb]{\smash{{\SetFigFont{5}{6.0}{\familydefault}{\mddefault}{\updefault}{\color[rgb]{0,0,0}$+-$}%
}}}}
\put(4636,-736){\makebox(0,0)[rb]{\smash{{\SetFigFont{5}{6.0}{\familydefault}{\mddefault}{\updefault}{\color[rgb]{0,0,0}$--$}%
}}}}
\put(4411,-736){\makebox(0,0)[rb]{\smash{{\SetFigFont{5}{6.0}{\familydefault}{\mddefault}{\updefault}{\color[rgb]{0,0,0}$-+$}%
}}}}
\put(1396,-779){\makebox(0,0)[lb]{\smash{{\SetFigFont{5}{6.0}{\familydefault}{\mddefault}{\updefault}{\color[rgb]{0,0,0}$--$}%
}}}}
\put(1081,-779){\makebox(0,0)[lb]{\smash{{\SetFigFont{5}{6.0}{\familydefault}{\mddefault}{\updefault}{\color[rgb]{0,0,0}$-+$}%
}}}}
\put(856,-779){\makebox(0,0)[lb]{\smash{{\SetFigFont{5}{6.0}{\familydefault}{\mddefault}{\updefault}{\color[rgb]{0,0,0}$++$}%
}}}}
\put(3016,-736){\makebox(0,0)[lb]{\smash{{\SetFigFont{5}{6.0}{\familydefault}{\mddefault}{\updefault}{\color[rgb]{0,0,0}$--$}%
}}}}
\put(2746,-736){\makebox(0,0)[lb]{\smash{{\SetFigFont{5}{6.0}{\familydefault}{\mddefault}{\updefault}{\color[rgb]{0,0,0}$-+$}%
}}}}
\put(2521,-736){\makebox(0,0)[lb]{\smash{{\SetFigFont{5}{6.0}{\familydefault}{\mddefault}{\updefault}{\color[rgb]{0,0,0}$++$}%
}}}}
\put(7966,-1366){\makebox(0,0)[lb]{\smash{{\SetFigFont{10}{12.0}{\familydefault}{\mddefault}{\updefault}{\color[rgb]{0,0,0}$w$}%
}}}}
\put(7291,-1501){\makebox(0,0)[lb]{\smash{{\SetFigFont{10}{12.0}{\familydefault}{\mddefault}{\updefault}{\color[rgb]{0,0,0}$(+,-)$}%
}}}}
\put(4951,-1366){\makebox(0,0)[lb]{\smash{{\SetFigFont{10}{12.0}{\familydefault}{\mddefault}{\updefault}{\color[rgb]{0,0,0}$a_2$}%
}}}}
\put(3781,-1366){\makebox(0,0)[lb]{\smash{{\SetFigFont{10}{12.0}{\familydefault}{\mddefault}{\updefault}{\color[rgb]{0,0,0}$-1$}%
}}}}
\put(5761,-1501){\makebox(0,0)[lb]{\smash{{\SetFigFont{10}{12.0}{\familydefault}{\mddefault}{\updefault}{\color[rgb]{0,0,0}$(-,+)$}%
}}}}
\put(5761,-1636){\makebox(0,0)[lb]{\smash{{\SetFigFont{10}{12.0}{\familydefault}{\mddefault}{\updefault}{\color[rgb]{0,0,0}$(+,-)$}%
}}}}
\put(6706,-1366){\makebox(0,0)[lb]{\smash{{\SetFigFont{10}{12.0}{\familydefault}{\mddefault}{\updefault}{\color[rgb]{0,0,0}$a_3$}%
}}}}
\put(4231,-1501){\makebox(0,0)[lb]{\smash{{\SetFigFont{10}{12.0}{\familydefault}{\mddefault}{\updefault}{\color[rgb]{0,0,0}$(-,+)$}%
}}}}
\put(1891,-1366){\makebox(0,0)[lb]{\smash{{\SetFigFont{10}{12.0}{\familydefault}{\mddefault}{\updefault}{\color[rgb]{0,0,0}$a_1$}%
}}}}
\put(991,-1501){\makebox(0,0)[lb]{\smash{{\SetFigFont{10}{12.0}{\familydefault}{\mddefault}{\updefault}{\color[rgb]{0,0,0}$(+,+)$}%
}}}}
\put(991,-1636){\makebox(0,0)[lb]{\smash{{\SetFigFont{10}{12.0}{\familydefault}{\mddefault}{\updefault}{\color[rgb]{0,0,0}$(-,+)$}%
}}}}
\put(2701,-1501){\makebox(0,0)[lb]{\smash{{\SetFigFont{10}{12.0}{\familydefault}{\mddefault}{\updefault}{\color[rgb]{0,0,0}$(-,+)$}%
}}}}
\end{picture}%

%% file: phaseappI.pdf_t
\begin{picture}(0,0)%
\includegraphics{phaseappI.pdf}%
\end{picture}%
\setlength{\unitlength}{4144sp}%
\begingroup\makeatletter\ifx\SetFigFont\undefined%
\gdef\SetFigFont#1#2#3#4#5{%
  \reset@font\fontsize{#1}{#2pt}%
  \fontfamily{#3}\fontseries{#4}\fontshape{#5}%
  \selectfont}%
\fi\endgroup%
\begin{picture}(2311,2284)(1077,-3673)
\put(2914,-2155){\makebox(0,0)[lb]{\smash{{\SetFigFont{5}{6.0}{\familydefault}{\mddefault}{\updefault}{\color[rgb]{0,0,0}$\sigma_B$}%
}}}}
\put(2276,-1658){\makebox(0,0)[lb]{\smash{{\SetFigFont{5}{6.0}{\familydefault}{\mddefault}{\updefault}{\color[rgb]{0,0,0}$\sigma_B$}%
}}}}
\put(2263,-3499){\makebox(0,0)[lb]{\smash{{\SetFigFont{5}{6.0}{\familydefault}{\mddefault}{\updefault}{\color[rgb]{0,0,0}$\sigma_B$}%
}}}}
\put(2699,-2920){\makebox(0,0)[lb]{\smash{{\SetFigFont{5}{6.0}{\familydefault}{\mddefault}{\updefault}{\color[rgb]{0,0,0}$\sigma_B$}%
}}}}
\put(1494,-2306){\makebox(0,0)[lb]{\smash{{\SetFigFont{5}{6.0}{\familydefault}{\mddefault}{\updefault}{\color[rgb]{0,0,0}$\sigma_B$}%
}}}}
\put(1494,-3161){\makebox(0,0)[lb]{\smash{{\SetFigFont{5}{6.0}{\familydefault}{\mddefault}{\updefault}{\color[rgb]{0,0,0}$\sigma_B$}%
}}}}
\put(3151,-2671){\makebox(0,0)[lb]{\smash{{\SetFigFont{8}{9.6}{\familydefault}{\mddefault}{\updefault}{\color[rgb]{0,0,0}$m_B^2$}%
}}}}
\put(3331,-1501){\makebox(0,0)[lb]{\smash{{\SetFigFont{8}{9.6}{\familydefault}{\mddefault}{\updefault}{\color[rgb]{0,0,0}$D_u$}%
}}}}
\put(3286,-3436){\makebox(0,0)[lb]{\smash{{\SetFigFont{8}{9.6}{\familydefault}{\mddefault}{\updefault}{\color[rgb]{0,0,0}$D_h$}%
}}}}
\put(1486,-1501){\makebox(0,0)[lb]{\smash{{\SetFigFont{8}{9.6}{\familydefault}{\mddefault}{\updefault}{\color[rgb]{0,0,0}$\lambda_B b_4$}%
}}}}
\put(1621,-1996){\makebox(0,0)[lb]{\smash{{\SetFigFont{8}{9.6}{\familydefault}{\mddefault}{\updefault}{\color[rgb]{0,0,0}$L$}%
}}}}
\end{picture}%

%% file: phaseappII.pdf_t
\begin{picture}(0,0)%
\includegraphics{phaseappII.pdf}%
\end{picture}%
\setlength{\unitlength}{4144sp}%
\begingroup\makeatletter\ifx\SetFigFont\undefined%
\gdef\SetFigFont#1#2#3#4#5{%
  \reset@font\fontsize{#1}{#2pt}%
  \fontfamily{#3}\fontseries{#4}\fontshape{#5}%
  \selectfont}%
\fi\endgroup%
\begin{picture}(2630,2284)(1436,-3763)
\put(3613,-2296){\makebox(0,0)[lb]{\smash{{\SetFigFont{5}{6.0}{\familydefault}{\mddefault}{\updefault}{\color[rgb]{0,0,0}$\sigma_B$}%
}}}}
\put(2175,-2299){\makebox(0,0)[lb]{\smash{{\SetFigFont{5}{6.0}{\familydefault}{\mddefault}{\updefault}{\color[rgb]{0,0,0}$\sigma_B$}%
}}}}
\put(3415,-3593){\makebox(0,0)[lb]{\smash{{\SetFigFont{5}{6.0}{\familydefault}{\mddefault}{\updefault}{\color[rgb]{0,0,0}$\sigma_B$}%
}}}}
\put(1860,-3023){\makebox(0,0)[lb]{\smash{{\SetFigFont{5}{6.0}{\familydefault}{\mddefault}{\updefault}{\color[rgb]{0,0,0}$\sigma_B$}%
}}}}
\put(4006,-2671){\makebox(0,0)[rb]{\smash{{\SetFigFont{8}{9.6}{\familydefault}{\mddefault}{\updefault}{\color[rgb]{0,0,0}$m_B^2$}%
}}}}
\put(3106,-1591){\makebox(0,0)[rb]{\smash{{\SetFigFont{8}{9.6}{\familydefault}{\mddefault}{\updefault}{\color[rgb]{0,0,0}$\lambda_B b_4$}%
}}}}
\put(1621,-3526){\makebox(0,0)[rb]{\smash{{\SetFigFont{8}{9.6}{\familydefault}{\mddefault}{\updefault}{\color[rgb]{0,0,0}$D_u$}%
}}}}
\put(2656,-1861){\makebox(0,0)[rb]{\smash{{\SetFigFont{8}{9.6}{\familydefault}{\mddefault}{\updefault}{\color[rgb]{0,0,0}$L$}%
}}}}
\put(2431,-3661){\makebox(0,0)[rb]{\smash{{\SetFigFont{8}{9.6}{\familydefault}{\mddefault}{\updefault}{\color[rgb]{0,0,0}$D_\nu$}%
}}}}
\put(4051,-3301){\makebox(0,0)[rb]{\smash{{\SetFigFont{8}{9.6}{\familydefault}{\mddefault}{\updefault}{\color[rgb]{0,0,0}$D_h$}%
}}}}
\end{picture}%

%% file: phaseappIII.pdf_t
\begin{picture}(0,0)%
\includegraphics{phaseappIII.pdf}%
\end{picture}%
\setlength{\unitlength}{4144sp}%
\begingroup\makeatletter\ifx\SetFigFont\undefined%
\gdef\SetFigFont#1#2#3#4#5{%
  \reset@font\fontsize{#1}{#2pt}%
  \fontfamily{#3}\fontseries{#4}\fontshape{#5}%
  \selectfont}%
\fi\endgroup%
\begin{picture}(2457,2284)(1564,-3673)
\put(2127,-2219){\makebox(0,0)[lb]{\smash{{\SetFigFont{5}{6.0}{\familydefault}{\mddefault}{\updefault}{\color[rgb]{0,0,0}$\sigma_B$}%
}}}}
\put(2127,-2984){\makebox(0,0)[lb]{\smash{{\SetFigFont{5}{6.0}{\familydefault}{\mddefault}{\updefault}{\color[rgb]{0,0,0}$\sigma_B$}%
}}}}
\put(2900,-1625){\makebox(0,0)[lb]{\smash{{\SetFigFont{5}{6.0}{\familydefault}{\mddefault}{\updefault}{\color[rgb]{0,0,0}$\sigma_B$}%
}}}}
\put(2808,-3479){\makebox(0,0)[lb]{\smash{{\SetFigFont{5}{6.0}{\familydefault}{\mddefault}{\updefault}{\color[rgb]{0,0,0}$\sigma_B$}%
}}}}
\put(3597,-3136){\makebox(0,0)[lb]{\smash{{\SetFigFont{5}{6.0}{\familydefault}{\mddefault}{\updefault}{\color[rgb]{0,0,0}$\sigma_B$}%
}}}}
\put(3597,-2281){\makebox(0,0)[lb]{\smash{{\SetFigFont{5}{6.0}{\familydefault}{\mddefault}{\updefault}{\color[rgb]{0,0,0}$\sigma_B$}%
}}}}
\put(1666,-3436){\makebox(0,0)[rb]{\smash{{\SetFigFont{8}{9.6}{\familydefault}{\mddefault}{\updefault}{\color[rgb]{0,0,0}$D_u$}%
}}}}
\put(1621,-1501){\makebox(0,0)[rb]{\smash{{\SetFigFont{8}{9.6}{\familydefault}{\mddefault}{\updefault}{\color[rgb]{0,0,0}$D_h$}%
}}}}
\put(4006,-2671){\makebox(0,0)[rb]{\smash{{\SetFigFont{8}{9.6}{\familydefault}{\mddefault}{\updefault}{\color[rgb]{0,0,0}$m_B^2$}%
}}}}
\put(3556,-1501){\makebox(0,0)[rb]{\smash{{\SetFigFont{8}{9.6}{\familydefault}{\mddefault}{\updefault}{\color[rgb]{0,0,0}$\lambda_B b_4$}%
}}}}
\put(3286,-1861){\makebox(0,0)[rb]{\smash{{\SetFigFont{8}{9.6}{\familydefault}{\mddefault}{\updefault}{\color[rgb]{0,0,0}$L$}%
}}}}
\end{picture}%

%% file: ellipse.pdf_t
\begin{picture}(0,0)%
\includegraphics{ellipse.pdf}%
\end{picture}%
\setlength{\unitlength}{4144sp}%
\begingroup\makeatletter\ifx\SetFigFont\undefined%
\gdef\SetFigFont#1#2#3#4#5{%
  \reset@font\fontsize{#1}{#2pt}%
  \fontfamily{#3}\fontseries{#4}\fontshape{#5}%
  \selectfont}%
\fi\endgroup%
\begin{picture}(8082,2011)(1249,-4076)
\put(7246,-2401){\makebox(0,0)[rb]{\smash{{\SetFigFont{8}{9.6}{\familydefault}{\mddefault}{\updefault}{\color[rgb]{0,0,0}No phase}%
}}}}
\put(8956,-3661){\makebox(0,0)[rb]{\smash{{\SetFigFont{8}{9.6}{\familydefault}{\mddefault}{\updefault}{\color[rgb]{0,0,0}$(+)$}%
}}}}
\put(9316,-3121){\makebox(0,0)[rb]{\smash{{\SetFigFont{8}{9.6}{\familydefault}{\mddefault}{\updefault}{\color[rgb]{0,0,0}$m_B^2$}%
}}}}
\put(8641,-2221){\makebox(0,0)[rb]{\smash{{\SetFigFont{8}{9.6}{\familydefault}{\mddefault}{\updefault}{\color[rgb]{0,0,0}$\lambda_Bb_4$}%
}}}}
\put(5941,-3121){\makebox(0,0)[lb]{\smash{{\SetFigFont{8}{9.6}{\familydefault}{\mddefault}{\updefault}{\color[rgb]{0,0,0}$m_B^2$}%
}}}}
\put(5176,-2176){\makebox(0,0)[lb]{\smash{{\SetFigFont{8}{9.6}{\familydefault}{\mddefault}{\updefault}{\color[rgb]{0,0,0}$\lambda_Bb_4$}%
}}}}
\put(5401,-2491){\makebox(0,0)[lb]{\smash{{\SetFigFont{8}{9.6}{\familydefault}{\mddefault}{\updefault}{\color[rgb]{0,0,0}$(+)$}%
}}}}
\put(4681,-2491){\makebox(0,0)[lb]{\smash{{\SetFigFont{8}{9.6}{\familydefault}{\mddefault}{\updefault}{\color[rgb]{0,0,0}$(-)$}%
}}}}
\put(3241,-2401){\makebox(0,0)[lb]{\smash{{\SetFigFont{8}{9.6}{\familydefault}{\mddefault}{\updefault}{\color[rgb]{0,0,0}No phase}%
}}}}
\put(2971,-3121){\makebox(0,0)[lb]{\smash{{\SetFigFont{8}{9.6}{\familydefault}{\mddefault}{\updefault}{\color[rgb]{0,0,0}$m_B^2$}%
}}}}
\put(2206,-2176){\makebox(0,0)[lb]{\smash{{\SetFigFont{8}{9.6}{\familydefault}{\mddefault}{\updefault}{\color[rgb]{0,0,0}$\lambda_Bb_4$}%
}}}}
\put(1531,-3661){\makebox(0,0)[lb]{\smash{{\SetFigFont{8}{9.6}{\familydefault}{\mddefault}{\updefault}{\color[rgb]{0,0,0}$(-)$}%
}}}}
\put(2431,-2491){\makebox(0,0)[lb]{\smash{{\SetFigFont{8}{9.6}{\familydefault}{\mddefault}{\updefault}{\color[rgb]{0,0,0}$(+)$}%
}}}}
\put(1621,-2491){\makebox(0,0)[lb]{\smash{{\SetFigFont{8}{9.6}{\familydefault}{\mddefault}{\updefault}{\color[rgb]{0,0,0}$(-)$}%
}}}}
\put(5446,-3661){\makebox(0,0)[lb]{\smash{{\SetFigFont{8}{9.6}{\familydefault}{\mddefault}{\updefault}{\color[rgb]{0,0,0}$(+)$}%
}}}}
\put(2521,-3661){\makebox(0,0)[lb]{\smash{{\SetFigFont{8}{9.6}{\familydefault}{\mddefault}{\updefault}{\color[rgb]{0,0,0}$(-)$}%
}}}}
\put(4276,-3481){\makebox(0,0)[lb]{\smash{{\SetFigFont{8}{9.6}{\familydefault}{\mddefault}{\updefault}{\color[rgb]{0,0,0}$(-)$}%
}}}}
\put(4681,-3661){\makebox(0,0)[lb]{\smash{{\SetFigFont{8}{9.6}{\familydefault}{\mddefault}{\updefault}{\color[rgb]{0,0,0}$(+)$}%
}}}}
\put(8011,-3661){\makebox(0,0)[rb]{\smash{{\SetFigFont{8}{9.6}{\familydefault}{\mddefault}{\updefault}{\color[rgb]{0,0,0}$(+)$}%
}}}}
\put(8821,-2491){\makebox(0,0)[rb]{\smash{{\SetFigFont{8}{9.6}{\familydefault}{\mddefault}{\updefault}{\color[rgb]{0,0,0}$(+)$}%
}}}}
\put(8101,-2491){\makebox(0,0)[rb]{\smash{{\SetFigFont{8}{9.6}{\familydefault}{\mddefault}{\updefault}{\color[rgb]{0,0,0}$(-)$}%
}}}}
\put(7921,-4021){\makebox(0,0)[lb]{\smash{{\SetFigFont{8}{9.6}{\familydefault}{\mddefault}{\updefault}{\color[rgb]{0,0,0}(c) $\phi_+ < x_6^B$}%
}}}}
\put(1801,-4021){\makebox(0,0)[lb]{\smash{{\SetFigFont{8}{9.6}{\familydefault}{\mddefault}{\updefault}{\color[rgb]{0,0,0}(a) $x_6^B < \phi_-$}%
}}}}
\put(4591,-4021){\makebox(0,0)[lb]{\smash{{\SetFigFont{8}{9.6}{\familydefault}{\mddefault}{\updefault}{\color[rgb]{0,0,0}(b) $\phi_- < x_6^B < \phi_+$}%
}}}}
\end{picture}%